\documentclass[12pt]{article}
\usepackage[utf8]{inputenc}

\usepackage{amsmath, amssymb, mathtools, bm, bbm}
\usepackage{graphicx}
\usepackage{subcaption}
\usepackage{multicol}
\usepackage{floatrow}
\usepackage{caption}
\usepackage{comment}
\usepackage[mathscr]{eucal}
\usepackage{multirow}
\usepackage{enumitem}
\usepackage{diagbox}
\usepackage{times}
\usepackage{natbib}
\usepackage{xcolor}
\usepackage{setspace}
\usepackage[plain,noend]{algorithm2e}

\definecolor{darkpowderblue}{rgb}{0.0, 0.2, 0.6}
\definecolor{goldenpoppy}{rgb}{0.99, 0.76, 0.0}
\definecolor{cardinal}{rgb}{0.77, 0.12, 0.23}

\usepackage{hyperref}
\hypersetup{
    colorlinks=true,
    linkcolor=cardinal,
    citecolor=darkpowderblue,
    urlcolor=cyan,
}
\usepackage{cleveref}

\newtheorem{theorem}{Theorem}[section]   
\newtheorem{lemma}[theorem]{Lemma}       
\newtheorem{proof}[theorem]{Proof}
\newtheorem{proposition}[theorem]{Proposition}
\newtheorem{corollary}[theorem]{Corollary}

\newtheorem{definition}[theorem]{Definition}

\newtheorem{remark}[theorem]{Remark}
\newtheorem{algo}{Algorithm}

\def\wtD{\widetilde{\Delta}}

\def\wtT{\widetilde{T}}
\def\whT{\widehat{T}}
\def\cD{\check{\Delta}}
\def\wtO{\widetilde{O}}
\newcommand{\E}{\mathbb{E}}
\newcommand{\pr}{\text{pr}}
\newcommand{\var}{\text{var}}

\newcommand*\samethanks[1][\value{footnote}]{\footnotemark[#1]}
\addtocontents{toc}{\protect\setcounter{tocdepth}{-1}}

\begin{document}
\title{ \bf Nonparametric Inference for Balance in Signed Networks}

\author{Xuyang Chen\thanks{Xuyang Chen and Yinjie Wang are co-first authors and have made equal contributions.} \thanks{Department of Applied Mathematics and Computational Science, University of Pennsylvania. }\quad Yinjie Wang\samethanks[1] \thanks{Department of Statistics, University of Chicago. }\quad Weijing Tang\thanks{Department of Statistics and Data Science, Carnegie Mellon University.}}

\date{}

\maketitle

\begin{abstract}
In many real-world networks, relationships often go beyond simple dyadic presence or absence; they can be positive, like friendship, alliance, and mutualism, or negative, characterized by enmity, disputes, and competition. To understand the formation mechanism of such \textit{signed networks}, the social \textit{balance theory} sheds light on the dynamics of positive and negative connections. In particular, it characterizes the proverbs, ``a friend of my friend is my friend'' and ``an enemy of my enemy is my friend''. In this work, we propose a nonparametric inference approach for assessing empirical evidence for the balance theory in real-world signed networks. We first characterize the generating process of signed networks with node exchangeability and propose a nonparametric sparse signed graphon model. Under this model, we construct confidence intervals for the population parameters associated with balance theory and establish their theoretical validity. Our inference procedure is as computationally efficient as a simple normal approximation but offers higher-order accuracy. By applying our method, we find strong real-world evidence for balance theory in signed networks across various domains, extending its applicability beyond social psychology.
\end{abstract}

\vspace{0.5em}

\section{Introduction}
\subsection{Motivation and Overview}

In many real-world complex systems, the relationships among subjects often extend beyond simple presence or absence. Apart from positive connections like friendship, alliance, and mutualism, relationships between two entities can also be negative, characterized by enmity, dispute, and competition. These multifaceted relationships among a group of subjects can be represented as \textit{signed networks}. A signed network consists of two elements, nodes, which are the basic units of the system, and edges, representing the positive or negative connections between the nodes. Such signed networks are ubiquitous in diverse engineering and scientific disciplines, such as biology, computer science, economics, and sociology, among others.
Examples include social networks where friendships and animosities coexist \citep{leskovec2010signed}, ecological networks depicting symbiotic and predatory relationships among species \citep{saiz2017evidence}, 
international relation networks indicating alliances and militarized disputes among countries \citep{kirkley2019balance}, and protein-protein interaction networks describing both positive and negative associations \citep{huttlin2021dual}. 

\begin{figure}[htbp]
  \centering
  \begin{subfigure}[t]{0.48\textwidth}
    \centering
    \includegraphics[height=1.8cm]{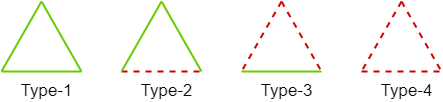}
    \caption{Four types of triangles}
    \label{fig:triangle}
  \end{subfigure}
  \hfill
  \begin{subfigure}[t]{0.48\textwidth}
    \centering
    \includegraphics[height=2.1cm]{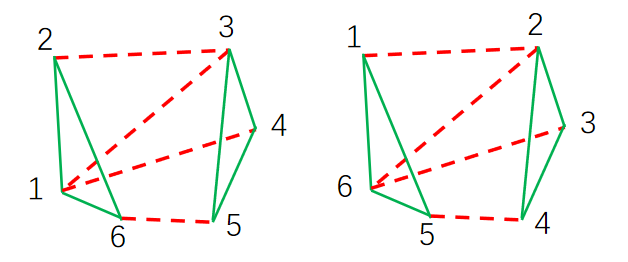}
    \caption{Node indices shuffling}
    \label{fig:shuffle}
  \end{subfigure}
  \caption{%
    (a) Illustration of the four types of triangles in signed networks, where solid green lines indicate positive edges, and dashed red lines indicate negative edges. Types 1 and 3 are balanced, Type 4 is weakly balanced, and Type 2 is unbalanced. (b) Illustration that randomly shuffling the node indices does not affect the balance pattern of the network.
  }
  \label{fig:combined}
\end{figure}

Such signed networks have substantially different properties and principles from unsigned ones. 
One fundamental theory that elucidates the connectivity in signed networks is the \textit{balance theory} \citep{heider1946attitudes,cartwright1956structural}. 
Originated from social psychology, the balance theory posits that individuals strive for psychological consistency in their relationships and beliefs with others in their social environment. 
Specifically, balance theory focuses on triangular configurations where three individuals are connected to each other. These triangles, due to the nature of the signed relationships, can be categorized into four types (see Figure~\ref{fig:triangle}). 
A triangle achieves a balanced state when it contains an even number of negative edges. Balanced triangles align with common proverbs, ``a friend of my friend is my friend'' and ``an enemy of my enemy is my friend'', corresponding to type-1 and type-3 triangles respectively. 
In contrast, an unbalanced state arises when relationships deviate from this pattern, leading to psychological discomfort. For example, in a type-2 triangle, where an individual has two friends who hold negative sentiments towards each other, the individual may reconcile two friends to restore balance. 

Balance theory posits that triangles in a signed network tend to be balanced, which provides valuable insights into the dynamics of positive and negative connections. Existing work has demonstrated that properly leveraging the balance theory in modeling signed networks can improve goodness of fit, interpretability, and node and link prediction~\citep{derr2018signed,tang2023population,hsieh2012low, chiang2014prediction, zhang2022signed}. However, balance theory is not universally applicable to all signed networks and improper application may lead to misleading results.  Therefore, a crucial first step is to assess the empirical evidence for balance theory in real-world signed networks to investigate its applicability.

To this end, it requires appropriate extensions of the concept of balance from individual triangles to the entire signed network to measure the overall degree of balance. \citet{harary1953notion} introduced structural balance theory, where a signed network is balanced if all its triangles are balanced. 
\citet{davis1967clustering}  proposed a weaker version, saying that a signed network is weakly balanced if no type-2 triangles occur, \textit{i.e.},  only triangles with two positive edges are impermissible and ``the enemy of my enemy can be my enemy''.
However, both structural and weak balance, which require all triangles to follow certain patterns, may be overly stringent for practical applications, especially for large signed networks. 
A natural quantity of interest is the count of balanced or weakly balanced triangles in an observed signed network. To account for noise in the observed networks, several studies have investigated the asymptotic properties of the normalized count under different balance-free models (\textit{i.e.}, a generating process that does not encode an explicit preference for balanced over unbalanced triangles)~\citep{facchetti2011computing,leskovec2010signed,feng2022testing,Jin2024-lk}. 
Yet, beyond balance-free models, when the balance pattern is present, uncertainty quantification of normalized counts remains unclear.

In this work, we introduce a nonparametric inference method to measure the degree of balance in observed signed networks. The inference procedure constructs confidence intervals for the expected proportion of balanced (or weakly balanced) triangles. 
Notably, the validity of our inference method holds under various degrees of balance. Thus, our inference approach provides insights into not only whether or not a network is balance-free but also \textit{how balanced (or weakly balanced) a signed network is}. 
Specifically, we consider a broad class of node-exchangeable random network models for signed networks, in which the distribution of the network is invariant under relabeling of node indices. Our inferential target, i.e., the proportion of balanced (or weakly balanced) triangles, is a global network statistic that remains unchanged under node relabeling (see \Cref{fig:shuffle}), which makes node exchangeability a natural modeling assumption for this purpose. At the same time, adopting a node-exchangeable nonparametric framework avoids assuming specific parametric generative models, which is particularly preferred given the challenges in accurately modeling complex signed networks. Building on a novel characterization of infinitely node-exchangeable random network models for signed networks, we propose a nonparametric graphon model that is flexible to accommodate sparsity in both edge density and sign proportion. This model family includes many commonly used statistical network models as special cases~\citep{nowicki2001estimation,airoldi2008mixed,nickel2008random,hoff2002latent,lei2021network}.  Under the proposed model, we establish a non-asymptotic approximation to the distribution of a studentized ratio of the normalized subgraph count, and construct valid confidence intervals for the population-level degree of balance.

We also apply our inference method to signed networks across various domains, including international relations in political science, protein-protein interactions in biology, and both small- and large-scale social networks in social science. 
Our results reveal strong empirical evidence for balance theory, extending its applicability beyond its root domain in social psychology.

\subsection{Related Works}

The uncertainty quantification of the normalized count of balanced triangles has been studied under several balance-free models~\citep{facchetti2011computing,leskovec2010signed,feng2022testing}. 
Here, \textit{balance-free} refers to the absence of an explicit mechanism favoring balanced triangles. Different works operationalize this idea differently.
For example, 
\citet{facchetti2011computing} considered a model in which the edge signs are independently and identically distributed Bernoulli variables, with the probability equal to the fraction of negative edges in the observed network. \citet{leskovec2010signed} reshuffed edge signed via a hypergeometric distribution that preserves the total number of negative edges. 
\citet{feng2022testing} further proposed a stratified model, where sign reshuffling occurs randomly across edges within the same embeddedness levels (\textit{i.e.}, the number of triangles that an edge is part of), acknowledging the distinct behaviors of positive and negative edges. 
With a different focus, \citet{Jin2024-lk} studied two-sample testing for binary networks by comparing counts of balanced and unbalanced even-length cycles under degree-corrected mixture models. Since we assess the degree of balance by analyzing balanced triangles, which are odd-length cycles, the results of \citet{Jin2024-lk} are not applicable.
Overall, there is no consensus of balance-free models in the existing literature.
Although the asymptotic properties of the normalized count under these balance-free models are useful for assessing whether or not a signed network is balance-free, they provide limited insight when the balance or weak balance is present in a signed network.

Our inference method is also closely related to the line of work on network moment inference~\citep{10.1214/15-AOS1338,10.1214/11-AOS904, thompson2016using, chen2019bootstrap, levin2019bootstrapping,green2022bootstrapping,DBLP:journals/corr/abs-2004-06615, 10.1093/biomet/asac032}. While several pioneering network bootstrap methods have been proposed for unsigned networks \citep{levin2019bootstrapping, green2022bootstrapping,10.1214/15-AOS1338}, their theoretical justifications cannot be directly applied to our setting. Among these methods, \citet{levin2019bootstrapping} developed a bootstrap procedure under the random dot-product graph (RDPG) model for inferring network quantities that can go beyond subgraph counts. However, it is unclear how to generalize the RDPG model for signed networks. Under the graphon model for unsigned networks, \citet{10.1214/15-AOS1338} and \citet{green2022bootstrapping} developed the node subsampling and the node resampling method respectively. 
\citet{DBLP:journals/corr/abs-2004-06615} established an explicit Edgeworth expansion formula that provides a higher-order accurate approximation of the distribution of a studentized network moment, compared to a simple normality approximation. 
Compared to bootstrap-based methods, the empirical Edgeworth expansion is computationally more efficient. 
Motivated by these advantages, we extend the Edgeworth expansion method to our setting, leading to an inference method sharing the computation efficiency and higher-order accuracy. Establishing the theoretical validity of this extension, however, requires nontrivial new analysis. Our balance statistics is the ratio of two empirical network moments. Unlike settings involving a single empirical moment with a deterministic normalization in \citep{DBLP:journals/corr/abs-2004-06615}, the denominator in our setting is itself random and estimated from the data, which introduces additional sources of randomness. Characterizing the higher-order distribution of this ratio therefore requires carefully handling estimation errors from both moments and their interaction, which substantially complicates the analysis despite the similarity in high-level proof strategy.

\noindent \textbf{Notation}\ \  We use ``$\buildrel d \over =$'' to denote that two random objects have the same distribution. For two sequences $\{a_n\}_{n=1}^\infty$ and $\{b_n\}_{n=1}^\infty$, we write $b_n=O(a_n)$ or $b_n\preceq a_n$ if there exist constants $C,N>0$ such that $|b_n| \le C|a_n|$ for all $n\ge N$; $b_n=o(a_n)$ or $b_n\prec a_n$ if $\lim_{n\to\infty}b_n/a_n=0$; $b_n = \omega (a_n)$ if $a_n=o(b_n)$; and $b_n\asymp a_n$ if both $b_n\preceq a_n$ and $a_n\preceq b_n$. We define $b_n=\wtO_{p,k}(a_n)$ if there exist constants $C_1,C_2,N>0$ such that $\pr(|b_n|\ge C_1\,|a_n|)\le C_2n^{-k}$ for all $n\ge N$. For a bounded function $f$ defined on $\mathbb{R}$, its supremum norm is defined as $\Vert f\Vert_{\infty} = \sup_{t \in \mathbb{R}} |f(t)|.$ The binomial coefficient is denoted by $\binom{n}{k}=n!/(k!(n-k)!)$.

\section{Methodology}
\subsection{Node-Exchangeable Random Network Model for Signed Networks}

Node-exchangeable random network models represent a wide class of probabilistic models for network data, including the stochastic block model \citep{nowicki2001estimation}, mixed-membership block model \citep{airoldi2008mixed}, random dot-product graph model \citep{nickel2008random}, and latent space model \citep{hoff2002latent}. 
The concept of node exchangeability views nodes as random samples and assumes that the network distribution is invariant under permutations of node labels. 
This invariance makes node exchangeability a natural assumption for studying the proportion of balanced triangles, since reshuffling node indices does not alter this global network quantity.
For unsigned (binary) networks, the generating process for the infinitely exchangeable network model has been well studied \citep{orbanz2014bayesian} and serves as the theoretical foundation. 
Specifically, the Aldous-Hoover theorem \citep{aldous1981representations, hoover1979relations} says that any infinitely exchangeable random binary network can be generated by first sampling the nodal variables $\{X_i:i\geq 1\}$ independently from a uniform distribution on $[0,1]$ and then independently of all others connecting each pair of nodes with probability $W(X_i, X_j)$, where $W:[0,1]^2 \rightarrow [0,1]$ is a symmetric function.  
However, an analogous characterization for signed networks remains unclear. Specifically, in an undirected signed network, each edge can take one of the three values $\{-1,0,1\}$. Denote the corresponding symmetric adjacency matrix for a signed network with $n$ nodes by $A=[A_{ij}]_{n\times n}\in \{-1,0,1\}^{n \times n}$, where $A_{ij} = 1$ indicates a positive edge between the nodes $i$ and $j$, $A_{ij} = -1$ indicates a negative edge, and $A_{ij} = 0$ indicates no edge.

We first characterize the generating process for infinitely node-exchangeable random network models for signed networks. Consider an infinite random two-way three-value symmetric array $\boldsymbol{A}=[A_{ij}]_{i,j \in \mathbb{N}}$ with $A_{ij}\in \{-1,0,1\}$ is row-column jointly exchangeable if 
\[[A_{ij}]_{i,j \in \mathbb{N}}  \overset{d}{=} [A_{\pi(i) \pi(j)}]_{i,j \in \mathbb{N}}\] 
holds for any permutation mapping $\pi$ on the finite indices, \textit{i.e.},  for some $1\le j < j'$, $\pi(i)=i$ if $i\notin \{j, j'\}$, $\pi(j)=j'$ and $\pi(j')=j$. 
The following proposition states its equivalent representation and the corresponding generating process, with the proof provided in the Supplemental Materials.

\vspace{2mm}

\begin{proposition}
\label{prop: exchangeability}
A random two-way array $\boldsymbol{A}=[A_{ij}]_{i,j \in \mathbb{N}}$ with $A_{ij}\in \{-1,0,1\}$  is row-column jointly exchangeable if and only if there exist two symmetric measurable functions $\widetilde{F}: [0,1]^2 \rightarrow [0,1]$ and $\widetilde{G}: [0,1]^2 \rightarrow [0,1]$ such that
\[[A_{ij}]_{i,j \in \mathbb{N}}  \overset{d}{=} [\mathbb{I}\{Y_{ij}<\widetilde{F}(X_i, X_j)\}(-1)^{\mathbb{I}\{Y_{ij}>{\widetilde{F}(X_i, X_j)}\widetilde{G}(X_i,X_j)\}}]_{i,j \in \mathbb{N}}, \]
where $X_i$ and $Y_{ij}$, for $i,j \in \mathbb{N}$, are $i.i.d.$ uniform random variables on $[0,1]$.
Here, $\mathbb{I}\{\cdot\}$ is the indicator function that equals $1$ if the event holds and $0$ otherwise. 
Therefore, the random two-way three-value array $\boldsymbol{A}$ can be generated by the following process.  
\begin{enumerate}
\item Generate $X_i \buildrel i.i.d. \over \sim U[0,1]$ for $i \ge 1$;
\item For each pair $(i,j)$, independently sample $|A_{ij}|=1$ with the probability $\widetilde{F}(X_i, X_j)$ and otherwise $A_{ij}=0$;
\item For each pair $(i,j)$ with  $|A_{ij}|= 1$, independently of all others, sample $A_{ij}=-1$ with the probability $\widetilde{G}(X_i, X_j)$ and otherwise $A_{ij}=1$.
\end{enumerate}
\end{proposition}

\vspace{2mm}

Proposition 1 shows that the distribution of $\boldsymbol{A}$ is fully determined by specifying the functions $\widetilde{F}(\cdot, \cdot)$ and $\widetilde{G}(\cdot, \cdot)$. 
Specifically, conditional on the node latent variables, the probabilities of observing a positive edge, a negative edge, or no edge are governed by two functions evaluated at the latent variables.

The representation in Proposition 1 describes an infinitely node-exchangeable array. For a finite signed adjacency matrix $A=[A_{ij}]_{n\times n}$, a direct finite-sample version is
\[[A_{ij}]_{n\times n}  \overset{d}{=} [\mathbb{I}\{Y_{ij}<\widetilde{F}_n(X_i, X_j)\}(-1)^{\mathbb{I}\{Y_{ij}>{\widetilde{F}_n(X_i, X_j)}\widetilde{G}_n(X_i,X_j)\}}]_{n\times n}, \]
where $\widetilde{F}_n$ and $\widetilde{G}_n$ may depend on $n$ arbitrarily. 
To decouple network size from structural patterns, it is standard to introduce explicit scaling parameters.
Following this line of work, we focus on a factorized form $\widetilde{F}_n = \rho_n F$ and $\widetilde{G}_n=s_n G$, where $F: [0,1]^2 \rightarrow [0,1]$ and $G: [0,1]^2 \rightarrow [0,1]$ are two bivariate functions independent of $n$ and $\rho_n$ and $s_n$ control, respectively, the network density and the prevalence of negative edges. Such scaling has been widely used in sparse graphon models \citep{klopp2017oracle, lei2021network, wolfe2013nonparametric, xu2014edge}. We propose the following nonparametric graphon model for undirected signed networks with $n$ nodes.

\vspace{2mm}
\begin{definition}[Graphon Model for Signed Networks $\mathcal{A}(n, F, G, \rho_n, s_n)$]
\label{def: graphon}
For $1\le i \le n $, let $X_i\sim U[0,1]$ be the latent variables independently sampled from the uniform distribution. Conditional on the latent variables of two nodes $i$ and $j$, an edge between two nodes is independently drawn with probability $\rho_n F(X_i, X_j)$, \textit{i.e.}, 
$$
\pr(|A_{ij}|=1 \mid X_i, X_j) =\rho_n F(X_i, X_j),\quad \text{for all }1\le i < j \le n; 
$$
then for each edge (\textit{i.e.} $|A_{ij}|=1$), independently of all others, it takes the negative sign with probability $s_n G(X_i, X_j)$ and the positive sign otherwise, \textit{i.e.},
$$
\pr(A_{ij}=-1\mid |A_{ij}|=1, X_i, X_j)=s_n G(X_i, X_j),\quad \text{for all }1\le i < j \le n.
$$
Besides, $A_{ij}=A_{ji}$ for all $i \neq j$ and $A_{ii}=0$.
We write $A \sim \mathcal{A}(n, F, G, \rho_n, s_n)$ to denote a signed network with $n$ nodes generated from the above procedure.
\end{definition}
\vspace{2mm}

For identifiability, we constrain the integral of $F$ and $G$ over $[0,1]^2$ being constants, while allowing $\rho_n$ and $s_n$ to vary with the network size $n$.
Throughout this work, we assume $\rho_n = o(1)$, which reflects the empirical sparsity commonly observed in many large real-world networks.  
Moreover, empirical evidence indicates that negative edges are typically less frequent than positive ones in signed networks (see \Cref{ppi and social networks}).  
Accordingly, we assume $\|s_n G\|_{\infty} < 1/2$, which implies that conditional on edge existence, negative signs are less likely than positive ones.

\vspace{2mm}
\begin{remark}
The proposed graphon model $\mathcal{A}(n,F,G,\rho_n,s_n)$ is flexible to accommodate varying degrees of balance through the choice of the function $G$.  
When $G(x,y)\equiv c$ is constant, the probability that an edge takes a negative sign does not depend on the latent node variables.  
In this case, the model does not encode any structural preference toward balanced over unbalanced triangles; balanced configurations arise only due to random sign fluctuations.  
This setting refers to a \emph{balance-free} scenario. 
In contrast, structured choices of $G$ can induce balanced configurations.  
For example, letting $G(x,y)=0.9$ when $(x-0.5)(y-0.5)>0$ and $G(x,y)=0.1$ otherwise implies that, conditional on edge existence, edges within the same latent block are more likely to be positive, while edges across blocks are more likely to be negative.  
This construction induces a network that strongly favors balanced triangles.
More generally, intermediate choices of $G$ give rise to networks exhibiting varying degree of balance (see \Cref{subsec: simu varying degree} for an example).
\end{remark}

\subsection{Population Parameters for Balance and Inference Procedure}
To measure the overall degree of balance, we aim to infer the expected proportion of balanced triangles among all triangles in a signed network. 
Let $A \sim \mathcal{A}(n, F, G, \rho_n, s_n)$ denote a signed network and  $A_{i,j,k}$ be the adjacency matrix of the subgraph induced by nodes $\{i, j, k\}$. We define the population-level proportion of balanced triangles as 
\[w_n=\pr(h_{\blacktriangle}(A_{i,j,k})=1\mid h_{\Delta}(A_{i,j,k})=1),\] where $h_{\Delta}(A_{i,j,k})$ and $h_{\blacktriangle}(A_{i,j,k})$ are the indicators that $A_{i,j,k}$ forms a triangle and a balanced triangle, respectively. 
A larger value of $w_n$ indicates stronger evidence for balance theory. Motivated by weak balance theory, we also introduce more fine-grained population parameters that describe the expected proportion of each type of triangle. Specifically, we define
\[w_{n,t} = \pr(h_{t}(A_{i,j,k})=1 \mid h_{\Delta}(A_{i,j,k})=1), \quad \text{for }1\le t \le 4,\]
where $h_{t}(A_{i,j,k})$ is the indicator that $A_{i,j,k}$ forms a type-$t$ triangle (see Figure~\ref{fig:triangle} for an illustration of the four triangle types). In particular, a smaller value of $w_{n,2}$ indicates stronger evidence of the weak balance theory.
Note that $w_n$ and $w_{n,t}$ are well-defined, meaning they are invariant to the specific node indices $\{i, j, k\}$ due to node exchangeability.

Next, we develop a computationally efficient inference procedure for population parameters, with its theoretical validity established under the proposed graphon model in Section~\ref{Theoretical}. For the sake of space, we present the inference
procedure for $w_n$ in the main text. The procedure for $w_{n,t}$ is analogous and deferred to the Supplementary Materials~\ref{sec: results for wnt}. 

We start with considering the normalized count of balanced triangles, which naturally serves as a point estimator for $w_n$.
Specifically, define  the following two sample network moments
\[\widehat{U}_n=\sum_{1 \le i_1 < i_2 < i_3 \le n}h_{\blacktriangle}(A_{i_1,i_2,i_3}) / \binom{n}{3},\quad 
\widehat{V}_n=\sum_{1 \le i_1 < i_2 < i_3 \le n}h_{\Delta}(A_{i_1,i_2,i_3})/\binom{n}{3},
\]
and their population counterparts $u_n=\E\{h_{\blacktriangle}(A_{1,2,3})\}$ and $v_n=\E\{h_{\Delta}(A_{1,2,3})\}$, respectively. 
By definition, the target parameter satisfies $w_n=u_n/v_n$, and its natural estimator is $\widehat{U}_n/\widehat{V}_n$. 

To quantify the uncertainty of this estimator, we study its studentized form. 
This approach is motivated by the fact that studentization often leads to higher-order accurate distributional approximations in both i.i.d.\ data \citep{wasserman2006all} and unsigned network data \citep{DBLP:journals/corr/abs-2004-06615}. 
However, unlike settings involving a single empirical moment with deterministic normalization~\citep{DBLP:journals/corr/abs-2004-06615}, both the numerator $\widehat{U}_n$ and  denominator $\widehat{V}_n$ here are random and estimated by the same data. 
This additional layer of dependence is what make the analysis substantially more involved, despite relying on a similar high-level strategy.
Applying studentization therefor requires careful characterization of the leading variance term and the construction of a consistent variance estimator. Readers primarily interested in the final inference procedure may safely skip the technical derivation and proceed directly to the variance estimator in~\eqref{estvaiance}.

Note that the stochastic variation of $\widehat{U}_n$ and $\widehat{V}_n$ arises from both the latent variables $\{X_i\}_{i=1}^n$ and the edge generation mechanism, where each $A_{ij}\in\{0, \pm 1\}$ follows a categorical distribution conditional on $\{X_i\}_{i=1}^n$. 
To disentangle these two sources of randomness and identify the leading variance term, we  first condition on the latent variables. Specifically, we introduce the conditional expectations:
$$U_n=\E(\widehat{U}_n\mid X_1,\ldots X_n),\qquad V_n=\E(\widehat{V}_n\mid X_1,\ldots X_n),$$
which serve as intermediate quantities capturing the variation induced by the latent variables alone. 
The leading variances of $U_n$ and $V_n$ can then be analyzed using standard Hoeffding decompositions of U-statistics~\citep{hoeffding1992class}.
Define the shorthand $h_{\blacktriangle}(X_{1},X_{2},X_{3})=\E\{h_{\blacktriangle}(A_{1,2,3}) \mid X_{1},X_{2},X_{3} \}$. The Hoeffding decomposition of $U_n$ takes the form
$$
U_{n}=u_{n} + 3\sum_{i=1}^{n}g_{1}(X_{i})/n+6\sum_{1\le i < j \le n}g_{2}(X_{i},X_{j})/\{n(n-1)\}+\widetilde{O}_{p, 2}(n^{-3/2} \rho_n^{3} \log^{3/2} n)
,$$
where 
\begin{align*}
    g_{1}(x_{1}) & =\E\{h_{\blacktriangle}(X_{1},X_{2},X_{3})\mid X_{1}=x_{1}\}-u_{n},\\
    g_{2}(x_{1},x_{2})& =\E\{h_{\blacktriangle}(X_{1},X_{2},X_{3})\mid X_{1}=x_{1},X_{2}=x_{2}\}-u_{n}-\sum_{1\le i \le 2}g_{1}(x_{i}).
\end{align*}
An analogous decomposition holds for $ V_n$, with corresponding functions $ f_1 $ and $ f_2$ in place of $g_1$ and $g_2$, respectively.
The leading variance term arises from the linear components of the Hoeffding decompositions combined with a first-order Taylor expansion of the ratio. We define 
\[q_1(X_i)=g_1(X_i)/v_n-u_n f_1(X_i)/v_n^2\quad \text{and}\quad \xi_1^2=\E\{q_1^2(X_1)\}.\]
The following proposition show that $\sigma_n^2:=9\xi_1^2/n$ is the dominating term of $\var(\widehat{U}_n/\widehat{V}_n)$.

\vspace{3mm}
\begin{proposition}
\label{proposition:variance}

Assume: 
\begin{itemize}
    \item[(a)]  $F(\cdot, \cdot)$, $G(\cdot, \cdot)>C_1$ for some constant $C_1>0$;
    \item[(b)] (Non-degeneracy) $\xi_1 > C_2$ for some constant $C_2>0$;
    \item[(c)] Either
\begin{itemize}
    \item[(c1)] (Vanishing negative proportion) $s_n \prec 1$, $\rho_n^3 s_n^2 \succeq n^{-2}\log^{8}n$, and $\rho_n \succ n^{-1}s_n$; or
    \item[(c2)] (Constant negative proportion) $s_n \asymp 1$ with $\|s_n G\|_\infty<1/2$, and $\rho_n^3 \succeq n^{-2}\log^{8}n$.
\end{itemize}
\end{itemize}
we have
\begin{align*}
\var(\widehat{U}_n/\widehat{V}_n-u_n/v_n)=\sigma_n^2+o(n^{-1}).
\end{align*}
\end{proposition}
\vspace{2mm}

The role and interpretation of Conditions (a)–(c) will be discussed together with the main theorem in the next section.  Guided by Proposition~\ref{proposition:variance}, we construct the studentized statistic 
\(\widehat{T}_n\coloneqq(\widehat{U}_n/\widehat{V}_n-u_n/v_n)/\widehat{S}_n\), where $\widehat{S}_n^2$ is an analytically tractable variance estimator of $\sigma_n^2$: 
\begin{align}
\label{estvaiance}
n\widehat{S}^2_n=\frac{9}{n}\sum_{i=1}^n\left[
\sum_{\substack{1\le i_{1}< i_{2}\le n\\i_{1},i_{2}\neq i}}\left\{h_{\blacktriangle}(A_{i,i_{1},i_{2}})/\widehat{V}_n-\widehat{U}_n h_{\Delta}(A_{i,i_{1},i_{2}})/\widehat{V}^2_n \right\}/\binom{n-1}{2}
\right]^2.
\end{align}

\section{Theoretical Results}\label{Theoretical}
Our inference procedure is based on an accurate approximation to the distribution function of the studentized statistic $\widehat{T}_n$, denoted by $\mathcal{F}_{\widehat{T}_n}$. 
While a normal approximation can be used, it can be inaccurate in finite samples, especially when the network size is relatively small. To address this, we apply Edgeworth expansion technique~\citep{edgeworth1883xlii, wallace1958asymptotic, bentkus1997edgeworth, hall2013bootstrap}, which refines the normal approximation by incorporating higher-order correction terms and yields a more accurate approximation of $\mathcal{F}_{\widehat{T}_n}$.

Edgeworth expansion has been widely applied to studentized or standardized U-statistics  \citep{callaert1980edgeworth, callaert1981order, bickel1986edgeworth, DBLP:journals/corr/abs-2004-06615}. 
A technical challenge in applying Edgeworth expansions to network data is that standard regularity conditions, such as Cram\'er's condition or non-lattice condition~\citep{bickel1986edgeworth,angst2017weak,mattner2017optimal,song2020uniform}, are violated under stochastic block models. This is undesirable in our context, as block models are a fundamental and widely used class of generative models for networks. When the network is not overly dense, \citet{DBLP:journals/corr/abs-2004-06615} observed that the randomness induced by edge formation provides a \textit{self-smoothing} effect to bypass the Cram\'er's condition. 
To accommodate denser regimes while preserving the same approximation accuracy, we adopt a smoothing approach in \cite{Shao2022-ar}. We add an auxiliary Gaussian perturbation $\delta_T\sim N(0, c_\delta n^{-1}\log n)$ with a sufficiently large constant $c_\delta$ and use $\widehat{T}_n+\delta_T$ rather than $\widehat{T}_n$ for inference.  This perturbation plays a purely technical role. Since its variance vanishes as $n \to \infty$, it has no asymptotic impact on inference, and its finite-sample effect is negligible in practice (see Supplementary Materials~\ref{sec: full simulation results}). 

Before presenting our Edgeworth expansion formula, we introduce some notation for simplicity. Define $p_1(X_i)=f_1(X_i)/v_n$, $q_1\left(X_i\right)=g_1\left(X_i\right) / v_n-u_n p_1(X_i) / v_n$, and $q_2(X_i,X_j)=g_2(X_i,X_j)/v_n-u_nf_2(X_i,X_j)/v_n^2.$ 
The following theorem characterizes the Edgeworth expansion approximation for $\mathcal{F}_{\widehat{T}_n+\delta_T}$, with a deviation bounded by 
\[\mathcal{E}(n,s_n,\rho_n):=n^{-1}\rho_n^{-3/2}\log^{3/2} n (1 + s_n \log n) + n^{-1} s_n^{-1/2} \rho_n^{-1/2}.\]

\vspace{2mm}
\begin{theorem}\label{thm1} 
Define the population Edgeworth expansion as \[
\mathcal{G}_n(x)=\Phi(x)+\phi(x)\left\{a(x^2/3+1/6)+b(x^2+1)-3cx^2\right\}/\sqrt{n},
\]
where $\phi$ and $\Phi$ denote the density and distribution functions of the standard normal distribution, respectively, and $a,b,c$ are given by \begin{align*}
  a=\E\{q_1^3(X_1)\}/\xi_1^3,\;b=\E\{q_1(X_1)q_1(X_2)q_2(X_1,X_2)\}/\xi_1^3,\;c=\E\{q_1(X_1)p_1(X_1)\}/\xi_1.
\end{align*}
Under Conditions (a)-(c) in Proposition~\ref{proposition:variance}, we have
$$
\|\mathcal{F}_{\widehat{T}_n+\delta_T}-\mathcal{G}_n\|_{\infty}=O(\mathcal{E}(n,s_n,\rho_n)).
$$
\end{theorem}

\vspace{2mm}

Condition (a) ensures that the probabilities of having an edge and a negative sign are of the order $\rho_n$ and $s_n$, respectively. This condition is used to obtain lower bounds on key conditional variances appearing in the linear leading terms of the studentized statistic. We note that this uniform boundedness away from zero is stronger than strictly necessary and is imposed here primarily to simplify the presentation. In Supplementary Material~\ref{sec:appendix-Lemma3}, we provide a weaker alternative that only requires the probabilities of observing the specific signed motifs contributing to our statistic to be bounded away from zero. 
Condition (b) is a standard non-degeneracy assumption that ensures the asymptotic variance of the limiting Gaussian distribution is strictly positive, and similar assumptions have been imposed in related work~\citep{DBLP:journals/corr/abs-2004-06615,Shao2022-ar}. 
Condition (c) specifies the feasible range for the network density and the proportion of negative edges. 
Specifically, $s_n^2\,\rho_n^3$ characterizes the order of the probability of observing a type-3 balanced triangle. The lower bounds imposed on $\rho_n$ in (c1)-(c2) ensure that such triangles occur with sufficient frequency to control higher-order remainder terms. We allow two regimes for $s_n$: in (c1), the proportion of negative edges vanishes, while in (c2) it remains bounded away from zero.  Both regimes rule out a near-symmetric sign regime in which positive and negative signs occur with nearly equal probability. In such cases, the conditional variance of the studentized statistic may degenerate due to the numerator–denominator dependence. 
Finally, if the density condition in (c1) or (c2) is strengthened to $\rho_n^3=\omega\left(n^{-1}\right)$, our Edgeworth expansion yields higher-order accuracy compared to a simple normal approximation. This sparsity condition $\rho_n^3=\omega\left(n^{-1}\right)$ aligns with those made in the existing literature on network moment inference~\citep{10.1214/15-AOS1338,10.1214/11-AOS904,DBLP:journals/corr/abs-2004-06615}.

The above population Edgeworth expansion still involves coefficients $a,b$, and $c$ that depend on the unknown underlying data-generating distribution. To obtain a feasible approximation, we construct an empirical Edgeworth expansion by replacing these unknown coefficients with their empirical counterparts. As an illustration, consider the coefficient $a=\E\{q_1^3(X_1)\}/\xi_1^3$. Define \[\widehat{q}_1(X_i)=\sum_{\substack{1\le i_1 < i_2 \le n\\ i \neq i_1, i_2}}\frac{\widehat{V}_n h_{\blacktriangle}(A_{i,\,i_1,i_2})-\widehat{U}_n h_{\Delta}(A_{i,\,i_1,i_2})}{\binom{n-1}{2}\widehat{V}^2_n},\]
and let $\widehat{\xi}_1^2=\frac{1}{n}\sum_{i=1}^n\widehat{q}_1^2(X_i)$ and $\widehat{\E}\{q_1^3(X_1)\}=\frac{1}{n}\sum_{i=1}^n\widehat{q}_1^3(X_i)$. We then define the empirical estimator $\widehat{a}=\widehat{\E}\{q_1^3(X_1)\}/\widehat{\xi}_1^3$.
The estimators $\widehat{b}$ and $\widehat{c}$ are constructed in an analogous manner with explicit forms provided in Supplemental Materials~\ref{defineabchat}. 
The following theorem establishes the approximation error of the empirical Edgeworth expansion to the CDF of $\widehat{T}_n+\delta_T$.

\vspace{2mm}
\begin{theorem}\label{thm2}
Define the empirical Edgeworth expansion as
$$
\widehat{\mathcal{G}}_n(x)=\Phi(x)+\phi(x)\left\{\widehat{a}(x^2/3+1/6)+\widehat{b}(x^2+1){-}3\widehat{c}x^2\right\}/\sqrt{n},
$$
where $\widehat{a}$, $\widehat{b}$, $\widehat{c}$ are the empirical counterparts of $a$, $b$, $c$. 
Under the conditions of Theorem~\ref{thm1}, we have
\begin{align*}
\|\mathcal{F}_{\widehat{T}_n + \delta_T}-\widehat{\mathcal{G}}_n\|_{\infty}=\wtO_{p,1}(\mathcal{E}(n,s_n,\rho_n)).
\end{align*}
\end{theorem}

\vspace{2mm}

Theorem~\ref{thm2} establishes that the empirical Edgeworth expansion approximates the CDF of \( \widehat{T}_n \) up to an error of order \( \mathcal{E}(n, s_n, \rho_n) \) with high probability (at least $1-n^{-1}$). Consequently, this result can be used to construct a Cornish-Fisher confidence interval~\citep{cornish1938moments} for $w_n$ that adjusts the normal quantiles to account for the higher-order terms in the distribution approximation, with the following coverage probability.

\vspace{2mm}
\begin{corollary}\label{thm3}
    For any \( \alpha \in (0,1) \), we define the adjusted quantile \( \widehat{q}_{\widehat{T}_{n;\alpha}} \) as
$$
\widehat{q}_{\widehat{T}_{n;\alpha}}:=z_{\alpha}-\{\widehat{a}(z_{\alpha}^2/3+1/6)+\widehat{b}(z_{\alpha}^2+1){-}3\widehat{c}z_{\alpha}^2\}/\sqrt{n} - \delta_T,
$$
where $z_{\alpha}:=\Phi^{-1}(\alpha)$ is the $\alpha$-th quantile of the standard normal distribution. Under the conditions of Theorem~\ref{thm1}, the two-sided confidence interval for $w_n$,
    \begin{align}
      \label{eq: CI for wn}
      (\widehat{U}_n/\widehat{V}_n-\widehat{q}_{\widehat{T}_{n;1-\alpha/2}}\widehat{S}_n , \ \widehat{U}_n/\widehat{V}_n-\widehat{q}_{\widehat{T}_{n;\alpha/2}}\widehat{S}_n).
    \end{align}
    achieves a \( 1 - \alpha + O\big(\mathcal{E}(n, s_n, \rho_n)\big) \) coverage probability.
\end{corollary}
\vspace{2mm}

\begin{remark}
  Our method always produces a confidence interval with length identical to that of the normal approximation (see the derivation in Supplementary Materials~\ref{sec: proof of CI length}). Compared to bootstrap-based methods, our procedure is computationally efficient and incurs negligible additional overhead beyond that of a simple normal approximation.
\end{remark}

\vspace{2mm}

\begin{remark}
Note that the Edgeworth expansion approximation in Theorem~\ref{thm2} can also be used to perform  both two-sided and one-sided hypothesis tests. To illustrate, consider testing a \textit{balance-free} null value with the hypothesis
\[H_0:w_n=c_n \quad vs \quad H_1:w_n> c_n.\]
Here, ``balance-free” refers to the absence of an explicit mechanism favoring balance. This notion has been operationalized in different ways in the literature, leading to different choices of the null value $c_n$. Our framework is flexible enough to accommodate these alternatives. For example, the null model in \citet{facchetti2011computing} assumes that the edge signs are i.i.d.\ Bernoulli variables, with probability equal to the fraction of negative edges in the observed network. Our null hypothesis accommodates this scenario by setting $c_n=(1-s_n)^3 + 3 s_n^2 ( 1- s_n)$. Similarly, in~\citet{Jin2024-lk}, the balance-free null model assumes equal probabilities for positive and negative edges, which our null hypothesis reflects by setting $c_n=0.5$. Using our empirical Edgeworth expansion formula, we evaluate the empirical p-value $\widehat{p}:=\widehat{\mathcal{G}}_n((\widehat{U}_n/\widehat{V}_n-c_n)/\widehat{S}_n)$. Following \cite{DBLP:journals/corr/abs-2004-06615}, given a significance level $\alpha$, we reject the null hypothesis $H_0$ if $\widehat{p} < \alpha$. The higher-order accuracy of our Edgeworth expansion allows us to control the Type-I error at $\alpha+O(\mathcal{E}(n,\,s_n,\,\rho_n))$ and the type-II error at $o(1)$ when $|w_n-c_n|=\omega(n^{-1/2})$. 
\end{remark}
\vspace{2mm}

\begin{remark} 
The inference procedure described above and its theoretical validity can be adapted to analyze each of the four types of triangles within the network, thereby providing detailed insights into specific patterns. For example, a lower proportion of type-2 triangles suggests strong evidence of weak balance in the network. For each type of triangle, we construct a studentized ratio of sample network moments and establish an Edgeworth expansion to approximate the CDF of this studentized ratio. The corresponding approximation error bounds are provided in the Supplemental Material~\ref{sec: results for wnt}. 
\end{remark}

\section{Simulation Studies}
\subsection{Comparative Analysis of CDF Approximations and their Computational Efficiency}
\label{subs: cdf}
We begin by assessing the accuracy and computational efficiency of approximations for the CDF of the studentized ratio of network moments. We compare our empirical Edgeworth expansion approach with two methods - the standard normal approximation and bootstrap-based node resampling method. Note that the node resampling method \citep{green2022bootstrapping}, originally developed for unsigned network moments, cannot be directly applied in our context. We adapt it to enable inference for the expected proportion of each type of triangle, with a detailed procedure provided in the Supplementary Material~\ref{sec: node resampling}.

\begin{figure}[!t]
\includegraphics[width=0.99\textwidth]{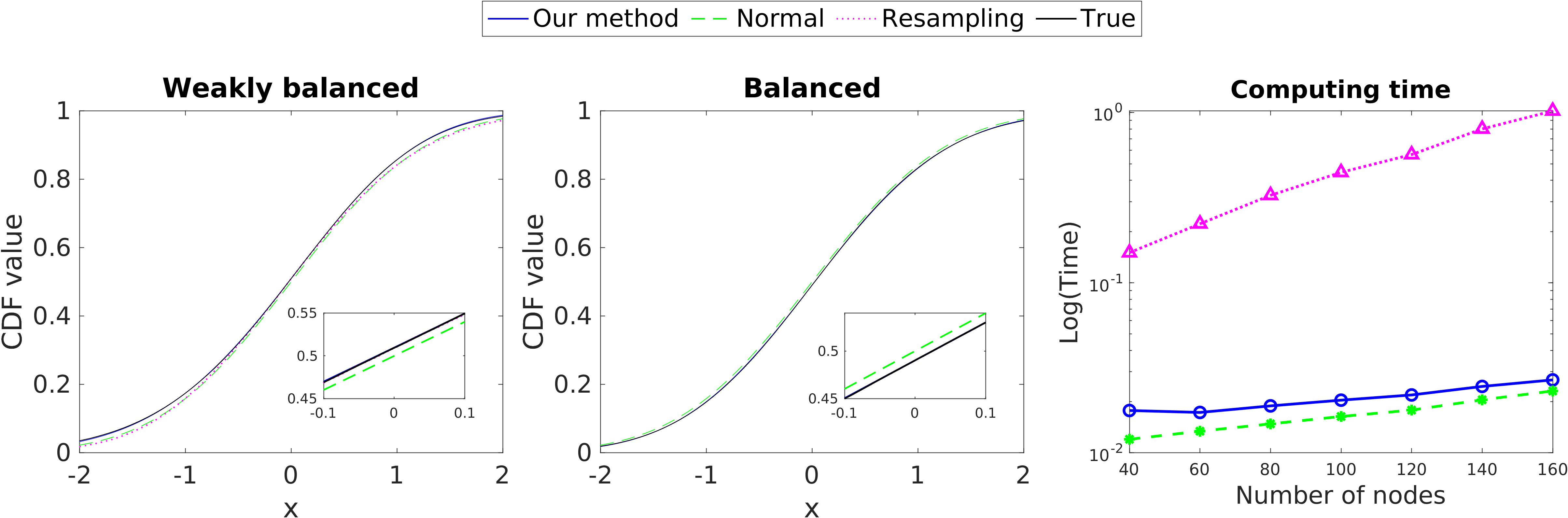}
\caption{The true CDF of the studentized ratio of network moments, along with its approximations (including our empirical Edgeworth expansion formula, standard normal approximation, and the bootstrap-based node-resampling method), is displayed on the left for type-2 triangles and in the middle for balanced triangles, respectively. On the right, we provide a comparison of the computing time for each approximation method.
} 
\label{fig.CDF}
\end{figure}
\begin{table}[h]
\centering
\begin{tabular}{|l|c|c|c|}
    \hline
    CDF distance & Our method & Normal & Resampling \\
    \hline
    Weakly balanced    & \bf{0.0028}  & 0.0171  & 0.0227  \\
    \hline
    Balanced  & \bf{0.0019}  & 0.0116  & \diagbox{}{} \\
    \hline
\end{tabular}
\caption{Kolmogorov--Smirnov (sup-norm) distances between the true cumulative distribution function and its approximations.}
\label{tab: CDF distance}
\end{table}

To compare their approximations to the true CDF, we generate signed networks with $n = 160$ nodes from the graphon model defined by the functions $\rho_nF(x,y)=0.8$ and $s_nG(x,y)=2\cos(x^2+y^2)/3+0.3$. We generate a large sample of $N=10^6$ networks from the graphon model to evaluate the true CDF of ${\whT_n}$. The adapted node-resampling method approximates the CDF by resampling $B=10^5$ networks from the estimated graphon functions. For the sake of space, we present the results for the population parameters for balance (\textit{i.e.}, $w_n$) and weak balance (\textit{i.e.}, $w_{n,2}$) in the main text, while the full results for other types of triangles are provided in the Supplementary Materials.
The left and middle panels of Figure~\ref{fig.CDF} suggest our method closely aligns with the true CDFs, offering a better approximation compared to the normal approximation and the resampling method. Table~\ref{tab: CDF distance} reports the Kolmogorov--Smirnov (sup-norm) distance between the true distribution and its approximations.

To further compare computational efficiency, we vary the network size from $40$ to $160$. The right panel of Figure~\ref{fig.CDF} summarizes the logarithm of computing time relative to the network size. It shows that our Edgeworth expansion approach and the normal approximation are computationally more efficient than the node-resampling method. The former two use explicit formulas, where the primary computational cost is a few matrix multiplications to calculate the studentized statistics, while the latter necessitates a large number of bootstrap replications. Overall, our method, with a computational cost identical to the simple normal approximation, provides higher-order accuracy, as demonstrated in Figure~\ref{fig.CDF}.

\subsection{Impact of Network Size and Sparsity}
\begin{figure}[!ht]
\centering
\begin{subfigure}{0.38\textwidth}
    \includegraphics[width=\textwidth]{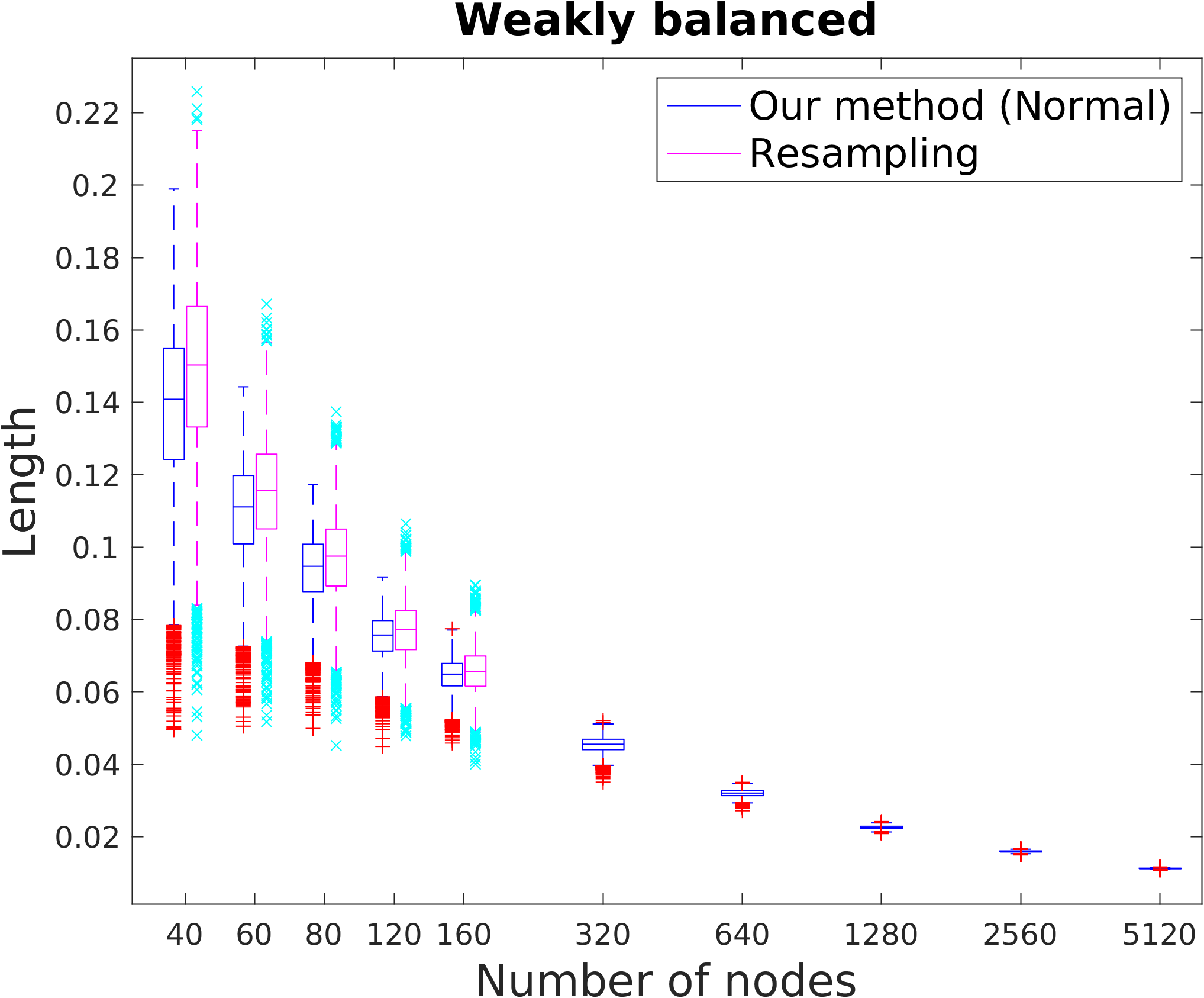}
\end{subfigure}
\hspace{0.02\textwidth}
\begin{subfigure}{0.39\textwidth}
    \includegraphics[width=\textwidth]{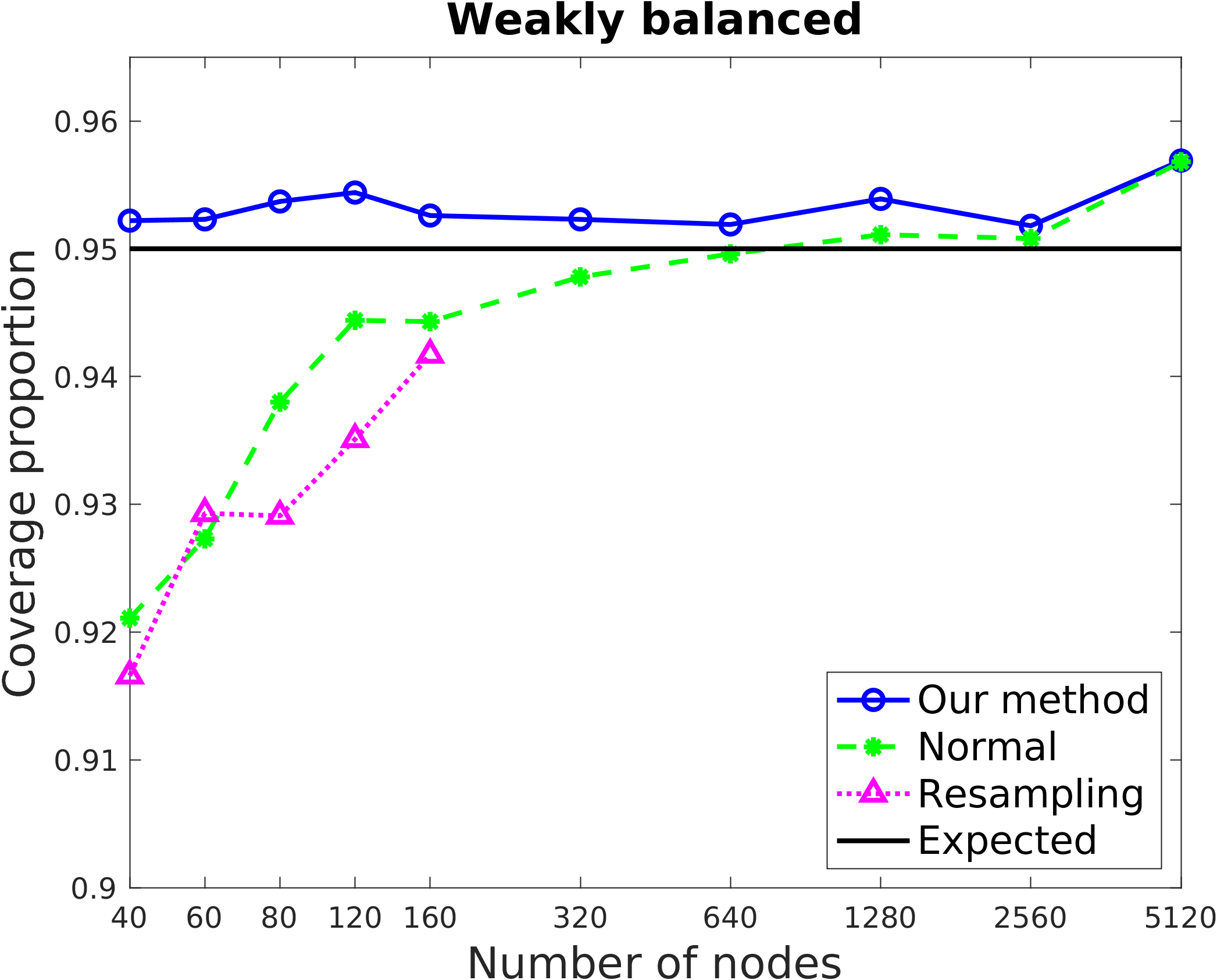}
\end{subfigure}
\begin{subfigure}{0.38\textwidth}
    \includegraphics[width=\textwidth]{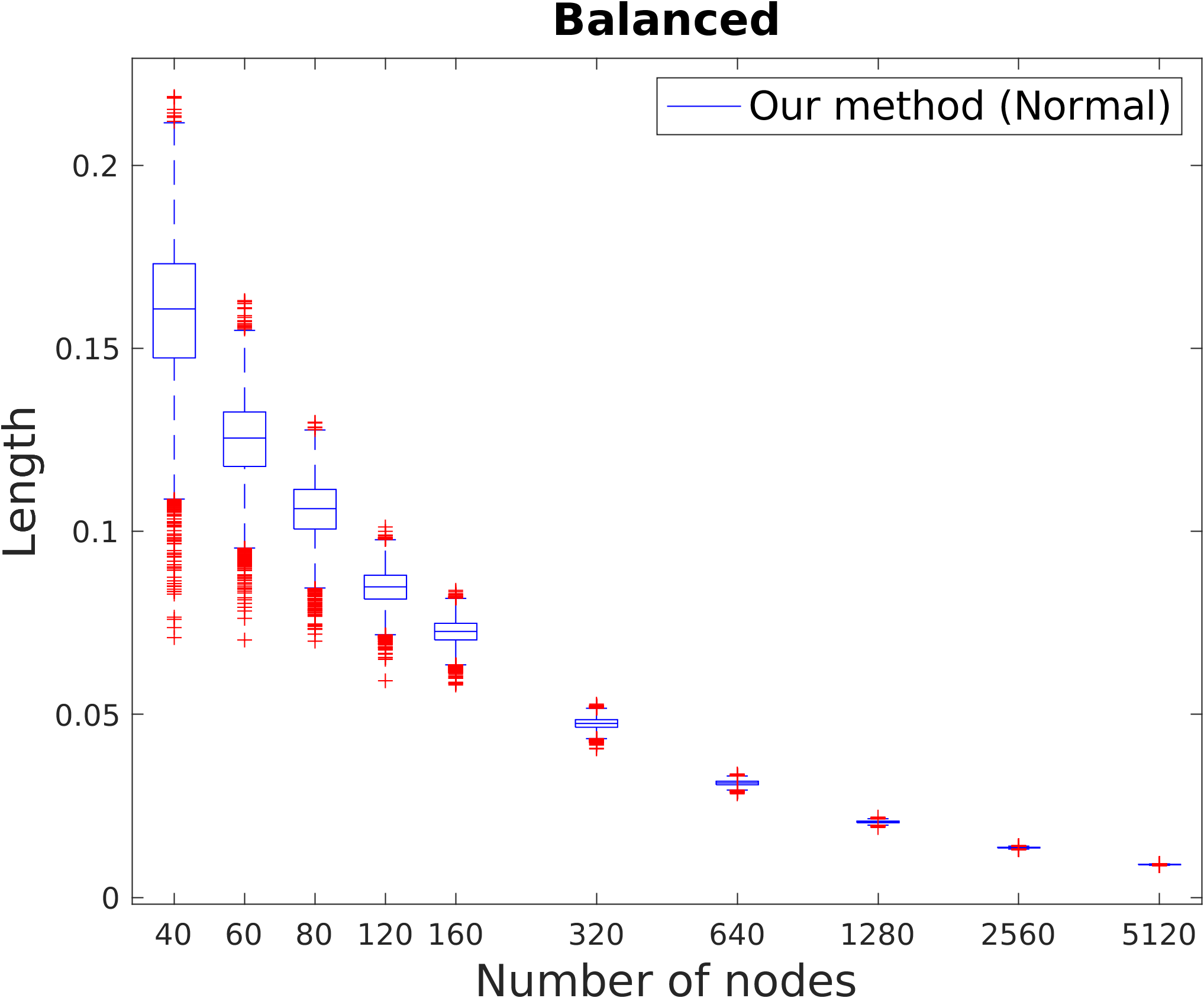}
\end{subfigure}
\hspace{0.02\textwidth}
\begin{subfigure}{0.39\textwidth}
    \includegraphics[width=\textwidth]{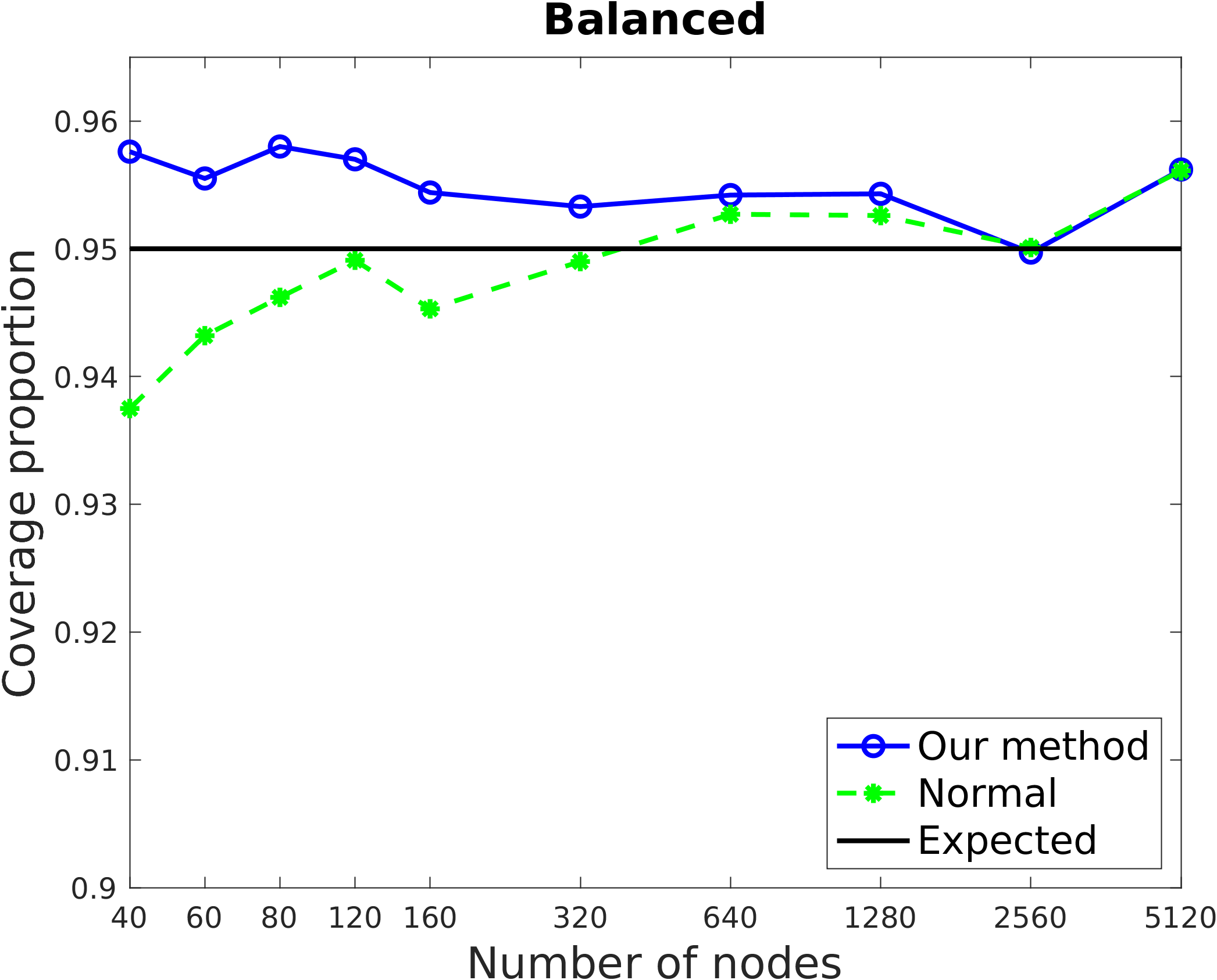}
\end{subfigure}
\caption{The length and coverage proportion of 95\% confidence intervals for (weakly) balanced triangles in relation to varying network size are presented from left to right, respectively.}
\label{fig.largen}
\end{figure}

\begin{figure}[!ht]
\centering
\begin{subfigure}{0.38\textwidth}
    \includegraphics[width=\textwidth]{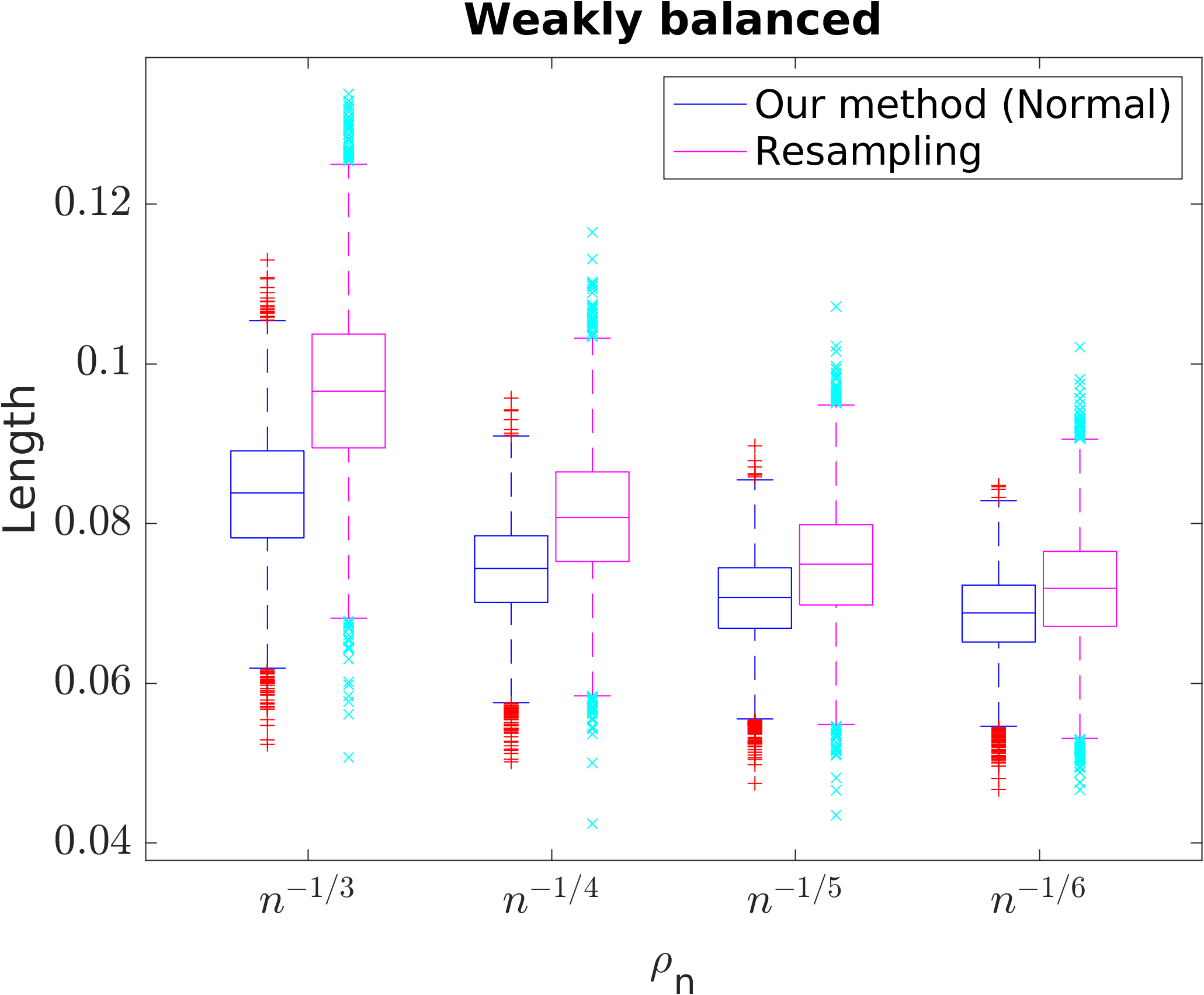}
\end{subfigure}
\hspace{0.02\textwidth}
\begin{subfigure}{0.39\textwidth}
    \includegraphics[width=\textwidth]{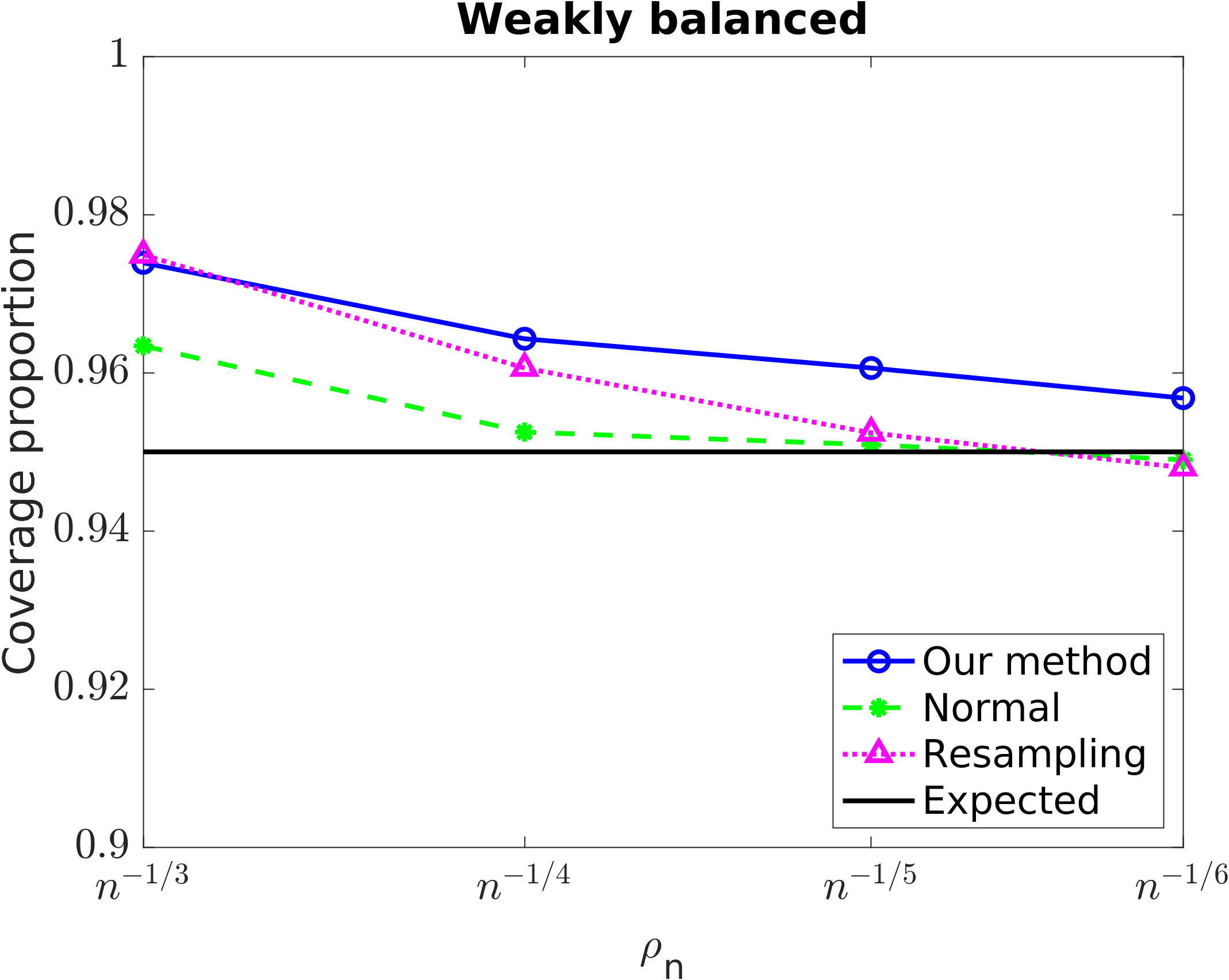}
\end{subfigure}
\begin{subfigure}{0.38\textwidth}
    \includegraphics[width=\textwidth]{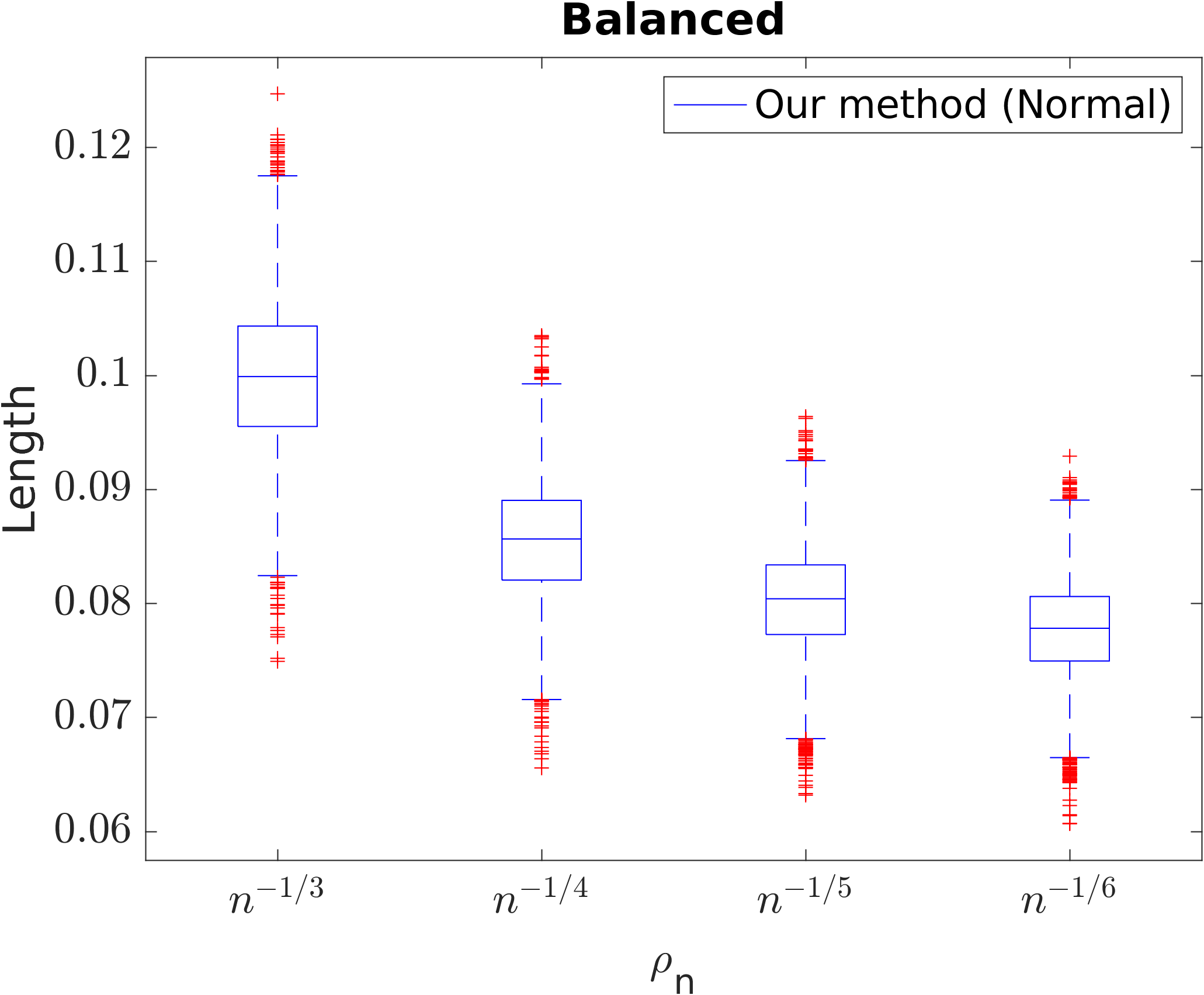}
\end{subfigure}
\hspace{0.02\textwidth}
\begin{subfigure}{0.39\textwidth}
    \includegraphics[width=\textwidth]{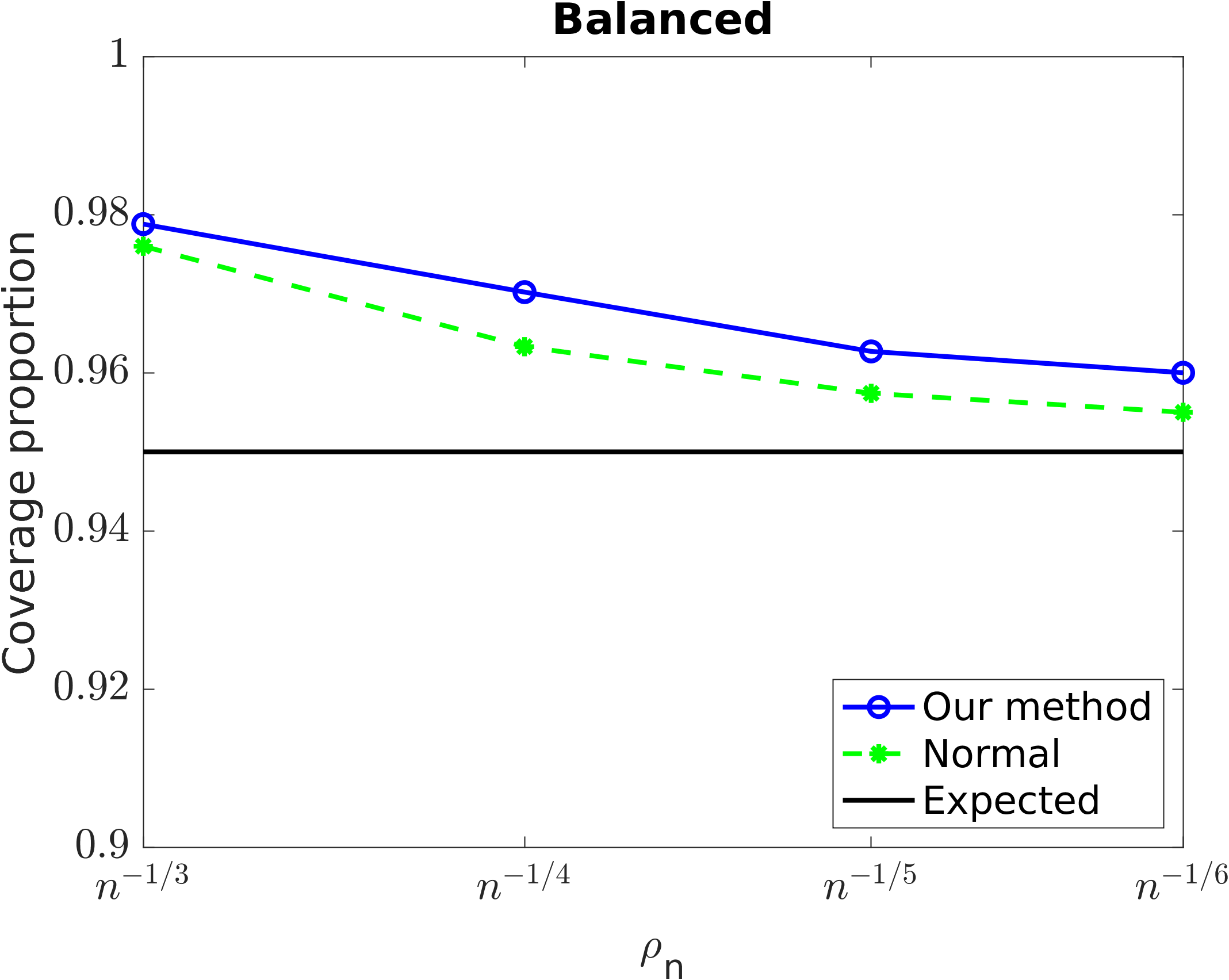}
\end{subfigure}
\caption{The length and coverage proportion of 95\% confidence intervals for (weakly) balanced triangles in relation to varying network sparsity are presented from left to right, respectively.}
\label{fig.rhon}
\end{figure}

In this subsection, we evaluate the coverage proportion and the length of confidence intervals (CIs) constructed by each method and investigate the impact of the network size and sparsity on their effectiveness. We generate networks using the same graphon model described in Section~\ref{subs: cdf}. Hereafter, we conduct $10,000$ replications for each setting.

We consider a wide range of network sizes, including small networks with $n\in \{40, 60, 80, 120, 160\}$ and larger networks with $n \in\{320,640,1280,2560,5120\}$. For large sizes $n\in\{320,640,\cdots,5120\}$, due to the high computational cost of the node-resampling method, we only compare our method with the standard normal approximation. Figure~\ref{fig.largen} summarizes the performance of 95\% CIs for $w_n$ and $w_{n,2}$ with respect to varying network sizes.
Our method consistently achieves proper coverage across different network sizes.
For smaller networks, our method obtains higher coverage than the normal and node resampling methods, which highlights the advantage of achieving higher-order approximation to the CDF. 
The CI length for all methods decreases as the network size increases, with our method and the normal approximation slightly outperforming the resampling method in terms of shorter lengths.

Furthermore, we set $n=160$ and vary $\rho_n$ with a range of $\{n^{-1/3}, n^{-1/4},n^{-1/5}, n^{-1/6}\}$, which leads to the network density ranging across 18.4\%, 27.2\%, 35.2\% and 42.4\%. Figure~\ref{fig.rhon} summarizes the effectiveness of 95\% CIs for $w_n$ and $w_{n,2}$ with respect to varying network sparsity. 
We can see that all methods achieve reasonable coverage proportions across varying network sparsity, and the CI length of our method and the normal approximation consistently remains shorter than that of the resampling method. 
As the network density increases, the CI length decreases and the coverage proportion approaches the nominal level 95\%. This is attributed to the increasing number of observed signs as the network becomes denser.

\subsection{Varying Degree of Balance} \label{subsec: simu varying degree}
Finally, we investigate the performance of our inference method across various extent of balance. To this end, we generate networks with $n=160$ nodes from the graphon model defined by $s_n G(x,y;\beta)=1/(1+\exp(\beta (x-0.4)(y-0.4))$. Here, $\beta$ controls the overall degree of balance. 
\begin{figure}[!ht]
\centering
\begin{subfigure}{0.38\textwidth}
    \includegraphics[width=\textwidth]{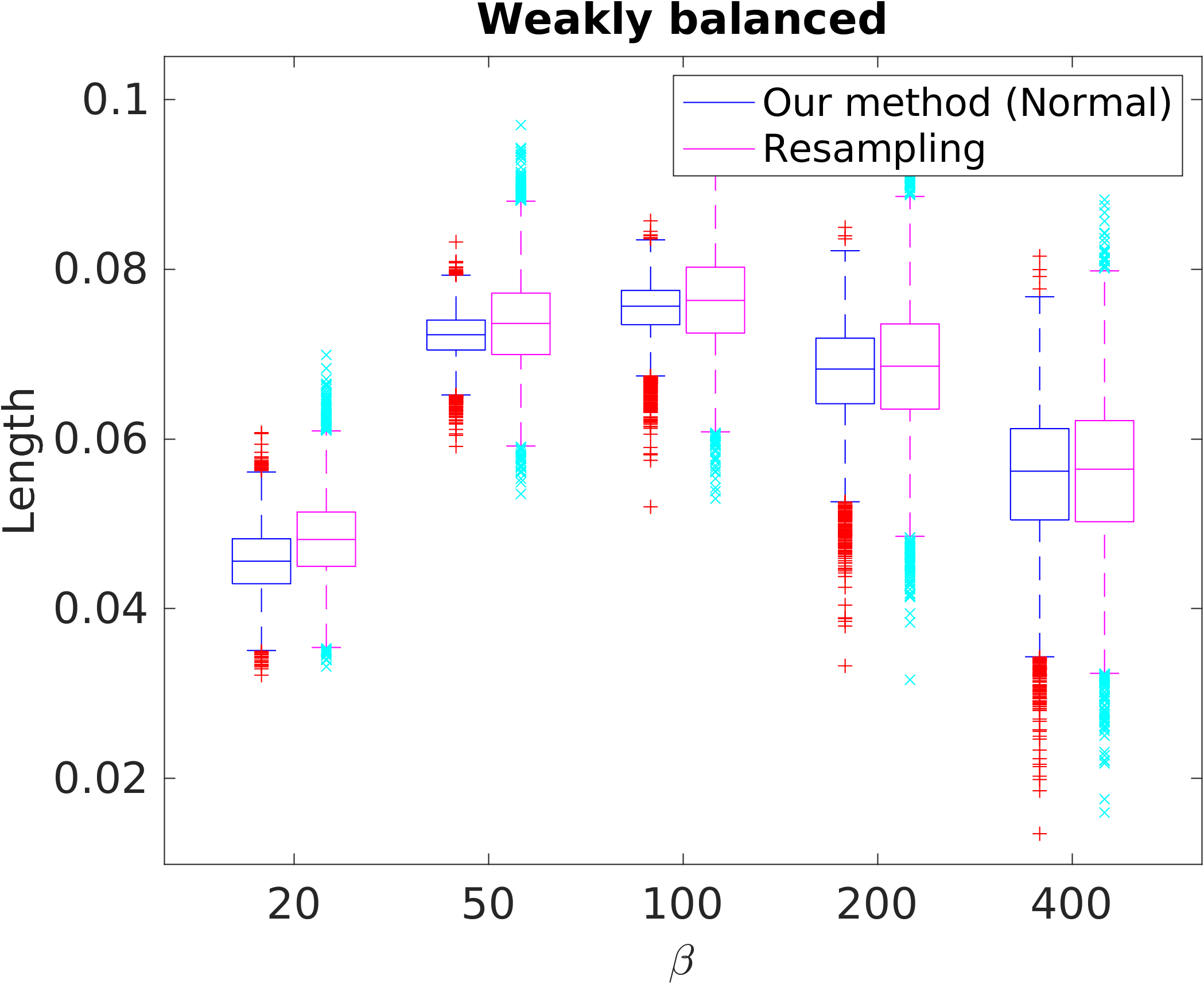}
\end{subfigure}
\hspace{0.02\textwidth}
\begin{subfigure}{0.39\textwidth}
    \includegraphics[width=\textwidth]{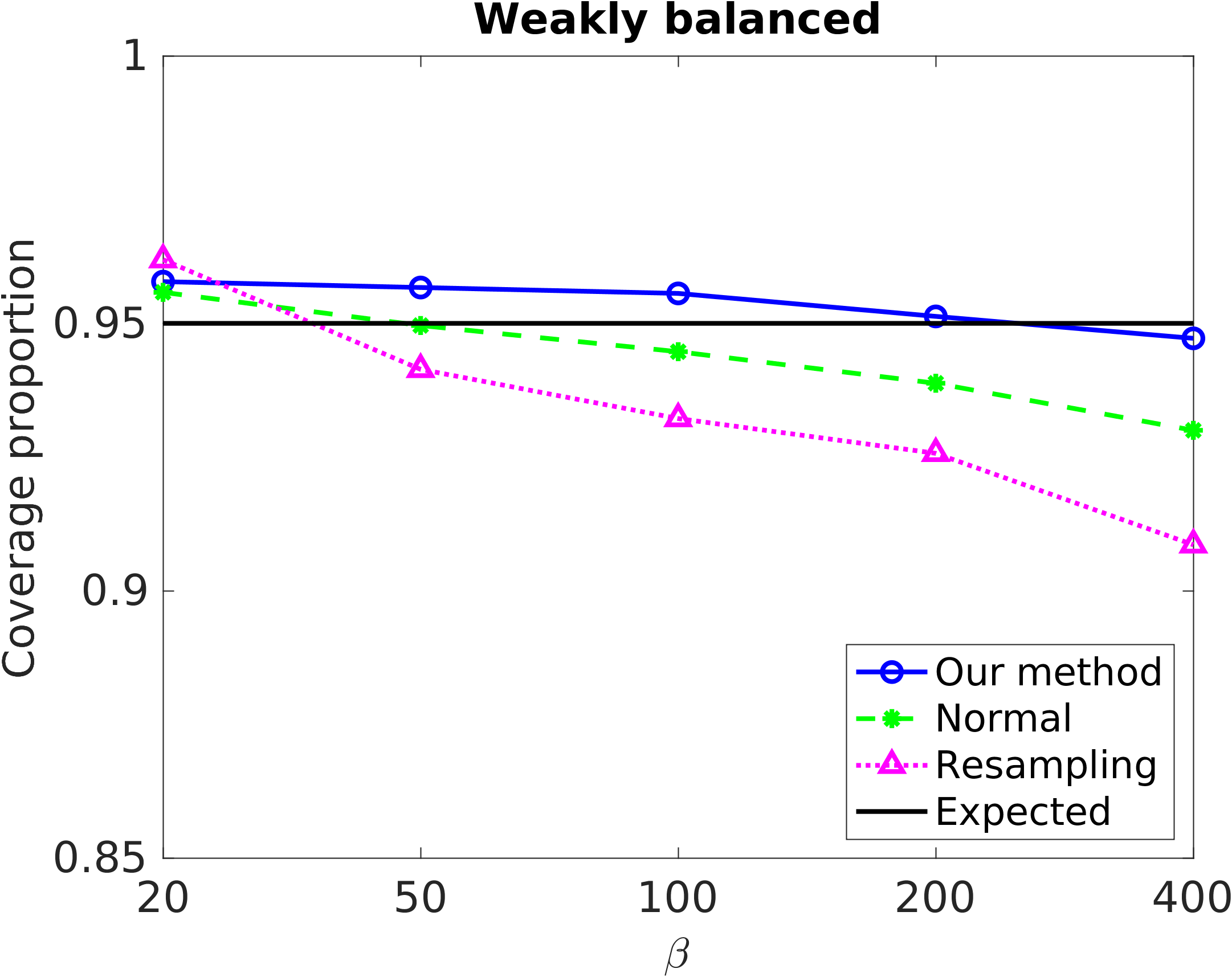}
\end{subfigure}
\begin{subfigure}{0.38\textwidth}
    \includegraphics[width=\textwidth]{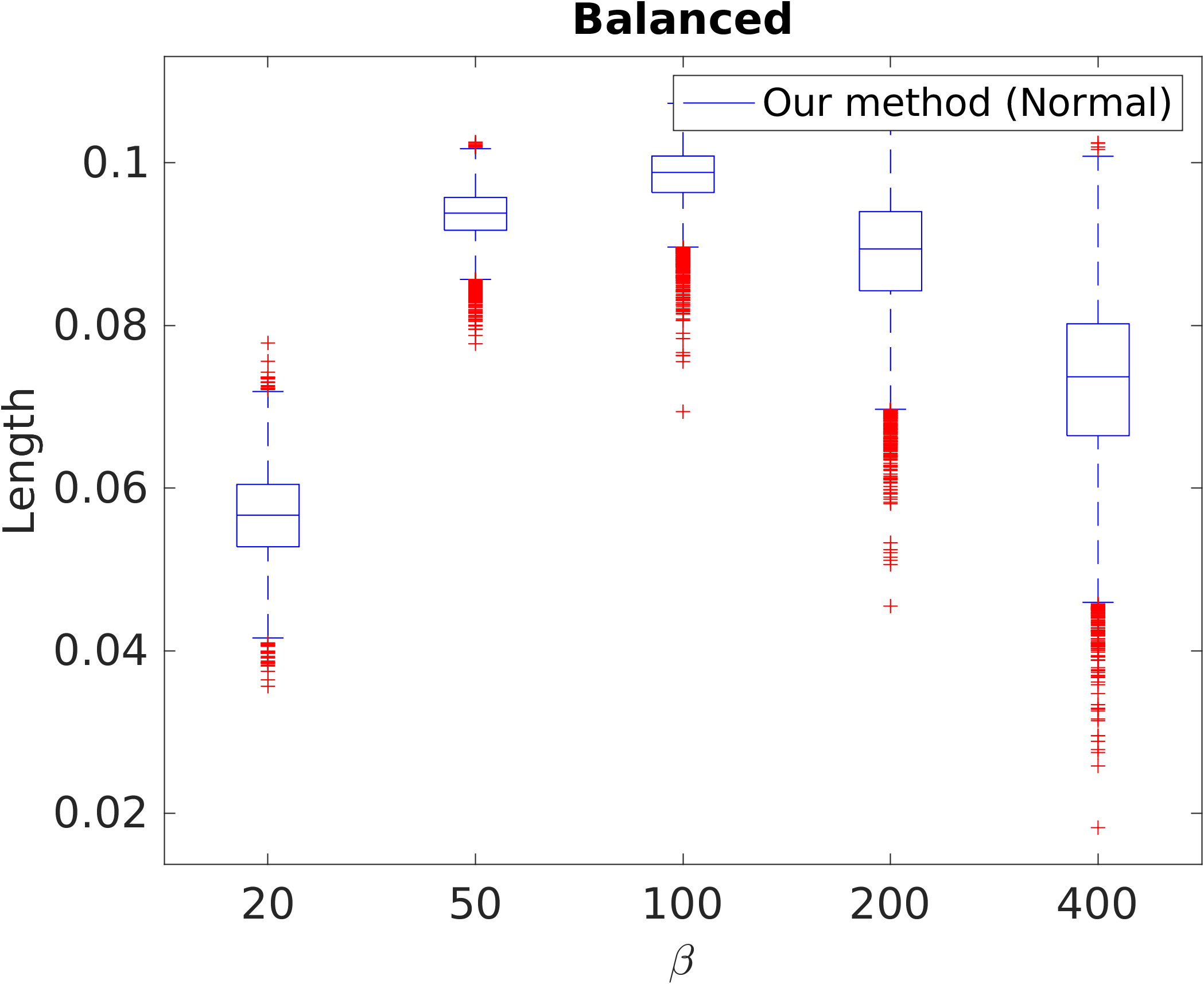}
\end{subfigure}
\hspace{0.02\textwidth}
\begin{subfigure}{0.39\textwidth}
    \includegraphics[width=\textwidth]{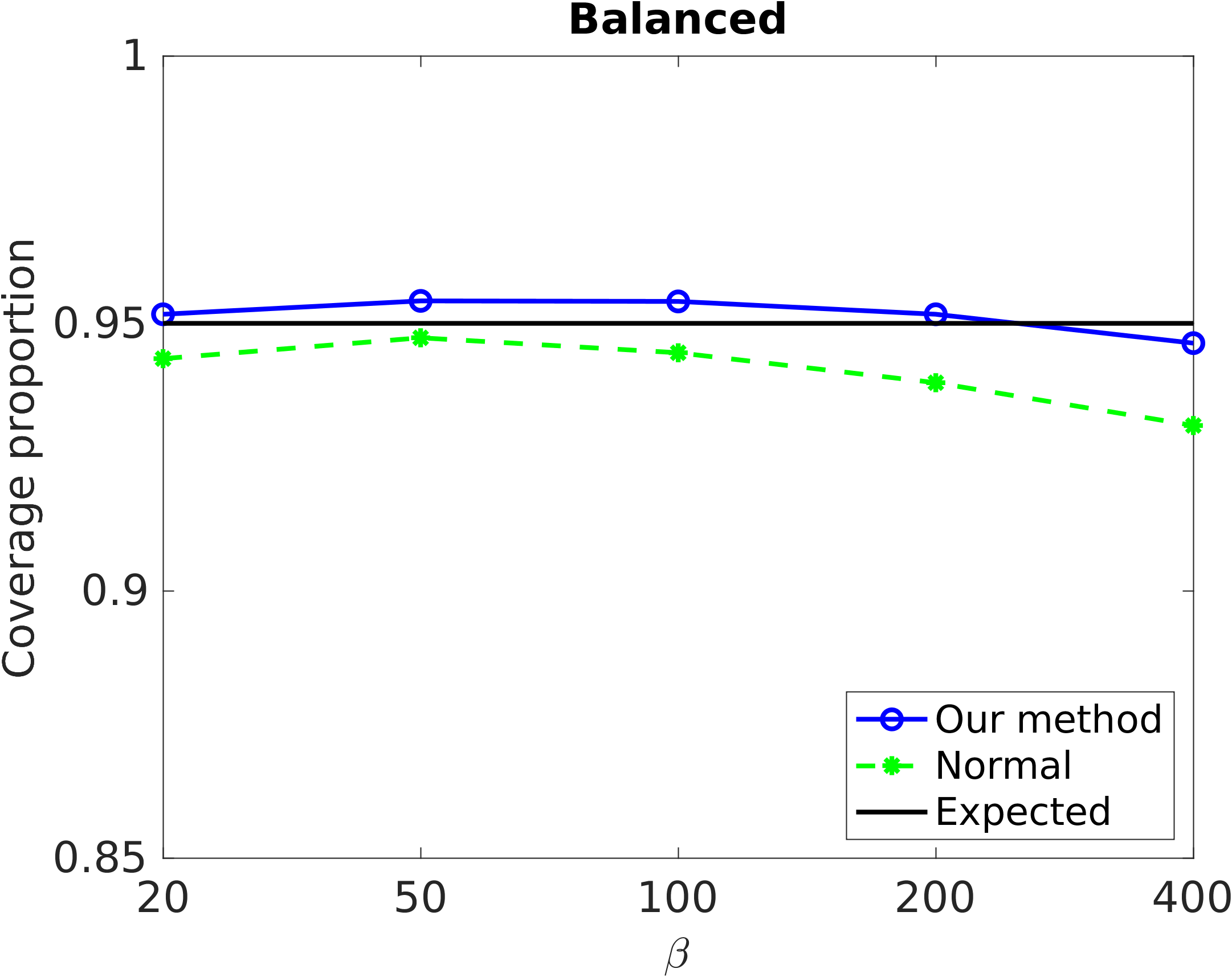}
\end{subfigure}
\caption{The length and coverage proportion of 95\% confidence intervals of (weakly) balanced triangles in relation to varying degree of balance are presented from left to right, respectively.}
\label{fig.balance}
\end{figure}
Specifically, when $\beta=0$, each edge is equally likely to be positive or negative, considered as a balance-free case. As $\beta$ approaches infinity, the edge signs follow a block model with membership probabilities $\pi=(0.4, 0.6)$. In particular, edges tend to have positive signs within blocks and negative signs between them, leading to balanced triangles. Therefore, a larger $\beta$ corresponds to a greater degree of balance.

We vary $\beta$ from $\{20,50,100,200,400\}$, with the proportion of balanced triangles increasing from $59.0\%$ to $92.6\%$. 
The proportion of each type of triangle is provided in Figure~\ref{fig:balance_summary}, showing an increase in balanced triangles (types 1 and 3) and a decrease in unbalanced triangles (types 2 and 4). 
As indicated in Figure~\ref{fig.balance}, our method consistently maintains proper coverage proportions across varying extent of balance. Notably, when $\beta$ is large, \textit{i.e.}, a great extent of balance is present, our method achieves more stable coverage proportion compared to both the normal approximation and the resampling method.  
We further examine the behavior of coverage in more extreme regimes of balance and network density in the Supplementary Materials~\ref{sec: full simulation results}, where we clarify the source of the downward trends observed in some settings.

\section{Real-world Data Examples}
\label{sec: real_world_example}
\subsection{Problem Setup}
In this section, we apply the proposed method to analyze the evidence of balance and weak balance in real-world signed networks across various fields. 
A higher expected proportion of balanced triangles, \textit{i.e.}, $w_n$, indicates stronger evidence of balance, while a lower expected proportion of type-2 triangles, \textit{i.e.}, $w_{n,2}$, suggests the weak balance. 
To measure the degree of balance in an observed signed network, we establish 95\% confidence intervals for $w_n$ and $w_{n,2}$, and compare them with balance-free baselines. Specifically, we consider two balance-free models to compute the corresponding expected proportions as baselines. The first model assumes that each edge has an equal probability of being positive or negative, setting baseline values as $0.5$ for balanced triangle and $0.25$ for type-2 triangle. We refer to them as ``Baseline 50\%'' and ``Baseline 25\%'' respectively below. This baseline aligns with the concept of the population-level balance introduced in~\citet{tang2023population}, where a signed network is considered population-level balanced if the expected proportion of balanced triangles exceeds $0.5$. The second model adjusts for the proportion of negative edges~\citep{facchetti2011computing}. Here, edge signs are independently and identically distributed Bernoulli variables, with the probability equal to the fraction of negative edges in the observed network. We refer to the second baseline as ``Adjusted Baseline''. Details on calculating these baselines and the results for other types of triangles are provided in the Supplemental Materials.

\subsection{International Relation Network}
We apply our method to international relations networks constructed from Correlates of War data \citep{gibler2008international, palmer2022mid5}, spanning 1910–1945 and involving 217 distinct countries across the full time period.
\begin{figure}[!t]
\centering
\begin{subfigure}{0.45\textwidth}
    \includegraphics[width=\textwidth]{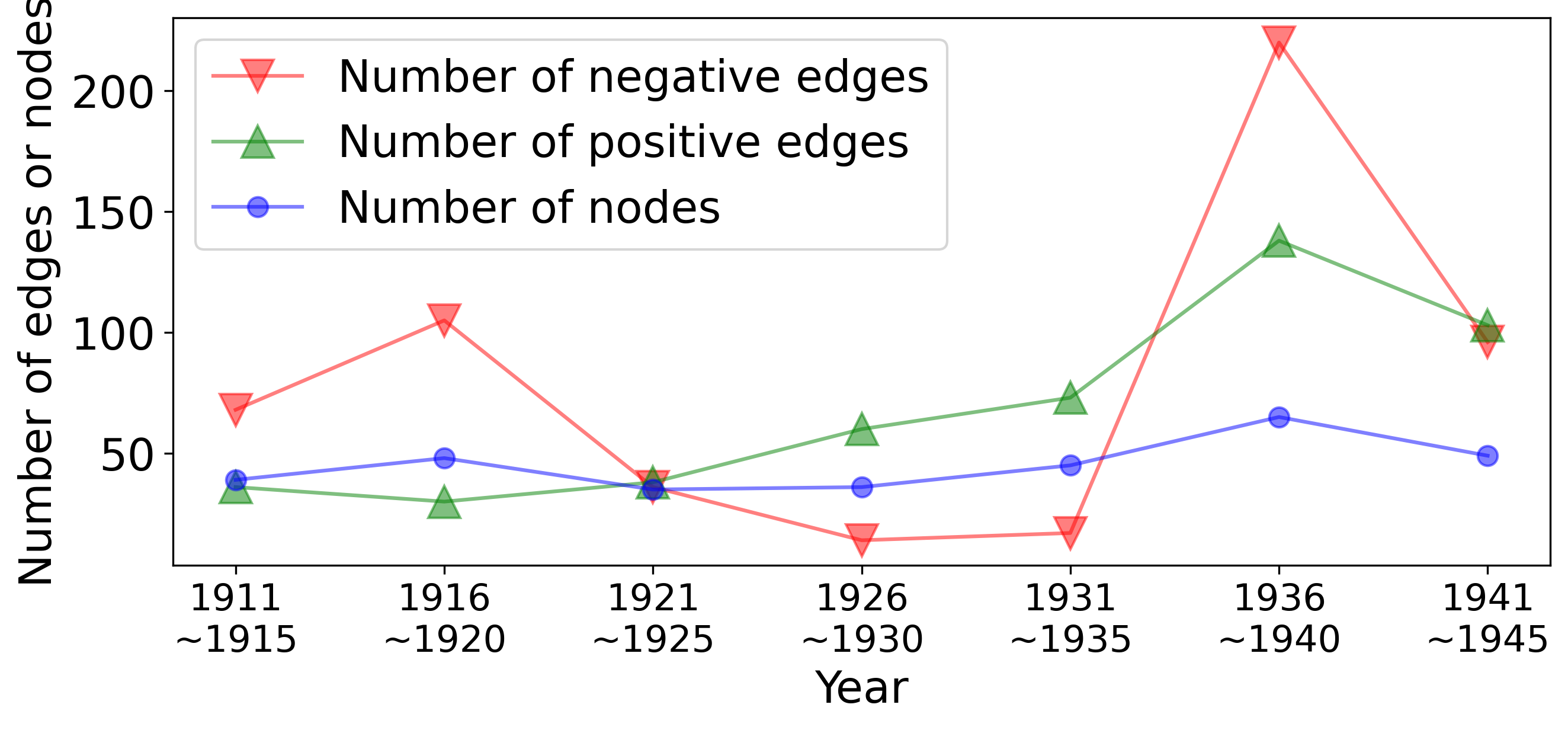}
    \caption{Number of nodes and edges in networks}
    \label{woc:first}
\end{subfigure}
\begin{subfigure}{0.45\textwidth}
    \includegraphics[width=\textwidth]{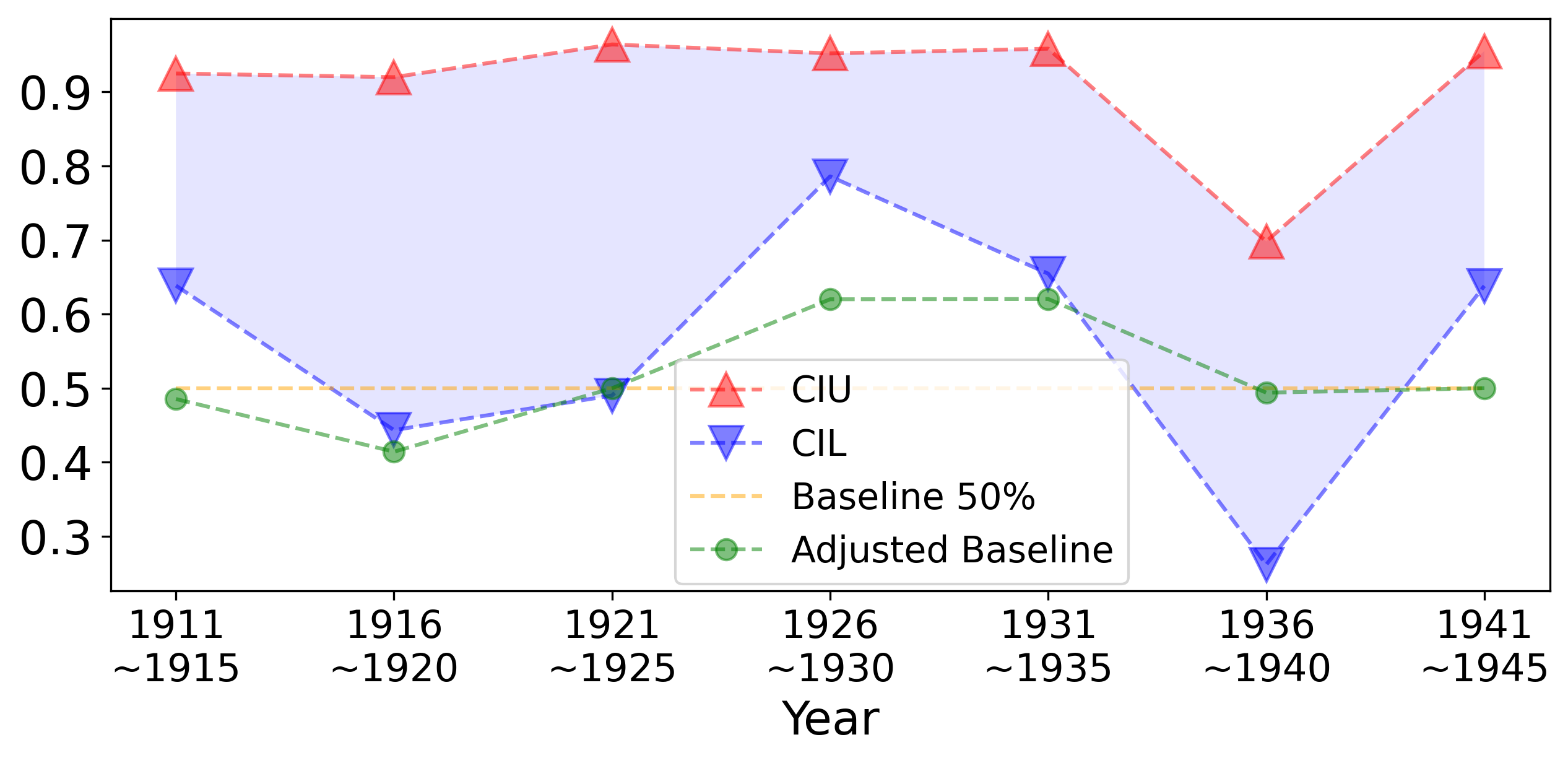}
    \caption{95\% confidence intervals for $w_n$}
    \label{woc:second}
\end{subfigure}
\begin{subfigure}{0.45\textwidth}
    \includegraphics[width=\textwidth]{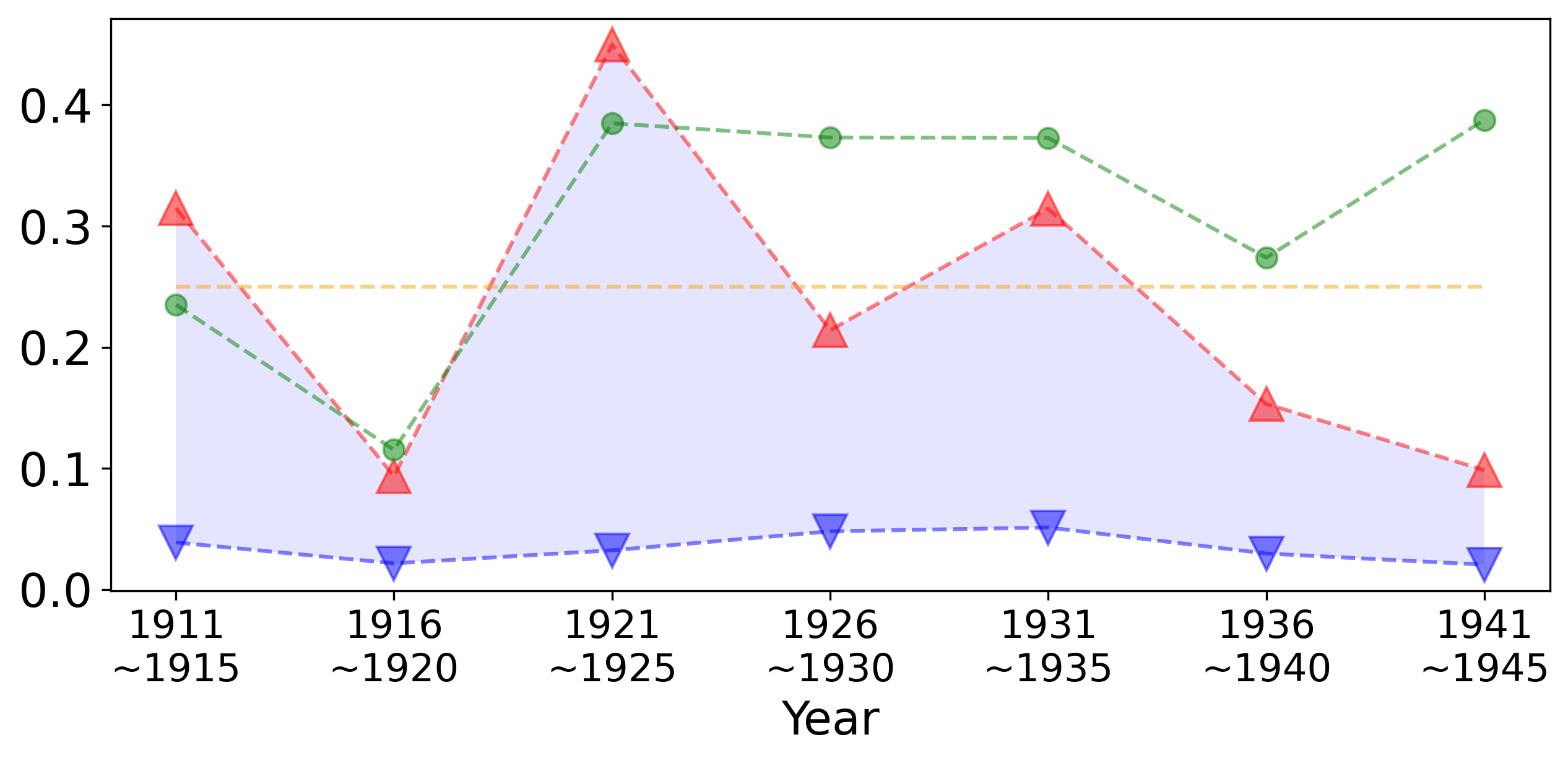}
    \caption{95\% confidence intervals for $w_{n,2}$}
    \label{woc:third}
\end{subfigure}
\begin{subfigure}{0.45\textwidth}
    \includegraphics[width=\textwidth]{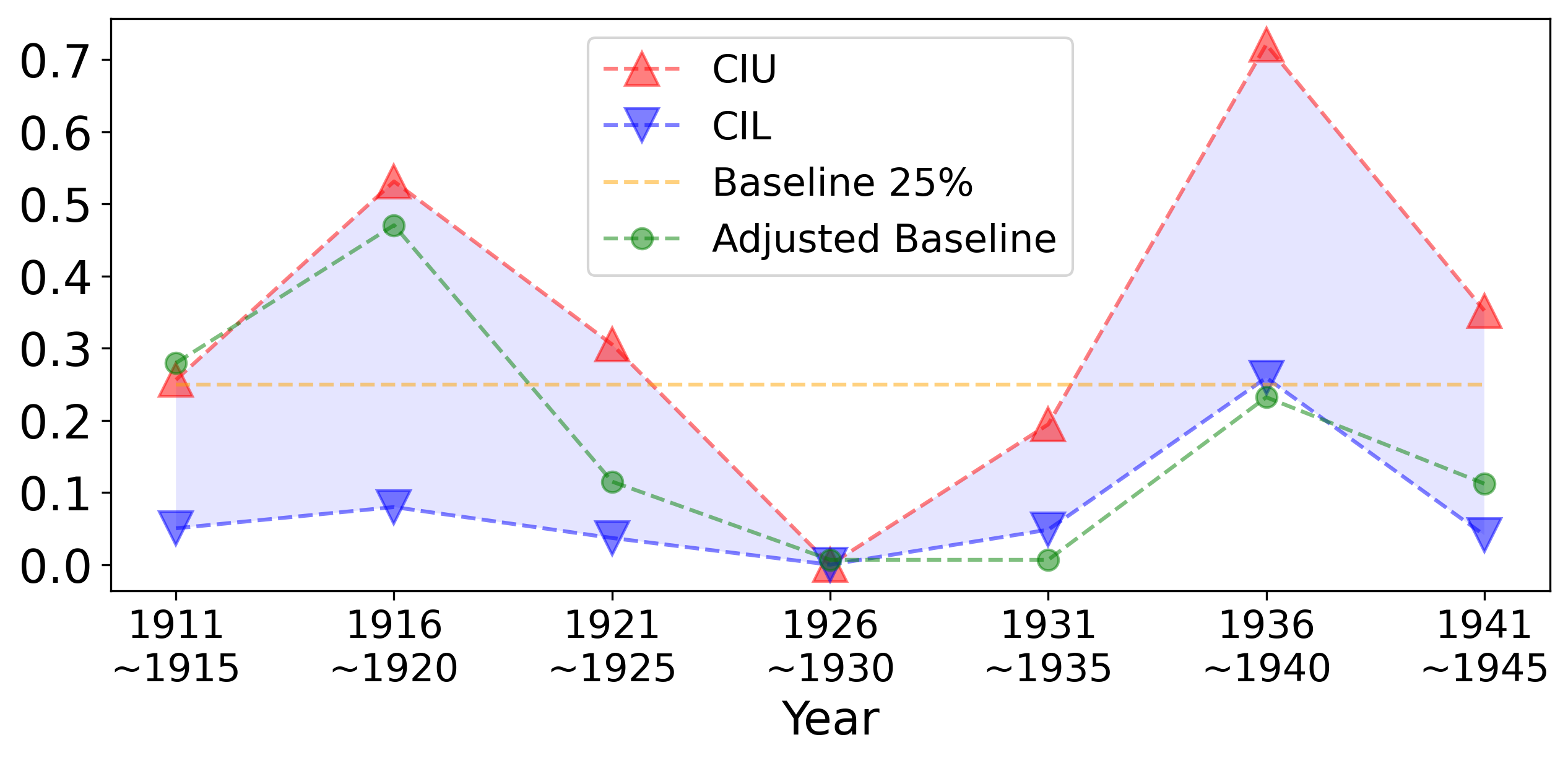}
    \caption{95\% confidence intervals for $w_{n,4}$}
    \label{woc:fourth}
\end{subfigure}
\caption{(a) The network size and the number of positive and negative edges for each five-year signed network from 1911 to 1945; 
(b) The 95\% confidence intervals for the expected proportion of balanced triangles, i.e., $w_{n}$; (c) The 95\% confidence intervals for the type-2 triangle, i.e.,  $w_{n,2}$;  (d) The 95\% confidence intervals for the type-4 triangle, i.e., $w_{n,4}$. For (b)-(d), the red upper triangles correspond to the upper bound of confidence intervals (CIU); the blue lower triangles correspond to the lower bound of confidence intervals (CIL); and the yellow and green dashed lines represent baselines under two balance-free models.
}
\label{woc}
\end{figure}
For each five-year period, we construct a signed network. We define a positive edge for a pair of countries if the duration of alliance agreements exceeds that of militarized disputes; a negative edge if the duration of disputes exceeds that of alliance agreements; and no edge if there was no relationship. Figure~\ref{woc:first} reports the network size and the numbers of positive and negative edges for each five-year signed network. In particular, the number of negative edges substantially increases during World War I~(1914-1918) and II~(1939-1945).

We summarize the confidence intervals for the expected proportion of balanced triangles in Figure~\ref{woc:second}. 
Compared to two baselines, the 95\% confidence intervals for $w_n$ are above the baseline values for all time frames except for the period from 1936 to 1940. This finding provides strong evidence supporting the structural balance theory over most time frames, suggesting that countries tend to form balanced triadic relationships in the international relation network. However, during the 1936-1940 period, we observe a noticeable deviation from the general trend. The confidence interval for the type-4 triangle is notably high, as shown in Figure~\ref{woc:fourth}, indicating that hostility often exists among  countries during this time frame. This observation may be attributed to the full-scale outbreak of World War II, with increasing tensions and the emergence of conflicts. 
While the network lacks strong structural balance during this time, it still maintains a degree of weak balance when considering only type-2 triangles as unbalanced. As illustrated in Figure~\ref{woc:third}, the confidence interval for the proportion of type-2 triangles is much below the baseline values.

\subsection{Protein-Protein Interaction Network}

\begin{figure}[!t]
\centering
\begin{subfigure}{0.45\textwidth}
    \includegraphics[width=\textwidth]{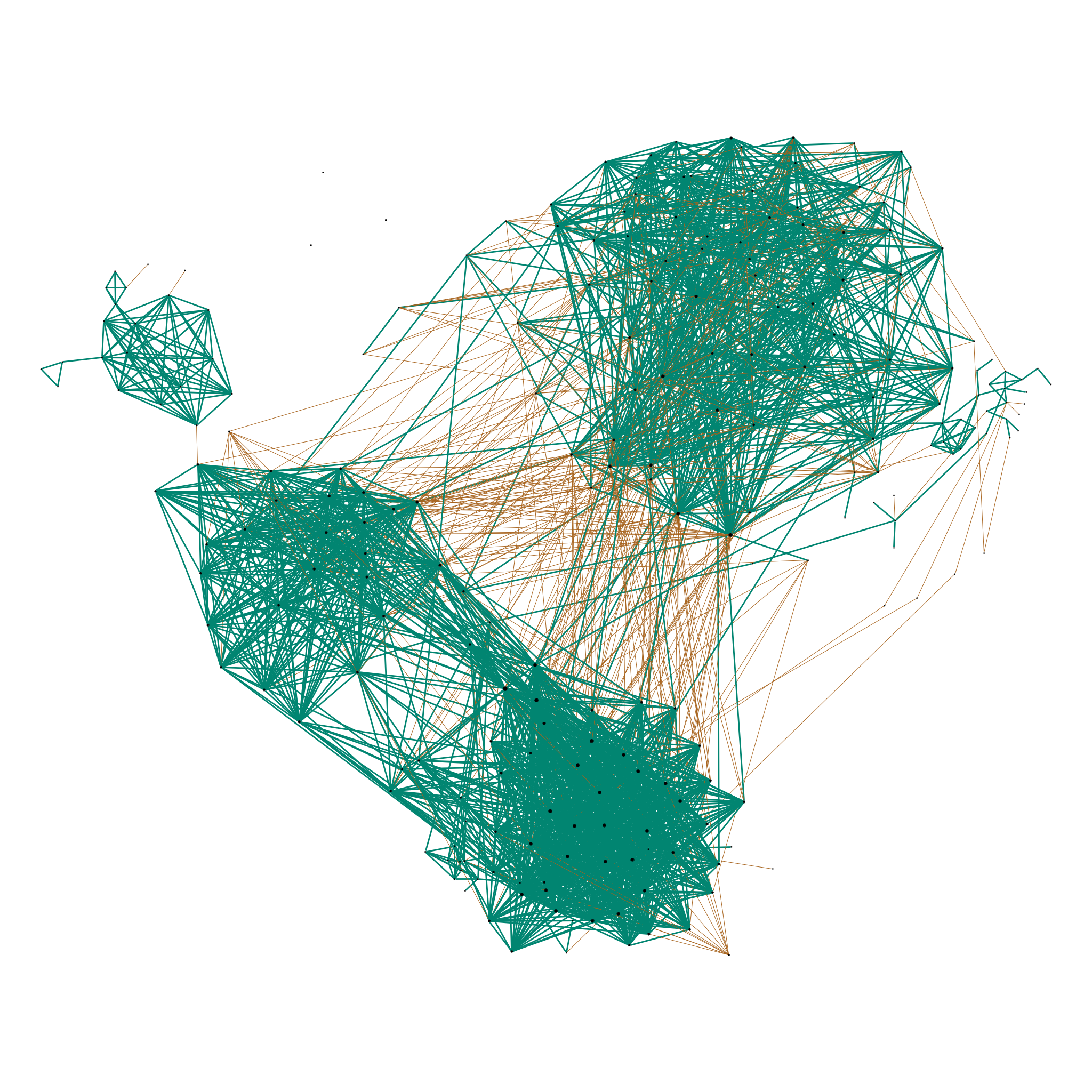}
    \caption{Protein-protein interaction network}
    \label{fig:ppi}
\end{subfigure}
\begin{subfigure}{0.45\textwidth}
    \includegraphics[width=\textwidth]{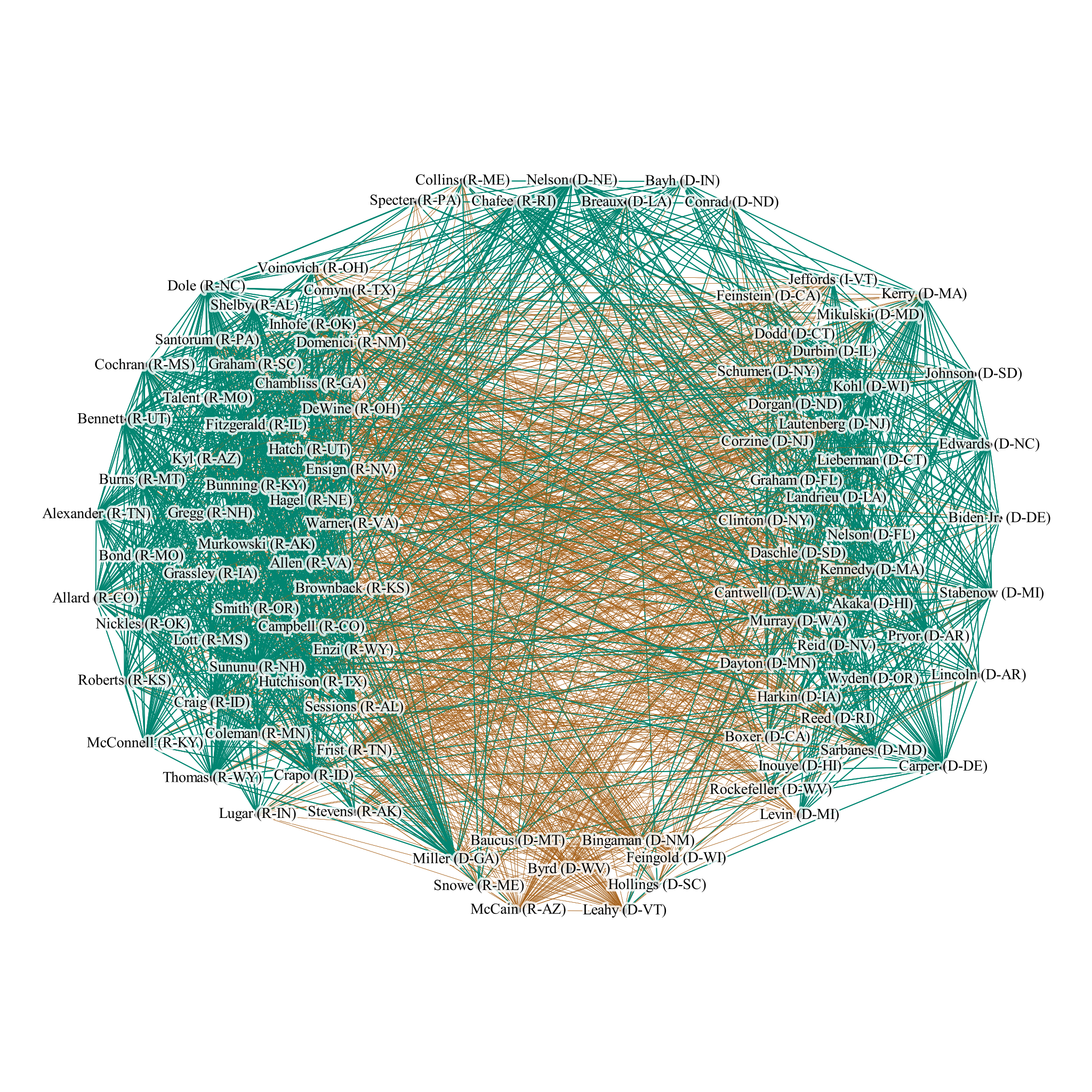}
    \caption{$108$th U.S.\ Senate network}
    \label{fig:senate}
\end{subfigure}
\caption{Visualization of two networks: (a) the functional similarity network among interacting proteins, where proteins with more than $25$ edges are plotted, and (b) the network among U.S. senators. In both networks, green edges represent positive interactions and red edges represent negative interactions.}
\end{figure}

Linking physical and functional associations among proteins is useful for understanding how their interactions relate to cellular function and viability. In particular, \citet{huttlin2021dual} integrated BioPlex networks~\citep{huttlin2015bioplex} in 293T and HCT116 cells with genome-wide CRISPR co-essentiality profiles from Project Achilles \citep{meyers2017computational}. 
They calculated the functional similarity between fitness profiles for each interacting protein pair using Spearman's correlation coefficient, which can be either positive or negative. 
Their findings indicate that these positive and negative correlations reflect different structural relationships, highlighting the importance of differentiating between signs of functional similarity.

To investigate this further, we construct a signed network among proteins by retaining only significant correlations between fitness profiles. Specifically, we assign a positive edge between a pair of interacting proteins if the p-value associated with Spearman's correlation is less than 0.05 and the correlation is positive. We assign a negative edge if the significant correlation is negative. No edge is assigned for non-significant correlations or non-interacting pairs. 
We then apply our inference method to this signed network and summarize the results in Table~\ref{ppi and social networks}. 
The confidence interval for $w_n$ is significantly above the baseline value, providing strong evidence of network balance. Compared to the international relation network, the confidence intervals in this PPI network are much shorter due to its larger network size. 

We visualize the signed PPI network in Figure~\ref{fig:ppi}, where we observe several clusters of nodes. There are more connections within clusters and fewer edges between clusters. Particularly, the edges between the upper-right cluster and the others are mostly negative, while edges within or between other clusters are mostly positive. This observation aligns with the population-level balance structure identified by \citet{tang2023population}, where node communities can be merged into two groups where edges are predominantly positive within groups and negative between groups. 
Notably, \citet{huttlin2021dual} observed that positively correlated proteins typically belong to the same complexes, whereas negatively correlated proteins belong to different complexes and exhibit antagonistic interactions. 
The balance structure provides insight into the functional organization of protein interactions, supporting the observation that proteins within the same complex tend to cooperate, while those in different complexes may have opposing functions.

\begin{table}[!ht]

    \centering
    \small
    \begin{tabular}{|c|c|c|c|}
    \hline
        Datasets &  Protein-Protein & U.S.Senate & Wikipedia Elections\\ \hline
        Nodes &  8777 & 100 & 7115 \\ \hline
        Edges  &  $6.05\times 10^{-4}$ & {$0.40$} & $3.98\times 10^{-3}$ \\ \hline
        Negative Signs  &  0.31 & {0.42} & 0.22 \\ \hline
        Type-1 & 0.891& 0.490 & 0.659 \\ \hline
        Type-2 & 0.019& 0.031 & 0.230 \\ \hline
        Type-3 & 0.089& 0.410 & 0.096  \\ \hline
        Type-4 & 0.001& 0.069 & 0.015 \\ \hline
        CI for $w_{n}$ & (0.968,0.989) & {(0.792, 0.943)} & (0.728, 0.780)\\ \hline
        Adjusted Baseline for $w_n$ & 0.529 & {0.502}& 0.587\\ \hline
    \end{tabular}
\caption{Characteristics of real-world signed networks, including the number of nodes, the proportion of edges, the proportion of negative signs, the proportion of type-$t$ triangles for $1 \le t \le 4$, 95\% confidence intervals for $w_n$, and the adjusted balance-free baseline for $w_n$.}
\label{ppi and social networks}
\end{table}

\subsection{Social Networks}
Finally, we apply our method to two social networks where balance theory originates: a small-scale U.S.\ Senate network~\citep{fowler2006legislative} and a large-scale Wikipedia Elections network~\citep{leskovec2010signed}. 
The first signed network among U.S.\ senators is constructed using Fowler’s Senate bill co-sponsorship data~\citep{fowler2006legislative}. A positive edge is assigned between two senators if they co-sponsored bills significantly more frequently than expected, suggesting a political alliance. Conversely, a negative edge is assigned if they co-sponsored bills significantly less frequently than expected. Detailed information on the network construction is provided in the Supplemental Materials. This signed network consists of 100 senators as nodes, with 1,164 positive edges and 835 negative edges. The second large-scale signed network among Wikipedia users is based on their votes for or against the promotion of other users to administrator status~\citep{leskovec2010signed}. We analyze an undirected version of the dataset, which includes 7,115 Wikipedia users, with 78,440 positive edges and 22,253 negative edges.

We apply the inference method to these two social networks, and our results in Table~\ref{ppi and social networks} show strong evidence of balance in both cases. The confidence intervals for the expected proportion of balanced triangles are all above the balance-free baselines. In particular, as shown in Figure~\ref{fig:senate}, the U.S.\ Senate network consists of two major clusters, which mostly align with the Democratic and Republican parties. Within each cluster, most connections are positive; in contrast,  the connections between the two clusters are mostly negative. This may be attributed to that senators within each cluster co-sponsor bills proposed by their  members and in most cases do not support bills from the other cluster. This partition structure aligns with the concept of balance theory.

\section{Discussions}
While our Edgeworth expansion provides refined distributional approximations for inference on the balance parameter, its practical utility for hypothesis testing depends critically on the choice of a scientifically meaningful balance-free null model. For instance, the ``Adjusted baseline'' null used in Section~\ref{sec: real_world_example} only matches the overall proportion of negative edges but does not account for other structural properties of the observed network, such as degree heterogeneity or the dependence of edge signs on local structure. Incorporating other conditional nulls~\citep{feng2022testing} would require characterizing the conditional distribution of balance statistics given observed network features, which would involve substantially different technical tools and raise additional challenges for power analysis. 
Developing inference procedures under such conditional null models is therefore an important direction for future work.

Our empirical findings also highlight the importance of sign information and the necessity of incorporating balance into statistical modeling for future work. Notably, in the protein-protein interaction network, we observed a high proportion of balanced triangles. We conjecture that this observation may be common in network data constructed by computing pairwise correlations from nodal measurements, such as neuron-neuron and gene-gene interactions \citep{zalesky2012connectivity,xia2013brainnet,greene2015understanding, wang2018network}. Exploring the application of balance theory for more efficient statistical analysis in these contexts is an intriguing direction for future work. 

Finally, this work focuses on inference for the degree of balance within time-static signed networks. More broadly, however, balance theory is often interpreted dynamically: systems tend to avoid conflict and evolve toward more balanced configurations.
This perspective suggests developing inference tools for balance in the context of dynamic signed networks as another natural direction for future research.

\noindent\textbf{Code availability.} Code for this work is publicly available at \url{https://github.com/weijtang/Nonparametric-Inference-for-Balance}.

\noindent \textbf{Acknowledgements.}
The authors thank Yuan Zhang for helpful and insightful discussions. Tang was partially supported by NSF Grant DMS-2412853 and DMS-2310668.

\bibliographystyle{plainnat}
\bibliography{reference}

\newpage
\appendix

\setcounter{equation}{0}
\setcounter{figure}{0}
\setcounter{table}{0}
\setcounter{section}{0}
\setcounter{algocf}{0}
\setcounter{page}{1}

\allowdisplaybreaks

\begin{center}
{\Large Supplemental Material for ``Nonparametric Inference for Balance in Signed Networks''}
\end{center}

\addtocontents{toc}{\protect\setcounter{tocdepth}{3}}

\tableofcontents
\newpage

\section{Proof of Proposition~\ref{prop: exchangeability}}

\begin{proof}[ of Proposition~\ref{prop: exchangeability}]
By Aldous-Hoover Theorem \citep{aldous1981representations, hoover1979relations}, there exists a measurable function $M:[0,1]^3 \rightarrow \{-1,0,1\}$ such that 
$$[A_{ij}]_{i,j \in \mathbb{N}} \buildrel d \over = [M(X_{i}, X_{j}, Y_{ij})]_{i,j \in \mathbb{N}},$$
where $(X_i)_{i \in \mathbb{N}}$ and $(Y_{ij})_{i,j \in \mathbb{N}}$ are independent random variables, each uniformly distributed on $[0,1]$. 
Let $\mathbb{I}\{\cdot\}$ be the indicator function of the event, we have
\begin{align*}
\pr(A_{ij}=1 \mid X_i, X_j)=&\E(\mathbb{I}\{ M(X_i, X_j, Y_{ij})=1 \} \mid X_i, X_j)\\
=&\int_{0}^{1} \mathbb{I}\{M(X_i, X_j, u)=1 \}du\\
=&\int_{0}^{1} \frac{1}{2}M(X_i, X_j, u) \left[M(X_i, X_j, u)+1 \right]du
\end{align*}
and
$$\pr(A_{ij}=-1 \mid X_i, X_j)=\int_{0}^{1} \frac{1}{2}M(X_i, X_j, u) \left[M(X_i, X_j, u)-1 \right]du.$$
Thus $$\pr(A_{ij}\neq 0 \mid X_i, X_j)=\int_{0}^{1} M^2(X_i, X_j, u) du.$$
We define
$$\widetilde{F}(x, y)=\int_{0}^{1} M^2(x, y, u) du\quad\text{and}\quad \widetilde{G}(x, y)=\frac{ \int_{0}^{1} \frac{1}{2}M(x, y, u) \left[M(x,y, u)-1 \right]du }{\int_{0}^{1} M^2(x, y, u) du}.$$
The distribution of $[A_{ij}]_{i,j \in \mathbb{N}}$ can be characterized as
$$[A_{ij}]_{i,j \in \mathbb{N}} \buildrel d \over = [\mathbb{I}\{Y_{ij}<\widetilde{F}(X_i, X_j)\}(-1)^{\mathbb{I}\{Y_{ij}>\widetilde{F}(X_i, X_j)\widetilde{G}(X_i,X_j)\}}]_{i,j \in \mathbb{N}},$$
which completes the proof.
\end{proof}

\section{Proof of Equal Length of Confidence Intervals}
\label{sec: proof of CI length}

The difference between the normal-approximation confidence interval \(\text{CI}^{normal}\) and the Cornish-Fisher confidence interval \(\text{CI}^{our}\) arises from a shift in the approximate distribution function, which does not affect interval length. Specifically, recall that
$$\text{CI}^{normal}=
(\widehat{U}_n/\widehat{V}_n-z_{1-\alpha/2}\widehat{S}_n , \ \widehat{U}_n/\widehat{V}_n-z_{\alpha/2}\widehat{S}_n)
$$
and
$$\text{CI}^{our}=
(\widehat{U}_n/\widehat{V}_n-\widehat{q}_{\widehat{T}_{n;1-\alpha/2}}\widehat{S}_n , \ \widehat{U}_n/\widehat{V}_n-\widehat{q}_{\widehat{T}_{n;\alpha/2}}\widehat{S}_n),
$$
where
$$
\widehat{q}_{\widehat{T}_{n;\alpha}}=z_{\alpha}-\{\widehat{a}(z_{\alpha}^2/3+1/6)+\widehat{b}(z_{\alpha}^2+1){-}3\widehat{c}z_{\alpha}^2\}/\sqrt{n} - \delta_T, \quad z_{\alpha}:=\Phi^{-1}(\alpha),
$$
and $\widehat{a}$, $\widehat{b}$ and $\widehat{c}$ are given by (\ref{defineabchat}). Here, $z_{\alpha}$ is the $\alpha$-th quantile of the standard normal distribution. 
The length of our confidence interval is then
$$
\text{length}(\text{CI}^{\text{our}})
= \Big( \widehat{q}_{\widehat{T}_{n;1-\alpha/2}} - \widehat{q}_{\widehat{T}_{n;\alpha/2}} \Big) \widehat{S}_n.
$$
Note that for the standard normal distribution, $z_{1-\alpha/2} = -z_{\alpha/2}$, which implies $z_{1-\alpha/2}^2 = z_{\alpha/2}^2$.
Thus, the squared terms and constants in the expansions for $\widehat{q}_{\widehat{T}_{n;1-\alpha/2}}$ and $\widehat{q}_{\widehat{T}_{n;\alpha/2}}$ are the same, and the higher-order correction terms cancel out in their difference:
$$
\widehat{q}_{\widehat{T}_{n;1-\alpha/2}} - \widehat{q}_{\widehat{T}_{n;\alpha/2}}
= z_{1-\alpha/2} - z_{\alpha/2}.
$$
Therefore,
$
\text{length}(\text{CI}^{\text{our}})
= (z_{1-\alpha/2} - z_{\alpha/2}) \widehat{S}_n
= \text{length}(\text{CI}^{\text{normal}}).
$
Compared to the standard normal approximation, our method achieves a high-order accuracy in coverage probability.

\section{Proof Sketch of the Main Lemma and Theorems}
\label{sec: proof of lemma sketch}
In this section, we provide the proof sketch of the main lemma (Lemma~\ref{lemma}) stated later in Section~\ref{sec: proof of lemma}, as well as of Theorem~\ref{thm1}. Lemma~\ref{lemma} paves the way for proving Theorem~\ref{thm1}. The high-level proof strategy is organized into the following steps and we highlight the connection and difference with~\citet{DBLP:journals/corr/abs-2004-06615} in each step:

\medskip
\noindent\textbf{Decomposition of the studentized statistic.}
We express the studentized statistic $\widehat{T}_n$ as a product of a normalized ``signal” term and a variance-estimation correction term: 
\begin{align}
\widehat{T}_n=&\frac{\widehat{U}_n/\widehat{V}_n-u_n/v_n}{\widehat{S}_n}=\frac{\widehat{U}_n/\widehat{V}_n-u_n/v_n}{\sigma_n}\cdot\frac{\sigma_n}{\widehat{S}_n}\nonumber\\
=&\frac{\widehat{U}_n/\widehat{V}_n-u_n/v_n}{\sigma_n}\cdot(1+\frac{\widehat{S}_n^2-\widehat{\sigma}_n^2}{\sigma_n^2}+\frac{\widehat{\sigma}_n^2-\sigma_n^2}{\sigma_n^2})^{-1/2}\label{eq:taylor},
\end{align}
where $\widehat{\sigma}_n^2$ defined in~(\ref{lemma13defineinterm}) is an intermediate term between $\widehat{S}_n^2$ and $\sigma_n^2$. In particular, the leading term $\sigma_n^2$ of variance, together with the design of the variance estimator $\widehat{S}_n^2$ is obtained by analyzing the behavior of $\widehat{U}_n/\widehat{V}_n-u_n/v_n$. 

\medskip
\noindent\textbf{Expansion of the ratio estimator.}
The analysis begins with a second-order Taylor expansion of $f(x,y)=x/y$ around $(u_n,v_n)$, combined with Hoeffding decompositions for the noiseless U-statistics $(U_n,V_n)$ and linearizations of the observation noise $(\widehat{U}_n-U_n,\widehat{V}_n-V_n)$. 
Because Taylor expansion introduces additional cross-terms, analyzing the leading term in variance decomposition is more involved than~\citet{DBLP:journals/corr/abs-2004-06615}.
It turns out that the leading term is given by
\[\operatorname{Var}(\widehat{U}_n/\widehat{V}_n-u_n/v_n)=\sigma_n^2+o(n^{-1}), \quad \text{with }\sigma_n\asymp n^{-1/2}.\] 
Moreover, this expansion produces an explicit decomposition of $(\widehat{U}_n/\widehat{V}_n-u_n/v_n)/\sigma_n$ into bounded canonical U-statistic components plus a key edge-noise term $\check{\Delta}_n$.

\medskip
\noindent\textbf{Asymptotic normality of the edge-noise term.}
Conditional on the latent variables $W=(X_1, \cdots, X_n)$, we will establish the asymptotic normality of the edge-noise term $\check{\Delta}_n$ using a Berry–Esseen bound, after showing that \[\operatorname{Var}(\check{\Delta}_n\mid W)\asymp n^{-1}\rho_n^{-1}s_n.\] 
Obtaining a nondegenerate lower bound for this conditional variance is nontrivial in our setting. The presence of additional randomness in edge signs, together with cross terms arising from the Taylor expansion, can lead to cancellations and potential degeneracy. Conditions (a) and (c) play a crucial technical role in ruling out this behavior. The asymptotic normality of $\check{\Delta}_n$ is a key ingredient for the subsequent Edgeworth expansion.

\medskip
\noindent\textbf{Control estimation error in $\widehat{S}_n^2$.}
To justify replacing $\widehat{S}_n$ by $\sigma_n$, we control two sources of variance discrepancy: $(\widehat{S}_n^2-\widehat{\sigma}_n^2)/\sigma_n^2$ and $(\widehat{\sigma}_n^2-\sigma_n^2)/\sigma_n^2$. 
The randomness in the former arises from observation error. Although we similarly apply the Schudy–Sviridenko polynomial concentration inequality (Lemma~\ref{hyperth}) to bound $(\widehat{S}_n^2-\widehat{\sigma}_n^2)/\sigma_n^2$, the analysis is technically more delicate than in unsigned settings: the additional randomness induced by edge signs complicates the underlying hypergraph structure and requires a more careful application of the concentration argument. 
To bound $(\widehat{\sigma}_n^2-\sigma_n^2)/\sigma_n^2$, whose randomness comes from the latent variables $W$, we expand the difference into tractable sample averages and bound each term accordingly. 
Taken together, these results control the studentization error and complete the proof of Lemma~\ref{lemma}.

\medskip
\noindent\textbf{Proof strategy for Theorem~\ref{thm1}.}
Theorem~\ref{thm1} establishes an upper bound on the Kolmogorov--Smirnov distance between the distribution of $\widehat{T}_n+\delta_T$ and the population Edgeworth expansion $\mathcal{G}_n$, i.e.,
\[
\|\mathcal{F}_{\widehat{T}_n+\delta_T}-\mathcal{G}_n\|_\infty.
\]
Once we have obtained the key Lemma~\ref{lemma}, we follow the same strategy as in~\citet{DBLP:journals/corr/abs-2004-06615} to bound this distance.
By applying a second-order Taylor expansion to the second term in Equation~\ref{eq:taylor} and using Lemma~\ref{lemma}, we can decompose $\widehat{T}_n$ into three components:
(i) a structured term $\widetilde{T}_n$ consisting of centering, linear, and quadratic terms from the expansion;
(ii) the edge-noise term $\check{\Delta}_n$; and
(iii) a remainder term.

By Lemma~\ref{lemma}~(3), the edge-noise term $\check{\Delta}_n$ can be approximated by an asymptotically normal variable $\widetilde{\Delta}_n$, which yields the bound
\[
\|\mathcal{F}_{\widetilde{T}_n+\check{\Delta}_n+\delta_T}
-\mathcal{F}_{\widetilde{T}_n+\widetilde{\Delta}_n+\delta_T}\|_\infty.
\]
Combining $\widetilde{\Delta}_n$ with $\widetilde{T}_n$ leads to the population Edgeworth expansion $\mathcal{G}_n$ and the bound
\[
\|\mathcal{F}_{\widetilde{T}_n+\widetilde{\Delta}_n+\delta_T}
-\mathcal{G}_n\|_\infty.
\]
Finally, the remainder term controls
\[
\|\mathcal{F}_{\widehat{T}_n+\delta_T}
-\mathcal{F}_{\widetilde{T}_n+\check{\Delta}_n+\delta_T}\|_\infty.
\]
Applying the triangle inequality to these three bounds yields the claimed result. 
See Lemma~\ref{lemma.CDF} for explicit upper bounds on each term.

\medskip
\noindent\textbf{Comparison with related work.}
We summarize the main differences between our analysis and that of~\citep{DBLP:journals/corr/abs-2004-06615}.
First, our target statistic involves the ratio of two sample network moments $\widehat{U}_n/\widehat{V}_n-u_n/v_n$, which introduces additional cross-terms through Taylor expansion and complicates the derivation of lower bounds for $\check{\Delta}_n$.
Second, the signed-network setting introduces extra randomness that makes variance estimation more delicate; for instance, bounding $(\widehat{S}_n^2-\widehat{\sigma}_n^2)/\sigma_n^2$ requires a more refined hypergraph-based argument to apply the Schudy--Sviridenko polynomial concentration inequality (Lemma~\ref{hyperth}).

\section{Proof of Proposition~\ref{proposition:variance} and the Main Lemma}
\label{sec: proof of lemma}
\subsection{The Main Lemma}

This section states the key lemma referenced in Section~\ref{sec: proof of lemma sketch} and provides its proof. The variance characterization in Proposition~\ref{proposition:variance} is incorporated as Part~(1) of this lemma (Lemma~\ref{lemma}).

\smallskip
\noindent \textbf{Notation.} We say $X_{n}^{(i)}=\wtO_{p,k}(a_n)$ holds uniformly for $1 \le i \le n$ if there exist universal constants $C_1, C_2,$ and $ N>0$ such that $\pr(|X_{n}^{(i)}|\ge C_1\,a_n)\le C_2n^{-k}$ for $1 \le i \le n$ and $n\ge N$.
We define \[M(n, s_n, \rho_n) = n^{-1}\,\rho_n^{-\frac{3}{2}}\,\log n ( 1+ s_n \log n).\]

\vspace{2mm}
\begin{lemma}\label{lemma} Assume: 
\begin{itemize}
    \item[(a)]  Either
\begin{itemize}
    \item[(a1)] $F(\cdot, \cdot)$, $G(\cdot, \cdot)>C_1$ for some constant $C_1>0$; or
    \item[(a2)] $\|F\|_{\infty}$ and $\|G\|_{\infty}$ are bounded, and\[\mathbb{E}\left[G(X_1,X_2)F(X_1,X_2)F(X_3,X_1)F(X_3,X_2)F(X_4,X_1)F(X_4,X_2)\right]>0;\]
\end{itemize}
    \item[(b)] (Non-degeneracy) $\xi_1 = \sqrt{n} \sigma_n / 3 > C_2$ for some constant $C_2>0$;
    \item[(c)] Either
\begin{itemize}
    \item[(c1)] (Vanishing negative proportion) $s_n \prec 1$, $\rho_n^3 s_n^2 \succeq n^{-2}\log^{8}n$, and $\rho_n \succ n^{-1}s_n$; or
    \item[(c2)] (Constant negative proportion) $s_n \asymp 1$ with $\|s_n G\|_\infty<1/2$, and $\rho_n^3 \succeq n^{-2}\log^{8}n$.
\end{itemize}
\end{itemize}
Then we have the following results:
\begin{enumerate}[label={(\arabic*)}]
\item $\var(\widehat{U}_n/\widehat{V}_n-u_n/v_n)=\sigma_n^2+o(n^{-1})$ with $\sigma_n \asymp n^{-1/2}$.
\item The decomposition for $(\widehat{U}_n/\widehat{V}_n-u_{n}/v_{n})/\sigma_n$ is
\begin{align*}
&-\frac{3}{\sqrt{n}\xi_1}\E\Big[p_1(X_1)q_1(X_1)\Big] +\frac{1}{\sqrt{n}\xi_1}\sum_{i=1}^n q_1(X_i)+\frac{1}{n^{3/2}\xi_1}\sum_{1 \le i \neq j \le n}q_2(X_i,X_j)\\
	&  -\frac{3}{n^{3/2}\xi_1}\sum_{1\le {i\neq j}\le n}p_1(X_i)q_1(X_j) +\check{\Delta}_n + \wtO_{p,2}(n^{-1}\log^{\frac{3}{2}} n + M(n, s_n, \rho_n)),
\end{align*}
where $\check{\Delta}_n$ is defined in~(\ref{eq: delta check def}). In addition, the first four terms are in the order of $\wtO_{p,2}(\log^{\frac{1}{2}}n)$ and $\check{\Delta}_n=\wtO_{p,2}(\rho_n^{-\frac{1}{2}}\,n^{-\frac{1}{2}}\,\log^{\frac{1}{2}}n)$.
\item Conditional on $W=(X_1,\cdots,X_n)$, the random variable $\check{\Delta}_n$ is asymptotically normal as follows,
\begin{equation}\label{lemma:distance}
\Vert \mathcal{F}_{\check{\Delta}_n|W}(u)-\mathcal{F}_{N(0,n^{-1}\,\rho_n^{-1}\,s_n\,\sigma_W^2)}(u)\Vert _{\infty}=\widetilde{O}_{p, k}(n^{-1}\,\rho_n^{-\frac{1}{2}}\,s_n^{-\frac{1}{2}}),
\end{equation}
for any $k$, where $\sigma_W \overset{\mathrm{w.h.p.}}{\asymp} 1$ is defined in~(\ref{equ:sigmaw}).
\vspace{1mm}
\item $(\widehat{S}_n^2-\widehat{\sigma}_n^2)/\sigma^2_n=\wtO_{p,1}({M(n,s_n,\rho_n)})$, where $\widehat{\sigma}_n^2$ is defined in~(\ref{lemma13defineinterm}).

\item We have the decomposition for $(\widehat{\sigma}^2_n-\sigma^2_n)/\sigma^2_n$:
\begin{align*}
\frac{\widehat{\sigma}^2_n-\sigma^2_n}{\sigma^2_n}&=\frac{1}{n\xi_{1}^2}\sum_{i=1}^{n}(q_{1}^2(X_{i})-\xi_{1}^2)+\frac{4}{n\xi_{1}^2}\sum_{i=1}^{n}g_{\delta,1}(X_i)\\
&\ \ \ -\frac{6}{n}\sum_{i=1}^n p_1(X_i)-\frac{6}{n\xi_1^2}\E[p_1(X_1)q_1(X_1)]\sum_{i=1}^n q_1(X_i) +\wtO_{p,1}(n^{-1}\,\log\,n).
\end{align*}
where $g_{\delta,1}(x)=\E\Big[q_1(X_2)q_2(X_1,X_2)\vert X_1=x\Big].$
\end{enumerate}
\end{lemma}

\begin{remark}
Condition (a1) is Condition (a) in Proposition~\ref{proposition:variance}. It ensures the probabilities of observing an edge and a negative sign are exactly of the order $\rho_n$ and $s_n$, respectively, which in turn yields a uniform lower bound on the conditional variance $\mathrm{Var}(\check{\Delta}_n \mid W)$ and prevents degeneracy of the studentized statistic. We note that this uniform boundedness away from zero is stronger than strictly necessary and is imposed for proof simplicity rather than being intrinsic to balance inference itself. We introduce Condition (a2) as a weaker alternative condition, which can be equivalently rewritten as 
\[\pr\left(h_{\Delta}\left(A_{1,2,3}\right) h_{\Delta}\left(A_{1,2,4}\right)=1, A_{12}=-1\right)>0.\]
\begin{figure}[!htbp]
  \centering
  \includegraphics[width=0.3\textwidth]{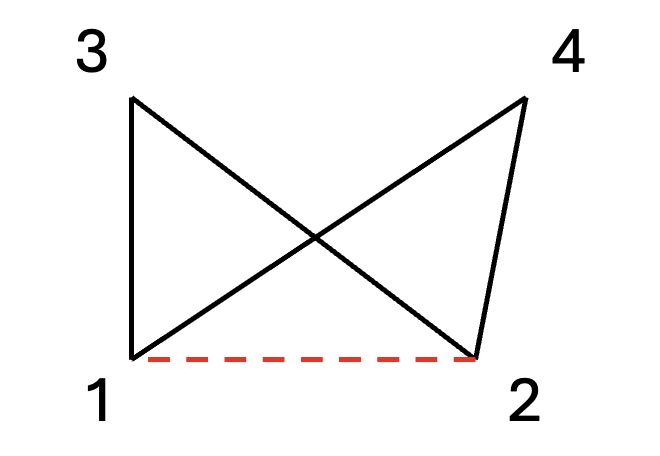} 
  \caption{Illustration of a signed motif whose probability is required to be nonzero under Condition (a2). Black solid edges indicate the presence of an edge regardless of sign, and the red dashed edge indicate a negative edge.}
  \label{example2}
\end{figure}
It requires that the probabilities of observing the signed motifs shown in Figure \ref{example2} are nonzero, which prevents triangles with negative edges from being isolated. Under this weaker condition, the leading order of $\mathrm{Var}(\check{\Delta}_n \mid W)$ remains lower bounded by $n^{-1}\rho_n^{-1}s_n$, but now with high probability $1-\exp(-c n)$, rather than almost surely as under Condition (a1).
\end{remark}
\vspace{2mm}

We organize the proof of Lemma~\ref{lemma} as follows. Section~\ref{sec:appendix-notation} introduces the notation and preliminary results. Sections~\ref{sec:appendix-Lemma1}–\ref{sec:appendix-Lemma5} then establish Parts~(1)–(5) of Lemma~\ref{lemma}, respectively.

\subsection{Notation and Preliminary Lemmas}
\label{sec:appendix-notation}
Before proceeding to the proof of Lemma~\ref{lemma}, we introduce some notation that will be used throughout the proof. 

Under our graphon model for signed networks (Definition~\ref{def: graphon}), conditional on the latent variables $\{X_i\}_{i=1}^n$, the edge $A_{ij}$ independently follows the distribution as follows:
$$
A_{ij}=
\begin{cases}
0,& \text{with probability $a_{ij}=1-\rho_n\,F(X_{i},X_{j})$};\\
-1,& \text{with probability 
$b_{ij}=\rho_n\,s_n\,F(X_{i},X_{j})\,G(X_{i},X_{j})$};\\
1,& \text{with probability $c_{ij}=\rho_n\,F(X_{i},X_{j})\,(1-s_n\,G(X_{i},X_{j}))$}.
\end{cases}
$$
We define
$
\eta_1(A_{ij})=|A_{ij}|-b_{ij}-c_{ij},
$
$
\eta_2(A_{ij})=A_{ij}(A_{ij}-1)/2-b_{ij}
$
and
$
\eta_3(A_{ij})=A_{ij}(1+A_{ij})/2-c_{ij}.
$
The triangle indicator functions $h_1$, $h_3$ and $h_{\Delta}$ can be written as
\begin{align}
    h_1(A_{i_1,i_2,i_3}) &=\prod_{\{i,j\}\subset \{i_1,i_2,i_3\}} [c_{ij}+\eta_3(A_{ij})], \nonumber\\
    h_{\triangle}(A_{i_1,i_2,i_3}) &=\prod_{\{i,j\}\subset \{i_1,i_2,i_3\}} [b_{ij}+c_{ij}+\eta_1(A_{ij})], \nonumber \\
    \text{and } \quad \quad \quad \quad \quad \quad h_3(A_{i_1,i_2,i_3}) &=\sum_{\{i,j\}\subset \{i_1,i_2,i_3\}} [c_{ij}+\eta_3(A_{ij})]\prod_{\substack{k\in \{i_1,i_2,i_3\}/\{i,j\}\\l\in\{i,j\} }} [b_{kl}+\eta_2(A_{kl})]. \label{h13}
\end{align}
We denote $h_t(X_{i_1},X_{i_2},X_{i_3})=\E[h_t(A_{i_1,i_2,i_3})\mid X_{i_1},X_{i_2},X_{i_3}]$ for $t=1, 2, 3, 4$, $\blacktriangle$ and $\triangle$. 
The conditional expectation of $h_1(A_{i_1,i_2,i_3})$ can be bounded through $\E[h_1(A_{i_1,i_2,i_3})\mid X_{i_1},X_{i_2},X_{i_3}]=c_{i_1 i_2}c_{i_1 i_3}c_{i_2 i_3}\preceq \rho^3_n$. Similarly, we have $\E[h_3(A_{i_1,i_2,i_3})\mid X_{i_1},X_{i_2},X_{i_3}]\preceq s_n^2 \rho^3_n$ and $\E[h_{\Delta}(A_{i_1,i_2,i_3})\mid X_{i_1},X_{i_2},X_{i_3}]\preceq \rho^3_n$. 

We summarize the properties of $\eta_\ell(A_{ij})$ for $1\le \ell \le 3$ that will be used frequently in Lemma~\ref{lemma: h property}.
\vspace{2mm}
\begin{lemma}\label{lemma: h property} Let $W=(X_1, \cdots, X_n)$, we have the following properties: 
\begin{enumerate}
    \item $\eta_1(A_{ij})$, $\eta_2(A_{ij})$ and $\eta_3(A_{ij})$ are all zero-mean and independent across different tuples $(i, j)$;
    \item we have \begin{align*}
    \var(\eta_1(A_{ij})\mid W) & = a_{ij} (b_{ij} + c_{ij}) \quad \text{and} \quad \E[|\eta_1(A_{ij})|^3 \mid W]  = a_{ij}(b_{ij}+c_{ij})[(b_{ij}+c_{ij})^2 + a_{ij}],\\
    \var(\eta_2(A_{ij}) \mid W) & = b_{ij}(a_{ij} + c_{ij}) \quad \text{and} \quad  \E[|\eta_2(A_{ij})|^3 \mid W]  = b_{ij}(a_{ij}+c_{ij})[(a_{ij}+c_{ij})^2 + b_{ij}^2],\\
    \var(\eta_3(A_{ij}) \mid W) & = c_{ij}(a_{ij}+ b_{ij}) \quad \text{and} \quad \E[|\eta_3(A_{ij})|^3 \mid W]  = c_{ij}(a_{ij}+b_{ij})[(a_{ij}+b_{ij})^2 + c_{ij}^2]. 
\end{align*}
\end{enumerate}
\end{lemma}
\vspace{2mm}
\begin{proof}[of Lemma~\ref{lemma: h property}]
Note that $A_{ij}$ are independent for different tuples $(i, j)$. Therefore $\eta_1(A_{ij})$, $\eta_2(A_{ij})$ and $\eta_3(A_{ij})$ are independent across different tuples $(i, j)$. 

By straightforward calculations,  we have,  for $\eta_1(A_{ij})$, $\var(\eta_1(A_{ij})) = a_{ij}(b_{ij} + c_{ij})^2 + b_{ij}(1-b_{ij} - c_{ij})^2 + c_{ij}(1-b_{ij} - c_{ij})^2 = a_{ij} (b_{ij} + c_{ij})$. Similarly, $\E[|\eta_1(A_{ij})|^3] = a_{ij}(b_{ij} + c_{ij})^3 + b_{ij}(1-b_{ij}-c_{ij})^3 + c_{ij}(1-b_{ij}-c_{ij})^2 = a_{ij}(b_{ij} + c_{ij})[(b_{ij} + c_{ij})^2 + a_{ij}^2]$. 

For $\eta_2(A_{ij})$, we have $\var(\eta_2(A_{ij})) = a_{ij}b^2_{ij} + b_{ij}(1-b_{ij})^2 + c_{ij} b_{ij}^2 = b_{ij}(a_{ij} + c_{ij})$ and $\E[|\eta_2(A_{ij})|^3] = b_{ij} (a_{ij} + c_{ij}) [(a_{ij}+ c_{ij})^2 + b_{ij}^2]$. 

For $\eta_3(A_{ij})$, we have $\var(\eta_3(A_{ij})) = a_{ij} c_{ij}^2 + b_{ij} c_{ij}^2 + c_{ij} (1-c_{ij})^2 = c_{ij}(a_{ij} + b_{ij})$ and $\E[|\eta_3(A_{ij})|^3] = c_{ij}(a_{ij} + b_{ij})[(a_{ij} + b_{ij})^2 + c^2_{ij}]$.
\end{proof}

\medskip
To prove Lemma~\ref{lemma}, we need the following characterization for the noiseless U-statistics and the edgewise observational errors conditional on $W=(X_1, \cdots, X_n)$. We postpone the lengthy proof to Section~\ref{subsec: proof U and V decomposition}.
\vspace{2mm}

\begin{lemma} \label{lemma: U and V decomposition}
	Assuming the conditions in Lemma~\ref{lemma},  we have the following decompositions:
\begin{enumerate}[label=(\arabic*)]
	\item For noiseless U-statistics, we have 
	\begin{align}
		U_n-u_n= \underbrace{\frac{3}{n} \sum_{i=1}^n g_1\left(X_i\right)}_{\mathcal{U}_n^{(1)}}+\underbrace{\frac{6}{n(n-1)} \sum_{1 \leq i<j \leq n} g_2\left(X_i, X_j\right)}_{\mathcal{U}_n^{(2)}}+\wtO_{p,2}\left(\rho_n^3 n^{-\frac{3}{2}} \log ^{\frac{3}{2}} n\right), \label{eq: Un decom}
	\end{align}
	where $\mathcal{U}_n^{(1)} = \wtO_{p,2}(\rho_n^3 n^{-\frac{1}{2}} \log ^{\frac{1}{2}} n)$ and $\mathcal{U}_n^{(2)} = \wtO_{p,2}(\rho_n^3 n^{-1} \log n)$. Similarly, for $V_n-v_n$, we have
	\begin{align}
		V_n-v_n = \underbrace{\frac{3}{n} \sum_{i=1}^n f_1\left(X_i\right)}_{\mathcal{V}_n^{(1)}}+\underbrace{\frac{6}{n(n-1)} \sum_{1 \leq i<j \leq n} f_2\left(X_i, X_j\right)}_{\mathcal{V}_n^{(2)}}+\wtO_{p,2}\left(\rho_n^3 n^{-\frac{3}{2}} \log ^{\frac{3}{2}} n\right), \label{eq: Vn decom}
	\end{align}
	where $\mathcal{V}_n^{(1)} = \wtO_{p,2}(\rho_n^3 n^{-\frac{1}{2}} \log ^{\frac{1}{2}} n)$ and $\mathcal{V}_n^{(2)} = \wtO_{p,2}(\rho_n^3 n^{-1} \log n)$.	
	\item Conditional on $W=(X_1, \cdots, X_n)$, we have 
	\begin{align}
		\widehat{U}_n - U_n = &   \underbrace{\frac{1}{\binom{n}{2}} \sum_{1 \leq k_1<k_2 \leq n} \mathring{B}_{k_1 k_2} \eta_2\left(A_{k_1 k_2}\right)}_{B_n} + \underbrace{\frac{1}{\binom{n}{2}} \sum_{1 \leq k_1<k_2 \leq n} \mathring{C}_{k_1 k_2} \eta_3\left(A_{k_1 k_2}\right)}_{C_n} \nonumber \\
		&  + \wtO_{p,2}(n^{-3/2} \rho_n^{3/2} \log^{1/2} n(1 + s_n \log^{3/2}n)), \label{eq: Uhat_n decom}
	\end{align}
	where   $\mathring{B}_{k_1 k_2}=\frac{3}{n-2} \sum_{\substack{1 \leq i \leq n \\ i \neq k_1, k_2}}\left(c_{i k_1} b_{i k_2}+b_{i k_1} c_{i k_2}\right)$ and $\mathring{C}_{k_1 k_2}=\frac{3}{n-2}  \sum_{\substack{1 \leq i \leq n \\ i \neq k_1, k_2}} (b_{i k_1} b_{i k_2}+c_{i k_1} c_{i k_2})$. In addition, the leading terms have orders
	\[B_n = \wtO_{p,2}(\rho_n^{\frac{5}{2}} s_n^{\frac{3}{2}}n^{-1} \log^{\frac{1}{2}} n)\quad \text{ and } \quad C_n = \wtO_{p,2}( \rho_n^{\frac{5}{2}} n^{-1} \log^{\frac{1}{2}} n).\]
	Similarly, for $\widehat{V}_n - V_n$, we have conditional on $W$
	\begin{align}
		\widehat{V}_n - V_n = &  \underbrace{\frac{1}{\binom{n}{2}} \sum_{1 \leq k_1<k_2 \leq n} \mathring{D}_{k_1 k_2} \eta_1\left(A_{k_1 k_2}\right)}_{D_n} + \wtO_{p,2}(\rho_n^{\frac{3}{2}}n^{-\frac{3}{2}}   \log^{\frac{1}{2}} n),\label{eq: Vhat_n decom}
	\end{align}
	where $\mathring{D}_{k_1 k_2}=\dfrac{3}{n-2} \sum_{\substack{1 \leq i \leq n \\ i \neq k_1, k_2}} (b_{i k_1}+c_{i k_1})(b_{i k_2}+c_{i k_2})$, and $D_n = \wtO_{p,2}( \rho_n^{\frac{5}{2}} n^{-1} \log^{\frac{1}{2}} n)$. 
\end{enumerate}
\end{lemma}

\subsection{Proof of Lemma~\ref{lemma}~(1)}
\label{sec:appendix-Lemma1}
This part of the lemma characterize the variance of $\widehat{U}_n/\widehat{V}_n-u_n/v_n$. We apply a second-order Taylor expansion of the function $f(x,y)=\dfrac{x}{y}$ at at point $(u_{n},v_{n})$.

\begin{align*}
	\dfrac{\widehat{U}_{n}}{\widehat{V}_{n}}-\dfrac{u_{n}}{v_{n}} = &\frac{1}{v_n}(\widehat{U}_{n}-u_{n})-\frac{u_n}{v^2_n}(\widehat{V}_{n}-v_{n})-\frac{1}{v^2_n}(\widehat{U}_{n}-u_{n})(\widehat{V}_{n}-v_{n})+\frac{u_n}{v^3_n}(\widehat{V}_{n}-v_{n})^2\\
& + O(\frac{1}{v_n^3}(\widehat{U}_{n}-u_{n})(\widehat{V}_{n}-v_{n})^2+ \frac{u_n}{v_n^4}(\widehat{V}_{n}-v_{n})^3).
\end{align*}
Combine (\ref{eq: Un decom})-(\ref{eq: Vhat_n decom}) in Lemma~\ref{lemma: U and V decomposition}, we have
\begin{align*}
	\widehat{U}_{n}-u_{n}=&\mathcal{U}_n^{(1)}+ \mathcal{U}_n^{(2)}+B_n +C_n + \wtO_{p,2}(\rho_n^{\frac{3}{2}}n^{-\frac{3}{2}}   \log^{\frac{1}{2}} n(\rho_n^{3/2}\log n+  1 + s_n \log^{\frac{3}{2}}n))\\
	 =&\wtO_{p,2}(\rho_n^{3}n^{-\frac{1}{2}}   \log^{\frac{1}{2}} n)
\end{align*}
and
\begin{align*}
	\widehat{V}_{n}-v_{n}=&\mathcal{V}_n^{(1)}+ \mathcal{V}_n^{(2)}+D_n + \wtO_{p,2}(\rho_n^{\frac{3}{2}}n^{-\frac{3}{2}}   \log^{\frac{1}{2}} n(\rho_n^{3/2}\log n+1))\\
	=&\wtO_{p,2}(\rho_n^{3}n^{-\frac{1}{2}}   \log^{\frac{1}{2}} n).
\end{align*}
Plugging them back into the Taylor expansion, we get
\begin{align}
\dfrac{\widehat{U}_{n}}{\widehat{V}_{n}}-\dfrac{u_{n}}{v_{n}} = &\frac{1}{v_n}(\mathcal{U}_n^{(1)}+ \mathcal{U}_n^{(2)}+B_n +C_n)-\frac{u_n}{v^2_n}(\mathcal{V}_n^{(1)}+ \mathcal{V}_n^{(2)}+D_n) -\frac{1}{v^2_n}(\mathcal{U}_n^{(1)}+ \mathcal{U}_n^{(2)})(\mathcal{V}_n^{(1)}+ \mathcal{V}_n^{(2)}) \nonumber\\
& +\frac{u_n}{v^3_n}(\mathcal{V}_n^{(1)}+ \mathcal{V}_n^{(2)})^2 +\wtO_{p,2}(n^{-\frac{3}{2}}\log^{\frac{3}{2}} n + \rho_n^{-\frac{3}{2}} n^{-\frac{3}{2}}\log^{\frac{1}{2}} n(1 + s_n\log^{3/2}n)) \quad \nonumber\\ 
=& \underbrace{\frac{1}{v_n}(\mathcal{U}_n^{(1)}+ \mathcal{U}_n^{(2)})-\frac{u_n}{v^2_n}(\mathcal{V}_n^{(1)}+ \mathcal{V}_n^{(2)}) -\frac{1}{v^2_n}(\mathcal{U}_n^{(1)}+ \mathcal{U}_n^{(2)})(\mathcal{V}_n^{(1)}+ \mathcal{V}_n^{(2)})+\frac{u_n}{v^3_n}(\mathcal{V}_n^{(1)}+ \mathcal{V}_n^{(2)})^2}_{I_1} \nonumber\\
& + \underbrace{\frac{1}{v_n}(B_n +C_n) - \frac{u_n}{v_n^2}D_n}_{I_2}+\wtO_{p,2}(n^{-\frac{3}{2}}\log^{\frac{3}{2}} n + \rho_n^{-\frac{3}{2}} n^{-\frac{3}{2}}\log^{\frac{1}{2}} n(1 + s_n\log^{3/2}n)) \label{eq: combined taylor},
\end{align}
where the last equality holds because the second term in the error bound is always dominated by the other two. 
Next, we analyze the leading term of the variance of $\widehat{U}_n/\widehat{V}_n-u_n/v_n$. Given the order of leading terms in Lemma~\ref{lemma: U and V decomposition}, we have 
\begin{align*}
	I_1 = \frac{1}{v_n}\mathcal{U}_n^{(1)} - \frac{u_n}{v_n^2}\mathcal{V}_n^{(1)} + \wtO_{p,2}(n^{-1}\log n) \quad \text{ and } \quad I_2 = \wtO_{p,2}(\rho_n^{-\frac{1}{2}} n^{-1} \log^{\frac{1}{2}} n).
\end{align*}
It follows that 
\begin{align*}
	&\dfrac{\widehat{U}_{n}}{\widehat{V}_{n}}-\dfrac{u_{n}}{v_{n}} \\
 = &\frac{1}{v_n}\mathcal{U}_n^{(1)} - \frac{u_n}{v_n^2}\mathcal{V}_n^{(1)} + \wtO_{p,2}(n^{-1}\log n + \rho_n^{-\frac{1}{2}} n^{-1} \log^{\frac{1}{2}} n + n^{-\frac{3}{2}} \rho_n^{-\frac{3}{2}} \log^{\frac{1}{2}} n (1 + s_n \log^{\frac{3}{2}} n))\\
	= &\underbrace{\frac{3}{n}\sum_{i=1}^{n}\left(\frac{1}{v_n}g_1(X_i)-\frac{u_n}{v_n^2}f_1(X_i)\right)}_{Y_1} + \wtO_{p,2}(n^{-1}\log n + \rho_n^{-\frac{1}{2}} n^{-1} \log^{\frac{1}{2}} n + n^{-\frac{3}{2}} \rho_n^{-\frac{3}{2}} \log^2 n),
\end{align*}
where $Y_1=\wtO_{p,2}(n^{-\frac{1}{2}}\,\log^{\frac{1}{2}}n)$. We have
\begin{align*}
v_n
&=\frac{1}{\binom{n}{3}} \sum_{1 \le i < j < k \le n} 
\mathbb{E}\!\big[(b_{ij}+c_{ij})(b_{ik}+c_{ik})(b_{jk}+c_{jk})\big]\\
&=\frac{\rho_n^3}{\binom{n}{3}} \sum_{1 \le i < j < k \le n} 
\mathbb{E}\!\big[F(X_i,X_j)\,F(X_i,X_k)\,F(X_k,X_j)\big]\\
&=\rho_n^3\,\mathbb{E}\!\big[F(X_1,X_2)\,F(X_1,X_3)\,F(X_2,X_3)\big] \asymp \rho_n^3.
\end{align*}

Define $Y_2:=\widehat{U}_n/\widehat{V}_n-u_n/v_n - Y_1 = \wtO_{p,2}(n^{-1}\log n + \rho_n^{-\frac{1}{2}} n^{-1} \log^{\frac{1}{2}} n + n^{-\frac{3}{2}} \rho_n^{-\frac{3}{2}} \log^2 n)$. 
Note that 
\begin{align*}
	\var(Y_1)=\frac{9}{n^2}\sum_{i=1}^n \E\left[\left(\frac{1}{v_n}g_1(X_i)-\frac{u_n}{v_n^2}f_1(X_i)\right)^2\right]=\frac{9}{n} \xi_1^2:= \sigma_n^2.
\end{align*}

Since $g_{1}(\cdot)\preceq \rho_n^3$ and $f_{1}(\cdot)\preceq \rho_n^3$, we have $\xi_1^2=\E\left[\left(\dfrac{1}{v_n}g_1(X_i)-\dfrac{u_n}{v_n^2}f_1(X_i)\right)^2\right] \preceq 1$. By Condition (b), we have $\xi_1 \asymp 1$ and then $\sigma_n \asymp n^{-1/2}$.  

It remains to prove that $\sigma_n^2$ is the leading term of the variance. 
Note that $|\widehat{U}_n/\widehat{V}-u_{n}/v_{n}| \le 2$, \textit{i.e.,} $Y_2$ is upper bounded by a universal constant. Then we bound the variance
\begin{align*}
	\var(Y_2) & \le \E(Y_2^2) \\
	=& \int Y_2^2 \cdot 1(|Y_2| \ge C(n^{-1}\log n + \rho_n^{-\frac{1}{2}} n^{-1} \log^{\frac{1}{2}} n + n^{-\frac{3}{2}} \rho_n^{-\frac{3}{2}} \log^2 n ) )d\mathcal{F}_{Y_2}\\
 &+ \int Y_2^2 \cdot 1(|Y_2| < C(n^{-1}\log n + \rho_n^{-\frac{1}{2}} n^{-1} \log^{\frac{1}{2}} n + n^{-\frac{3}{2}} \rho_n^{-\frac{3}{2}} \log^2 n ))d\mathcal{F}_{Y_2} \\ 
	\preceq &  n^{-2} + n^{-2} \log^2 n + n^{-2} \rho_n^{-1} \log n + n^{-3} \rho_n^{-3} \log^4 n\\
 =&O(n^{-2} \log^2 n + n^{-2} \rho_n^{-1} \log n + n^{-3} \rho_n^{-3} \log^4 n).
\end{align*}
Similarly, we have $\E(Y_1Y_2) \le \E|Y_1Y_2| \preceq n^{-2}+n^{-\frac{3}{2}}\log^{\frac{3}{2}} n + \rho_n^{-\frac{1}{2}} n^{-\frac{3}{2}} \log n   + n^{-2} \rho_n^{-\frac{3}{2}} \log^{5/2} n= O(n^{-\frac{3}{2}}\log^{\frac{3}{2}} n + \rho_n^{-\frac{1}{2}} n^{-\frac{3}{2}} \log n   + n^{-2} \rho_n^{-\frac{3}{2}} \log^{5/2} n)$. By combining them, we~have
\begin{align*}
&\var(\widehat{U}_n/\widehat{V}_n-u_n/v_n) \\
=&\var(Y_1+Y_2)=\var(Y_1)+\var(Y_2)+2\E(Y_1Y_2) \quad \quad (\text{by } \E(Y_1)=0 ) \\
=&\sigma_n^2+O(n^{-\frac{3}{2}} \log^{\frac{3}{2}}n + n^{-2} \rho_n^{-1} \log n + n^{-\frac{3}{2}} \rho_n^{-\frac{1}{2}} \log n + n^{-3} \rho_n^{-3} \log^4 n + n^{-2} \rho_n^{-\frac{3}{2}} \log^{\frac{5}{2}} n)\\
=&\sigma_n^2+o(n^{-1}), \quad \quad (\text{by }\rho_n^3 s_n^2 \succeq n^{-2} \log^8 n) 
\end{align*}
which indicates that $\sigma_n^2 \asymp n^{-1}$ is the leading term.

\subsection{Proof of Lemma~\ref{lemma}~(2)}
\label{sec:appendix-Lemma2}

By (\ref{eq: combined taylor}) and $\sigma_n^2 \asymp n^{-1}$, we have 
\[(\widehat{U}_n/\widehat{V}_n-u_n/v_n)/\sigma_n = I_1 / \sigma_n + I_2/\sigma_n + \wtO_{p,2}(n^{-1}\log^{\frac{3}{2}} n + M(n, s_n, \rho_n)).\]
We further simplify the first term $I_1$. Note that, given the order of leading terms in Lemma~\ref{lemma: U and V decomposition}~(1), we have 
\begin{align*}
\frac{1}{v^2_n}(\mathcal{U}_n^{(1)}+ \mathcal{U}_n^{(2)})(\mathcal{V}_n^{(1)}+ \mathcal{V}_n^{(2)})=&\frac{1}{v^2_n}\left(\frac{3}{n}\sum_{i=1}^{n}g_{1}(X_{i})+\wtO_{p,2}(\rho^3_n\,n^{-1}\,\log\,n)\right)\left(\frac{3}{n}\sum_{i=1}^{n}f_{1}(X_{i})+\wtO_{p,2}(\rho^3_n\,n^{-1}\,\log\,n)\right)\\
=&\frac{1}{v^2_n}\frac{9}{n^2}\sum_{i=1}^{n}g_{1}(X_{i}) \sum_{i=1}^{n}f_{1}(X_{i})+\wtO_{p,2}(n^{-\frac{3}{2}}\,\log^{\frac{3}{2}}n)
\end{align*}
and
\begin{align*}
\frac{u_n}{v^3_n}(\mathcal{V}_n^{(1)}+ \mathcal{V}_n^{(2)})^2=&\frac{u_n}{v^3_n}\left(\frac{3}{n}\sum_{i=1}^{n}f_{1}(X_{i})+\wtO_{p,2}(\rho^3_n\,n^{-1}\,\log\,n)
\right)^2\\
=& \frac{u_n}{v^3_n}\frac{9}{n^2}\left(\sum_{i=1}^{n}f_{1}(X_{i})\right)^2+\wtO_{p,2}(n^{-\frac{3}{2}}\,\log^{\frac{3}{2}}n).
\end{align*}
By plugging them into $I_1$, we have
\begin{align}
I_1=&\dfrac{3}{n\,v_n}\sum\limits_{i=1}^{n}g_{1}(X_{i})+\dfrac{6}{n(n-1)\,v_n}\sum\limits_{1\le i < j \le n}g_{2}(X_{i},X_{j})\nonumber \\
& -\dfrac{3\,u_n}{n\,v^2_n}\sum\limits_{i=1}^{n}f_{1}(X_{i})-\dfrac{6\,u_n}{n(n-1)\,v^2_n}\sum\limits_{1\le i < j \le n}f_{2}(X_{i},X_{j}) \nonumber\\
&- \dfrac{1}{v^2_n}\dfrac{9}{n^2}\sum\limits_{i=1}^{n}g_{1}(X_{i})\cdot \sum\limits_{i=1}^{n}f_{1}(X_{i})+\dfrac{u_n}{v^3_n}\dfrac{9}{n^2}\left(\sum\limits_{i=1}^{n}f_{1}(X_{i})\right)^2+\wtO_{p,2}(n^{-\frac{3}{2}}\,\log^{\frac{3}{2}}n) \nonumber\\
=&\dfrac{3}{n}\sum\limits_{i=1}^{n}(\dfrac{1}{v_n}g_1(X_i)-\dfrac{u_n}{v^2_n}f_1(X_i))+\dfrac{6}{n(n-1)}\sum\limits_{1 \le i < j \le n}(\dfrac{1}{v_n}g_2(X_i,X_j)-\dfrac{u_n}{v^2_n}f_2(X_i,X_j)) \nonumber\\
&-\dfrac{9}{n^2}\sum\limits_{i=1}^n \dfrac{f_1(X_i)}{v_n}\cdot \sum\limits_{i=1}^n(\dfrac{1}{v_n}g_1(X_i)-\dfrac{u_n}{v^2_n}f_1(X_i))+\wtO_{p,2}(n^{-\frac{3}{2}}\,\log^{\frac{3}{2}}n) \nonumber\\
:=&\frac{3}{n}\sum_{i=1}^n q_1(X_i)+\frac{6}{n(n-1)}\sum_{1 \le i < j \le n}q_2(X_i,X_j)-\frac{9}{n^2}\sum_{i=1}^n p_1(X_i)\sum_{i=1}^n q_1(X_i) \nonumber\\
&+\wtO_{p,2}(n^{-\frac{3}{2}}\,\log^{\frac{3}{2}}n), \label{eq: lemma1.2}
\end{align}
where we define 
$q_1(X_i):=g_1(X_i)/v_n-u_nf_1(X_i)/v^2_n$, $q_2(X_i,X_j):=g_2(X_i,X_j)/v_n-u_nf_2(X_i,X_j)/v^2_n$, and $p_1(X_i):={f_1(X_i)}/{v_n}$.

Since $p_1(X_i)q_1(X_j)$ and $q_2(X_i, X_j)$ are canonical and bounded by constants, by Lemma~\ref{applemma1}, we have $$\frac{1}{n^2}\sum_{1\le i\neq j\le n} p_1(X_i)q_1(X_j)=\wtO_{p,2}(n^{-1}\log n)\quad \text{ and }\quad \frac{1}{n^2}\sum_{1 \le i < j \le n}q_2(X_i,X_j) = \wtO_{p,2}(n^{-1}\log n).$$
Therefore, under Condition (b), we have 
\begin{align} \label{eq: q2}
	\frac{2}{n(n-1)}\sum_{1 \le i < j \le n}q_2(X_i,X_j)- \frac{1}{n^2}\sum_{1 \le {i \neq j}\le n}q_2(X_i,X_j) = \wtO_{p,2}(n^{-2}\log n).
\end{align}
In addition, by Lemma~\ref{applemma2} and $|p_1(X_i)q_1(X_i) - \E[p_1(X_i)q_1(X_i)]| \preceq 1$, we have
\begin{align} \label{eq: p1q1}
\frac{1}{n}\sum_{i=1}^np_1(X_i)q_1(X_i)=\E\Big[p_1(X_1)q_1(X_1)\Big]+\wtO_{p,2}(n^{-1/2}\log^{1/2}n).
\end{align}
By combining (\ref{eq: lemma1.2})-(\ref{eq: p1q1}), we can simplify $I_1/\sigma_n$ as 
\begin{align*}
	I_1/\sigma_n=& -\frac{3}{\sqrt{n}\xi_1}\E\Big[p_1(X_1)q_1(X_1)\Big] +\frac{1}{\sqrt{n}\xi_1}\sum_{i=1}^n q_1(X_i)+\frac{1}{n^{3/2}\xi_1}\sum_{1 \le i \neq j \le n}q_2(X_i,X_j)\\
	&  -\frac{3}{n^{3/2}\xi_1}\sum_{1\le {i\neq j}\le n}p_1(X_i)q_1(X_j) +\wtO_{p,2}(n^{-1}\,\log^{\frac{3}{2}}n).
\end{align*}
Define 
\begin{align}
	\check{\Delta}_n=I_2 / \sigma_n = \frac{1}{v_n \sigma_n}(B_n +C_n) - \frac{u_n}{v_n^2 \sigma_n}D_n=\wtO_{p,2}(\rho_n^{-\frac{1}{2}} n^{-\frac{1}{2}} \log^{\frac{1}{2}} n). \label{eq: delta check def}
\end{align}
Thus, we have derived the decomposition for 
\begin{align*}
	(\widehat{U}_n/\widehat{V}_n-u_n/v_n)/\sigma_n= & -\frac{3}{\sqrt{n}\xi_1}\E\Big[p_1(X_1)q_1(X_1)\Big] +\frac{1}{\sqrt{n}\xi_1}\sum_{i=1}^n q_1(X_i)+\frac{1}{n^{3/2}\xi_1}\sum_{1 \le i \neq j \le n}q_2(X_i,X_j)\\
	&  -\frac{3}{n^{3/2}\xi_1}\sum_{1\le {i\neq j}\le n}p_1(X_i)q_1(X_j) +\check{\Delta}_n + \wtO_{p,2}(n^{-1}\log^{\frac{3}{2}} n + M(n, s_n, \rho_n)).
\end{align*}

\subsection{Proof of Lemma~\ref{lemma}~(3)}
\label{sec:appendix-Lemma3}

Note that the randomness of $\check{\Delta}_n$ comes from two parts, the latent variables $W=(X_1, \cdots, X_n)$ and the Bernoulli random variables $A_{ij} \mid X_{i}, X_{j}$ conditional on $W$. 
Throughout this proof, 
we first treat $W$ as fixed by conditioning on them and focus on the randomness from Bernoulli random variables.

We prove the asymptotic normality for $\check{\Delta}_n$ through the Berry-Esseen bound in Lemma~\ref{berryessentheroem}, which requires to estimate the variance of $\check{\Delta}_n$. 
By the definition of $B_n$, $C_n$ and $D_n$ in Lemma~\ref{lemma: U and V decomposition}, we have
\begin{align*}
&\check{\Delta}_n=\frac{1}{v_n \sigma_n}\frac{1}{\binom{n}{2}}\sum_{1\le k_1 < k_2 \le n}\Big(\mathring{B}_{k_1 k_2}\eta_{2}(A_{k_1 k_2})+\mathring{C}_{k_1 k_2}\eta_{3}(A_{k_1 k_2})- \frac{u_n}{v_n}\mathring{D}_{k_1 k_2}\eta_{1}(A_{k_1 k_2})\Big).
\end{align*}
Because $\E[\check{\Delta}_n|W]=0$, we compute the variance of $\check{\Delta}_n$ given $W$ as
\begin{align*}
&\var(\check{\Delta}_n |W)\\
=& \frac{1}{\binom{n}{2}^2 v_n^2\sigma_n^2}\sum_{1\le k_1 < k_2 \le n}\E\left[\mathring{B}_{k_1 k_2}\eta_{2}(A_{k_1 k_2})+\mathring{C}_{k_1 k_2}\eta_{3}(A_{k_1 k_2})- \frac{u_n}{v_n}\mathring{D}_{k_1 k_2}\eta_{1}(A_{k_1 k_2})\right]^2\\
=& \frac{1}{\binom{n}{2}^2v_n^2\sigma_n^2}\sum_{1\le k_1 < k_2 \le n}  \Bigg( a_{k_1 k_2} \bigg( \underbrace{\mathring{B}_{k_1 k_2} (-b_{k_1 k_2})+\mathring{C}_{k_1 k_2} (-c_{k_1 k_2}) - \frac{u_n}{v_n} \mathring{D}_{k_1 k_2} (-b_{k_1 k_2} -c_{k_1 k_2})}_{\mathring{\theta}_{k_1 k_2}}\bigg)^2 \\
&\ \ \quad +b_{k_1 k_2} \bigg( \mathring{B}_{k_1 k_2}(1-b_{k_1 k_2})+\mathring{C}_{k_1 k_2}(-c_{k_1 k_2}) - \frac{u_n}{v_n} \mathring{D}_{k_1 k_2}(1-b_{k_1 k_2} - c_{k_1 k_2}) \bigg)^2\\
&\ \ \quad +c_{k_1 k_2} \bigg( \mathring{B}_{k_1 k_2}(-b_{k_1 k_2})+\mathring{C}_{k_1 k_2}(1-c_{k_1 k_2}) - \frac{u_n}{v_n} \mathring{D}_{k_1 k_2}(1-b_{k_1 k_2} - c_{k_1 k_2}) \bigg)^2 \Bigg)\\
=& \frac{1}{\binom{n}{2}^2v_n^2\sigma_n^2}\sum_{1\le k_1 < k_2 \le n}  \Bigg( a_{k_1 k_2} \mathring{\theta}_{k_1 k_2}^2 +b_{k_1 k_2} \bigg( \mathring{\theta}_{k_1 k_2} + \mathring{B}_{k_1 k_2} - \frac{u_n}{v_n} \mathring{D}_{k_1 k_2} \bigg)^2\\
&\ \ \quad \quad\quad\quad\quad\quad\quad\quad\quad\quad+c_{k_1 k_2} \bigg(\mathring{\theta}_{k_1 k_2}+ \mathring{C}_{k_1 k_2}- \frac{u_n}{v_n} \mathring{D}_{k_1 k_2} \bigg)^2 \Bigg)\\
=&\frac{1}{\binom{n}{2}^2v_n^2\sigma_n^2}\sum_{1\le k_1 < k_2 \le n}\Bigg(-\mathring{\theta}_{k_1 k_2}^2+b_{k_1 k_2}(\mathring{B}_{k_1 k_2}- \frac{u_n}{v_n} \mathring{D}_{k_1 k_2})^2+c_{k_1 k_2}(\mathring{C}_{k_1 k_2}- \frac{u_n}{v_n} \mathring{D}_{k_1 k_2})^2\Bigg).
\end{align*}
By the definition of $\mathring{B}_{k_1 k_2}$, $\mathring{C}_{k_1 k_2}$, and $\mathring{D}_{k_1 k_2}$ in Lemma~\ref{lemma: U and V decomposition} and $1-u_n/v_n \preceq s_n$, we can compute
\begin{align}
\mathring{B}_{k_1 k_2}- \frac{u_n}{v_n} \mathring{D}_{k_1 k_2} & \asymp \rho_n^2s_n -\rho_n^2 \preceq \rho_n^2, \label{eq: B o order}\\
	\mathring{C}_{k_1 k_2}- \frac{u_n}{v_n} \mathring{D}_{k_1 k_2} &= \mathring{C}_{k_1 k_2}- \frac{u_n}{v_n} (\mathring{B}_{k_1 k_2}+\mathring{C}_{k_1 k_2})=(1-\frac{u_n}{v_n})\mathring{C}_{k_1 k_2}-\frac{u_n}{v_n}\mathring{B}_{k_1 k_2} \nonumber \\
	& \preceq \rho_n^2 s_n. \label{eq: C o order}
\end{align}
Thus we have the following upper bounds:
\begin{align}
	\mathring{\theta}_{k_1 k_2}= -b_{k_1 k_2}(\mathring{B}_{k_1 k_2}- \frac{u_n}{v_n} \mathring{D}_{k_1 k_2})-c_{k_1 k_2}(\mathring{C}_{k_1 k_2}- \frac{u_n}{v_n} \mathring{D}_{k_1 k_2}) \preceq \rho_n^3s_n
\end{align}
and
\begin{align}
	b_{k_1 k_2}(\mathring{B}_{k_1 k_2}- \frac{u_n}{v_n} \mathring{D}_{k_1 k_2})^2+c_{k_1 k_2}(\mathring{C}_{k_1 k_2}- \frac{u_n}{v_n} \mathring{D}_{k_1 k_2})^2 \preceq \rho_n^5s_n.
\end{align}
It follows that $\var(\check{\Delta}_n |W)\preceq n^{-1}\rho_n^{-1}s_n$. Now we prove its lower bound is also in the order of $n^{-1} \rho_n^{-1} s_n$. We discuss two admissible regimes under Conditions (c1) and (c2) separately.

\vspace{2mm}

\noindent \textbf{(i) Vanishing $s_n$.} When Condition (c1) holds ($s_n \prec 1$), given  $F, G > C$ for some constant $C$, we have
\[
\mathring{B}_{k_1 k_2}- \frac{u_n}{v_n} \mathring{D}_{k_1 k_2} \asymp \rho_n^2s_n -\rho_n^2 \asymp \rho_n^2.
\]
Therefore
\[
b_{k_1 k_2}(\mathring{B}_{k_1 k_2}- \frac{u_n}{v_n} \mathring{D}_{k_1 k_2})^2+c_{k_1 k_2}(\mathring{C}_{k_1 k_2}- \frac{u_n}{v_n} \mathring{D}_{k_1 k_2})^2 \asymp s_n \rho_n (\rho_n^2)^2 + \rho_n (s_n \rho_n^2)^2 \asymp s_n \rho_n^5.
\]

Then we have $\operatorname{Var}(\check{\Delta}_n \mid W)\asymp n^{-1}\rho_n^{-1}s_n$. 
If we replace the Condition (a1) by the moment Condition (a2)
\[
\mathbb{E}\!\Big[ G(X_1,X_2)\,F(X_1,X_2)\,F(X_3,X_1)\,F(X_3,X_2)\,F(X_4,X_1)\,F(X_4,X_2)  \Big]>0,
\]
and assume $\|F\|_{\infty}$ and $\|G\|_{\infty}$ are bounded, the same conclusion holds with high probability. 
In this situation, it suffices to prove that
\begin{align*}
&\frac{1}{\binom{n}{2}^2v_n^2\sigma_n^2} (\frac{u_n}{v_n})^2 \sum_{1\le k_1<k_2\le n} b_{k_1k_2}\,\mathring{D}_{k_1k_2}^2\\
=&\,\frac{s_n \rho_n^5}{\binom{n}{2}^2v_n^2\sigma_n^2} (\frac{u_n}{v_n})^2 \sum_{\substack{i,j,k_1,k_2\\ \text{distinct}}}
G(X_{k_1},X_{k_2})\,F(X_{k_1},X_{k_2})\,F(X_i,X_{k_1})\,F(X_i,X_{k_2})\,F(X_j,X_{k_1})\,F(X_j,X_{k_2})
\end{align*}
is of order of $n^{-1} \rho_n^{-1} s_n$ with high probability, since all remaining terms in $\var(\check{\Delta}_n |W)$ are $O(n^{-1} \rho_n^{-1} s_n^2)=o(n^{-1} \rho_n^{-1} s_n)$.
In fact, by Hoeffding’s inequality for bounded U-statistics,
\[
n^{-4}\sum_{\substack{i,j,k_1,k_2\\ \text{distinct}}}
G(X_{k_1},X_{k_2})\,F(X_{k_1},X_{k_2})\,F(X_i,X_{k_1})\,F(X_i,X_{k_2})\,F(X_j,X_{k_1})\,F(X_j,X_{k_2}) > C_1
\]
with probability at least $1-\exp(-C_2 n)$ for some constants $C_1>0$ and $C_2>0$.

\vspace{2mm}
\noindent \textbf{(ii) Constant-order $s_n$.} When Condition (c2) holds ($s_n \|G\|_{\infty} < 1/ 2$). Notice that
\begin{align*}
    &(\mathring{B}_{k_1 k_2}- \frac{u_n}{v_n} \mathring{D}_{k_1 k_2})^2 + (\mathring{C}_{k_1 k_2}- \frac{u_n}{v_n} \mathring{D}_{k_1 k_2})^2\\
    =&2 \frac{u_n^2}{v_n^2} \mathring{D}_{k_1 k_2}^2 - 2 \frac{u_n}{v_n} \mathring{D}_{k_1 k_2}^2 + \mathring{B}_{k_1 k_2}^2 + \mathring{C}_{k_1 k_2}^2\\
    =&2 \mathring{D}_{k_1 k_2}^2 (\frac{u_n}{v_n} - \frac{1}{2})^2 + \frac{1}{2} ( \mathring{B}_{k_1 k_2}- \mathring{C}_{k_1 k_2})^2,
\end{align*}
we have 
\begin{align*}
	&b_{k_1 k_2}(\mathring{B}_{k_1 k_2}- \frac{u_n}{v_n} \mathring{D}_{k_1 k_2})^2+c_{k_1 k_2}(\mathring{C}_{k_1 k_2}- \frac{u_n}{v_n} \mathring{D}_{k_1 k_2})^2 \\
    \succeq & \min\{b_{k_1 k_2}, c_{k_1 k_2}\} \left[ (\mathring{B}_{k_1 k_2}- \frac{u_n}{v_n} \mathring{D}_{k_1 k_2})^2 + (\mathring{C}_{k_1 k_2}- \frac{u_n}{v_n} \mathring{D}_{k_1 k_2})^2 \right] \\
    = & \rho_nF(X_{k_1},X_{k_2})\min\{s_nG(X_{k_1},X_{k_2}),1-s_nG(X_{k_1},X_{k_2})\} \left[ 2 \mathring{D}_{k_1 k_2}^2 (\frac{u_n}{v_n} - \frac{1}{2})^2 + \frac{1}{2} ( \mathring{B}_{k_1 k_2}- \mathring{C}_{k_1 k_2})^2 \right]\\
    \ge& 2(\frac{u_n}{v_n} - \frac{1}{2})^2\rho_nF(X_{k_1},X_{k_2})\min\{s_nG(X_{k_1},X_{k_2}),1-s_nG(X_{k_1},X_{k_2})\} \mathring{D}_{k_1 k_2}^2.
\end{align*}
Recall that
\begin{align*}
\mathring{D}_{k_1 k_2}=&\frac{3}{n-2} \sum_{\substack{1 \leq i \leq n \\ i \neq k_1, k_2}} (b_{i k_1}+c_{i k_1})(b_{i k_2}+c_{i k_2})=\frac{3 \rho_n^2}{n-2} \sum_{\substack{1 \leq i \leq n \\ i \neq k_1, k_2}} F(X_i, X_{k_1})F(X_i, X_{k_2}).
\end{align*}
It follows that
\begin{align*}
&\sum_{1\le k_1 < k_2 \le n}\Bigg(b_{k_1 k_2}(\mathring{B}_{k_1 k_2}- \frac{u_n}{v_n} \mathring{D}_{k_1 k_2})^2+c_{k_1 k_2}(\mathring{C}_{k_1 k_2}- \frac{u_n}{v_n} \mathring{D}_{k_1 k_2})^2\Bigg)\\
\ge&2(\frac{u_n}{v_n} - \frac{1}{2})^2\sum_{1\le k_1 < k_2 \le n}\rho_nF(X_{k_1},X_{k_2})\min\{s_nG(X_{k_1},X_{k_2}),1-s_nG(X_{k_1},X_{k_2})\} \mathring{D}_{k_1 k_2}^2\\
=&\dfrac{9\rho_n^5}{(n-2)^2}(\frac{u_n}{v_n} - \frac{1}{2})^2\sum_{1\le k_1 \neq k_2 \le n}\Big[F(X_{k_1},X_{k_2})\min\{s_nG(X_{k_1},X_{k_2}),1-s_nG(X_{k_1},X_{k_2})\}\\
&\qquad\qquad\qquad\qquad\qquad\qquad\cdot\Big(\sum_{\substack{1 \leq i \leq n \\ i \neq k_1, k_2}} F(X_{i},X_{k_1})F(X_{i},X_{k_2})\Big)^2\Big]\\
\ge&\dfrac{9\rho_n^5}{(n-2)^2}(\frac{u_n}{v_n} - \frac{1}{2})^2\sum_{\text{distinct }i,j,k_1,k_2}F(X_{k_1},X_{k_2})F(X_{i},X_{k_1})F(X_{i},X_{k_2})F(X_{j},X_{k_1})F(X_{j},X_{k_2})\\
&\qquad\qquad\qquad\qquad\qquad\qquad\quad\cdot\min\{s_nG(X_{k_1},X_{k_2}),1-s_nG(X_{k_1},X_{k_2})\}\\
\ge&\dfrac{9\rho_n^5s_n}{(n-2)^2}(\frac{u_n}{v_n} - \frac{1}{2})^2\sum_{\text{distinct }i,j,k_1,k_2}F(X_{k_1},X_{k_2})F(X_{i},X_{k_1})F(X_{i},X_{k_2})F(X_{j},X_{k_1})F(X_{j},X_{k_2})\\
&\qquad\qquad\qquad\qquad\qquad\qquad\quad\cdot G(X_{k_1},X_{k_2}),\\
\succeq&n^{2} \rho_n^5 s_n,
\end{align*}
where we use Condition (c2) to guarantee that $|\dfrac{u_n}{v_n}-\dfrac{1}{2}|\ge\epsilon$ for some constant $\epsilon$, and the lower bound in Condition (a1), namely $F,G>C_1$ for some $C_1>0$. In fact, by Hoeffding's inequality for bounded U-statistics again, Condition (a1) can be replaced by (a2), and the above result still holds with high probability.
Therefore, the variance is in the order 
\[\var(\check{\Delta}_n |W) \overset{\mathrm{w.h.p.}}{\asymp} n^{-1}\,\rho_n^{-1}\,s_n.\]

Combining the upper and lower bounds of $\var(\check{\Delta}_n |W)$, with high probability we have
\begin{align}
\label{equ:sigmaw}
\sigma_{W}:=(n\,\rho_n\,s_n^{-1})^{\frac{1}{2}}\sqrt{\var(\check{\Delta}_n |W)}\asymp 1.
\end{align}
We proceed under $\sigma_{W}\overset{\mathrm{w.h.p.}}{\asymp} 1$. By Lemma~\ref{berryessentheroem}, there exists a universal constant $C$ such that 
\begin{align*}
&\Vert \mathcal{F}_{\dfrac{(n\,\rho_n\,s_n^{-1})^{\frac{1}{2}}\check{\Delta}_n}{\sigma_{W}}|W}(u)-\mathcal{F}_{N(0,1)}(u)\Vert _{\infty}\\
\le & C\left\{0+\sum_{1\le i < j \le n}\frac{(n\,\rho_n\,s_n^{-1})^{\frac{3}{2}}}{(\sigma_W\,\binom{n}{2}v_n \sigma_n)^3} (a_{i j}|\mathring{\theta}_{i j}|^3+b_{i j}|\mathring{B}_{i j}- \frac{u_n}{v_n}\mathring{D}_{i j}-\mathring{\theta}_{i j}|^3+c_{i j}|\mathring{C}_{i j}- \frac{u_n}{v_n}\mathring{D}_{i j}-\mathring{\theta}_{i j}|^3)\,\right\}\\
\preceq &n^2\cdot \frac{n^{\frac{3}{2}}\,s_n^{-\frac{3}{2}}\,\rho_n^{\frac{3}{2}}}{n^6\rho_n^9 n ^{-\frac{3}{2}}}\cdot s_n\,\rho_n^{7} \quad (\text{by }  (\ref{eq: B o order})-(\ref{equ:sigmaw}))\\
= & n^{-1}s_n^{-\frac{1}{2}}\,\rho_n^{-\frac{1}{2}}.
\end{align*}
By taking a variable transformation, it follows that for any $k$,
\begin{align*}
	\Vert \mathcal{F}_{\check{\Delta}_n|W}(u)-\mathcal{F}_{N(0,n^{-1}\,\rho_n^{-1}\,s_n\,\sigma_W^2)}(u)\Vert _{\infty}=\widetilde{O}_{p, k}(n^{-1}\,\rho_n^{-\frac{1}{2}}\,s_n^{-\frac{1}{2}}),
\end{align*}
where $\sigma_W \asymp 1$.
Here, we have completed the proof of Lemma~\ref{lemma}~(3).

\subsection{Proof of Lemma~\ref{lemma}~(4)}
\label{sec:appendix-Lemma4}

We define the intermediate term $\widehat{\sigma}^2_n$ as follow:
\begin{align}
\frac{n\widehat{\sigma}^2_n}{9}&=\frac{1}{n}\sum_{i=1}^n\left\{\frac{1}{\binom{n-1}{2}}\sum_{\substack{1\le i_{1}< i_{2}\le n\\i_{1},i_{2}\neq i}}\left(\frac{\E[h_{\blacktriangle}(A_{i,i_{1},i_{2}})\mid X_{i},X_{i_1},X_{i_2}]}{V_n}-\frac{\E[h_{\Delta}(A_{i,i_{1},i_{2}})\mid X_{i},X_{i_1},X_{i_2}]}{V^2_n}U_n\right)\right\}^2 \nonumber\\
&=\frac{1}{n}\sum_{i=1}^n\left\{\frac{1}{\binom{n-1}{2}}\sum_{\substack{1\le i_{1}< i_{2}\le n\\i_{1},i_{2}\neq i}}\left(\frac{h_{\blacktriangle}(X_{i},X_{i_1},X_{i_2}))}{V_n}-\frac{h_{\Delta}(X_{i},X_{i_1},X_{i_2})}{V^2_n}U_n\right)\right\}^2. \label{lemma13defineinterm}
\end{align}
To bound 
\[
\widehat{\delta}_n:=\dfrac{\widehat{S}^2_n-\widehat{\sigma}^2_n}{\sigma^2_n}=(\dfrac{n\widehat{S}^2_n}{9}-\dfrac{n\widehat{\sigma}^2_n}{9})/\dfrac{n\sigma^2_n}{9},
\]
similar to the proof of Lemma~\ref{lemma} (3), we first treat $W = (X_{1}, \cdots, X_n)$,  as fixed by conditioning them, and focus on the randomness from the second layer $A_{ij} \mid X_{i}, X_{j}$. 
Recall that
$$
\frac{n\widehat{S}^2_n}{9}=\frac{1}{n}\sum_{i=1}^n\left(
\frac{1}{\binom{n-1}{2}}\sum_{\substack{1\le i_{1}< i_{2}\le n\\i_{1},i_{2}\neq i}}\left(\frac{h_{\blacktriangle}(A_{i,i_{1},i_{2}})}{\widehat{V}_n}-
\frac{h_{\Delta}(A_{i,i_{1},i_{2}})}{\widehat{V}^2_n}\widehat{U}_n\right)
\right)^2.
$$
For notation simplicity, we define the following shorthand:
$$
\widehat{A}_i=\frac{1}{\binom{n-1}{2}}\sum_{\substack{1\le i_{1}< i_{2}\le n\\i_{1},i_{2}\neq i}}\left(\frac{h_{\blacktriangle}(A_{i,i_{1},i_{2}})}{\widehat{V}_n}-
\frac{h_{\Delta}(A_{i,i_{1},i_{2}})}{\widehat{V}^2_n}\widehat{U}_n\right),
$$
$$
A_i=\frac{1}{\binom{n-1}{2}}\sum_{\substack{1\le i_{1}< i_{2}\le n\\i_{1},i_{2}\neq i}}\left(\frac{h_{\blacktriangle}(X_{i},X_{i_1},X_{i_2})}{\widehat{V}_n}-
\frac{h_{\Delta}(X_{i},X_{i_1},X_{i_2})}{\widehat{V}^2_n}\widehat{U}_n\right),
$$
and
$$
\widehat{a}_i=\frac{1}{\binom{n-1}{2}}\sum_{\substack{1\le i_{1}< i_{2}\le n\\i_{1},i_{2}\neq i}}\left(\frac{h_{\blacktriangle}(X_{i},X_{i_1},X_{i_2})}{V_n}-
\frac{h_{\Delta}(X_{i},X_{i_1},X_{i_2})}{V^2_n}U_n\right).
$$
It follows that $n\widehat{\sigma}^2_n/9=\sum_{i=1}^n \widehat{a}_i^2 / n$ and 
\begin{align}
\label{lemma13decompAwidehat}
\frac{n\widehat{S}^2_n}{9}=\frac{1}{n}\sum_{i=1}^n\widehat{A}^2_i=\frac{1}{n}\sum_{i=1}^n(\widehat{A}_i-A_i+A_i)^2=\frac{1}{n}\sum_{i=1}^n(\widehat{A}_i-A_i)^2+\frac{2}{n}\sum_{i=1}^n(\widehat{A}_i-A_i)A_i+\frac{1}{n}\sum_{i=1}^n A_i^2.
\end{align}
We want to bound $\frac{1}{n}\sum_{i=1}^n(\widehat{A}_i-A_i)^2+\frac{2}{n}\sum_{i=1}^n(\widehat{A}_i-A_i)A_i$ next. To proceed, we claim the following decomposition for any fixed $i$ in Lemma~\ref{apppro2}, the proof of which is given in Section~\ref{subsec: proof of apppro2}.

\smallskip
\begin{lemma}\label{apppro2}  Assuming the conditions in Lemma~\ref{lemma}, we have the following decompositions:
\begin{enumerate}
	\item For any fixed $1\le i \le n$, we have 
\begin{align*}
&\frac{1}{\binom{n-1}{2}}\sum_{\substack{1 \le i_1 < i_2 \le n\\i_1,i_2\neq i}}\left(h_{3}(A_{i,i_1, i_2})-h_{3}(X_i,X_{i_1}, X_{i_2})\right)\\
=&\frac{1}{n-1}\sum_{\substack{1\le j \le n:\\j \neq i}}\widehat{\Theta}_{ij}^{(1)} \eta_3(A_{ij})+\frac{1}{n-1}\sum_{\substack{1\le j \le n:\\j \neq i}}\widehat{\Theta}_{ij}^{(2)}\eta_2(A_{ij})+\wtO_{p,2}(n^{-1} \rho_n^{3/2} s_n \log^2 n),
\end{align*}
where $\widehat{\Theta}_{ij}^{(1)}$ and $\widehat{\Theta}_{ij}^{(2)}$, defined in (\ref{proplema2deftheta12}), are functions of $X_k$, $1 \le k \le n$, so here we treat them as constants. Additionally, 
$\widehat{\Theta}_{ij}^{(1)}\asymp s_n^2 \rho_n^2$, and
$\widehat{\Theta}_{ij}^{(2)}\asymp s_n \rho_n^2$.
\item For any fixed $1\le i \le n$, we have
\begin{align*}
\frac{1}{\binom{n-1}{2}}\sum_{\substack{1 \le i_1 < i_2 \le n\\i_1,i_2\neq i}}&\left(h_1(A_{i,i_1, i_2})-h_1(X_i,X_{i_1}, X_{i_2})\right) \\
=&\frac{1}{n-1}\sum_{\substack{1\le j \le n:\\j \neq i}}\widehat{\Theta}_{ij}^{(3)} \eta_3(A_{ij})+\wtO_{p,2}(n^{-1}\,\rho_n^{\frac{3}{2}}\,\log n ),\\
\frac{1}{\binom{n-1}{2}}\sum_{\substack{1 \le i_1 < i_2 \le n\\i_1,i_2\neq i}}&\left(h_{\Delta}(A_{i,i_1, i_2})-h_{\triangle}(X_i,X_{i_1}, X_{i_2})\right) \\
=&\frac{1}{n-1}\sum_{\substack{1\le j \le n:\\j \neq i}}\widehat{\Theta}_{ij}^{(4)}\eta_1(A_{ij})+\wtO_{p,2}(n^{-1}\,\rho_n^{\frac{3}{2}}\,\log n ),
\end{align*}
where $\widehat{\Theta}_{ij}^{(3)}$ and $\widehat{\Theta}_{ij}^{(4)}$, defined in (\ref{proplema2deftheta34}), are functions of $X_k$, $1 \le k \le n$. Additionally, 
$\widehat{\Theta}_{ij}^{(3)}\asymp \rho_n^2$, and
$\widehat{\Theta}_{ij}^{(4)}\asymp \rho_n^2$.
\end{enumerate}
\end{lemma}
\smallskip
\noindent
By Lemma~\ref{apppro2}, we have 
\begin{equation}
\begin{aligned}
\widehat{A}_i-A_i=&\frac{1}{\binom{n-1}{2}}\frac{1}{\widehat{V}_n}\sum_{\substack{1\le i_{1}< i_{2}\le n\\i_{1},i_{2}\neq i}}\{h_{\blacktriangle}(A_{i,i_{1},i_{2}})-\E[h_{\blacktriangle}(A_{i,i_{1},i_{2}})]\}\\
&-\frac{1}{\binom{n-1}{2}}\frac{\widehat{U}_n}{\widehat{V}^2_n}\sum_{\substack{1\le i_{1}< i_{2}\le n\\i_{1},i_{2}\neq i}}\{h_{\Delta}(A_{i,i_{1},i_{2}})-\E[h_{\Delta}(A_{i,i_{1},i_{2}})]\}\\
=&\frac{1}{\widehat{V}_n}\Bigg(\frac{1}{n-1}\sum_{\substack{1\le j \le n\\j\neq i}}\widehat{\Theta}^{(1)}_{ij}\eta_3(A_{ij})+\frac{1}{n-1}\sum_{\substack{1\le j \le n\\j\neq i}}\widehat{\Theta}^{(2)}_{ij}\eta_2(A_{ij})\\
&\quad \quad+\frac{1}{n-1}\sum_{\substack{1\le j \le n\\j\neq i}}\widehat{\Theta}^{(3)}_{ij}\eta_3(A_{ij})+\wtO_{p,2}(\rho^3_n\,M(n,s_n,\rho_n))\Bigg)\\
&-\frac{\widehat{U}_n}{\widehat{V}^2_n}\left(\frac{1}{n-1}\sum_{\substack{1\le j \le n\\j\neq i}}\widehat{\Theta}^{(4)}_{ij}\eta_1(A_{ij})+\wtO_{p,2}(\rho^3_n\,M(n,s_n,\rho_n))\right)\\
:=& \frac{1}{\widehat{V}_n} R_i^{(1)} -\frac{\widehat{U}_n}{\widehat{V}^2_n} R_i^{(2)},
\end{aligned}
\label{hatA-Abound}
\end{equation}
where
$\widehat{\Theta}_{ij}^{(1)}\asymp s_n^2 \rho_n^2$, $\widehat{\Theta}_{ij}^{(2)}\asymp s_n \rho_n^2$, 
$\widehat{\Theta}_{ij}^{(3)}\asymp \rho_n^2$,
$\widehat{\Theta}_{ij}^{(4)}\asymp \rho_n^2$, $R_i^{(1)} := \frac{1}{n-1}\sum_{\substack{1\le j \le n\\j\neq i}}\widehat{\Theta}^{(1)}_{ij}\eta_3(A_{ij})+\frac{1}{n-1}\sum_{\substack{1\le j \le n\\j\neq i}}\widehat{\Theta}^{(2)}_{ij}\eta_2(A_{ij})+\frac{1}{n-1}\sum_{\substack{1\le j \le n\\j\neq i}}\widehat{\Theta}^{(3)}_{ij}\eta_3(A_{ij})+\wtO_{p,2}(\rho^3_n\,M(n,s_n,\rho_n))$, and $R_i^{(2)} := \frac{1}{n-1}\sum_{\substack{1\le j \le n\\j\neq i}}\widehat{\Theta}^{(4)}_{ij}\eta_1(A_{ij})+\wtO_{p,2}(\rho^3_n\,M(n,s_n,\rho_n))$.

By Lemma~\ref{applemma2}, there exist universal constants $C_1$, $C_2$, and $C_3$ such that, for each $i$, $\pr(|R_i^{(1)}| > C_1 \rho_n^{\frac{5}{2}}\,n^{-\frac{1}{2}}\,\log^{\frac{1}{2}}n) < C_2 n^{-2}$ holds for $n > C_3$. 
Define the event $\mathfrak{A}_1 := \{A: R_i^{(1)} \le C_1 \rho_n^{\frac{5}{2}}\,n^{-\frac{1}{2}}\,\log^{\frac{1}{2}}n \text{ for any } i\}$. It follows that $\pr(\mathfrak{A}_1^\complement) \le \sum_{i=1}^n \pr(|R_i^{(1)}| > C_1 \rho_n^{\frac{5}{2}}\,n^{-\frac{1}{2}}\,\log^{\frac{1}{2}}n) \le n\cdot C_2 n^{-2} = C_2 n^{-1}$ for $n > C_3$. 
Note that, by (\ref{eq: Vhat_n decom}), we have $\widehat{V}_n-V_n=\wtO_{p,2}(\rho^{\frac{5}{2}}_n\,n^{-1}\,\log^{\frac{1}{2}}n + \rho_n^{\frac{3}{2}} n^{-\frac{3}{2}}\log^{\frac{1}{2}}n)$ and $V_n \asymp \rho^3_n$. Under Condition (c), there exist constants $C_4$, $C_5$ and $C_6$ such that, for $n > C_6$, the event $\mathfrak{A}_2 := \{A: \widehat{V}_n \ge C_4 \rho^3_n\}$ holds with probability at least $1-C_5 n^{-2}$. 
It follow that for $n > \max\{C_3, C_6 \}$, we have
\begin{align*}
&\pr(R_i^{(1)} / \widehat{V}_n \le C_1 / C_4 n^{-1/2} \rho_n^{-1/2} \log^{1/2} n, \text{ for any }1 \le i \le n)\\
\ge & \pr(\mathfrak{A}_1 \cap \mathfrak{A}_2) \ge 1 - \pr(\mathfrak{A}_1^\complement)- \pr(\mathfrak{A}_2^\complement) \ge 1 - (C_2+C_5)n^{-1}.
\end{align*}
Following a similar procedure, we obtain the same error bound for $\widehat{U}_n R_i^{(2)}/\widehat{V}_n^2$. This implies that 
\begin{align}
\label{hatA-Abound2}
|\widehat{A}_i - A_i| =\wtO_{p,1}( n^{-1/2}\,\rho_n^{-1/2}\,\log^{1/2}n )
\end{align}
holds uniformly for $1\le i \le n$. 
It follows that
\begin{align}
\label{final(hatA-A)2}
\frac{1}{n}
\sum_{i=1}^n (\widehat{A}_i-A_i)^2=\wtO_{p,1}(n^{-1}\,\rho_n^{-1}\,\log\,n).
\end{align}

Next, we analyze the interaction term $\frac{2}{n}\sum_{i=1}^n(\widehat{A}_i-A_i)A_i$. Note that
\begin{align}
\dfrac{2}{n}\sum_{i=1}^n(\widehat{A}_i-A_i)A_i \nonumber
=&\frac{2}{n(n-1)\widehat{V}_n}\sum_{1 \le i \neq j \le n}A_i\widehat{\Theta}^{(1)}_{ij}\eta_3(A_{ij})+ A_i\widehat{\Theta}^{(2)}_{ij}\eta_2(A_{ij})+A_i\widehat{\Theta}^{(3)}_{ij}\eta_3(A_{ij}) \nonumber\\
&-\frac{2\widehat{U}_n}{n(n-1)\widehat{V}^2_n}\sum_{1 \le i \neq j \le n}A_i\widehat{\Theta}^{(4)}_{ij}\eta_1(A_{ij})+\wtO_{p,1}(M(n,s_n,\rho_n)). \label{A(hatA-A)}
\end{align}
We rewrite $A_i$ as $\frac{1}{\widehat{V}_n}A^{(1)}_{i} - \frac{\widehat{U}_n}{\widehat{V}^2_n}A^{(2)}_i$, where $A^{(1)}_{i}:=\sum_{\substack{1\le i_{1}< i_{2}\le n\\i_{1},i_{2}\neq i}}h_{\blacktriangle}(X_i, X_{i_1}, X_{i_2})/\binom{n-1}{2}$ and $A^{(2)}_{i}:=\sum_{\substack{1\le i_{1}< i_{2}\le n\\i_{1},i_{2}\neq i}}h_{\Delta}(X_i, X_{i_1}, X_{i_2})/\binom{n-1}{2}$. Since $h_{\blacktriangle}(X_i, X_{i_1}, X_{i_2})$ and $h_{\Delta}(X_i, X_{i_1}, X_{i_2}) \preceq \rho_n^3$,  we have $A^{(1)}_{i}$ and $ A^{(2)}_{i} \preceq \rho_n^3$. Note that the randomness of $A^{(1)}_{i}$ and $A^{(2)}_{i}$ comes from $W$. By Lemma~\ref{applemma2}, we have
\begin{align*}
&\frac{2}{n(n-1)\widehat{V}_n}\sum_{1 \le i \neq j \le n}A_i\widehat{\Theta}^{(1)}_{ij}\eta_3(A_{ij})\\
=& \frac{1}{n(n-1)\widehat{V}_n^2} \sum_{1 \le i \neq j \le n} A_i^{(1)} \widehat{\Theta}^{(1)}_{ij}\eta_3(A_{ij}) - \frac{\widehat{U}_n}{n(n-1)\widehat{V}_n^3} \sum_{1 \le i \neq j \le n} A_i^{(2)} \widehat{\Theta}^{(1)}_{ij}\eta_3(A_{ij})\\
=& \wtO_{p,2}(n^{-1}\,\rho^{-\frac{1}{2}}_n\,\log^{\frac{1}{2}}n).
\end{align*}
Similarly, we can bound the remaining terms in (\ref{A(hatA-A)}), which leads to
\begin{align}
\label{finalA(hatA-A)}
\frac{2}{n}\sum_{i=1}^n(\widehat{A}_i-A_i)A_i=\wtO_{p,2}(n^{-1}\,\rho^{-\frac{1}{2}}_n\,\log^{\frac{1}{2}}n)+\wtO_{p,1}(M(n,s_n,\rho_n)).
\end{align}
We further decompose the last term $\sum_{i=1}^{n}A_i^2$: 
$$
\sum_{i=1}^{n}A_i^2=\sum_{i=1}^n(\widehat{a}_i+A_{i}-\widehat{a}_i)^2=\sum_{i=1}^{n}\widehat{a}_i^2+2\sum_{i=1}^{n}(A_{i}-\widehat{a}_i)\widehat{a}_i+\sum_{i=1}^{n}(A_{i}-\widehat{a}_i)^2,
$$
where
$$
A_{i}-\widehat{a}_i=\frac{1}{\binom{n-1}{2}}(\frac{1}{\widehat{V}_n}-\frac{1}{V_n})\sum_{\substack{1\le i_{1}< i_{2}\le n\\i_{1},i_{2}\neq i}}h_{\blacktriangle}(X_{i},X_{i_1},X_{i_2})-\frac{1}{\binom{n-1}{2}}(\frac{\widehat{U}_n}{\widehat{V}^2_n}-\frac{U_n}{V^2_n})\sum_{\substack{1\le i_{1}< i_{2}\le n\\i_{1},i_{2}\neq i}}h_{\Delta}(X_{i},X_{i_1},X_{i_2}).
$$
First, to bound $A_{i}-\widehat{a}_i$, we estimate $\dfrac{1}{\widehat{V}_n}-\dfrac{1}{V_n}$ and $\dfrac{\widehat{U}_n}{\widehat{V}^2_n}-\dfrac{U_n}{V^2_n}$ using earlier results:
$$
\dfrac{1}{\widehat{V}_n}-\dfrac{1}{V_n}=\frac{V_n-\widehat{V}_n}{V_n \widehat{V}_n}=\wtO_{p,2}(n^{-1}\,\rho^{-\frac{7}{2}}_n\,\log^{\frac{1}{2}}n + \rho_n^{-3} n^{-\frac{1}{2}} M(n, s_n, \rho_n)),
$$
$$
\frac{\widehat{U}_n}{\widehat{V}^2_n}-\frac{U_n}{V^2_n}=\frac{(\widehat{U}_n-U_n) V_n^2+U_n(V_n-\widehat{V}_n)(V_n+\widehat{V}_n)}{\widehat{V}_n^2 V_n^2}=\wtO_{p,2}(n^{-1}\,\rho^{-\frac{7}{2}}_n\,\log^{\frac{1}{2}}n + \rho_n^{-3} n^{-\frac{1}{2}} M(n, s_n, \rho_n)).
$$
Given that $h_{\blacktriangle}(X_{i},X_{i_1},X_{i_2}) \preceq \rho_n^3$ and $h_{\Delta}(X_{i},X_{i_1},X_{i_2})\preceq \rho_n^3$, we have
\begin{align}
\label{A-abound}
A_{i}-\widehat{a}_i=\wtO_{p,2}(n^{-1}\,\rho^{-\frac{1}{2}}_n\,\log^{\frac{1}{2}}n +n^{-\frac{1}{2}} M(n, s_n, \rho_n)),
\end{align}
holds uniformly for any $i$.
Given that $\widehat{a}_i \preceq 1$, it follows that
\begin{align}
\frac{1}{n}\sum_{i=1}^{n}A_i^2&=\frac{1}{n}\sum_{i=1}^{n}\widehat{a}_i^2+\frac{2}{n}\sum_{i=1}^{n}(A_{i}-\widehat{a}_i)\widehat{a}_i+\frac{1}{n}\sum_{i=1}^{n}(A_{i}-\widehat{a}_i)^2 \nonumber\\
&=\frac{1}{n}\sum_{i=1}^{n}\widehat{a}_i^2+\widetilde{O}_{p}(n^{-1}\,\rho^{-\frac{1}{2}}_n\,\log^{\frac{1}{2}}n).\label{Ahata} 
\end{align}
Finally, the combination of (\ref{lemma13decompAwidehat}), (\ref{final(hatA-A)2}), (\ref{finalA(hatA-A)}) and (\ref{Ahata}) yields the following results:
\begin{align}
\frac{1}{n}\sum_{i=1}^{n}\widehat{A}_i^2&=\widetilde{O}_{p}(n^{-1}\,\rho^{-1}_n\,\log\,n)+\wtO_{p,1}(M(n,s_n,\rho_n))+\frac{1}{n}\sum_{i=1}^{n}A_i^2 \nonumber\\
&=\wtO_{p,1}(M(n,s_n,\rho_n))+\frac{1}{n}\sum_{i=1}^{n}\widehat{a}_i^2. \label{lemma13final}
\end{align}
It follows that
\[
\widehat{\delta}_n=
\dfrac{\dfrac{1}{n}\sum\limits_{i=1}^{n}\widehat{A}_i^2-\dfrac{1}{n}\sum\limits_{i=1}^{n}\widehat{a}_i^2}{{n\sigma^2_n}/{9}}=
\wtO_{p,1}(M(n,s_n,\rho_n)),
\]
which completes the proof of Lemma~\ref{lemma} (4).

\subsection{Proof of Lemma~\ref{lemma}~(5)}
\label{sec:appendix-Lemma5}

Next, we bound the relative difference between the intermediate term and the variance $\delta_n :={(\widehat{\sigma}^2_n-\sigma_n^2)}/{\sigma^2_n}.$ Define
\begin{align*}
a_i=\frac{1}{\binom{n-1}{2}}\sum_{\substack{1\le i_{1}< i_{2}\le n\\i_{1},i_{2}\neq i}}\left(\frac{h_{\blacktriangle}(X_{i},X_{i_1},X_{i_2})}{v_n}-
\frac{h_{\Delta}(X_{i},X_{i_1},X_{i_2})}{v^2_n}u_n\right),
\end{align*}
and \[\widetilde{a}_i = \frac{\E[h_{\blacktriangle}(X_{i},X_{i_1},X_{i_2})|X_i]}{v_n} -\frac{\E[h_{\Delta}(X_{i},X_{i_1},X_{i_2})|X_i]u_n}{v^2_n}=\frac{g_1(X_i)}{v_n} - \frac{u_n f_1(X_i)}{v_n^2}:=q_1(X_i).\]
We decompose ${n\widehat{\sigma}^2_n}/{9}$ as follows,
\begin{align}
\dfrac{n\widehat{\sigma}^2_n}{9}=\frac{1}{n}\sum_{i=1}^n\widehat{a}_i^2=\frac{1}{n}\sum_{i=1}^{n}(a_i-\widetilde{a}_i+\widetilde{a}_i)^2+\frac{2}{n}\sum_{i=1}^{n}(\widehat{a}_i-a_{i})(a_i-\widetilde{a}_i+\widetilde{a}_i)+\frac{1}{n}\sum_{i=1}^{n}(\widehat{a}_i-a_{i})^2,\label{eq: decompose sigma hat}
\end{align}
where 
\begin{align*}
\widehat{a}_i-a_i=\frac{1}{\binom{n-1}{2}}(\frac{1}{V_n}-\frac{1}{v_n})\sum_{\substack{1\le i_{1}< i_{2}\le n\\i_{1},i_{2}\neq i}}h_{\blacktriangle}(X_{i},X_{i_1},X_{i_2})-\frac{1}{\binom{n-1}{2}}(\frac{U_n}{V^2_n}-\frac{u_n}{v^2_n})\sum_{\substack{1\le i_{1}< i_{2}\le n\\i_{1},i_{2}\neq i}}h_{\Delta}(X_{i},X_{i_1},X_{i_2}).
\end{align*}
For $1\le i \le n$, by Hoeffding's decomposition and Lemma~\ref{applemma1}, we have 
\begin{align}
&\frac{1}{\binom{n-1}{2}}\sum\limits_{\substack{1 \le i_1 < i_2 \le n\\i_1 , i_2\neq i}}h_{\blacktriangle}(X_{i},X_{i_1},X_{i_2})\nonumber\\
=&u_n + g_1(X_i)+\frac{2}{n-1}\sum_{\substack{1 \le j \le n\\ j\neq i}}g_2(X_i, X_j)+\frac{2}{\binom{n-1}{2}}\sum_{\substack{1 \le j_1 < j_2 \le n\\ j_1, j_2 \neq i}}g_3(X_i, X_{j_1},X_{j_2})\nonumber\\
=&u_n + g_1(X_i)+\frac{2}{n-1}\sum_{\substack{1 \le j \le n\\ j\neq i}}g_2(X_i, X_j)+\wtO_{p,2}(n^{-1} \rho_n^3 \log\,n) \label{lemma:blackhde higher order}\\
=& u_n + g_1(X_i)+\wtO_{p,2}(n^{-1/2}\rho_n^3\,\log^{1/2}n).\label{lemma:blackhde}
\end{align}
Similarly, we have 
\begin{align}
&\frac{1}{\binom{n-1}{2}}\sum\limits_{\substack{1 \le i_1 < i_2 \le n\\i_1 , i_2\neq i}}h_{\Delta}(X_{i},X_{i_1},X_{i_2})\nonumber\\
=&v_n + f_1(X_i)+\frac{2}{n-1}\sum_{\substack{1 \le j \le n\\ j\neq i}}f_2(X_i, X_j)+\wtO_{p,2}(n^{-1}\rho_n^3  \log\,n) \label{lemma:hde higher order}\\
=& v_n + f_1(X_i)+\wtO_{p,2}(n^{-1/2}\rho_n^3\,\log^{1/2}n).\label{lemma:hde}
\end{align}
Given $U_n-u_n$ and $V_n-v_n=\wtO_{p,2}(n^{-\frac{1}{2}}\,\rho_n^3\,\log^{\frac{1}{2}}n)$ in Lemma~\ref{lemma: U and V decomposition} and $U_n \asymp V_n \asymp \rho_n^3$, it follows that
\begin{align}
\frac{1}{V_n} - \frac{1}{v_n}& = \frac{v_n - V_n}{V_n v_n} =\frac{v_n - V_n}{v^2_n} + \wtO_{p,2}(n^{-1} \rho_n^{-3}\log n)\nonumber\\
& = -\frac{1}{v^2_n}\cdot \frac{3}{n }\sum_{i=1}^n f_1(X_i) + \wtO_{p,2}(n^{-1} \rho_n^{-3}\log n) = \wtO_{p,2}(n^{-1/2} \rho_n^{-3}\log^{1/2} n),\label{eq: v inverse diff}
\end{align}
and
\begin{align}
\frac{U_n}{V^2_n} - \frac{u_n}{v^2_n}&= \frac{U_n v_n^2 - V_n^2 u_n}{V_n^2 v_n^2}= \frac{2u_n}{v_n^3} (v_n - V_n) + \frac{1}{v_n^2}(U_n - u_n) + \wtO_{p,2}(n^{-1} \rho_n^{-3}\log n)\nonumber\\
& = -\frac{2u_n}{v^3_n}\cdot \frac{3}{n }\sum_{i=1}^n f_1(X_i) +\frac{1}{v^2_n}\cdot \frac{3}{n }\sum_{i=1}^n g_1(X_i) +\wtO_{p,2}(n^{-1} \rho_n^{-3}\log n)\nonumber\\
& = \wtO_{p,2}(n^{-1/2} \rho_n^{-3}\log^{1/2} n). \label{eq: u v inverse diff}
\end{align}
By combining (\ref{lemma:blackhde}), (\ref{lemma:hde}), (\ref{eq: v inverse diff}), and (\ref{eq: u v inverse diff}), it follows that 
\begin{align}
	\widehat{a}_i-a_i=&  -\frac{1}{v^2_n}\cdot \frac{3}{n }\sum_{j=1}^n f_1(X_j)  \left( u_n + g_1(X_i)\right)\nonumber \\
	 &\ + \left( \frac{2u_n}{v^3_n}\cdot \frac{3}{n }\sum_{j=1}^n f_1(X_j) -\frac{1}{v^2_n}\cdot \frac{3}{n }\sum_{j=1}^n g_1(X_j) \right) \left(v_n + f_1(X_i) \right)  + \wtO_{p,2}(n^{-1} \log n)\nonumber\\
	 = & -\frac{3}{n } \sum_{j=1}^n q_1(X_j) - q_1(X_i) \frac{3}{n } \sum_{j=1}^n p_1(X_j)-p_1(X_i) \frac{3}{n } \sum_{j=1}^n q_1(X_j)+ \wtO_{p,2}(n^{-1} \log n)\nonumber\\
	 =&\wtO_{p,2}(n^{-1/2} \log^{1/2} n),\label{lemma4ahat-a}
\end{align}
uniformly for any $i$. By (\ref{lemma:blackhde higher order}) and (\ref{lemma:hde higher order}), we have
\begin{align}
\label{aitildea1a2}
	a_i-\widetilde{a}_i = & \frac{1}{v_n} \left(u_n + g_1(X_i)+\frac{2}{n-1}\sum_{\substack{1 \le j \le n\\ j\neq i}}g_2(X_i, X_j)\right)\\
	& - \frac{u_n}{v_n^2} \left(v_n + f_1(X_i)+\frac{2}{n-1}\sum_{\substack{1 \le j \le n\\ j\neq i}}f_2(X_i, X_j) \right) -q_1(X_i)+\wtO_{p,2}(n^{-1}  \log\,n)\\
	= & \frac{2}{n-1}\sum_{\substack{1 \le j \le n\\ j\neq i}}q_2(X_i, X_j)+\wtO_{p,2}(n^{-1}  \log\,n)\\
	= & \wtO_{p,2}(n^{-1/2} \log^{1/2} n),
\end{align}
uniformly for any $i$.
Again, for any $1 \le i \le n$, the constants involved in the above high probability statements are independent of the realization $W$ and the sample size $n$.  Then we have $\sum\limits_{i=1}^{n}(\widehat{a}_i-a_{i})^2 / n=\wtO_{p,1}(n^{-1}\,\log\,n)$,  $\sum\limits_{i=1}^{n}(a_i-\widetilde{a}_i)^2 / n=\wtO_{p,1}(n^{-1}\,\log\,n)$, and $\sum\limits_{i=1}^{n}(\widehat{a}_i-a_{i})(a_i-\widetilde{a}_i) / n=\wtO_{p,1}(n^{-1}\,\log\,n)$.
By plugging them into (\ref{eq: decompose sigma hat}), we have 
\begin{align*}
	\dfrac{n\widehat{\sigma}^2_n}{9}& =\frac{1}{n}\sum_{i=1}^{n}(a_i-\widetilde{a}_i+\widetilde{a}_i)^2+\frac{2}{n}\sum_{i=1}^{n}(\widehat{a}_i-a_{i})(a_i-\widetilde{a}_i+\widetilde{a}_i)+\frac{1}{n}\sum_{i=1}^{n}(\widehat{a}_i-a_{i})^2\\
	& = \frac{1}{n}\sum_{i=1}^{n}\widetilde{a}_i^2 + \frac{2}{n}\sum_{i=1}^{n}(a_i-\widetilde{a}_i)\widetilde{a}_i+\frac{2}{n}\sum_{i=1}^{n}(\widehat{a}_i-a_{i})\widetilde{a}_i+\wtO_{p,1}(n^{-1}\,\log\,n)\\
	& = \frac{1}{n}\sum_{i=1}^{n}q_1^2(X_i) + \frac{4}{n(n-1)}\sum_{\substack{1 \le i \neq j \le n}}q_2(X_i, X_j)q_1(X_i)- \frac{6}{n^2}\left(\sum_{i=1}^{n}q_1(X_i)\right)^2\\
	& \ \ \ \  - \frac{6}{n^2}\sum_{i=1}^{n}  q_1^2(X_i)  \sum_{j=1}^n p_1(X_j)-\frac{6}{n^2}\sum_{i=1}^{n}p_1(X_i)q_1(X_i) \sum_{j=1}^n q_1(X_j)+\wtO_{p,1}(n^{-1}\,\log\,n).
\end{align*}
Furthermore, by Lemma~\ref{applemma1}, we have $\sum_{i=1}^{n}q_1(X_i)/n = \wtO_{p,2}(n^{-1/2}\,\log^{1/2}\,n)$ and $\sum_{i=1}^{n}p_1(X_i)/n = \wtO_{p,2}(n^{-1/2}\,\log^{1/2}\,n)$, and by Lemma~\ref{applemma2}, we have $\sum_{i=1}^{n}  q_1^2(X_1)/n=\E[q_1^2(X_i)]+\wtO_{p,2}(n^{-1/2}\,\log^{1/2}\,n)$ and $\sum_{i=1}^{n}p_1(X_i)q_1(X_i)/n = \E[p_1(X_1)q_1(X_1)]+\wtO_{p,2}(n^{-1/2}\,\log^{1/2}\,n)$. Therefore, we can further simplify the decomposition for $n\widehat{\sigma}^2_n/9$:
\begin{align*}
	\dfrac{n\widehat{\sigma}^2_n}{9}& = \frac{1}{n}\sum_{i=1}^{n}q_1^2(X_i) + \frac{4}{n(n-1)}\sum_{\substack{1 \le i \neq j \le n}}q_2(X_i, X_j)q_1(X_i)\nonumber\\
	& \ \ \ \  - \frac{6\xi^2_1}{n}  \sum_{j=1}^n p_1(X_j)-\frac{6}{n}\E[p_1(X_1)q_1(X_1)] \sum_{j=1}^n q_1(X_j)+\wtO_{p,1}(n^{-1}\,\log\,n).
\end{align*}
By applying Hoeffding's decomposition, it follows that $$\frac{2}{n(n-1)}\sum_{1\le i\neq j\le n}q_1(X_i)q_2(X_i,X_j)=\frac{2}{n}\sum_{i=1}^ng_{\delta,1}(X_i)+\wtO_{p,2}(n^{-1}\log n),$$
where $g_{\delta,1}(x)=\E\Big[q_1(X_2)q_2(X_1,X_2)\vert X_1=x\Big].$ 
Therefore, we have
\begin{align*}
\delta_n=\frac{\widehat{\sigma}^2_n-\sigma^2_n}{\sigma^2_n}&=\frac{1}{n\xi_{1}^2}\sum_{i=1}^{n}(q_{1}^2(X_{i})-\xi_{1}^2)+\frac{4}{n\xi_{1}^2}\sum_{i=1}^{n}g_{\delta,1}(X_i)\\
&-\frac{6}{n}\sum_{i=1}^n p_1(X_i)-\frac{6}{n\xi_1^2}\E[p_1(X_1)q_1(X_1)]\sum_{i=1}^n q_1(X_i) +\wtO_{p,1}(n^{-1}\,\log\,n).
\end{align*}
Here, we have completed the proof of Lemma~\ref{lemma}.

\subsection{Proof of Lemma~\ref{lemma: U and V decomposition}} \label{subsec: proof U and V decomposition}

\begin{proof}[of Lemma~\ref{lemma: U and V decomposition}~(1)]
	Based on Hoeffding's decomposition \citep{hoeffding1992class}, we have \begin{align*}
&U_{n}-u_{n} \\
=&\frac{1}{\binom{n}{3}}\sum_{1\le i_{1}< i_{2} < i_{3} \le n}h_{\blacktriangle}(X_{i_{1}},X_{i_{2}},X_{i_{3}})-u_n\\
=&\underbrace{\frac{3}{n} \sum_{i=1}^n g_1\left(X_i\right)}_{\mathcal{U}_n^{(1)}}+\underbrace{\frac{6}{n(n-1)} \sum_{1 \leq i<j \leq n} g_2\left(X_i, X_j\right)}_{\mathcal{U}_n^{(2)}}+\underbrace{\frac{6}{n(n-1)(n-2)}\sum_{1\le i_{1} < i_{2} < i_{3} \le n} g_{3}(X_{i_{1}},X_{i_{2}},X_{i_{3}})}_{\mathcal{U}_n^{(3)}},
\end{align*}
where $g_{1}(x_{1}):=\E[h_{\blacktriangle}(X_{1},X_{2},X_{3})|X_{1}=x_{1}]-u_{n}$, 
$
g_{2}(x_{1},x_{2}):=\E[h_{\blacktriangle}(X_{1},X_{2},X_{3})|X_{1}=x_{1},X_{2}=x_{2}]-u_{n}-\sum_{1\le i \le 2}g_{1}(x_{i})
$, and 
$
g_{3}(x_{1},x_{2},x_{3}):=\E[h_{\blacktriangle}(X_{1},X_{2},X_{3})|X_{1}=x_{1},X_{2}=x_{2},,X_{3}=x_{3}]-u_{n}-\sum_{k'=1}^{2}\sum_{1 \le i_{1} < 
\cdots < i_{k'} \le 3}g_{k'}(x_{i_{1}}\cdots x_{i_{k'}})
$. 
We have $g_{1}(x_{1})\preceq \rho_n^3$ due to $u_n \preceq \rho_n^3$ and $$h_{\blacktriangle}(X_{i_1},X_{i_2},X_{i_3})=\E[h_1(A_{i_1,i_2,i_3})+h_3(A_{i_1,i_2,i_3})\mid X_{i_1},X_{i_2},X_{i_3}]\preceq \rho_n^3.$$
Similarly, we have $g_2(x_1,x_2)\preceq \rho_n^3$ and $g_{3}(x_{1},x_{2},x_{3})\preceq \rho_n^3$. By Lemma~\ref{applemma1}, each term of Hoeffding's decomposition for $U_n-u_n$ can be bounded as follow:
\[\mathcal{U}_n^{(1)} = \wtO_{p,2}(\rho^3_n\,n^{-\frac{1}{2}}\,\log^{\frac{1}{2}}n), \ \ \mathcal{U}_n^{(2)} = \wtO_{p,2}(\rho^3_n\,n^{-1}\,\log n), \text{ and }\mathcal{U}_n^{(3)} = \wtO_{p,2}(\rho^3_n\,n^{-\frac{3}{2}}\,\log^{\frac{3}{2}}n).\]
Similarly, denoting $\frac{3}{n}\sum_{i=1}^{n}f_{1}(X_{i})$ and $\frac{6}{n(n-1)}\sum_{1\le i < j \le n}f_{2}(X_{i},X_{j})$ as the linear and quadratic terms in the Hoeffding's decomposition for $V_n-v_n$, we have
\begin{align}
V_{n}-v_{n}=&\frac{1}{\binom{n}{3}}\sum_{1\le i_{1}< i_{2} <i_{3} \le n}h_{\Delta}(X_{i_{1}},X_{i_{2}},X_{i_{3}})-v_n \nonumber\\
=&\underbrace{\frac{3}{n} \sum_{i=1}^n f_1\left(X_i\right)}_{\mathcal{V}_n^{(1)}}+\underbrace{\frac{6}{n(n-1)} \sum_{1 \leq i<j \leq n} f_2\left(X_i, X_j\right)}_{\mathcal{V}_n^{(2)}}+\wtO_{p,2}(\rho^3_n\,n^{-\frac{3}{2}}\,\log^{\frac{3}{2}}n) \nonumber
\end{align}
where $\mathcal{V}_n^{(1)} = \wtO_{p,2}(\rho_n^3 n^{-\frac{1}{2}} \log ^{\frac{1}{2}} n)$ and $\mathcal{V}_n^{(2)} = \wtO_{p,2}(\rho_n^3 n^{-1} \log n)$.
\end{proof}

\medskip
\begin{proof}[of Lemma~\ref{lemma: U and V decomposition}~(2)]
	To derive the decomposition of $\widehat{U}_n - U_n$ and $\widehat{V}_n - V_n$, it suffices to analyze $\frac{1}{\binom{n}{3}}\sum_{1 \le i_1 < i_2 < i_3 \le n}[h(A_{i_1, i_2, i_3})-h(X_{i_1},X_{i_2},X_{i_3})]$, with the function $h$ being any of $h_1$, $h_3$ and $h_{\Delta}$. 
	The following lemma provides the decomposition for each term.
	\medskip 
	\begin{lemma}\label{apppro1} Assuming the conditions in Lemma~\ref{lemma}, we have the following decompositions:
	\begin{align*}
		\dfrac{1}{\binom{n}{3}}\sum\limits_{1 \le i_1 < i_2 < i_3 \le n}[h_1(A_{i_1, i_2, i_3})-h_1(X_{i_1},X_{i_2},X_{i_3})]& = C_{n}^{(1)}+\wtO_{p,2}(\rho_n^{\frac{3}{2}}n^{-\frac{3}{2}}   \log^{\frac{1}{2}} n),\\
		\dfrac{1}{\binom{n}{3}}\sum\limits_{1 \le i_1 < i_2 < i_3 \le n}[h_3(A_{i_1, i_2, i_3})-h_3(X_{i_1},X_{i_2},X_{i_3})]& =B_n+C_{n}^{(2)}+\wtO_{p,2}(s_n \rho_n^{\frac{3}{2}}n^{-\frac{3}{2}}  \log^{2} n),\\
		\dfrac{1}{\binom{n}{3}}\sum\limits_{1 \le i_1 < i_2 < i_3 \le n}[h_{\triangle}(A_{i_1, i_2, i_3})-h_{\triangle}(X_{i_1},X_{i_2},X_{i_3})]& =D_n+\wtO_{p,2}(\rho_n^{\frac{3}{2}}n^{-\frac{3}{2}}   \log^{\frac{1}{2}} n),
	\end{align*}
	where we define
	\begin{align*}
	C_n^{(1)} & =\frac{1}{\binom{n}{2}}\sum_{1\le k_1 < k_2 \le n}\mathring{C}_{k_1 k_2}^{(1)}\eta_{3}(A_{k_1 k_2})\text{\ with }
\mathring{C}_{k_1 k_2}^{(1)}=\frac{3}{n-2}\sum_{\substack{1\le i\le n\\i\neq k_1,k_2}}c_{i\,k_1}c_{i\,k_2}, \\
C_n^{(2)} & =\frac{1}{\binom{n}{2}}\sum_{1\le k_1 < k_2 \le n}\mathring{C}_{k_1 k_2}^{(2)}\eta_{3}(A_{k_1 k_2})\text{\ with }
\mathring{C}_{k_1 k_2}^{(2)}=\frac{3}{n-2}\sum_{\substack{1\le i\le n\\i\neq k_1,k_2}}b_{i\,k_1}b_{i\,k_2}, \\
B_n & =\frac{1}{\binom{n}{2}}\sum_{1\le k_1 < k_2 \le n}\mathring{B}_{k_1 k_2}\eta_{2}(A_{k_1 k_2})\text{\ with }
\mathring{B}_{k_1 k_2}=\frac{3}{n-2}\sum_{\substack{1\le i\le n\\i\neq k_1,k_2}}(c_{i\,k_1}b_{i\,k_2}+b_{i\,k_1}c_{i\,k_2}),\\
	D_n & =\frac{1}{\binom{n}{2}}\sum_{1\le k_1 < k_2 \le n}\mathring{D}_{k_1 k_2}\eta_{1}(A_{k_1 k_2})\text{\ with } 
\mathring{D}_{k_1 k_2}=-\frac{3}{n-2}\sum_{\substack{1\le i\le n\\i\neq k_1,k_2}}(c_{i\,k_1}+b_{i\,k_1})(b_{i\,k_2}+c_{i\,k_2}).
\end{align*}
In addition, we have $C_n^{(1)} = \wtO_{p,2}(\rho_n^{\frac{5}{2}}\,n^{-1}\,\log^{\frac{1}{2}}n)$, $C_n^{(2)} = \wtO_{p,2}(s_n^{2}\,\rho_n^{\frac{5}{2}}\,n^{-1}\,\log^{\frac{1}{2}}n)$, $B_n = \wtO_{p,2}(s_n^{\frac{3}{2}}\,\rho_n^{\frac{5}{2}}\,n^{-1}\,\log^{\frac{1}{2}}n)$, and $D_n = \wtO_{p,2}(\rho_n^{\frac{5}{2}}\,n^{-1}\,\log^{\frac{1}{2}}n)$.
\end{lemma}
\medskip
Lemma~\ref{lemma: U and V decomposition}~(2) is a direct result of the above lemma by combining the decomposition of $h_1$ and $h_3$. The proof of Lemma~\ref{apppro1} is given in Section~\ref{subsec: proof of apppro1}.
\end{proof}

\subsection{Proof of Lemma~\ref{apppro1}} \label{subsec: proof of apppro1}
We start by proving the decomposition for $h=h_1$ or $h_\Delta$, as these two cases are similar and more straightforward. Their proof provides an overview of the technique, which is then extended to the case where $h=h_3$. 

\noindent \underline{\bf Proof of Lemma~\ref{apppro1} when $h=h_1$ or $h_\Delta$}:
To provide a unified notation for the cases where $h=h_1$ or $h_\Delta$, we define $\eta_{ij}=\eta_{3}(A_{ij})$ and $\alpha_{ij}=c_{ij}$ when $h=h_1$; and $\eta_{ij}=\eta_{1}(A_{ij})$ and $\alpha_{ij}=c_{ij}+b_{ij}$ when $h=h_\Delta$ for $1 \le i < j \le n$.
For both cases, by Lemma~\ref{lemma: h property}, we have $\eta_{ij}=\eta_{ji},\, \E[\eta_{ij}]=0,\, \E[|\eta_{ij}|]\asymp \rho_n, \, \E[\eta_{ij}^2]\asymp \rho_n$, and $\E[|\eta_{ij}^3|]\asymp \rho_n$. In addition, we have $\alpha_{ij}=\alpha_{ji}$ and $\alpha_{ij} \asymp \rho_n$.
Therefore, it suffices to prove for a unified $h_{i_1 i_2 i_3}:=(\alpha_{i_1 i_2}+\eta_{i_1 i_2})(\alpha_{i_1 i_3}+\eta_{i_1 i_3})(\alpha_{i_2 i_3}+\eta_{i_2 i_3})$ with the above properties. Note that 
$\E[h_{i_1 i_2 i_3}\mid X_{i_1},X_{i_2},X_{i_3}] = \alpha_{i_1 i_2} \alpha_{i_1 i_3}\alpha_{i_2 i_3}$. 
It follows that
\begin{align}
\sum\limits_{1\le i_1 < i_2 < i_3 \le n}h_{i_1 i_2 i_3}=\sum\limits_{1\le i_1 < i_2 < i_3 \le n}\E[h_{i_1 i_2 i_3}\mid X_{i_1},X_{i_2},X_{i_3}]+\sum_{1\le k_1 < k_2 \le n}\Theta_{k_1 k_2} \eta_{k_1 k_2}+R_1+R_2,
\nonumber
\end{align}
where we define
$
\Theta_{k_1 k_2}=\sum_{\substack{1\le i \le n\\ i \neq k_1,k_2}}\alpha_{i k_1} \alpha_{i k_2}$, 
\[R_1=\sum\limits_{\substack{1\le k_1 < k_2 < k_3 \le n\\or 1\le k_2 < k_1 < k_3 \le n\\or 1\le k_2 < k_3 < k_1 \le n}}\alpha_{k_2 k_3}\eta_{k_1 k_2}\eta_{k_1 k_3},
\quad \text{ and } \quad R_2=\sum\limits_{1 \le k_1 < k_2 < k_3 \le n}\eta_{k_1 k_2}\eta_{k_1 k_3}\eta_{k_2 k_3}.\]

Next, we apply a variant of Berstein inequality in Lemma~\ref{applemma2} to bound the linear terms and apply Lemma~\ref{hyperth} (Theorem 1.3 of \cite{schudy2011bernstein}) to bound the remainder terms $R_i$. 

For the linear term, we have 
\[\frac{1}{\binom{n}{3}}\sum_{1\le k_1 < k_2 \le n}\Theta_{k_1 k_2} \eta_{k_1 k_2} = \frac{1}{\binom{n}{2}}\sum_{1\le k_1 < k_2 \le n}\frac{3}{n-2}\Theta_{k_1 k_2} \eta_{k_1 k_2}.\]
Given that $3\Theta_{k_1 k_2}/(n-2)\asymp \rho_n^2$ and $\E[\eta_{ij}^2]\asymp \rho_n$, by Lemma~\ref{applemma2}, we have \[\frac{1}{\binom{n}{3}}\sum_{1\le k_1 < k_2 \le n}\Theta_{k_1 k_2} \eta_{k_1 k_2} = \wtO_{p,2}(n^{-1}\rho_n^{5/2} \log^{1/2}n ).\] 

For each remainder term, we first need to define the hypergraph associated with the multilinear polynomial $f$ over a set of random variables $Y$, then bound the corresponding $\var(f(Y))$ and $\Xi_{\mathfrak{r}}$.
\begin{enumerate}
	\item For $R_1=\sum\limits_{\substack{1\le k_1 < k_2 < k_3 \le n\\or 1\le k_2 < k_1 < k_3 \le n\\or 1\le k_2 < k_3 < k_1 \le n}}\alpha_{k_2 k_3}\eta_{k_1 k_2}\eta_{k_1 k_3}$, we define the hypergraph as follows. We define the node set as $\mathcal{V}(H)=\{\eta_{ij}|i\neq j,\,i\in [n] , j\in [n]\}$, where $\eta_{ij}$ and $\eta_{ji}$ are the same random variable, the hyperedge set as $\mathcal{H}(H)=\{\mathfrak{h}=(\eta_{is},\eta_{it})|i\neq s,s\neq t,i \neq t, i,s,t \in [n]\}$, and the weight on the hyperedge $(\eta_{is},\eta_{it})$ as $\mathfrak{W}_{\mathfrak{h}}=\alpha_{k_2 k_3} \asymp \rho_n$. The condition~(\ref{hyperineqcondition}) in Lemma~\ref{hyperth} holds with $L=1$ because $\E[|\eta_{ij}|^k] \le \E[|\eta_{ij}|^{k-1}] \le k \E[|\eta_{ij}|^{k-1}]$. Next, we bound  \[\Xi_{\mathfrak{r}}=\max\limits_{S\subset[N]:|S|=\mathfrak{r}}
\sum\limits_{\substack{\mathfrak{h}\in \mathcal{H}(H): \\ S\subseteq \mathcal{V}(\mathfrak{h})}}\,
|\mathfrak{W}_{\mathfrak{h}}|
\prod\limits_{\mathfrak{v}\in 
\mathcal{V}(\mathfrak{h})\backslash
S}\E[|Y_{\mathfrak{v}}|], \] where
$\mathfrak{r}$ represents the number of nodes in $S$ and $1\le \mathfrak{r}\le 2$. We analyze each case below.

(1)  When $\mathfrak{r}=1$ and $|S|=1$, the summation is over $2(n-2)$ different hyperedges $\mathfrak{h}\in \mathcal{H}(H)$ such that $S\subseteq \mathcal{V}(\mathfrak{h})$. Note that, with one random variable chosen in $S$, there is only one random variable left in $\mathcal{V}(\mathfrak{h})$. Thus, we have $\prod\limits_{\mathfrak{v}\in 
\mathcal{V}(\mathfrak{h})\backslash
S}\E[|Y_{\mathfrak{v}}|]=\E[|\eta_{ij}|] \asymp \rho_n$ for some $(i,j)$. It follows that
$$
\Xi_{1}=\max\limits_{S\subset[N]:\mathfrak{r}=1}
\sum\limits_{\substack{\mathfrak{h}\in \mathcal{H}(H): \\ S\subseteq \mathcal{V}(\mathfrak{h})}}\,
|\mathfrak{W}_{\mathfrak{h}}|
\prod\limits_{\mathfrak{v}\in 
\mathcal{V}(\mathfrak{h})\backslash
S}\E[|Y_{\mathfrak{v}}|] \asymp 2(n-2)\cdot \rho_n \cdot \rho_n \asymp n\,\rho_n^2.
$$

(2) When $\mathfrak{r}=2$ and $|S|=2$, given $S$ fixed, the hyperedge $\mathfrak{h}$ such that $S\subseteq \mathcal{V}(\mathfrak{h})$ is fully determined and $\mathcal{V}(\mathfrak{h})\backslash S=\emptyset$. 
Therefore, we have
$$
\Xi_{2}=\max\limits_{S\subset[N]:\mathfrak{r}=2}
\sum\limits_{\substack{\mathfrak{h}\in \mathcal{H}(H): \\ S\subseteq \mathcal{V}(\mathfrak{h})}}\,
|\mathfrak{W}_{\mathfrak{h}}|
\prod\limits_{\mathfrak{v}\in 
\mathcal{V}(\mathfrak{h})\backslash
S}\E[|Y_{\mathfrak{v}}|] = \max\limits_{\substack{S\subset[N]:\mathfrak{r}=2\\ \mathfrak{h}\in \mathcal{H}(H): S\subseteq \mathcal{V} (\mathfrak{h})} }
|\mathfrak{W}_{\mathfrak{h}}|  \asymp \rho_n.
$$
The bound for the variance is  given by
\begin{align*}
&\var(R_1\mid W)\\
=&\sum\limits_{\substack{1\le k_1 < k_2 < k_3 \le n\\or 1\le k_2 < k_1 < k_3 \le n\\or 1\le k_2 < k_3 < k_1 \le n}}\alpha^2_{k_2 k_3}\var(\eta_{k_1 k_2})\var(\eta_{k_1 k_3})\asymp n^3\cdot \rho_n^2 \cdot \rho_n\cdot \rho_n\asymp n^3\, \rho_n^4.
\end{align*}
Here, the first equality holds because the covariance between dependent terms equals zero, \textit{i.e.}, $cov(\eta_{k_1k_2}\eta_{k_1k_3}, \eta_{k_1k_3}\eta_{k_1k_4} \mid W)=0$ for $k_2 \neq k_4$.	

By Lemma~\ref{hyperth}, there exists a universal constant $C>0$ such that 
\begin{align*}
&\pr(|\frac{R_1}{\binom{n}{3}}|\ge C_0 n^{-\frac{3}{2}} \rho_n^{2} \log^{\frac{1}{2}}n \mid W)\\
\le&e^2 \cdot \max \Bigg\{\exp\left\{-\frac{C_0^2 n^{-3} \rho^4_n \log n}{C^2\,\var(\frac{R_1}{\binom{n}{3}}|W)}\right\},\exp\left\{-\frac{C_0 n^{-\frac{3}{2}} \rho_n^{2} \log^{\frac{1}{2}}n}{C^2 \frac{\Xi_{1}}{\binom{n}{3}}}\right\},\\ 
& \quad \quad \quad \quad \exp\left\{-\left(\frac{C_0 n^{-\frac{3}{2}} \rho_n^{2} \log^{\frac{1}{2}}n}{C^2\,\frac{\Xi_2}{\binom{n}{3}}}\right)^{1/2}\right\} \Bigg\} \\
\le& e^2 \cdot \max\Bigg\{\exp\left\{-\frac{C_0^2 n^{-3} \rho^4_n \log n}{C^2 n^{-3} \rho_n^4}\right\},\exp\left\{-\frac{C_0 n^{-\frac{3}{2}} \rho_n^{2} \log^{\frac{1}{2}}n}{C^2\,n^{-2}\rho_n^2}\right\}, \\
& \quad \quad \quad \quad  \exp\left\{-\left(\frac{C_0 n^{-\frac{3}{2}}  \rho_n^{2} \log^{\frac{1}{2}}n}{C^2 n^{-3} \rho_n }\right)^{1/2}\right\}\Bigg\} \\
\le & e^2\cdot \exp\left\{-\log n\,\max \{\frac{C_0^2}{C^2}, (\frac{C_0}{C^2})^{1/2} \}\right\} \le \frac{e^2}{n^2},
\end{align*}
where the second last inequality holds as $\rho_n \succeq n^{-3/2} \log^{3/2}n$ and the last inequality holds by choosing $C_0 > \max (\sqrt{2}C, 4 C^2)$.
Note that the constants $C_0$ and $e$ are universal, and the above inequality holds for any realization $W$. Thus, we have $R_1 /\binom{n}{3} = \wtO_{p,2}(n^{-3/2} \rho_n^2 \log^{1/2}n)$. 

\item For $R_2=\sum\limits_{1 \le k_1 < k_2 < k_3 \le n}\eta_{k_1 k_2}\eta_{k_1 k_3}\eta_{k_2 k_3}$, we similarly define a hypergraph with the node set being $\mathcal{V}(H)=\{\eta_{ij}|i\neq j,\,i\in [n] , j\in [n]\}$ and the hyperedge set being $\mathcal{H}(H)=\{\mathfrak{h}=(\eta_{is},\eta_{it}, \eta_{st})|i\neq s,s\neq t,i \neq t, i,s,t \in [n]\}$. The weight on the hyperedge $\mathfrak{h}$ is $\mathfrak{W}_{\mathfrak{h}}=1$. Similarly, for $1\le \mathfrak{r}\le 3$, we bound  \[\Xi_{\mathfrak{r}}=\max\limits_{S\subset[N]:|S|=\mathfrak{r}}
\sum\limits_{\substack{\mathfrak{h}\in \mathcal{H}(H): \\ S\subseteq \mathcal{V}(\mathfrak{h})}}\,
\prod\limits_{\mathfrak{v}\in 
\mathcal{V}(\mathfrak{h})\backslash
S}\E[|Y_{\mathfrak{v}}|].\] 

(1)  When $\mathfrak{r}=1$ and $|S|=1$, the summation is over $3(n-2)$ different hyperedges $\mathfrak{h}\in \mathcal{H}(H)$ such that $S\subseteq \mathcal{V}(\mathfrak{h})$. Note that, with one random variable chosen in $S$, there are two random variables left in $\mathcal{V}(\mathfrak{h})$. Thus, we have $\prod\limits_{\mathfrak{v}\in 
\mathcal{V}(\mathfrak{h})\backslash
S}\E[|Y_{\mathfrak{v}}|]=\E[|\eta_{it}||\eta_{st}|] \asymp \rho_n^2$ for some $(i,s,t)$. It follows that
$$
\Xi_{1}=\max\limits_{S\subset[N]:\mathfrak{r}=1}
\sum\limits_{\substack{\mathfrak{h}\in \mathcal{H}(H): \\ S\subseteq \mathcal{V}(\mathfrak{h})}}\,
\prod\limits_{\mathfrak{v}\in 
\mathcal{V}(\mathfrak{h})\backslash
S}\E[|Y_{\mathfrak{v}}|] \asymp 2(n-2)\cdot 1 \cdot \rho_n^2 = n\,\rho_n^2.
$$

(2) When $\mathfrak{r}=2$ and $|S|=2$, with two random variables chosen in $S$, there is only one random variable left in $\mathcal{V}(\mathfrak{h})$ and its indices are determined.  
Therefore, we have
$$
\Xi_{2}=\max\limits_{S\subset[N]:\mathfrak{r}=2}
\sum\limits_{\substack{\mathfrak{h}\in \mathcal{H}(H): \\ S\subseteq \mathcal{V}(\mathfrak{h})}}\,
\prod\limits_{\mathfrak{v}\in 
\mathcal{V}(\mathfrak{h})\backslash
S}\E[|Y_{\mathfrak{v}}|] \asymp \rho_n.
$$

(3) When $\mathfrak{r}=3$ and $|S|=3$, given $S$ fixed, the hyperedge $\mathfrak{h}$ such that $S\subseteq \mathcal{V}(\mathfrak{h})$ is fully determined and $\mathcal{V}(\mathfrak{h})\backslash S=\emptyset$. 
Therefore, we have $\Xi_{3} = 1$.

Similarly, the bound for the variance is  given by
\begin{align*}
&\var(R_2\mid W)\\
=&\sum\limits_{\substack{1\le k_1 < k_2 < k_3 \le n\\or 1\le k_2 < k_1 < k_3 \le n\\or 1\le k_2 < k_3 < k_1 \le n}}\var(\eta_{k_1 k_2})\var(\eta_{k_1 k_3})\var(\eta_{k_2 k_3})\asymp  n^3\, \rho_n^3.
\end{align*}

By Lemma~\ref{hyperth}, there exists a universal constant $C>0$ such that 
\begin{align*}
&\pr(|\frac{R_2}{\binom{n}{3}}|\ge C_0 n^{-\frac{3}{2}} \rho_n^{\frac{3}{2}} \log^{\frac{1}{2}}n \mid W)\\
\le&e^2 \cdot \max \Bigg\{\exp\left\{-\frac{C_0^2 n^{-3} \rho^3_n \log n}{C^3\,\var(\frac{R_2}{\binom{n}{3}}|W)}\right\},\exp\left\{-\frac{C_0 n^{-\frac{3}{2}} \rho_n^{\frac{3}{2}} \log^{\frac{1}{2}}n}{C^3 \frac{\Xi_{1}}{\binom{n}{3}}}\right\},\\ 
& \quad \quad \quad \quad \exp\left\{-\left(\frac{C_0 n^{-\frac{3}{2}} \rho_n^{\frac{3}{2}} \log^{\frac{1}{2}}n}{C^3\,\frac{\Xi_2}{\binom{n}{3}}}\right)^{1/2}\right\}, \exp\left\{-\left(\frac{C_0 n^{-\frac{3}{2}} \rho_n^{\frac{3}{2}} \log^{\frac{1}{2}}n}{C^3\,\frac{\Xi_3}{\binom{n}{3}}}\right)^{1/3}\right\} \Bigg\} \\
\le& e^2 \cdot \max\Bigg\{\exp\left\{-\frac{C_0^2 n^{-3} \rho^3_n \log n}{C^3 n^{-3} \rho_n^3}\right\},\exp\left\{-\frac{C_0 n^{-\frac{3}{2}} \rho_n^{\frac{3}{2}} \log^{\frac{1}{2}}n}{C^3\,n^{-2}\rho_n^2}\right\}, \\
& \quad \quad \quad \quad  \exp\left\{-\left(\frac{C_0 n^{-\frac{3}{2}}  \rho_n^{\frac{3}{2}} \log^{\frac{1}{2}}n}{C^3 n^{-3} \rho_n }\right)^{1/2}\right\}, \exp\left\{-\left(\frac{C_0 n^{-\frac{3}{2}} \rho_n^{\frac{3}{2}} \log^{\frac{1}{2}}n}{C^3\,n^{-3}}\right)^{1/3}\right\} \Bigg\} \\
\le & e^2\cdot \exp\left\{-\log n\,\max \{\frac{C_0^2}{C^3}, \frac{C_0}{C^3}, (\frac{C_0}{C^3})^{1/2}, (\frac{C_0}{C^3})^{1/3}\}\right\} \le \frac{e^2}{n^2}.
\end{align*}
where the second last inequality holds as $\rho_n \succeq n^{-1} \log^{5/3} n$ and the last inequality holds by choosing $C_0 > \max ((2 C^3)^{1/2}, 8 C^3)$.
Note that the constants $C_0$ and $e$ are universal, and the above inequality holds for any realization $W$. Thus, we have $R_2 /\binom{n}{3} = \wtO_{p,2}(n^{-3/2} \rho_n^{3/2} \log^{1/2}n)$.
\end{enumerate}
Thus, we have shown that $(R_1+R_2) /\binom{n}{3} = \wtO_{p,2}(\rho_n^{3/2}  n^{-3/2} \log^{1/2}n)$, which completes the decomposition of $h=h_1$ and $h_\Delta$.

\medskip
\noindent \underline{\bf Proof of Lemma~\ref{apppro1} when $h=h_3$}:
The technique used to bound the remainder term relies on the independence of the node variables associated with the hyperedges. However, in our analysis of the decomposition of  $h=h_3$ below, the remainder term will involve both $\eta_{2}(A_{ij})$ and $\eta_{3}(A_{ij})$, which are dependent for the same tuple $(i,j)$. To resolve this, we introduce a new variable for sign distribution that is independent of the edge distribution. The following notations will be used going forward. 
Define 
\begin{align*}
	B_{ij}=
\begin{cases}
1,& \text{with probability 
$\bar{s}_{ij}:=1 - s_n G(X_{i},X_{j})$};\\
0,& \text{with probability $s_{ij}:=s_n G(X_{i},X_{j})$},
\end{cases}
\end{align*}
which is independent with $\{|A_{ij}|: 1 \le i \neq j \le n\}$ conditional on $W=(X_1, \cdots, X_n)$.
Define the shorthands $\rho_{ij} := \rho_n F(X_{i},X_{j})$ and $\bar{\rho}_{ij} := 1- \rho_n F(X_{i},X_{j})$, and  the centered variables
\begin{align}
\label{defmuomega}
\mu_{ij} := B_{ij} - \bar{s}_{ij}, \quad \omega_{ij} :=\eta_1(A_{ij})= |A_{ij}| - \rho_{ij}.
\end{align}
The following properties for the centered variables are straightforward to verify; the proof is omitted here.  
\begin{lemma}
\label{secondlayerrandomness}
Assuming the conditions in Lemma~\ref{lemma}, we have the following properties: 
\begin{enumerate}
\item $\eta_2(A_{ij}) \buildrel d \over = -\rho_{ij} \mu_{ij} + s_{ij}\omega_{ij} - \omega_{ij} \mu_{ij}$ and $\eta_3(A_{ij}) \buildrel d \over = \rho_{ij} \mu_{ij} + \bar{s}_{ij}\omega_{ij} + \omega_{ij} \mu_{ij}$.
\item $\E[\mu_{ij} \mid W] = 0$, $\E[|\mu_{ij}| \mid W] \asymp s_n$ and $\E[\mu_{ij}^2 \mid W] \asymp s_n$.
\item $\E[\omega_{ij} \mid W] = 0$, $\E[|\omega_{ij}| \mid W] \asymp \rho_n$ and $\E[\omega_{ij}^2 \mid W] \asymp \rho_n$.
\end{enumerate}
\end{lemma}

Following~(\ref{h13}), we expand $\sum\limits_{1 \le i_1 < i_2 < i_3 \le n}h_3(A_{i_1,i_2,i_3})$ as 
\begin{align*}
&\sum\limits_{1 \le i_1 < i_2 < i_3 \le n}h_3(A_{i_1,i_2,i_3})\\
=&\sum\limits_{1 \le i_1 < i_2 < i_3 \le n}\sum_{\{i,j\}\subset \{i_1,i_2,i_3\}} [c_{ij}+\eta_3(A_{ij})]\prod_{\substack{k\in \{i_1,i_2,i_3\}/\{i,j\}\\l\in\{i,j\} }} [b_{kl}+\eta_2(A_{kl})]\\
=&\sum_{1 \le i_1 < i_2 < i_3 \le n}c_{i_1 i_2} b_{i_1 i_3} b_{i_2 i_3}+b_{i_1 i_2} c_{i_1 i_3} b_{i_2 i_3}+b_{i_1 i_2} b_{i_1 i_3} c_{i_2 i_3}\\
&+\frac{n-2}{3}\sum_{1\le k_1 < k_2 \le n}\mathring{C}^{(2)}_{k_1 k_2}\eta_3(A_{k_1 k_2})+\frac{n-2}{3}\sum_{1\le k_1 < k_2 \le n}\mathring{B}_{k_1 k_2}\eta_2(A_{k_1 k_2})+R\\
=&\sum_{1 \le i_1 < i_2 < i_3 \le n}h_3(X_{i_1},X_{i_2},X_{i_3})\\
&+\frac{n-2}{3}\sum_{1\le k_1 < k_2 \le n}\mathring{C}^{(2)}_{k_1 k_2}\eta_3(A_{k_1 k_2})+\frac{n-2}{3}\sum_{1\le k_1 < k_2 \le n}\mathring{B}_{k_1 k_2}\eta_2(A_{k_1 k_2})+R,
\end{align*}
where we define
\begin{align*}
\mathring{C}^{(2)}_{k_1 k_2}=\frac{3}{n-2}\sum_{\substack{1\le i \le n\\ k_1, k_2\neq i}}b_{i k_1}b_{i k_2}, \quad \quad \mathring{B}_{k_1 k_2}=\frac{3}{n-2}\sum_{\substack{1\le i \le n\\ k_1, k_2\neq i}}(c_{i k_1}b_{i k_2}+c_{i k_2}b_{i k_1}),
\end{align*}
and $R$ for the remainder term that involves quadratic and third-order terms. 
We first bound the two linear terms. By Lemma~\ref{applemma2}, along with $\mathring{C}^{(2)}_{k_1 k_2} \asymp \rho_n^2 s_n^2$ and $\mathring{B}_{k_1 k_2} \asymp \rho_n^2 s_n$, we have
\begin{align*}
\frac{n-2}{3 \binom{n}{3}}\sum_{1\le k_1 < k_2 \le n}\mathring{C}^{(2)}_{k_1 k_2}\eta_3(A_{k_1 k_2}) & = \frac{1}{\binom{n}{2}} \sum_{1\le k_1 < k_2 \le n}\mathring{C}^{(2)}_{k_1 k_2}\eta_3(A_{k_1 k_2}) = \wtO_{p,2}(n^{-1} \rho_n^{5/2} s_n^2 \log^{1/2} n),\\
\frac{n-2}{3 \binom{n}{3}}\sum_{1\le k_1 < k_2 \le n}\mathring{B}_{k_1 k_2}\eta_2(A_{k_1 k_2}) & = \frac{1}{\binom{n}{2}} \sum_{1\le k_1 < k_2 \le n}\mathring{B}_{k_1 k_2}\eta_2(A_{k_1 k_2}) = \wtO_{p,2}(n^{-1} \rho_n^{5/2} s_n^{3/2} \log^{1/2} n).
\end{align*}

Next, we bound the remainder term $R$, which can be expressed as
\begin{align*}
R = \sum_{\substack{0\le p, q \le 3 , \\ 2 \le p+q \le 6}} \underbrace{\sum_{(i_{1:p}; j_{1:p}; k_{1:q}; l_{1:q}) \in I_{p, q}} \Theta_{i_{1:p}; j_{1:p}; k_{1:q}; l_{1:q} } (\omega_{i_1 j_1} \cdots \omega_{i_p j_p})\cdot (\mu_{k_1 l_1} \cdots \mu_{k_q l_q})}_{R^{(p, q)}},
\end{align*}
where $i_{1:p}$ represents tuple $(i_1,\dots, i_p)$, and similarly for $j_{1:p}$, $k_{1:q}$, and $l_{1:q}$; $I_{p, q}$ denotes an index set for $(i_{1:p}; j_{1:p}; k_{1:q}; l_{1:q})$ such that $(\omega_{i_1 j_1} \cdots \omega_{i_p j_p})\cdot (\mu_{k_1 l_1} \cdots \mu_{k_q l_q})$ is one of the secondary terms (excluding the mean and the linear term) in the decomposition of $\sum_{1 \le i_1 < i_2 < i_3 \le n}h_3(A_{i_1,i_2,i_3})$; $\Theta_{i_{1:p}; j_{1:p}; k_{1:q}; l_{1:q} }$ is the weight of $(\omega_{i_1 j_1} \cdots \omega_{i_p j_p})\cdot (\mu_{k_1 l_1} \cdots \mu_{k_q l_q})$ in $R$, with the randomness arising from $W$. Notice that $\omega_{ij}$ and $\mu_{ij}$, as defined in (\ref{defmuomega}), are independent because their randomness derive from $|A_{ij}|$ and $B_{ij}$ respectively. 

We will prove $R^{(p, q)} = \wtO_{p,2}(n^{3/2} \rho_n^{3/2} s_n \log^2 n)$ for every $(p, q)$ using Lemma~\ref{hyperthquick} (a succinct version of Lemma~\ref{hyperth}) and Condition (a). Before that, we first need to establish a universal upper bound for $\Theta_{i_{1:p}; j_{1:p}; k_{1:q}; l_{1:q} }$ for every $(i_{1:p}; j_{1:p}; k_{1:q}; l_{1:q}) \in I_{p, q}$. Then we bound $\Xi_{\mathfrak{r}}$ as $\max_{1 \le \mathfrak{r} \le \min \{ 2, \mathfrak{q} - 1\}}\Xi_{\mathfrak{r}} \preceq \max\{n\rho_n^2 s_n, 1 \}$ and $\max_{\mathfrak{r} > \min \{ 2, \mathfrak{q} - 1\}} \Xi_{\mathfrak{r}} \preceq 1$ where $\mathfrak{q} = p+q$. Additionally, we prove $\var(R^{(p, q)} \mid W) \preceq n^3 \rho_n^3 s_n^2$.

Consider the case where $(p, q) = (2, 0)$. First, we need to identify the index set $I_{2, 0}$ and estimate the order of the weight $\Theta_{i_{1:2}; j_{1:2}; k_{1:0}; l_{1:0} }$. Recall that $\eta_3(A_{ij}) \sim \rho_{ij} \mu_{ij} + \bar{s}_{ij} \omega_{ij} + \omega_{ij} \mu_{ij}$ and $\eta_2(A_{ij}) \sim -\rho_{ij} \mu_{ij} + s_{ij} \omega_{ij} - \omega_{ij} \mu_{ij}$. By examining the term $(c_{ij} + \eta_3(A_{ij}))(b_{ik} + \eta_2(A_{ik}))(b_{jk} + \eta_2(A_{jk}))$, we find that $R^{(2, 0)}$ can be expressed as:
\begin{align*}
R^{(2, 0)}=C \sum_{\substack{1 \le i,j,k \le n\\ i \neq j, j \neq k, k \neq i}} ( s_{kj} s_{ki} c_{ij} + \bar{s}_{kj}s_{ki} b_{ij} ) \omega_{ki}\omega_{kj}.
\end{align*}
It is straightforward that the weight $s_{kj} s_{ki} c_{ij} + \bar{s}_{kj}s_{ki} b_{ij}$ is bounded by $O(\rho_n s_n^2)$. Next, we define a hypergraph and apply Lemma~\ref{hyperthquick} to bound $\Xi_{\mathfrak{r}}$. The node set is $\mathcal{V}(H)=\{\omega_{ij}|i\neq j,\,i\in [n] , j\in [n]\}$, where $\omega_{ij}$ and $\omega_{ji}$ represent the same random variable. The hyperedge set is $\mathcal{H}(H)=\{\mathfrak{h}=(\omega_{ki},\omega_{kj})|i\neq j,j\neq k,k \neq i, i,j,k \in [n]\}$, with the weight on a hyperedge $(\omega_{ki},\omega_{kj})$ bounded by $\mathfrak{W}_{\mathfrak{h}} \preceq \rho_n s_n^2$.
Notice that we condition on $W$ throughout this analysis. The condition~(\ref{hyperineqcondition}) in Lemma~\ref{hyperth} holds with $L=1$ because $\E[|\omega_{ij}|^k] \le \E[|\omega_{ij}|^{k-1}] \le k \E[|\omega_{ij}|^{k-1}]$. Recall that  \[\Xi_{\mathfrak{r}}=\max\limits_{S\subset[N]:|S|=\mathfrak{r}}
\sum\limits_{\substack{\mathfrak{h}\in \mathcal{H}(H): \\ S\subseteq \mathcal{V}(\mathfrak{h})}}\,
|\mathfrak{W}_{\mathfrak{h}}|
\prod\limits_{\mathfrak{v}\in 
\mathcal{V}(\mathfrak{h})\backslash
S}\E[|Y_{\mathfrak{v}}|], \] where
$\mathfrak{r}$ represents the number of nodes in $S$ and $1\le \mathfrak{r}\le 2$.

(1)  When $\mathfrak{r}=1$ and $|S|=1$, the summation is over $2(n-2)$ distinct hyperedges $\mathfrak{h}\in \mathcal{H}(H)$ such that $S\subseteq \mathcal{V}(\mathfrak{h})$. Note that with one random variable chosen in $S$, there remains only one random variable in $\mathcal{V}(\mathfrak{h})$. Consequently, we have $\prod\limits_{\mathfrak{v}\in 
\mathcal{V}(\mathfrak{h})\backslash
S}\E[|Y_{\mathfrak{v}}|]=\E[|\omega_{ij}|] \asymp \rho_n$ for some $(i,j)$. It follows that
$$
\Xi_{1}=\max\limits_{S\subset[N]:\mathfrak{r}=1}
\sum\limits_{\substack{\mathfrak{h}\in \mathcal{H}(H): \\ S\subseteq \mathcal{V}(\mathfrak{h})}}\,
|\mathfrak{W}_{\mathfrak{h}}|
\prod\limits_{\mathfrak{v}\in 
\mathcal{V}(\mathfrak{h})\backslash
S}\E[|Y_{\mathfrak{v}}|] \preceq 2(n-2)\cdot \rho_n s_n^2 \cdot \rho_n \asymp n \rho_n^2 s_n^2.
$$

(2) When $\mathfrak{r}=2$ and $|S|=2$, for a fixed $S$, the hyperedge $\mathfrak{h}$ such that $S\subseteq \mathcal{V}(\mathfrak{h})$ is uniquely determined, and $\mathcal{V}(\mathfrak{h})\backslash S=\emptyset$. 
Therefore, we have
$$
\Xi_{2}=\max\limits_{S\subset[N]:\mathfrak{r}=2}
\sum\limits_{\substack{\mathfrak{h}\in \mathcal{H}(H): \\ S\subseteq \mathcal{V}(\mathfrak{h})}}\,
|\mathfrak{W}_{\mathfrak{h}}|
\prod\limits_{\mathfrak{v}\in 
\mathcal{V}(\mathfrak{h})\backslash
S}\E[|Y_{\mathfrak{v}}|] = \max\limits_{\substack{S\subset[N]:\mathfrak{r}=2\\ \mathfrak{h}\in \mathcal{H}(H): S\subseteq \mathcal{V} (\mathfrak{h})} }
|\mathfrak{W}_{\mathfrak{h}}|  \preceq \rho_n s_n^2.
$$
We bound $\var(R^{(2,0)}\mid W)$ as
\begin{align*}
&\var(R^{(2,0)}\mid W)\\
\preceq &\sum\limits_{\substack{1 \le i,j,k \le n\\ i \neq j, j \neq k, k \neq i}} \rho_n^2 s_n^4 \var(\omega_{ki})\var(\omega_{kj})\asymp n^3\cdot \rho_n^2 s_n^4 \cdot \rho_n \cdot \rho_n\asymp n^3  \rho_n^4 s_n^4,
\end{align*}
where the first equality holds because the covariance between dependent terms equals zero, \textit{i.e.}, $cov(\omega_{k_1k_2}\omega_{k_1k_3}, \omega_{k_1k_3}\omega_{k_1k_4} \mid W)=0$ for $k_2 \neq k_4$.	
In summarize, we prove $\Xi_{1} \preceq n \rho_n^2 s_n$, $\Xi_{2} \preceq 1$, and $\var(R^{(2, 0)} \mid W) \preceq n^3 \rho_n^3 s_n^2$, which leads to $R^{(2, 0)} = \wtO_{p,2}(n^{3/2} \rho_n^{3/2} s_n 
\log^2 n)$.

The term $R^{(0, 2)}$ can be bounded similarly. We now present an example where both $\omega_{ij}$ and $\mu_{ij}$ appear in the node set of the corresponding hypergraph: $R^{(1, 1)}$. Note that $R^{(1, 1)}$ can be expressed as
\begin{align*}
R^{(1, 1)} = C \sum_{\substack{1 \le i,j,k \le n\\ i \neq j, j \neq k, k \neq i}} R^{(1, 1)}_{i,j,k} \omega_{ij} \mu_{jk},
\end{align*}
where $C$ is a constant and the weight $R^{(1, 1)}_{i,j,k}$ satisfies $R^{(1, 1)}_{i,j,k} \preceq \rho_n^2 s_n$. We define the node set as $\mathcal{V}(H)=\{\omega_{ij}, \mu_{ij}|i\neq j,\,i\in [n] , j\in [n]\}$, and the hyperedge set as $\mathcal{H}(H)=\{\mathfrak{h}=(\omega_{ij},\mu_{jk})|i\neq j,j\neq k,k \neq i, i,j,k \in [n]\}$. 

When $\mathfrak{r} = 1$, the random variable in $\mathcal{V}(\mathfrak{h})\backslash
S$ could be either $\mu_{i'j'}$ or $\omega_{i'j'}$, which implies that $\Xi_{1} \preceq n \cdot \rho_n^2 s_n \cdot \max \{\rho_n + s_n\} \preceq n \rho_n^2 s_n$. When $\mathfrak{r} = 2$, there are no random variables in $\mathcal{V}(\mathfrak{h})\backslash
S$. Thus, we have $\Xi_2 \preceq \rho_n^2 s_n$. By straightforward calculation, we obtain $\var(R^{(1, 1)} \mid W) \preceq n^3 \cdot (\rho_n^2 s_n)^2 \cdot \rho_n s_n \asymp n^3 \rho_n^5 s_n^3$.

We conclude that $\Xi_{1} \preceq n \rho_n^2 s_n$, $\Xi_{2} \preceq 1$, and $\var(R^{(1, 1)} \mid W) \preceq n^3 \rho_n^3 s_n^2$, thereby completing bound for $R^{(1, 1)}$.

Now we consider the case where $\mathfrak{q} = p + q = 3$. We first establish a bound for $R^{(2, 1)}$, which can be expressed as
\begin{align*}
R^{(2,1)} = \underbrace{C_1 \sum_{\substack{1 \le i,j,k \le n\\ i \neq j, j \neq k, k \neq i}} R^{(2, 1)}_{i, j, k; 1} \mu_{ij} \omega_{ik} \omega_{jk}}_{R^{(2, 1)}_1} + \underbrace{C_2 \sum_{\substack{1 \le i,j,k \le n\\ i \neq j, j \neq k, k \neq i}} R^{(2, 1)}_{i, j, k; 2} \mu_{ij} \omega_{ij} \omega_{ik}}_{R^{(2, 1)}_2},
\end{align*}
where $R^{(2, 1)}_{i, j, k; 1}$ and $ R^{(2, 1)}_{i, j, k; 2}$ are bounded above by $\rho_n s_n$. We will demonstrate the bound for $R^{(2, 1)}_1$, as the argument for $R^{(2, 1)}_2$ follows a similar line of reasoning.

When $\mathfrak{r} = 1$, there is at least one random variable $\omega_{i'j'}$ remaining in $\mathcal{V}(\mathfrak{h})\backslash
S$. Therefore, we have $\Xi_1 \preceq n \rho_n^2 s_n$. When $\mathfrak{r} \ge 2$, there is exactly one hyperedge $\mathfrak{h}$ such that $S\subseteq \mathcal{V}(\mathfrak{h})$, which implies that $\Xi_{\mathfrak{r}} \preceq 1 \cdot \rho_n s_n \asymp \rho_n s_n$. Additionally, the variance of $R^{(2, 1)}$ conditional on $W$ satisfies $\var(R^{(2, 1)}_1 \mid W) \preceq n^3 \cdot (\rho_n s_n)^2 \cdot \rho_n^2 s_n \asymp n^3 \rho_n^4 s_n^3$.

We conclude that $\max_{1 \le \mathfrak{r} \le 2} \Xi_{\mathfrak{r}} \preceq \max \{n\rho_n^2 s_n, 1 \}$, $\Xi_3 \preceq 1$, and variance is $\preceq n^3 \rho_n^3 s_n^2$. This completes the bound for $R^{(2,1)}_1$. The bound for $R^{(2,1)}_2$ is similar.

We now establish a bound for $R^{(3, 0)}$, which can be expressed as
\begin{align*}
R^{(3, 0)} = C \sum_{\substack{1 \le i,j,k \le n\\ i \neq j, j \neq k, k \neq i}} R^{(3, 0)}_{i,j,k} \omega_{ij} \omega_{ik} \omega_{jk},
\end{align*}
where $R^{(3, 0)}_{i,j,k} \preceq s_n^2$. We define node set as $\mathcal{V}(H)=\{\omega_{ij}|i\neq j,\,i\in [n] , j\in [n]\}$, and the hyperedge set as $\mathcal{H}(H)=\{\mathfrak{h}=(\omega_{ki}, \omega_{ij},\omega_{kj})|i\neq j,j\neq k,k \neq i, i,j,k \in [n]\}$.
When $\mathfrak{r}=1$, there are still two random variables $\omega_{i'j'}$ and $\omega_{j'k'}$, in $\mathcal{V}(\mathfrak{h})\backslash
S$. Therefore, we have $\Xi_{1} \preceq n \cdot s_n^2 \cdot \rho_n^2 \asymp n \rho_n^2 s_n^2$.
When $\mathfrak{r} \ge 2$, there is exactly one hyperedge $\mathfrak{h}$ such that $S\subseteq \mathcal{V}(\mathfrak{h})$, which implies that $\Xi_{\mathfrak{r}} \preceq 1 \cdot s_n^2 \cdot 1 \asymp s^2_n$. The variance can be bounded similarly: $\var(R^{(3,0)} \mid W) \preceq n^3 \cdot s_n^4 \cdot \rho_n^3 \asymp n^3 \rho_n^3 s_n^4$. This completes the bound for $R^{(3, 0)}$.

The remaining terms are similar to those previously bounded. We conclude that for all $R^{(p, q)}$, we have $\var(R^{(p, q)} \mid W) \preceq n^3 \rho_n^3 s_n^2$, $\max_{1 \le \mathfrak{r} \le \min \{ 2, \mathfrak{q} - 1\}}\Xi_{\mathfrak{r}} \preceq \max\{n\rho_n^2 s_n, 1 \}$, and $\max_{\mathfrak{r} > \min \{ 2, \mathfrak{q} - 1\}} \Xi_{\mathfrak{r}} \preceq 1$. We now examine the case where the variance reaches the maximum order $n^3 \rho_n^3 s_n^2$, specifically for $R^{(3, 2)}$. Note that $R^{(3, 2)}$ can be expressed as 
\begin{align*}
R^{(3, 2)} = C \sum_{\substack{1 \le i,j,k \le n\\ i \neq j, j \neq k, k \neq i}} R^{(3, 2)}_{i,j,k} \omega_{ij} \omega_{ik} \omega_{jk} \mu_{ij} \mu_{ik},
\end{align*}
where $R^{(3, 2)}_{i,j,k} \preceq 1$. 

When $\mathfrak{r} = 1$, there are at least two random variables $\omega_{i'j'}$ and $\omega_{j'k'}$, as well as one random variable $\mu_{i"j"}$ in $\mathcal{V}(\mathfrak{h})\backslash
S$. Therefore, we have $\Xi_1 \preceq n \rho_n^2 s_n$.

When $\mathfrak{r} = 2$, if $\omega_{i'j'}$ and $\mu_{i'j'}$ are both included in $S$, there are still $2(n-2)$ choices for hyperedge $\mathfrak{h}$ such that $S\subseteq \mathcal{V}(\mathfrak{h})$. In this case, $\sum_{\mathfrak{h}\in \mathcal{H}(H): S\subseteq \mathcal{V}(\mathfrak{h})}\,
|\mathfrak{W}_{\mathfrak{h}}|
\prod_{\mathfrak{v}\in 
\mathcal{V}(\mathfrak{h})\backslash
S}\E[|Y_{\mathfrak{v}}|] \preceq n \cdot \rho_n^2 s_n \asymp n \rho_n^2 s_n$. If $\omega_{i'j'}$ and $\mu_{i'j'}$ are not both included in $S$, there is only one hyperedge $\mathfrak{h}$ such that $S\subseteq \mathcal{V}(\mathfrak{h})$. In this case, $\sum_{\mathfrak{h}\in \mathcal{H}(H): S\subseteq \mathcal{V}(\mathfrak{h})}\,
|\mathfrak{W}_{\mathfrak{h}}|
\prod_{\mathfrak{v}\in 
\mathcal{V}(\mathfrak{h})\backslash
S}\E[|Y_{\mathfrak{v}}|] \preceq 1$. Thus, we have $\Xi_2 \preceq \max \{n \rho_n^2 s_n , 1\}$.

When $\mathfrak{r} \ge 3$, there is only one hyperedge $\mathfrak{h}$ such that $S\subseteq \mathcal{V}(\mathfrak{h})$, which implies that $\Xi_{\mathfrak{r}} \preceq 1$.

We conclude that $\max_{1 \le \mathfrak{r} \le 2} \preceq \max\{n\rho_n^2 s_n, 1 \}$, $\max_{\mathfrak{r} \ge 3} \Xi_{\mathfrak{r}} \preceq 1$, and $\var(R^{(3,2)} \mid W) \preceq n^3 \rho_n^3 s_n^2$, which completes the bound for $R^{(3, 2)}$. Consequently, we have $R/\binom{n}{3} = \wtO_{p,2}(n^{-3/2} \rho_n^{3/2} s_n \log^2 n)$, thereby completing the proof of Lemma~\ref{apppro1}.

\subsection{Proof of Lemma~\ref{apppro2}}
\label{subsec: proof of apppro2}

\begin{proof}[of Lemma~\ref{apppro2} (1)]
The proof is analogous to that of Lemma~\ref{apppro1}.  We start by decomposing $\sum\limits_{\substack{1 \le i_1 < i_2 \le n\\i_1,i_2\neq i}}h_{3}(A_{i,i_1, i_2})$:
\begin{align*}
&\sum\limits_{\substack{1 \le i_1 < i_2 \le n\\i_1,i_2\neq i}}h_{3}(A_{i,i_1, i_2})\\
=&\sum\limits_{\substack{1 \le i_1 < i_2 \le n\\i_1,i_2\neq i}}[(c_{i\,i_1}+\eta_3(A_{i\,i_1}))(b_{i\,i_2}+\eta_2(A_{i\,i_2}))(b_{i_1 i_2}+\eta_2(A_{i_1 i_2}))\\
&\quad\quad\quad\quad+(b_{i\,i_1}+\eta_2(A_{i\,i_1}))(b_{i\,i_2}+\eta_2(A_{i\,i_2}))(c_{i_1 i_2}+\eta_3(A_{i_1 i_2}))\\
&\quad\quad\quad\quad+(b_{i\,i_1}+\eta_2(A_{i\,i_1}))(c_{i\,i_2}+\eta_2(A_{i\,i_2}))(b_{i_1 i_2}+\eta_2(A_{i_1 i_2}))]\\
=&\sum\limits_{\substack{1 \le i_1 < i_2 \le n\\i_1,i_2\neq i}}h_{3}(X_{i},X_{i_1}, X_{i_2})+\frac{n-2}{2} \sum_{\substack{1 \le j \le n:\\ j\neq i}}\widehat{\Theta}^{(1)}_{ij}\eta_3(A_{ij})+\frac{n-2}{2} \sum_{\substack{1 \le j \le n:\\ j\neq i}}\widehat{\Theta}^{(2)}_{ij}\eta_2(A_{ij})\\
&+\sum\limits_{\substack{1 \le i_1 < i_2 \le n\\i_1,i_2\neq i}} (b_{i i_1} b_{i i_2} \eta_3(A_{i_1 i_2}) + b_{i i_1} c_{i i_2} \eta_2(A_{i_1 i_2})) +R\\
=&\sum\limits_{\substack{1 \le i_1 < i_2 \le n\\i_1,i_2\neq i}}h_{3}(X_{i},X_{i_1}, X_{i_2})+\frac{n-2}{2} \sum_{\substack{1 \le j \le n:\\ j\neq i}}\widehat{\Theta}^{(1)}_{ij}\eta_3(A_{ij})+\frac{n-2}{2} \sum_{\substack{1 \le j \le n:\\ j\neq i}}\widehat{\Theta}^{(2)}_{ij}\eta_2(A_{ij})\\
&+\wtO_{p,2}(n \rho_n^{5/2} s_n^{3/2} \log^{1/2} n) + R,
\end{align*}
where we define
\begin{align}
\label{proplema2deftheta12}
\widehat{\Theta}_{ij}^{(1)}:= \frac{2}{n-2} \sum_{\substack{1 \le k \le n:\\ k\neq i,j}}b_{ik}b_{jk}, \quad \widehat{\Theta}_{ij}^{(2)}:= \frac{2}{n-2} \sum_{\substack{1 \le k \le n:\\ k\neq i,j}}(c_{ik}b_{jk}+c_{jk}b_{ik}),
\end{align}

We bound the remainder term $R$, which can be expressed as
\begin{align*}
R = \sum_{\substack{0\le p, q \le 3 , \\ 2 \le p+q \le 6}} \underbrace{\sum_{\substack{(i_{1:p}; j_{1:p}; k_{1:q}; l_{1:q}) \in I_{p, q}\\ i \in \{ i_{1:p}, j_{1:p}, k_{1:q}, l_{1:q}\} } } \Theta_{i_{1:p}; j_{1:p}; k_{1:q}; l_{1:q} } (\omega_{i_1 j_1} \cdots \omega_{i_p j_p})\cdot (\mu_{k_1 l_1} \cdots \mu_{k_q l_q})}_{R^{(p, q)}},
\end{align*}
where $i_{1:p}$ denotes tuple $(i_1,\dots, i_p)$, and similarly for $j_{1:p}$, $k_{1:q}$, and $l_{1:q}$; $I_{p, q}$ represents the index set for $(i_{1:p}; j_{1:p}; k_{1:q}; l_{1:q})$ such that $(\omega_{i_1 j_1} \cdots \omega_{i_p j_p})\cdot (\mu_{k_1 l_1} \cdots \mu_{k_q l_q})$ is one of the secondary terms (excluding the mean and the linear term) in the decomposition of $\sum_{1 \le i_1 < i_2 < i_3 \le n}h_3(A_{i_1,i_2,i_3})$. Note that $i$ must be in $\{i_{1:p}, j_{1:p}, k_{1:q}, l_{1:q}\}$ in this context. $\Theta_{i_{1:p}; j_{1:p}; k_{1:q}; l_{1:q} }$ is the weight composed of $s_n$, $\rho_n$, $F(X_k, X_l)$, $G(X_k, X_l)$, $b_{kl}$ and $c_{kl}$, with randomness originating from $W$. Recall that $\omega_{kl}$ and $\mu_{kl}$, as defined in (\ref{defmuomega}), have their randomness derived from $|A_{kl}|$ and $B_{kl}$, respectively, rendering $\omega_{kl}$ and $\mu_{kl}$ independent. 

Similar to the approach used in the proof of Lemma~\ref{apppro1}, we will apply Lemma~\ref{hyperthquick} to demonstrate that $R^{(p, q)} = \wtO_{p,2}(n^{2} \rho_n^{3/2} s_n \log^2 n)$ for every pair $(p, q)$.
Specifically, for each $R^{(p, q)}$, it is sufficient to establish that $\var(R^{(p, q)} \mid W) \preceq n^2 \rho_n^3 s_n^2$, $\max_{1 \le \mathfrak{r} \le \min \{ 2, \mathfrak{q} - 1\}}\Xi_{\mathfrak{r}} \preceq \max\{n\rho_n^2 s_n, 1 \}$, and $\max_{\mathfrak{r} > \min \{ 2, \mathfrak{q} - 1\}} \Xi_{\mathfrak{r}} \preceq 1$, where $\mathfrak{q} = p+q$. Given these conditions, we obtain $R^{(p, q)} = \wtO_{p,2}(n \rho_n^{3/2} s_n \log^2 n)$ by applying Lemma~\ref{hyperthquick} and utilizing the condition $\rho_n^3 s_n^2 \succeq n^{-2} \log^8 n$.

The only difference lies in the form of $R^{(p,q)}$. For each pair $(p, q)$, $R^{(p, q)}$ can be decomposed into the sum of three terms: $R^{(p, q)} = R^{(p, q; 1)} + R^{(p, q; 2)} + R^{(p, q; 3)}$, where these three terms arise based on which index among $\{ i_{1:p}, j_{1:p}, k_{1:q}, l_{1:q}\} \}$ (with $|i_{1:p}, j_{1:p}, k_{1:q}, l_{1:q}\}| = 3$) equals $i$. However, we assert that these three terms all share the same bound.

We begin with $(p, q) = (2, 0)$. We have
\begin{align*}
R^{(2, 0)} =C_1 \underbrace{\sum_{\substack{1 \le i_1 \le i_2 \le n\\i_1, i_2 \neq i}} R^{(2,0; 1)}_{i_1, i_2} \omega_{i i_1} \omega_{i i_2}}_{R^{(2,0)}_1} + C_2 \underbrace{\sum_{\substack{1 \le i_1 \le i_2 \le n\\i_1, i_2 \neq i}} R^{(2,0; 2)}_{i_1, i_2}\omega_{i_1 i} \omega_{i_1 i_2}}_{R^{(2,0)}_2},
\end{align*}
where $R^{(2,0; 1)}_{i_1, i_2}$ and $R^{(2,0; 2)}_{i_1, i_2} \preceq \rho_n s_n^2$. 

We bound $R_1^{(2,0)}$ using Lemma~\ref{hyperthquick}. We define the node set as $\mathcal{V}(H)=\{\omega_{ii_1}|i_1\neq i, i_1\in [n]\}$, where $\omega_{ii_1}$ and $\omega_{i_1i}$ represent the same random variable. The hyperedge set is defined as $\mathcal{H}(H)=\{\mathfrak{h}=(\omega_{ii_1},\omega_{ii_2})|i_1\neq i_2,i_2\neq i,i \neq i_1, i_1,i_2 \in [n]\}$, and the weight on the hyperedge $(\omega_{ki},\omega_{kj})$ is denoted by $\mathfrak{W}_{\mathfrak{h}} \preceq \rho_n s_n^2$. The condition~(\ref{hyperineqcondition}) in Lemma~\ref{hyperth} holds with $L=1$ because $\E[|\omega_{ij}|^k] \le \E[|\omega_{ij}|^{k-1}] \le k \E[|\omega_{ij}|^{k-1}]$. We then proceed to bound $\Xi_{\mathfrak{r}}$ and the variance.

(1). When $\mathfrak{r}=1$, the summation is over at most $2(n-2)$ distinct hyperedges $\mathfrak{h}\in \mathcal{H}(H)$ for which $S\subseteq \mathcal{V}(\mathfrak{h})$. Note that with one random variable chosen in $S$, there is only one remaining random variable in $\mathcal{V}(\mathfrak{h})$. Consequently, we have $\prod\limits_{\mathfrak{v}\in 
\mathcal{V}(\mathfrak{h})\backslash
S}\E[|Y_{\mathfrak{v}}|] \asymp \rho_n$. Thus, it follows that $\Xi_1 \preceq n \cdot \rho_n s_n^2 \cdot \rho_n \asymp n \rho_n^2 s_n^2$.

(2) When $\mathfrak{r}=2$, the hyperedge $\mathfrak{h}$ for which $S\subseteq \mathcal{V}(\mathfrak{h})$ is uniquely determined, and $\mathcal{V}(\mathfrak{h})\backslash S=\emptyset$. 
Therefore, we have $\Xi_2 \preceq 1 \cdot \rho_n s_n^2 \cdot 1 \asymp \rho_n s_n^2$.

The variance can be bounded as $\var(R^{(2,0)}_1 \mid W) \preceq n^2 \rho_n^4 s_n^4$. We conclude that $\Xi_1 \preceq n \rho_n^2 s_n$, $\Xi_2 \preceq 1$, and $\var(R^{(2,0)}_1 \mid W) \preceq n^2 \rho_n^3 s_n^2$, which completes the bound for $R^{(2,0)}_1$.

Next, we bound $R_2^{(2,0)}$. We define the node set as $\mathcal{V}(H)=\{\omega_{ii_1} \omega_{i_1, i_2}|i_1\neq i, i \neq i_2, i_2 \neq i_1; i_1, i_2\in [n]\}$, the hyperedge set as $\mathcal{H}(H)=\{\mathfrak{h}=(\omega_{ii_1},\omega_{i_1 i_2})|i_1\neq i_2,i_2\neq i,i \neq i_1, i_1,i_2 \in [n]\}$, and the weight on the hyperedge $(\omega_{ki},\omega_{kj})$ as $\mathfrak{W}_{\mathfrak{h}} \preceq \rho_n s_n^2$. We have $\Xi_1 \preceq n \rho_n^2 s_n^2 \preceq n \rho_n^2 s_n$, $\Xi_2 \preceq \rho_n s_n^2 \preceq 1$, and $\var(R^{(2,0)}_2\mid W) \preceq n^2 \rho_n^4 s_n^4 \preceq n^2 \rho_n^3 s_n^2$, which completes the bound for $R_2^{(2,0)}$. Thus, we have $R^{(2,0)} = \wtO_{p,2}(n \rho_n^{3/2} s_n \log^2 n)$. The bound for $R^{(0,2)}$ can be obtained in a similar manner.

Now we provide an example where both $\omega$ and $\mu$ are present in the node set of the corresponding hypergraph: $R^{(1, 1)}$. Note that $R^{(1, 1)}$ can be expressed as

\begin{align*}
&R^{(1, 1)} \\
=& \underbrace{C_1 \sum_{\substack{1 \le i_1 \le i_2 \le n\\i_1, i_2 \neq i}} R^{(1, 1; 
 1)}_{i_1 i_2} \omega_{i i_1} \mu_{i_1 i_2}}_{R^{(1, 1; 1)}} +\underbrace{C_2 \sum_{\substack{1 \le i_1 \le i_2 \le n\\i_1, i_2 \neq i}} R^{(1, 1; 
 2)}_{i_1 i_2} \omega_{i i_1} \mu_{i i_2}}_{R^{(1, 1; 2)}} + \underbrace{C_3 \sum_{\substack{1 \le i_1 \le i_2 \le n\\i_1, i_2 \neq i}} R^{(1, 1; 
 3)}_{i_1 i_2} \omega_{i_1 i_2} \mu_{i i_2}}_{R^{(1, 1; 3)}},
\end{align*}
where weight $R^{(1, 1; 
 1)}_{i_1 i_2}$, $R^{(1, 1; 
 2)}_{i_1 i_2}$, $R^{(1, 1; 
 3)}_{i_1 i_2} \preceq \rho_n^2 s_n$. 

We focus on $R^{(1, 1; 1)}$ in this context. When $\mathfrak{r} = 1$, we have $\Xi_1 \preceq n(\rho_n^3 s_n + \rho_n^2 s_n^2) \preceq n \rho_n^2 s_n$. For $\mathfrak{r} \ge 2$,
we have $\Xi_{\mathfrak{r}} \preceq \rho_n^2 s_n \preceq 1$. Thus, $R^{(1, 1; 1)} = \wtO_{p,2}(n \rho_n^{3/2} s_n \log^2 n)$. The bounds for $R^{(1, 1; 2)}$ and $R^{(1, 1; 3)}$ can be obtained in a similar manner, which implies that $R^{(1, 1)} = \wtO_{p,2}(n \rho_n^{3/2} s_n \log^2 n)$.

The remaining terms follow similar bounds to those previously established. We conclude that for all $R^{(p, q)}$, $\var(R^{(p, q)} \mid W) \preceq n^2 \rho_n^3 s_n^2$, $\max_{1 \le \mathfrak{r} \le \min \{ 2, \mathfrak{q} - 1\}}\Xi_{\mathfrak{r}} \preceq \max\{n\rho_n^2 s_n, 1 \}$, and $\max_{\mathfrak{r} > \min \{ 2, \mathfrak{q} - 1\}} \Xi_{\mathfrak{r}} \preceq 1$. We now address the case where the variance achieves the maximum order $n^2 \rho_n^3 s_n^2$: $R^{(3, 2)}$. Note that $R^{(3, 2)}$ can be expressed as
\begin{align*}
&R^{(3, 2)} \\
=& \underbrace{C_1 \sum_{\substack{1 \le i_1 \le i_2 \le n\\i_1, i_2 \neq i}} R^{(3, 2; 
 1)}_{i_1 i_2} \omega_{i_1 i_2} \omega_{i_1 i} \omega_{i_2 i} \mu_{i_1 i_2} \mu_{i_1 i} 
 }_{R^{(3, 2; 1)}} +\underbrace{C_2 \sum_{\substack{1 \le i_1 \le i_2 \le n\\i_1, i_2 \neq i}} R^{(3, 2; 
 2)}_{i_1 i_2} \omega_{i i_1} \omega_{i i_2} 
 \omega_{i_1 i_2} \mu_{i i_1} \mu_{i i_2} }_{R^{(3, 2; 2)}} ,
\end{align*}
where weight $R^{(3, 2; 
 1)}_{i_1 i_2}$, $R^{(3, 2; 
 2)}_{i_1 i_2} \preceq 1$.

We focus on $R^{(3,2; 1)}$ in this context. When $\mathfrak{r} = 1$, there are at least two random variables $\omega$ and one random variable $\mu$ in $\mathcal{V}(\mathfrak{h})\backslash
S$. So we have $\Xi_1 \preceq n \rho_n^2 s_n$.

When $\mathfrak{r} = 2$, if we include $\omega_{i_1' i}$ and $\mu_{i_1' i}$ together in $S$, there are still $2(n-2)$ choices for hyperedge $\mathfrak{h}$ such that $S\subseteq \mathcal{V}(\mathfrak{h})$. In this case, $\sum_{\mathfrak{h}\in \mathcal{H}(H): S\subseteq \mathcal{V}(\mathfrak{h})}\,
|\mathfrak{W}_{\mathfrak{h}}|
\prod_{\mathfrak{v}\in 
\mathcal{V}(\mathfrak{h})\backslash
S}\E[|Y_{\mathfrak{v}}|] \preceq n \cdot \rho_n^2 s_n \asymp n \rho_n^2 s_n$. If we do not include $\omega_{i_1' i}$ and $\mu_{i_1' i}$ together in $S$, there is only one hyperedge $\mathfrak{h}$ such that $S\subseteq \mathcal{V}(\mathfrak{h})$. In this case, $\sum_{\mathfrak{h}\in \mathcal{H}(H): S\subseteq \mathcal{V}(\mathfrak{h})}\,
|\mathfrak{W}_{\mathfrak{h}}|
\prod_{\mathfrak{v}\in 
\mathcal{V}(\mathfrak{h})\backslash
S}\E[|Y_{\mathfrak{v}}|] \preceq 1$. Thus, we have $\Xi_2 \preceq \max \{n \rho_n^2 s_n , 1\}$.

When $\mathfrak{r} \ge 3$, there is only one hyperedge $\mathfrak{h}$ such that $S\subseteq \mathcal{V}(\mathfrak{h})$, which implies that $\Xi_{\mathfrak{r}} \preceq 1$.

We conclude that $\max_{1 \le \mathfrak{r} \le 2} \preceq \max\{n\rho_n^2 s_n, 1 \}$, $\max_{\mathfrak{r} \ge 3} \Xi_{\mathfrak{r}} \preceq 1$, and $\var(R^{(3,2)} \mid W) \preceq n^2 \rho_n^3 s_n^2$, which completes the bound for $R^{(3, 2; 1)}$. The bound for $R^{(3, 2; 2)}$ can be obtained in a similar manner. 

Therefore,
\begin{align*}
&\frac{1}{\binom{n-1}{2}}\sum_{\substack{1 \le i_1 < i_2 \le n\\i_1,i_2\neq i}}\left(h_{3}(A_{i,i_1, i_2})-h_{3}(X_i,X_{i_1}, X_{i_2})\right)\\
=&\frac{1}{n-1}\sum_{\substack{1\le j \le n:\\j \neq i}}\widehat{\Theta}_{ij}^{(1)} \eta_3(A_{ij})+\frac{1}{n-1}\sum_{\substack{1\le j \le n:\\j \neq i}}\widehat{\Theta}_{ij}^{(2)}\eta_2(A_{ij})+\wtO_{p,2}(n^{-1} \rho_n^{3/2} s_n \log^2 n),
\end{align*}
where $\widehat{\Theta}^{(1)}_{ij}$ and $\widehat{\Theta}^{(2)}_{ij}$ are defined in (\ref{proplema2deftheta12}).
Then we have the following results:
$
\widehat{\Theta}_{ij}^{(1)}\asymp s_n^2 \rho_n^2$ and $\widehat{\Theta}_{ij}^{(2)}\asymp s_n \rho_n^2
$,
which completes our proof of Lemma~\ref{apppro2} (1).
\end{proof}

\begin{proof}[of Lemma~\ref{apppro2} (2)]
Similar to the proof of Lemma~\ref{apppro1}, we define $\eta_{ij}=\eta_{3}(A_{ij})$ and $\alpha_{ij}=c_{ij}$ when $h=h_1$; and $\eta_{ij}=\eta_{1}(A_{ij})$ and $\alpha_{ij}=c_{ij}+b_{ij}$ when $h=h_\Delta$ for $1 \le i < j \le n$.
For both cases, by Lemma~\ref{lemma: h property}, we have $\eta_{ij}=\eta_{ji},\, \E[\eta_{ij}]=0,\, \E[|\eta_{ij}|]\asymp \rho_n, \, \E[\eta_{ij}^2]\asymp \rho_n$, and $\E[|\eta_{ij}^3|]\asymp \rho_n$. In addition, we have $\alpha_{ij}=\alpha_{ji}$ and $\alpha_{ij} \asymp \rho_n$.
Therefore, it suffices to prove for a unified $h_{i_1 i_2 i_3}:=(\alpha_{i_1 i_2}+\eta_{i_1 i_2})(\alpha_{i_1 i_3}+\eta_{i_1 i_3})(\alpha_{i_2 i_3}+\eta_{i_2 i_3})$ with the above properties. Note that 
$\E[h_{i_1 i_2 i_3}\mid X_{i_1},X_{i_2},X_{i_3}] = \alpha_{i_1 i_2} \alpha_{i_1 i_3}\alpha_{i_2 i_3}$. 
It follows that
\begin{align*}
&\sum_{\substack{1 \le i_1 < i_2 \le n\\i_1,i_2\neq i}}h_{i,i_1, i_2}\\
=&\sum_{\substack{1 \le i_1 < i_2 \le n\\i_1,i_2\neq i}}(\alpha_{i\,i_1}+\eta_{i\,i_1})(\alpha_{i\,i_2}+\eta_{i\,i_2})(\alpha_{i_1 i_2}+\eta_{i_1 i_2})\\
=&\sum_{\substack{1 \le i_1 < i_2 \le n\\i_1,i_2\neq i}}\E_{A|X_1,...,X_n}[h_{i,i_1, i_2}]+\frac{n-2}{2}\sum_{\substack{1 \le j \le n:\\ j\neq i}}\widehat{\Theta}_{ij}\eta_{ij}+ \sum_{\substack{1 \le i_1 < i_2 \le n\\i_1,i_2\neq i}}\alpha_{i i_1} \alpha_{i i_2} \eta_{i_1 i_2}  +R\\
=&\sum_{\substack{1 \le i_1 < i_2 \le n\\i_1,i_2\neq i}}\E_{A|X_1,...,X_n}[h_{i,i_1, i_2}]+\frac{n-2}{2}\sum_{\substack{1 \le j \le n:\\ j\neq i}}\widehat{\Theta}_{ij}\eta_{ij}+ \wtO_{p,2}(n \rho_n^{3/2} \log^{1/2} n)  +R,
\end{align*}
where we define
$
\widehat{\Theta}_{ij}:=\frac{2}{n-2}\sum_{\substack{1 \le k \le n\\ k\neq i,j}}\alpha_{ik}\alpha_{jk},
$ and 
\begin{align*}
R:=&\sum_{\substack{1 \le j_1 < j_2 \le n\\ j_1, j_2 \neq i}}\alpha_{i\,j_1}\alpha_{i\,j_2}\eta_{j_1 j_2} + \underbrace{\sum_{\substack{1\le k_1 < k_2 \le n\\ k_1, k_2 \neq i}}\alpha_{k_1 k_2}\eta_{i\,k_1}\eta_{i\,k_2}}_{R_1} + \underbrace{\sum_{\substack{1\le k_1, k_2 \le n\\k_1 \neq k_2,\, k_1, k_2 \neq i}}\alpha_{i\,k_2}\eta_{k_1 i}\eta_{k_1 k_2}}_{R_2} \\
&+ \underbrace{\sum_{\substack{1\le k_1 < k_2 \le n\\ k_1, k_2 \neq i}}\eta_{i\,k_1}\eta_{i\,k_2}\eta_{k_1 k_2}}_{R_3}.
\end{align*}

Now we use Lemma~\ref{hyperth} to bound the three reminders: $R_1$, $R_2$ and $R_3$.
\begin{enumerate}
\item For $R_1$ and $R_2$, the corresponding hypergraph consists of nodes $\eta_{ij}$, where each edge contains 2 nodes. We have $\Xi_1 \preceq n \rho_n^2$, $\Xi_2 \preceq \rho_n$ and $\var(R_1 \mid W), \var(R_2 \mid W) \preceq n^2 \rho_n^4 $. Denote both $R_1$ and $R_2$ as $R'$. By Lemma~\ref{hyperth}, there exists a universal constant $C>0$ such that 
\begin{align*}
&\pr(|R'|\ge C_0 n \rho_n^{2} \log n \mid W)\\
\le&e^2 \cdot \max \Bigg\{\exp\left\{-\frac{C_0^2 n^{2} \rho^4_n \log^2 n}{C^2\,\var(R'\mid W)}\right\},\exp\left\{-\frac{C_0 n \rho_n^{2} \log n}{C^2 \Xi_{1}}\right\},\\ 
& \quad \quad \quad \quad \exp\left\{-\left(\frac{C_0 n \rho_n^{2} \log n}{C^2\,\Xi_2}\right)^{1/2}\right\} \Bigg\} \\
\le&e^2 \cdot \max \Bigg\{\exp\left\{-\frac{C_0^2 n^{2} \rho^4_n \log^2 n}{C^2\, n^2 \rho_n^3 }\right\},\exp\left\{-\frac{C_0 n \rho_n^{2} \log n}{C^2 n \rho_n^2}\right\},\\ 
& \quad \quad \quad \quad \exp\left\{-\left(\frac{C_0 n \rho_n^{2} \log n}{C^2\, \rho_n }\right)^{1/2}\right\} \Bigg\} \\
\le & e^2\cdot \exp\left\{-\log n\,\max \{\frac{C_0^2}{C^2}, \frac{C_0}{C^2}, (\frac{C_0}{C^2})^{1/2} \}\right\} \le \frac{e^2}{n^2},
\end{align*}
where the second last inequality holds as $\rho_n \succeq n^{-1} \log^{3/2}n$ and the last inequality holds by choosing $C_0 > \max (\sqrt{2} C, 4 C^2)$.
Note that the constants $C_0$ and $e$ are universal, and the above inequality holds for any realization $W$. Thus, we have $R_1 /\binom{n}{3}$, $R_2 /\binom{n}{3}= \wtO_{p,2}(n^{-1} \rho_n^2 \log n)$.
\item For $R_3$, the corresponding hypergraph consists of nodes $\eta_{ij}$, where each edge contains 3 nodes. We have $\Xi_1 \preceq n \rho_n^2$, $\Xi_2 \preceq \rho_n$, $\Xi_3 \preceq 1$ and $\var(R_3 \mid W)\preceq n^2 \rho_n^3$. By Lemma~\ref{hyperth}, there exists a universal constant $C>0$ such that 
\begin{align*}
&\pr(|R_3|\ge C_0 n \rho_n^{3/2} \log n \mid W)\\
\le&e^2 \cdot \max \Bigg\{\exp\left\{-\frac{C_0^2 n^{2} \rho^3_n \log^2 n}{C^3\,\var(R'\mid W)}\right\},\exp\left\{-\frac{C_0 n \rho_n^{3/2} \log n}{C^3 \Xi_{1}}\right\},\\ 
& \quad \quad \quad \quad \exp\left\{-\left(\frac{C_0 n \rho_n^{3/2} \log n}{C^3\,\Xi_2}\right)^{1/2}\right\} , \exp\left\{-\left(\frac{C_0 n \rho_n^{3/2} \log n}{C^3\,\Xi_3}\right)^{1/3}\right\} \Bigg\}\\
\le&e^2 \cdot \max \Bigg\{\exp\left\{-\frac{C_0^2 n^{2} \rho^3_n \log^2 n}{C^3\, n^2 \rho_n^3 }\right\},\exp\left\{-\frac{C_0 n \rho_n^{3/2} \log n}{C^3 n \rho_n^2}\right\},\\ 
& \quad \quad \quad \quad \exp\left\{-\left(\frac{C_0 n \rho_n^{3/2} \log n}{C^3\, \rho_n }\right)^{1/2}\right\}, \exp\left\{-\left(\frac{C_0 n \rho_n^{3/2} \log n}{C^3}\right)^{1/3}\right\} \Bigg\} \\
\le & e^2\cdot \exp\left\{-\log n\,\max \{\frac{C_0^2}{C^3}, \frac{C_0}{C^3}, (\frac{C_0}{C^3})^{1/2}, (\frac{C_0}{C^3})^{1/3} \}\right\} \le \frac{e^2}{n^2},
\end{align*}
where the second last inequality holds as $\rho_n^3 \succeq n^{-2} \log^{2}n$ and the last inequality holds by choosing $C_0 > \max((2 C^3)^{1/2}, 8 C^3)$.
Note that the constants $C_0$ and $e$ are universal, and the above inequality holds for any realization $W$. Thus, we have $R_3 /\binom{n}{3}= \wtO_{p,2}(n^{-1} \rho_n^{3/2} \log n)$.
\end{enumerate}

By Lemma~\ref{applemma2},
\begin{align*}
\frac{1}{\binom{n}{2}}\sum_{\substack{1 \le j_1 < j_2 \le n\\ j_1, j_2 \neq i}}\alpha_{i\,j_1}\alpha_{i\,j_2}\eta_{j_1 j_2}=\wtO_{p,2}(n^{-1}\,\rho_n^{\frac{5}{2}}\,\log^{\frac{1}{2}}n).
\end{align*}
Above all, we have $R/\binom{n}{2}=\wtO_{p,2}(n^{-1}\,\rho_n^{\frac{3}{2}}\,\log n )$, which leads to
\begin{align*}
&\frac{1}{\binom{n-1}{2}}\sum_{\substack{1 \le i_1 < i_2 \le n\\i_1,i_2\neq i}}\left(h_1(A_{i,i_1, i_2})-h_1(X_i,X_{i_1}, X_{i_2})\right) \\
=&\frac{1}{n-1}\sum_{\substack{1\le j \le n:\\j \neq i}}\widehat{\Theta}_{ij}^{(3)} \eta_3(A_{ij})+\wtO_{p,2}(n^{-1}\,\rho_n^{\frac{3}{2}}\,\log n ),
\end{align*}
\begin{align*}
&\frac{1}{\binom{n-1}{2}}\sum_{\substack{1 \le i_1 < i_2 \le n\\i_1,i_2\neq i}}\left(h_{\Delta}(A_{i,i_1, i_2})-h_{\triangle}(X_i,X_{i_1}, X_{i_2})\right) \\
=&\frac{1}{n-1}\sum_{\substack{1\le j \le n:\\j \neq i}}\widehat{\Theta}_{ij}^{(4)}\eta_1(A_{ij})+\wtO_{p,2}(n^{-1}\,\rho_n^{\frac{3}{2}}\,\log n ).
\end{align*}
where
\begin{align}
\label{proplema2deftheta34}
\widehat{\Theta}^{(3)}_{ij}:=\frac{2}{n-2}\sum_{\substack{1 \le k \le n\\ k\neq i,j}}c_{ik}c_{jk}, \quad \widehat{\Theta}^{(4)}_{ij}:=\frac{2}{n-2}\sum_{\substack{1 \le k \le n\\ k\neq i,j}}(c_{ik} + b_{ik})(c_{jk} + b_{jk}).
\end{align}
It's straightforward to see that $\widehat{\Theta}^{(3)}_{ij} \asymp \rho_n^2$ and $\widehat{\Theta}^{(4)}_{ij} \asymp \rho_n^2$,
which completes this proof.
\end{proof}

\section{Proof of Theorem~\ref{thm1}}
\subsection{Preliminary Decomposition}

To approximate the distribution function $\mathcal{F}_{\whT_n+\delta_T}$, we build on the procedure of \citet{DBLP:journals/corr/abs-2004-06615}. Since the decompositions and error bounds in Lemma~\ref{lemma} differ, we provide and verify the detailed derivation below for readers' convenience. 

Recall that $M(n,s_n,\rho_n)=n^{-1} \rho_n^{-3/2} \log n (1 + s_n \log n)$. For presentation simplicity, we use shorthand for the terms in Lemma~\ref{lemma} as follows.
\begin{align*}
	\widehat{T}_n
=&(\frac{\widehat{U}_n/\widehat{V}_n-u_n/v_n}{\sigma_n})\cdot(1+\frac{\widehat{S}_n^2-\widehat{\sigma}_n^2}{\sigma_n^2}+\frac{\widehat{\sigma}_n^2-\sigma_n^2}{\sigma_n^2})^{-1/2}\\
= & \Big(\check{\Delta}_n-\frac{3}{\sqrt{n}\xi_1}\E\Big[p_1(X_1) q_1(X_1)\Big] +\underbrace{\dfrac{1}{\sqrt{n}\xi_1}\sum_{i=1}^nq_1(X_i)}_{W_n^{(1)}} + \underbrace{\dfrac{1}{n^{3/2}\xi_1}\sum_{1\le i\neq j \le n}q_2(X_i,X_j)}_{W_n^{(2)}}\\
& \ \ - \underbrace{\dfrac{3}{n^{3/2}\xi_1}\sum_{1\le i\neq j\le n}p_1(X_i)q_1(X_j)}_{W_n^{(3)}} + \wtO_{p,2}(n^{-1}\log^{\frac{3}{2}}n + M(n,s_n,\rho_n)) \Big)\\
& \cdot \Big(1+\check{\delta}_n + \wtO_{p,1}(M(n,s_n,\rho_n))\Big)^{-1/2},
\end{align*}
where  
\begin{align*}
	\check{\delta}_n = \underbrace{\dfrac{1}{n\xi_{1}^2}\sum_{i=1}^{n}(q_{1}^2(X_{i})-\xi_{1}^2)}_{\check{\delta}_{n,1}} + \underbrace{\dfrac{4}{n\xi_{1}^2}\sum_{i=1}^{n}g_{\delta,1}(X_i)}_{\check{\delta}_{n,2}} \underbrace{-\dfrac{6}{n}\sum_{i=1}^{n}p_1(X_i)}_{\check{\delta}_{n,3}}   \underbrace{-\dfrac{6}{n\xi_{1}^2}\E[p_1(X_1)q_1(X_1)]\sum_{i=1}^{n}q_1(X_i)}_{\check{\delta}_{n,4}}.
\end{align*}
By Lemma~\ref{lemma} (2), we have $\check{\Delta}_n=\wtO_{p,1}(\rho_n^{-1/2}n^{-1/2}\log^{1/2}n)$. 
By Lemma~\ref{applemma1} and~\ref{applemma2}, we bound other terms~by
\begin{align*}
	W_n^{(1)} =\wtO_{p,2}(\log^{1/2}{n}), \quad W_n^{(2)}&=\wtO_{p,2}(n^{-1/2}\log{n}), \quad W_n^{(3)}=\wtO_{p,2}(n^{-1/2}\log n),\\
	\check{\delta}_{n,1} =\wtO_{p,2}(n^{-1/2}\log^{1/2}n), \quad \check{\delta}_{n,2}&=\wtO_{p,2}(n^{-1/2}\log^{1/2}n), \quad \check{\delta}_{n,3}=\wtO_{p,2}(n^{-1/2}\log^{1/2}n),
\end{align*}
and $\check{\delta}_{n,4}=\wtO_{p,2}(n^{-1/2}\log^{1/2}n)$. Thus, it follows that $\check{\delta}_n=\wtO_{p,2}(n^{-1/2}\log^{1/2}n)$.  
By applying the second-order Taylor expansion with $x\coloneqq\check{\delta}_n+\wtO_{p,1}(M(n,s_n,\rho_n))$, we have $(1+x)^{-1/2}=1-\dfrac{1}{2}x+O(x^2)=1-\dfrac{1}{2}\check{\delta}_n+\wtO_{p,1}(M(n,s_n,\rho_n)+ n^{-1}\log n).$
Therefore, we have
\begin{align*}
\whT_n=&(-\frac{3}{\sqrt{n}\xi_1}\E\Big[p_1(X_1) q_1(X_1)\Big]+W_n^{(1)}+W_n^{(2)}-W_n^{(3)}+\check{\Delta}_n+\wtO_{p,1}(n^{-1}\log^{\frac{3}{2}}n + M(n,s_n,\rho_n)))\\
& \cdot(1-\dfrac{1}{2}\check{\delta}_n+\wtO_{p,1}(M(n,s_n,\rho_n)+ n^{-1}\log n))\\
=&-\frac{3}{\sqrt{n}\xi_1}\E\Big[p_1(X_1) q_1(X_1)\Big]+W_n^{(1)}+W_n^{(2)}-W_n^{(3)}-\dfrac{W_n^{(1)}\check{\delta}_n}{2}+\check{\Delta}_n\\
&+\wtO_{p,1}(M(n,\rho_n,s_n)\log^{\frac{1}{2}}n+ n^{-1}\log^{\frac{3}{2}}n).
\end{align*}
We further simplify  $W_n^{(1)}\check{\delta}_n$. Regarding the term $W_n^{(1)}\check{\delta}_{n,1}$, we have
\begin{align*}
W_n^{(1)}\check{\delta}_{n,1}&=\frac{1}{n^{3/2}\xi_1^3}\Big(\sum_{1\le i\le n}q_1(X_i)(q_1^2(X_i)-\xi_1^2)+\sum_{1\le i\neq j\le n}q_1(X_i)(q_{1}^2(X_{j})-\xi_{1}^2)\Big).
\end{align*}
By applying Lemma~\ref{applemma2}, the first term can be decomposed by
\begin{align*}
&\frac{1}{n^{3/2}\xi_1^3}\sum_{1\le i\le n}q_1(X_i)(q_1^2(X_i)-\xi_1^2)\\
=&\dfrac{1}{\sqrt{n}\xi_1^3}\E\Big[q_1^3(X_1)\Big]+\frac{1}{n^{3/2}\xi_1^3}\sum_{1\le i\le n}(q_1^3(X_i)-q_1(X_i)\xi_1^2-\E\Big[q_1^3(X_1)\Big])\\
=&\dfrac{1}{\sqrt{n}\xi_1^3}\E\Big[q_1^3(X_1)\Big]+\wtO_{p,2}(n^{-1}\log^{1/2}n).
\end{align*}
Since $h(X_i,X_j)=q_1(X_i)(q_{1}^2(X_{j})-\xi_{1}^2)$ is canonical, by Lemma~\ref{applemma1}, the second term is bounded by
\begin{align*}
L_{n,1}\coloneqq\frac{1}{n^{3/2}\xi_1^3}\sum_{1\le i\neq j\le n}q_1(X_i)(q_{1}^2(X_{j})-\xi_{1}^2)=\wtO_{p,2}(n^{-1/2}\log n).
\end{align*}
Regarding the term $W_n^{(1)}\check{\delta}_{n,2}$, we have
\begin{align*}
W_n^{(1)}\check{\delta}_{n,2}&=\dfrac{4}{n^{3/2}\xi_1^3}\Big(\sum_{i=1}^nq_1(X_i)g_{\delta,1}(X_i)+\sum_{1\le i\neq j\le n}q_1(X_i)g_{\delta,1}(X_j)\Big).
\end{align*}
For the first term, note that $\E\Big[q_1(X_i)g_{\delta,1}(X_i)\Big]=\E\Big[q_1(X_1)q_1(X_2)q_2(X_1,X_2)\Big]$, by Lemma~\ref{applemma2}, we have
\begin{align*}
\dfrac{4}{n^{3/2}\xi_1^3}\sum_{i=1}^nq_1(X_i)g_{\delta,1}(X_i)=\dfrac{4}{\sqrt{n}\xi_1^3}\E\Big[q_1(X_1)q_1(X_2)q_2(X_1,X_2)\Big]+\wtO_{p,2}(n^{-1}\log^{1/2}n).
\end{align*}
For the second term, as $h(X_i,X_j)=q_1(X_i)g_{\delta,1}(X_j)$ is canonical, by Lemma~\ref{applemma1}, we have
\begin{align*}
L_{n,2}\coloneqq\dfrac{4}{n^{3/2}\xi_1^3}\sum_{1\le i\neq j\le n}q_1(X_i)g_{\delta,1}(X_j)=\wtO_{p,2}(n^{-1/2}\log n).
\end{align*}
Similarly, we have 
\begin{align*}
	W_n^{(1)}\check{\delta}_{n,3}= & -\frac{6}{\sqrt{n}\xi_1}\E[p_1(X_1)q_1(X_1)]-\frac{6}{n^{3/2}\xi_1}\sum_{1\le i\neq j\le n}q_1(X_i)p_1(X_j)+\wtO_{p,2}(n^{-1}\log^{1/2}n)\\
	= & -\frac{6}{\sqrt{n}\xi_1}\E[p_1(X_1)q_1(X_1)]-2W_n^{(3)}+\wtO_{p,2}(n^{-1}\log^{1/2}n) , 
\end{align*}
and
$$W_n^{(1)}\check{\delta}_{n,4}=-\frac{6}{\sqrt{n}\xi_1}\E[p_1(X_1)q_1(X_1)]-\underbrace{\frac{6}{n^{3/2}\xi_1^3}\E[p_1(X_1)q_1(X_1)]\sum_{1\le i\neq j\le n}q_1(X_i)q_1(X_j)}_{L_{n,3}=\wtO_{p,2}(n^{-1/2}\log n)}+\wtO_{p,2}(n^{-1}\log^{1/2}n).$$
By combining the above results, it follows that 
\begin{align*}
W_n^{(1)}\check{\delta}_n&=\dfrac{1}{\sqrt{n}\xi_1^3}\E\Big[q_1^3(X_1)\Big]+\dfrac{4}{\sqrt{n}\xi_1^3}\E\Big[q_1(X_1)q_1(X_2)q_2(X_1,X_2)\Big]-\frac{12}{\sqrt{n}\xi_1}\E[p_1(X_1)q_1(X_1)]\\
&\ \ \ -2W_n^{(3)} +\underbrace{L_{n,1}+L_{n,2}-L_{n,3}}_{L_n=\wtO_{p,2}(n^{-1/2}\log n)}+\wtO_{p,2}(n^{-1}\log^{1/2}n).
\end{align*}
Plugging the decomposition for $W_n^{(1)}\check{\delta}_n$ into $\whT_n$, we have 
\begin{align*}
\whT_n=& \underbrace{\alpha_n + W_n^{(1)}+W_n^{(2)}-\dfrac{L_n}{2}}_{\widetilde{T}_n}+\,\check{\Delta}_n+\wtO_{p,1}(M(n,\rho_n,s_n)\log^{1/2}n\vee n^{-1}\log^{3/2}n),
\end{align*}
where $\alpha_n:=\dfrac{3}{\sqrt{n}\xi_1}\E\Big[p_1(X_1) q_1(X_1)\Big]-\dfrac{1}{2\sqrt{n}\xi_1^3}\E\Big[q_1^3(X_1)\Big]-\dfrac{2}{\sqrt{n}\xi_1^3}\E\Big[q_1(X_1)q_1(X_2)q_2(X_1,X_2)\Big]$. 
With the above decomposition for $\whT_n$, we divide $\Vert \mathcal{F}_{\widehat{T}_n+\delta_T}-\mathcal{G}_n\Vert_{\infty}$ into three parts and bound them separately in the following lemma.

\begin{lemma}\label{lemma.CDF} Denote $\widetilde{\Delta}_n\mid W\sim N(0,n^{-1}\rho_n^{-1}s_n\sigma_W^2)$ and define $\mathcal{E}(n,s_n,\rho_n)=M(n, s_n, \rho_n) \log^{\frac{1}{2}}n + n^{-1}\rho_n^{-\frac{1}{2}}s_n^{-\frac{1}{2}} = n^{-1}\rho_n^{-\frac{3}{2}}\log^{\frac{3}{2}} n( 1 + s_n \log n)+n^{-1}\rho_n^{-\frac{1}{2}}s_n^{-\frac{1}{2}}$.  Assuming the conditions in Theorem~\ref{thm1}, we have
\begin{itemize}
	\item[(a)] $\Vert \mathcal{F}_{\widetilde{T}_n+\check{\Delta}_n{+\delta_T}}-\mathcal{F}_{\widetilde{T}_n+\widetilde{\Delta}_n{+\delta_T}}\Vert_{\infty}=O(n^{-1}\rho_n^{-\frac{1}{2}}s_n^{-\frac{1}{2}})$;
	\item[(b)] With a sufficiently large $c_\delta$, $\Vert \mathcal{F}_{\widetilde{T}_n+\widetilde{\Delta}_n{+\delta_T}}-\mathcal{G}_n\Vert_{\infty}=O(n^{-1}\rho_n^{-1}s_n\log n{+n^{-1}\log n})$;
	\item[(c)] $\Vert \mathcal{F}_{\widehat{T}_n{+\delta_T}}-\mathcal{F}_{\widetilde{T}_n+\check{\Delta}_n{+\delta_T}}\Vert_{\infty}=O(\mathcal{E}(n,\rho_n,s_n))$.
\end{itemize}
\end{lemma}
The proof of Theorem~\ref{thm1} is completed by combining the bounds in Lemma~\ref{lemma.CDF} using the triangle inequality. 
We provide the proof of Lemma~\ref{lemma.CDF} in Section~\ref{subsec: proof of lemma.CDF (a)}-\ref{subsec: proof of lemma.CDF (c)}. Specifically, we prove Lemma~\ref{lemma.CDF} (a) and Lemma~\ref{lemma.CDF} (b), and then combine Lemma~\ref{lemma.CDF} (a)-(b) and Lemma~\ref{smoothing lemma} to prove Lemma~\ref{lemma.CDF} (c).

\subsection{Proof of Lemma~\ref{lemma.CDF} (a)}\label{subsec: proof of lemma.CDF (a)} 
Lemma~\ref{lemma} (3) provides a bound for the distance between $\mathcal{F}_{\check{\Delta}_n|W}$ and $\mathcal{F}_{\wtD_n|W}$ in (\ref{lemma:distance}). Denote $W_{bad}:=\{W: \Vert \mathcal{F}_{\check{\Delta}_n|W}-\mathcal{F}_{N(0,n^{-1}\,\rho_n^{-1}\,s_n\,\sigma_W^2)}\Vert _{\infty}> C_1\cdot n^{-1}\rho_n^{-\frac{1}{2}}s_n^{-\frac{1}{2}}\}$ for some constant $C_1$, then $\pr\Big(W_{bad}\Big)\le C_2\cdot n^{-1}$ for some constant $C_2$. For any $u\in \mathbb{R}$, we have
\begin{align*}
& \mathcal{F}_{\widetilde{T}_n+\check{\Delta}_n{+\delta_T}}(u)-\mathcal{F}_{\widetilde{T}_n+\widetilde{\Delta}_n{+\delta_T}}(u)\\
&=\pr\Big(\check{\Delta}_n\le u-\widetilde{T}_n{-\delta_T}\Big)-\pr\Big(\widetilde{\Delta}_n\le u-\widetilde{T}_n{-\delta_T}\Big)\\
&=\Big(\pr\Big(\check{\Delta}_n\le u-\widetilde{T}_n{-\delta_T}\Big|W_{bad}\Big)-\pr\Big(\widetilde{\Delta}_n\le u-\widetilde{T}_n{-\delta_T}\Big|W_{bad}\Big)\Big)\pr\Big(W_{bad}\Big)\\
&\ \ \ +\Big(\pr\Big(\check{\Delta}_n\le u-\widetilde{T}_n{-\delta_T}\Big|W_{bad}^\complement\Big)-\pr\Big(\widetilde{\Delta}_n\le u-\widetilde{T}_n{-\delta_T}\Big|W_{bad}^\complement\Big)\Big)\pr\Big(W_{bad}^\complement\Big)\\
&=O(n^{-1})+O(n^{-1}\rho_n^{-1/2}s_n^{-1/2})=O(n^{-1}\rho_n^{-1/2}s_n^{-1/2}).
\end{align*}
Therefore, we have $\Vert \mathcal{F}_{\widetilde{T}_n+\check{\Delta}_n}-\mathcal{F}_{\widetilde{T}_n+\widetilde{\Delta}_n}\Vert_{\infty}=O(n^{-1}\rho_n^{-1/2}s_n^{-1/2})$.

\subsection{Proof of Lemma~\ref{lemma.CDF} (b)}
\label{subsec: proof of lemma.CDF (b)}
We apply Lemma~\ref{Esseen's smoothing lemma} to prove this part, where we set $\gamma=n$ and bound the integral in (\ref{eq:Esseen}) by the following lemma, the proof of which is given in Section~\ref{subsec: proof of lemma.integral} later. Notice that by Condition (c), $\dfrac{1}{4}+\dfrac{\log\rho_n-\log s_n}{4\log n}>0.$

\begin{lemma}\label{lemma.integral} Assuming the conditions in Theorem~\ref{thm1} and $0<\epsilon\le {\min(\dfrac{1}{4},\dfrac{1}{4}+\dfrac{\log\rho_n-\log s_n}{4\log n})}$, we have the following results:
\begin{enumerate}
	\item[(a)] For any $lower\asymp n^{\epsilon}$ and $upper\asymp n$, we have
$$\int_{lower}^{upper} \mid \frac{Ch.f.(\mathcal{G}_n; t)}{t} \mid dt=O(n^{-1});$$
	\item[(b)] For any $\vert lower\vert\ge\sqrt{2/c_\delta}n^{1/2}$ and $upper\asymp n$, we have $$\int_{lower}^{upper} \mid \frac{Ch.f.(\mathcal{F}_{\widetilde{T}_n+\widetilde{\Delta}_n{+\delta_T}}; t)}{t} \mid dt=O(n^{-1}\log n);$$
	\item[(c)] For any $lower\asymp n^{\epsilon}$ and $\vert upper\vert\le c_0n^{1/2}$ where $c_0>0$ is small enough, we have
$$\int_{lower}^{upper} \mid \frac{Ch.f.(\mathcal{F}_{\widetilde{T}_n+\widetilde{\Delta}_n{+\delta_T}}; t)}{t} \mid dt=O(s_n\rho^{-1}_nn^{-1}\log{n} + n^{-\frac{3}{2}}\log ^{\frac{5}{2}}n);$$
	\item[(d)]  
	For any $upper\asymp n^{\epsilon}$, we have
$$\int_{0}^{upper} \mid \frac{Ch.f.(\mathcal{F}_{\widetilde{T}_n+\widetilde{\Delta}_n{+\delta_T}}; t)-Ch.f.(\mathcal{G}_n; t)}{t} \mid dt=O(s_n\rho^{-1}_nn^{-1}+n^{-1}\log n).$$
\end{enumerate}
\end{lemma}
By Lemma~\ref{lemma.integral} and taking $c_\delta\ge 2/c_0^2$ and $\epsilon=1/12$, we have 
\begin{align*}
    \int_{n^{\epsilon}}^{n} \mid \frac{Ch.f.(\mathcal{G}_n; t)}{t} \mid dt & =O(n^{-1}),\\
    \int_{\sqrt{2/c_\delta}n^{1/2}}^{n} \mid \frac{Ch.f.(\mathcal{F}_{\widetilde{T}_n+\widetilde{\Delta}_n{+\delta_T}}; t)}{t} \mid dt & =O(n^{-1}\log n), \\
    \int_{n^{\epsilon}}^{c_0n^{1/2}} \mid \frac{Ch.f.(\mathcal{F}_{\widetilde{T}_n+\widetilde{\Delta}_n{+\delta_T}}; t)}{t} \mid dt & =O(s_n\rho^{-1}_nn^{-1}\log{n} + n^{-\frac{3}{2}}\log ^{\frac{5}{2}}n),
\end{align*}
and
$$\int_{0}^{n^{\epsilon}} \mid \frac{Ch.f.(\mathcal{F}_{\widetilde{T}_n+\widetilde{\Delta}_n{+\delta_T}}; t)-Ch.f.(\mathcal{G}_n; t)}{t} \mid dt=O(n^{-1}\rho_n^{-1}s_n+n^{-1}\log n).$$
Thus, it follows that $$\int_{0}^\gamma\Big| \frac{Ch.f.(\mathcal{F}_{\widetilde{T}_n+\widetilde{\Delta}_n{+\delta_T}}; t)-Ch.f.(\mathcal{G}_n; t)}{t}\Big|dt=O(n^{-1}\rho_n^{-1}s_n\log n{+n^{-1}\log n}).$$
Similarly, we have $$\int_{-\gamma}^0\Big| \frac{Ch.f.(\mathcal{F}_{\widetilde{T}_n+\widetilde{\Delta}_n{+\delta_T}}; t)-Ch.f.(\mathcal{G}_n; t)}{t}\Big|dt=O(n^{-1}\rho_n^{-1}s_n\log n{+n^{-1}\log n}).$$
By Lemma~\ref{Esseen's smoothing lemma}, it follows that
\begin{align*}
\Vert \mathcal{F}_{\widetilde{T}_n+\widetilde{\Delta}_n{+\delta_T}}-\mathcal{G}_n\Vert_\infty&\le C_1\int_{-\gamma}^\gamma\Big| \frac{Ch.f.(\mathcal{F}_{\widetilde{T}_n+\widetilde{\Delta}_n{+\delta_T}}; t)-Ch.f.(\mathcal{G}_n; t)}{t}\Big|dt+\frac{C_2M}{\gamma}\\
&=O(n^{-1}\rho_n^{-1}s_n\log n {+n^{-1}\log n}).
\end{align*}

\subsection{Proof of Lemma~\ref{lemma.CDF} (c)}
\label{subsec: proof of lemma.CDF (c)}
We apply Lemma~\ref{smoothing lemma} to prove this part, where we set $X=\whT_n{+\delta_T}$, $Y=\wtT_n+\cD_n{+\delta_T}$, and $Z=X-Y=\wtO_{p,1}(\mathcal{E}(n,\rho_n,s_n))$. Denote $M:=\sup_{u\in \mathbb{R}}|\mathcal{G}_n'(u)|<\infty$. For any $u$ and $a>0$, we have
\begin{align*}
\mathcal{F}_Y(u+a)-\mathcal{F}_Y(u) &\le \underbrace{\left| \mathcal{F}_{\wtT_n+\cD_n{+\delta_T}}(u+a)-\mathcal{F}_{\wtT_n+\wtD_n{+\delta_T}}(u+a)\right|}_{\text{Bounded by Lemma~\ref{lemma.CDF} (a)}}+\underbrace{\left| \mathcal{F}_{\wtT_n+\wtD_n{+\delta_T}}(u+a)-\mathcal{G}_n(u+a)\right|}_{\text{Bounded by Lemma~\ref{lemma.CDF} (b)}} \\
&\ \ \ + \left|\mathcal{G}_n(u+a)-\mathcal{G}_n(u)\right| +\underbrace{\left|\mathcal{G}_n(u)-\mathcal{F}_{\wtT_n+\wtD_n{+\delta_T}}(u)\right|}_{\text{Bounded by Lemma~\ref{lemma.CDF} (b)}}\\
&\ \ \ + \underbrace{\left|\mathcal{F}_{\wtT_n+\wtD_n{+\delta_T}}(u)-\mathcal{F}_{\wtT_n+\cD_n{+\delta_T}}(u)\right|}_{\text{Bounded by Lemma~\ref{lemma.CDF} (a)}} \\
&\le M\cdot{a}+O(n^{-1}\rho_n^{-1/2}s_n^{-1/2}+ n^{-1}\rho_n^{-1}s_n\log n{+n^{-1}\log n}).
\end{align*}
Note that all terms in $O(\cdot)$ are bounded by $\mathcal{E}(n,\rho_n,s_n)$, by Lemma~\ref{smoothing lemma}, we have completed the proof.

\subsection{Proof of Lemma~\ref{lemma.integral}}
\label{subsec: proof of lemma.integral}
\begin{proof}[of Lemma~\ref{lemma.integral} (a)] Based on the expression of $\mathcal{G}_n$, it follows that $Ch.f.(\mathcal{G}_n; t)$ is the multiplication of $e^{-t^2/2}$ and a degree-3 polynomial of $t$. Note that the coefficients of $\mathcal{G}_n$ are bounded and that, for $0\le k \le 3$, there exists some constant $C_k$ such that $t^ke^{-t^2/2}\le C_ke^{-t^2/3}$ holds for any $n^{\epsilon}\preceq |t|\preceq n$. Therefore, we have 
\[
\int_{lower}^{upper} \mid \frac{Ch.f.(\mathcal{G}_n; t)}{t} \mid dt \preceq \int_{lower}^{\infty} e^{-\frac{t^2}{3}} dt\le\int_{lower}^{\infty} \frac{t}{lower} e^{-\frac{t^2}{3}} dt\preceq \frac{1}{n^{\epsilon}}e^{-\frac{n^{2\epsilon}}{3}}=O(n^{-1}),
\]
which completes the proof.
\end{proof}

\medskip
\begin{proof}[of Lemma~\ref{lemma.integral} (b)] Note that $\widetilde{T}_n$ is determined by $W$. Because $\widetilde{\Delta}_n\mid W\sim N(0,n^{-1}\rho_n^{-1}s_n\sigma_W^2)$  and $\delta_T\sim N(0,c_\delta n^{-1}\log n)$ is independent of $W$, we have
\begin{align}
Ch.f.(\mathcal{F}_{\widetilde{T}_n+\widetilde{\Delta}_n{+\delta_T}}; t) &=\E\Big[e^{\mathrm{i}t(\widetilde{T}_n+\widetilde{\Delta}_n{+\delta_T})}\Big]=\E\Big[\E\Big[e^{\mathrm{i}t(\widetilde{T}_n+\widetilde{\Delta}_n)}\Big| W\Big]{\E\Big[e^{\mathrm{i}t\delta_T}\Big]}\Big]\nonumber \\
&=\E\Big[e^{\mathrm{i}t\widetilde{T}_n}e^{-(\var(\wtD_n\mid W)+\var(\delta_T))t^2/2}\Big] \label{eq: chf_var}\\
&={n^{-c_\delta n^{-1}t^2/2}}\E\Big[e^{\mathrm{i}t\widetilde{T}_n}e^{-(n\rho_n/s_n)^{-1}\sigma_W^2t^2/2}\Big]. \label{eq: chf}
\end{align}
For $\vert lower\vert\ge\sqrt{2/c_\delta}n^{1/2}$ and $upper\asymp n$, 
\begin{align*}
\int_{lower}^{upper} \mid \frac{Ch.f.(\mathcal{F}_{\widetilde{T}_n+\widetilde{\Delta}_n+\delta_T}; t)}{t} \mid dt\le n^{-1}\int_{lower}^{upper} \mid \frac{1}{t} \mid dt=O(n^{-1}\log n),
\end{align*}
which completes the proof of Lemma~\ref{lemma.integral} (b).
\end{proof}

\medskip
\begin{proof}[of Lemma~\ref{lemma.integral} (c)] 
Note that $\widetilde{T}_n$ can be written in the following form:
\begin{align*}
\widetilde{T}_n
&=\alpha_n+W_n^{(1)}+W_n^{(2)}-\dfrac{L_n}{2}\\
&:=\alpha_n+\frac{1}{\sqrt{n}\xi_1}\sum_{i=1}^nq_1(X_i)+\frac{1}{\sqrt{n}(n-1){\xi_1}}\sum_{1 \le i < j \le n}\psi(X_i,X_j), 
\end{align*}
where $\E\Big[\psi(X_i,X_j)\Big]=0$ and $\Vert\psi\Vert_\infty\le M$ for some $M>0$. Note that $\displaystyle\sum_{i=1}^nq_1(X_i)=\dfrac{1}{n-1}\sum_{1\le i<j\le n}(q_1(X_i)+q_1(X_j))$, we define \[K_n=\sum_{1\le i<j\le n}q_1(X_i)+q_1(X_j)+\psi(X_i,X_j)\quad\text{and}\quad \kappa_n=\sqrt{n}(n-1)\xi_1.\]
It follows that $\widetilde{T}_n = \alpha_n + K_n / \kappa_n$. 
Recall that  
\begin{align*}
\sigma_{W}^2 
&=n\,\rho_n\,s_n^{-1} \var(\check{\Delta}_n |W) \\
&=\frac{n\,\rho_n\,s_n^{-1}}{(C_{n}^2v_n \sigma_n)^2}\sum_{1\le k_1 < k_2 \le n}\Big(-\mathring{\theta}_{k_1 k_2}^2+b_{k_1 k_2}(\mathring{B}_{k_1 k_2}-\frac{u_n}{v_n}\mathring{D}_{k_1 k_2})^2 +c_{k_1 k_2}(\mathring{C}_{k_1 k_2}-\frac{u_n}{v_n}\mathring{D}_{k_1 k_2})^2\Big).
\end{align*}
After plugging in the expression of $\mathring{B}_{k_1 k_2}$, $\mathring{C}_{k_1 k_2}$, $\mathring{D}_{k_1 k_2}$ and $\mathring{\theta}_{k_1 k_2}$, we note that $\sigma_{W}^2$ can be expressed as the sum of a U-statistic of degree $3$ and a U-statistic of degree $4$ by simple calculation.  By Hoeffding's decomposition, $\sigma_{W}^2$ can be decomposed as
\begin{align}\label{Hoeffding.sigma_W^2}
\sigma_{W}^2=\E\big[\sigma_W^2\big]+\frac{1}{n}\sum_{i=1}^ng_{\sigma_W^2}(X_i)+R_n,
\end{align}
where $\E[\sigma_W^2]\asymp 1$, $g_{\sigma_W^2}(x):= \E[\sigma_W^2 \mid X_1=x]- \E[\sigma_W^2]$ and $R_n$ denotes the remainder term. It follows that $\E[g_{\sigma_W^2}(X_i)]=0$ and all items in Hoeffding's decomposition are $\preceq 1$ with some universal constants. Then, by Lemma~\ref{applemma1}, we have  $R_n=\wtO_{p,1}(n^{-1}\log{n})$.

Now we are ready to study $Ch.f.(\mathcal{F}_{\widetilde{T}_n+\widetilde{\Delta}_n{+\delta_T}}; t)$. By (\ref{eq: chf})-(\ref{Hoeffding.sigma_W^2}), we have
\begin{align}
& \Big|Ch.f.(\mathcal{F}_{\widetilde{T}_n+\widetilde{\Delta}_n{+\delta_T}}; t)\Big|\le\Big|\E\Big[e^{\mathrm{i}t\widetilde{T}_n-(n\rho_n/s_n)^{-1}t^2\sigma_W^2/2}\Big]\Big| \nonumber\\
=&\Big|\E\Big[e^{\mathrm{i}tK_n/\kappa_n-(n\rho_n/s_n)^{-1}t^2\sigma_W^2/2}\Big]\Big| \quad\quad\quad \quad\quad\quad(\text{By $\widetilde{T}_n=\alpha_n+K_n/\kappa_n$ and $|e^{\mathrm{i}t \alpha_n}|=1$}) \nonumber\\
=&\Big|\E\Big[e^{\mathrm{i}tK_n/\kappa_n-(n\rho_n/s_n)^{-1}t^2(\E[\sigma_W^2]+\frac{1}{n}\sum_{1\le i\le n}g_{\sigma_W^2}(X_i))/2} \cdot e^{-(n\rho_n/s_n)^{-1}t^2R_n/2}\Big]\Big| \nonumber \\
\le & \underbrace{\Big|\E\Big[e^{\mathrm{i}tK_n/\kappa_n-\sum_{1\le i\le n}\widetilde{g}(X_i)}\Big]\Big|}_{J_1}+ \underbrace{\Big|\E\Big[e^{\mathrm{i}tK_n/\kappa_n-\sum_{1\le i\le n}\widetilde{g}(X_i)}\cdot \bar{R}_n\Big]\Big|}_{J_2},\label{eq: chf (c)}
\end{align}
where $\widetilde{g}(X_i):=(n^2\rho_n/s_n)^{-1}t^2(\E[\sigma_W^2]+g_{\sigma_W^2}(X_i))/2$ and $\bar{R}_n:=e^{-(n\rho_n/s_n)^{-1}t^2R_n/2}-1$.  Note that {$\widetilde{g}(X_i)\prec n^{-1/2}$} under Condition (c) and $t \preceq n^{1/2}$, where the constant involved in $\prec$ is universal and holds for any $1\le i\le n$. In addition, by $R_n=\wtO_{p,1}(n^{-1}\log{n})$ and Lemma~\ref{lemma.expop}, we have $\bar{R}_n=\wtO_{p,1}(t^2 s_n \rho_n^{-1}n^{-2}\log n)$ and there exists a constant $C_0$ such that $|\bar{R}_n|\le C_0 t^2 s_n \rho_n^{-1}n^{-1}$.

To bound the second term $J_2$ in (\ref{eq: chf (c)}), we discuss two events $W_1=\{W:\sum_{i=1}^n\widetilde{g}(X_i)\ge 0\}$  and $W_2=\{W: |\bar{R}_n | \le C_2 t^2 s_n \rho_n^{-1}n^{-2}\log n\}$. Then it follows that there exists a constant $C_1$ such that $\pr\Big(W_2^\complement\Big)\le C_1/n$. By the Bernstein inequality, there exists some constant $k_1>0$ such that
\begin{align*}
\pr\Big(W_1^\complement\Big)=\pr\Big(\frac{1}{n}\sum_{i=1}^ng_{\sigma_W^2}(X_i)<-\E\Big[\sigma_W^2\Big]\Big)\le e^{-k_1n}.
\end{align*}
Then we have
\begin{align*}
& \Big|\E\Big[e^{\mathrm{i}tK_n/\kappa_n-\sum_{1\le i\le n}\widetilde{g}(X_i)}\cdot \bar{R}_n \mid W_1\Big]\Big| \cdot \pr(W_1)\\
 \le & \E\Big[ e^{-\sum_{1\le i\le n}\widetilde{g}(X_i)} \vert\bar{R}_n \vert \mid W_1\Big]  \\
\le &\E\Big[ \vert\bar{R}_n \vert \mid W_1, W_2\Big] \pr(W_2) +\E\Big[ \vert\bar{R}_n \vert \mid W_1, W_2^\complement\Big] \pr(W_2^\complement) \\
\le & C_2 t^2 s_n \rho_n^{-1}n^{-2}\log n +C_0t^2 s_n \rho_n^{-1}n^{-1} \frac{C_1}{n} =O( t^2 s_n \rho_n^{-1}n^{-2}\log n),
\end{align*}
where the first inequality holds by $\pr(W_1)\le 1$ and the second inequality holds as $e^{-\sum_{1\le i\le n}\widetilde{g}(X_i)} \le 1 $ on the event $W_1$. 
We also have 
\begin{align*}
&	\Big|\E\Big[e^{\mathrm{i}tK_n/\kappa_n-\sum_{1\le i\le n}\widetilde{g}(X_i)}\cdot \bar{R}_n \mid W_1^\complement\Big]\Big|\cdot \pr(W_1^\complement) \\
\le &\E\Big[e^{-\sum_{1\le i\le n}\widetilde{g}(X_i)}\cdot \vert\bar{R}_n \vert \mid W_1^\complement\Big]\cdot e^{-k_1n} 
\\
\le & C_0 t^2 s_n \rho_n^{-1}n^{-1}e^{k_2n^{1/2}-k_1n}
\end{align*}
for some constant $k_2>0$, 
where the first inequality holds as $\pr(W_1^\complement) \le e^{-k_1n} $ and the second one holds due to $\widetilde{g}(X_i)\prec n^{-1/2}$. 
Combining the above together, it follows that
\begin{align*}
	J_2 & \le \Big|\E\Big[e^{\mathrm{i}tK_n/\kappa_n-\sum_{1\le i\le n}\widetilde{g}(X_i)}\cdot \bar{R}_n \mid W_1\Big]\Big|\cdot \pr(W_1) \\
	& \quad \quad + \Big|\E\Big[e^{\mathrm{i}tK_n/\kappa_n-\sum_{1\le i\le n}\widetilde{g}(X_i)}\cdot \bar{R}_n \mid W_1^\complement\Big]\Big|\cdot \pr(W_1^\complement)\\
	& =O( t^2 s_n \rho_n^{-1}n^{-2}\log n).
\end{align*}

To bound the first term $J_1$ in (\ref{eq: chf (c)}), we proceed by following arguments (2.17) - (2.20) in \cite{10.1214/aos/1176350170}. Let $\square_n(m)=\sum_{i=1}^m\sum_{j=i+1}^n\psi(X_i,X_j)$ with $1\le m \le n-1 $ to be specified later. By \citet[Corollary on page 1722]{dharmadhikari1968bounds}, it follows that $\E[|\square_n(m)|^r]=O((mn)^{r/2})$. Therefore, we have
\begin{align*}
J_1=&\E\Big[e^{\mathrm{i}t\Big(K_n-\square_n(m)\Big)/\kappa_n-\sum_{1\le i\le n}\widetilde{g}(X_i)} e^{\mathrm{i}t\square_n(m)/\kappa_n}\Big]\\
=&\E\Big[e^{\mathrm{i}t\Big(K_n-\square_n(m)\Big)/\kappa_n-\sum_{1\le i\le n}\widetilde{g}(X_i)}\sum_{v=0}^r\frac{(\mathrm{i}t\square_n(m)/\kappa_n)^v}{v!}\Big]+\text{Remainder}, 
\end{align*}
where $r \ge 2$ and the remainder term is bounded by
\begin{align*}
\text{Remainder}& \preceq \Big|\E\Big[e^{\mathrm{i}t\Big(K_n-\square_n(m)\Big)/\kappa_n-\sum_{1\le i\le n}\widetilde{g}(X_i)}(\mathrm{i}t\square_n(m)/\kappa_n)^{r+1}\Big] \Big|\\
	& \le \E\Big[ e^{-\sum_{1\le i\le n}\widetilde{g}(X_i)}(|t \square_n(m)|/\kappa_n)^{r+1} \Big] \\
	& \le\E\Big[ e^{-\sum_{1\le i\le n}\widetilde{g}(X_i)}(|t \square_n(m)|/\kappa_n)^{r+1}\mid W_1\Big]\pr(W_1)\\
	& \quad \quad \quad \quad +\E\Big[ e^{-\sum_{1\le i\le n}\widetilde{g}(X_i)}(|t\square_n(m)|/\kappa_n)^{r+1} \mid W_1^\complement\Big]\pr(W_1^\complement) \\
        & \le\E\Big[ (|t \square_n(m)|/\kappa_n)^{r+1}\mid W_1\Big]\pr(W_1)\\
	& \quad \quad \quad \quad + (|t|/\kappa_n)^{r+1}{\E\Big[|\square_n(m)|^{r+1}\mid W_1^\complement\Big]}\cdot {e^{k_2n^{1/2}-k_1n}} \\
	& \preceq \left({tm^{\frac{1}{2}}n^{-1}}\right)^{r+1} + \left({tmn^{-1/2}}\right)^{r+1}\cdot {e^{k_2n^{1/2}-k_1n}}\preceq \left({tm^{\frac{1}{2}}n^{-1}}\right)^{r+1}.
\end{align*}
Here, the second last inequality holds because $\E[|\square_n(m)|^{r+1}]=O((mn)^{(r+1)/2})$ and $|\square_n(m)|\preceq mn$.
We further expand $(\square_n(m))^v$ and bound the leading terms as follows:
\begin{align*}
&\Big|\E\Big[e^{\mathrm{i}t\Big(K_n-\square_n(m)\Big)/\kappa_n-\sum_{i=1}^n\widetilde{g}(X_i)}(\square_n(m))^v\Big]\Big|\\
=&\Big|\E\Big[e^{\mathrm{i}t\Big(K_n-\square_n(m)\Big)/\kappa_n-\sum_{i=1}^n\widetilde{g}(X_i)}(\sum_{i=1}^m\sum_{j=i+1}^n\psi(X_i,X_j))^v\Big]\Big|\\
\le&\sum_{\substack{1 \le j_1\le m \\ j_1<k_1 \le n}}\cdots \sum_{\substack{1 \le j_v\le m \\ j_v<k_v \le n}}\Big|\E\Big[e^{\mathrm{i}t\Big(K_n-\square_n(m)\Big)/\kappa_n-\sum_{i=1}^n\widetilde{g}(X_i)}\prod_{l=1}^v\psi(X_{j_l},X_{k_l})\Big]\Big|.
\end{align*}
For each term in the summation, we have
\begin{align*}
&\Big|\E\Big[e^{\mathrm{i}t\Big(K_n-\square_n(m)\Big)/\kappa_n-\sum_{i=1}^n\widetilde{g}(X_i)}\prod_{l=1}^v\psi(X_{j_l},X_{k_l})\Big]\Big|\\
=&\Big|\E\Big[\exp\Big\{\sum_{i=1}^m(\mathrm{i}tq_1(X_i)/\sqrt{n}-\widetilde{g}(X_i))+\sum_{i=m+1}^n(\mathrm{i}tq_1(X_i)/\sqrt{n}-\widetilde{g}(X_i))\\
&\ \ \ \ \quad \quad +\sum_{i=m+1}^{n-1}\sum_{j=i+1}^n\mathrm{i}t\psi(X_i,X_j)/\kappa_n \Big\}\prod_{l=1}^v\psi(X_{j_l},X_{k_l})\Big]\Big|\\
=&\Big|\E\Big[\exp\Big\{\sum_{i\in S}(\mathrm{i}tq_1(X_i)/\sqrt{n}-\widetilde{g}(X_i))\Big\}\Big]\Big|\\
\cdot&\Big|\E\Big[\exp\Big\{\sum_{i\in [m]\setminus S}(\mathrm{i}tq_1(X_i)/\sqrt{n}-\widetilde{g}(X_i))+\sum_{i=m+1}^n(\mathrm{i}tq_1(X_i)/\sqrt{n}-\widetilde{g}(X_i))\\
&\ \ \ \ \quad \quad +\sum_{i=m+1}^{n-1}\sum_{j=i+1}^n\mathrm{i}t\psi(X_i,X_j)/\kappa_n\Big\}\prod_{l=1}^v\psi(X_{j_l},X_{k_l})\Big]\Big|\\
\coloneqq& \mathcal{A}_{S}\cdot \mathcal{B}_{S^\complement},
\end{align*}
where $S=[m]\setminus \cup_{l=1}^v\{j_l, k_l\}$  and satisfies $|S|\ge m-2v$.

It follows that
\begin{align*}
	\mathcal{A}_{S}= \prod_{i\in S} \underbrace{\left|\E\Big[e^{(\mathrm{i}tq_1(X_i)/\sqrt{n}-\widetilde{g}(X_i))}\Big] \right|}_{\gamma_n(t)}=\left| \gamma_n(t) \right|^{|S|}.
\end{align*}
Note that, for $n^\epsilon\preceq|t|\le c_0n^{1/2}$, we have {$\widetilde{g}(X_i)\prec n^{-1/2}\prec\dfrac{t}{\sqrt{n}}$}. Thus, by the Taylor expansion, we have
\begin{align*}
&\gamma_n(t)\\
= &\E\Big[1+\Big(\frac{\mathrm{i}tq_1(X_i)}{\sqrt{n}\xi_1}-\widetilde{g}(X_i)\Big)+\frac{1}{2}\Big(\frac{\mathrm{i}tq_1(X_i)}{\sqrt{n}\xi_1}-\widetilde{g}(X_i)\Big)^2+o\big|\frac{\mathrm{i}tq_1(X_i)}{\sqrt{n}\xi_1}-\widetilde{g}(X_i)\big|^2\Big]\\
=&1-\E[\widetilde{g}(X_i)]-\frac{t^2\E[q_1^2(X_i)]}{2n\xi_1^2}+o\Big(\frac{t^2}{n}\Big)\\
=&1-\E[\widetilde{g}(X_i)]-\frac{t^2}{2n}+o\Big(\frac{t^2}{n}\Big) {\le} 1-\frac{t^2}{3n}\le \exp\Big(-\frac{t^2}{3n}\Big),
\end{align*}
where the second last inequality holds because $\E[\widetilde{g}(X_i)]=\left(n^2 \rho_n / s_n\right)^{-1} t^2 / 2\E\left[\sigma_W^2\right]\ge 0.$
Therefore, we have 
\begin{align*}
	\mathcal{A}_{S}\le e^{-\frac{t^2}{3n}|S|}\le e^{-\frac{t^2}{3n}(m-2v)} .
\end{align*}

To bound $\mathcal{B}_{S^\complement}$, we have
\begin{align*}
\mathcal{B}_{S^\complement}&\le\E\Big[\exp\{-\sum_{i\in [m]\setminus S}\widetilde{g}(X_i)-\sum_{i=m+1}^n\widetilde{g}(X_i)\}\Big|\prod_{l=1}^v\psi(X_{j_l},X_{k_l})\Big|\Big]\\
&=\E\Big[\exp\{-\sum_{i\in I}\widetilde{g}(X_i)\}\Big|\prod_{l=1}^v\psi(X_{j_l},X_{k_l})\Big|\Big]\\
&=\E\Big[\exp\{-(n^2\rho_n/s_n)^{-1}t^2/2\sum_{i\in I}(\E[\sigma_W^2]+g_{\sigma_W^2}(X_i))\}\Big|\prod_{l=1}^v\psi(X_{j_l},X_{k_l})\Big|\Big],
\end{align*}
where $I=([m]\setminus S) \cup[m+1:n]$ with $|I|\ge n-m$. 
By Bernstein inequality, there exists a constant $k_3$ such that
\begin{align*}
\pr\Big(\sum_{i\in I}(\E[\sigma_W^2]+g_{\sigma_W^2}(X_i))<0\Big)=\pr\Big(\frac{1}{I}\sum_{i\in I}g_{\sigma_W^2}(X_i)<-\E[\sigma_W^2]\Big)\le e^{-k_3|I|}.
\end{align*}
Thus, given that $\psi(X_{j_l},X_{k_l})$ is bounded by constant, by similarly discussing the event $W_1$, we have
\begin{align*}
\mathcal{B}_{S^\complement} \preceq(1+e^{{k_2n^{-1/2}|I|}-k_3|I|})\preceq 1.
\end{align*}
With the bounds for $\mathcal{A}_{S}$ and $B_{S^\complement}$, we have 
\begin{align*}
\Big|\E\Big[e^{\mathrm{i}t\Big(K_n-\square_n(m)\Big)/\kappa_n-\sum_{i=1}^n\widetilde{g}(X_i)}(\square_n(m))^v\Big]\Big| \le \sum_{\substack{1 \le j_1\le m \\ j_1<k_1 \le n}}\cdots \sum_{\substack{1 \le j_v\le m \\ j_v<k_v \le n}} e^{-\frac{t^2}{3n}(m-2v)} \preceq (mn)^v e^{-\frac{t^2}{3n}(m-2v)}.
\end{align*}
It follows that 
\begin{align*}
&\Big|\E\Big[e^{\mathrm{i}t\Big(K_n-\square_n(m)\Big)/\kappa_n-\sum_{1\le i\le n}\widetilde{g}(X_i)}\sum_{v=0}^r\frac{(\mathrm{i}t\square_n(m)/\kappa_n)^v}{v!}\Big]\Big|\\
\preceq &\sum_{v=0}^r({mn^{-\frac{1}{2}}t})^v e^{-\frac{t^2}{3n}(m-2v)} =e^{-\frac{t^2m}{3n}} \sum_{v=0}^r(mn^{-\frac{1}{2}}te^{\frac{2t^2}{3n}})^v.
\end{align*}
Therefore, combined with the remainder term and the upper bound for $J_1$, we have
\begin{align*}
\Big|Ch.f.(\mathcal{F}_{\widetilde{T}_n+\widetilde{\Delta}_n{+\delta_T}}; t)\Big|=O\left( t^2 s_n \rho_n^{-1}n^{-2}\log n + \left({tm^{\frac{1}{2}}n^{-1}}\right)^{r+1} + e^{-\frac{t^2m}{3n}} \sum_{v=0}^r(mn^{-\frac{1}{2}}te^{\frac{2t^2}{3n}})^v \right).
\end{align*}
We choose $m=6rn\log n /t^2$ with a constant $r \ge 2$ that satisfies $m \le n$ when $n^{\epsilon}\preceq |t|$. Then it follows that $\left({tm^{\frac{1}{2}}n^{-1}}\right)^{r+1} =O( n^{-\frac{r+1}{2}}\log ^{\frac{r+1}{2}}n)$ and $e^{-\frac{t^2m}{3n}} \sum_{v=0}^r\left(mn^{-\frac{1}{2}}te^{\frac{2t^2}{3n}}\right)^v =O(n^{-\frac{3r}{2}}\,\log^rn\,t^{-r})$ when $ |t| \le c_0 n^{1/2}$. Note that the latter term $O(n^{-\frac{3r}{2}}\,\log^rn\,t^{-r})$ is always dominated by the former $O( n^{-\frac{r+1}{2}}\log ^{\frac{r+1}{2}}n)$ when $r \ge 1$. Therefore, we have for any constant $r \ge 2$ 
\[\Big|Ch.f.(\mathcal{F}_{\widetilde{T}_n+\widetilde{\Delta}_n{+\delta_T}}; t)\Big|=O\left( t^2 s_n \rho_n^{-1}n^{-2}\log n + n^{-\frac{r+1}{2}}\log^{\frac{r+1}{2}}{n} \right).\]
For any $lower\asymp n^{\epsilon}$ and $upper\asymp n^{1/2}$, it follows that
\begin{align*}
\int_{lower}^{upper}\Big|\frac{Ch.f.(\mathcal{F}_{\widetilde{T}_n+\widetilde{\Delta}_n{+\delta_T}}; t)}{t}\Big|dt=O( s_n\rho^{-1}_nn^{-1}\log{n}+n^{-\frac{r+1}{2}}\log ^{\frac{r+3}{2}}n),
\end{align*}
which completes the proof of Lemma~\ref{lemma.integral} (c) by setting $r=2$.
\end{proof}

\smallskip
\begin{proof}[of Lemma~\ref{lemma.integral} (d)] In this part, we expand $Ch.f.(\mathcal{F}_{\widetilde{T}_n+\widetilde{\Delta}_n{+\delta_T}}; t)$ and compare  the difference with $Ch.f.(\mathcal{G}_n;t)$. By (\ref{eq: chf_var}) and Taylor expansion, we have
\begin{align}
& Ch.f.(\mathcal{F}_{\widetilde{T}_n+\widetilde{\Delta}_n{+\delta_T}}; t) \nonumber\\
=&\E\Big[e^{\mathrm{i}t\widetilde{T}_n}e^{-(\var(\wtD_n\mid W)+\var(\delta_T))t^2/2}\Big]\nonumber \\
=&\E\Big[e^{\mathrm{i}t\widetilde{T}_n}\Big(1-\frac{t^2}{2}((\var(\wtD_n\mid W)+\var(\delta_T)))+O\Big(t^4(\var(\wtD_n\mid W)\vee \var(\delta_T))^2\Big)\Big)\Big].\label{eq: chf Gn}
\end{align}
We first analyze the common multiplier $\E[e^{\mathrm{i}t\widetilde{T}_n}]$. Note that, 
\begin{align*}
	\widetilde{T}_n = \underbrace{W_n^{(1)}}_{\wtO_{p,2}(\log^{1/2}{n})}+ \underbrace{\alpha_n + W_n^{(2)}-\dfrac{L_n}{2}}_{\wtO_{p,2}(n^{-1/2}\log n)}.
\end{align*}
We keep the leading term $W_n^{(1)}$ and use Taylor expansion for the remaining terms:
\begin{align}
\E\Big[e^{\mathrm{i}t\widetilde{T}_n}\Big]=&\E\Big[e^{\mathrm{i}tW_n^{(1)}}\Big\{1+\mathrm{i}t\Big(\alpha_n+W_n^{(2)}-\dfrac{L_n}{2}\Big)-\frac{1}{2}t^2\Big(\alpha_n+W_n^{(2)}-\dfrac{L_n}{2}\Big)^2 \nonumber\\
&\quad \quad \quad \quad \quad +O\Big(t^3\vert\alpha_n+W_n^{(2)}-\dfrac{L_n}{2}\vert^3\Big)\Big\}\Big].\label{eq: Ttilde leading term}
\end{align}
The remainder term in (\ref{eq: Ttilde leading term}) can be bounded by 
\begin{align*}
	\left|\E\Big[e^{\mathrm{i}tW_n^{(1)}} O\Big(t^3\vert\alpha_n+W_n^{(2)}-\dfrac{L_n}{2}\vert^3 \Big)\Big] \right| \le t^3\E\Big[\vert\alpha_n+W_n^{(2)}-\dfrac{L_n}{2}\vert^3\Big] = O(t^3 n^{-3/2}\log^3n).
\end{align*}

Next, we analyze the intercept term and the linear (w.r.t\ $t$) term in (\ref{eq: Ttilde leading term}). The quadratic term can similarly be shown to be of a smaller order using the same procedure.

Note that $\E[e^{\mathrm{i}tW_n^{(1)}}]=\E\Big[e^{\mathrm{i}t\cdot \sum_{i=1}^n q_1(X_i)/(\sqrt{n}\xi_1)}\Big]=\prod_{i=1}^n\E\Big[e^{\mathrm{i}t\cdot q_1(X_i)/(\sqrt{n}\xi_1)}\Big]:= \left(\varphi_n(t)\right)^n$. Denote a polynomial of $t$ that involves only components with powers greater than $\ell$ by $\text{P}^{\ge \ell}(t)$. 
By applying Lemma~\ref{lemma: phin n-k} to $q_1(X_1)$, we have, for $k=0,1,2$,
\begin{align*}
\varphi_n^{n-k}(t) & =e^{-t^2/2}\Big(1-\frac{\mathrm{i}t^3}{6\sqrt{n}\xi_1^3}\E\Big[q_1^3(X_1)\Big]\Big)+O\left(n^{-1}\cdot \text{P}_k^{\ge 1}(t) e^{-t^2 / 12}\right),
\end{align*}
where $\text{P}_k^{\ge 1}(t)$ are fixed polynomials.

Therefore, we have the approximation of the intercept and linear terms
\begin{align*}
	\E\Big[e^{\mathrm{i}tW_n^{(1)}}\Big]= &\  e^{-t^2/2}\Big(1-\dfrac{\mathrm{i}t^3}{6\sqrt{n}\xi_1^3}\E\Big[q_1^3(X_1)\Big]\Big)+O\Big(n^{-1}\cdot \text{P}_k^{\ge 1}(t) e^{-t^2 / 12}\Big),\\
	\E\Big[e^{\mathrm{i}tW_n^{(1)}}\cdot\mathrm{i}t\alpha_n\Big] =&  \dfrac{\mathrm{i}t}{\sqrt{n}\xi_1^3}e^{-t^2/2}\Big({3\xi_1^2\E\Big[p_1(X_1)q_1(X_1)\Big]}-\dfrac{1}{2}\E\Big[q_1^3(X_1)\Big]\\
&\quad \quad \quad \quad \quad \quad -2\E\Big[q_1(X_1)q_1(X_2)q_2(X_1,X_2)\Big]\Big)\\
& +O\Big(n^{-1}\cdot \text{P}_k^{\ge 1}(t) e^{-t^2 / 12}\Big),\\
\E\Big[e^{\mathrm{i}tW_n^{(1)}}\cdot\mathrm{i}tW_n^{(2)}\Big]=&\frac{\mathrm{i}t}{n^{3/2}\xi_1}\E\Big[e^{\mathrm{i}t\cdot\sum_{i=1}^nq_1(X_i)/(\sqrt{n}\xi_1)}\cdot\sum_{1\le i\neq j \le n}q_2(X_i,X_j)
\Big] \\
=&\frac{\mathrm{i}tn(n-1)}{n^{3/2}\xi_1}\varphi_n^{n-2}(t)\E\Big[e^{\mathrm{i}t(q_1(X_1)+q_1(X_2))/(\sqrt{n}\xi_1)}\cdot q_2(X_1,X_2)\Big] \\
=&\frac{\mathrm{i}t(n-1)}{\sqrt{n}\xi_1}\varphi_n^{n-2}(t)\E\Big[q_2(X_1,X_2)+\frac{\mathrm{i}t(q_1(X_1)+q_1(X_2))}{\sqrt{n}\xi_1}q_2(X_1,X_2)\\
&-\frac{t^2(q_1(X_1)+q_1(X_2))^2}{2n\xi_1^2}q_2(X_1,X_2)+O\Big(n^{-3/2}t^3\Big)\Big]\\
=&-\frac{\mathrm{i}t^3}{\sqrt{n}\xi_1^3}e^{-t^2/2}\E\Big[q_1(X_1)q_1(X_2)q_2(X_1,X_2)\Big] +O\Big(n^{-1}\cdot \text{P}_k^{\ge 1}(t) e^{-t^2 / 12}\Big),\\
\E\Big[e^{\mathrm{i}tW_n^{(1)}}\cdot\mathrm{i}tL_{n,1}\Big]=&\frac{\mathrm{i}t}{n^{3/2}\xi_1^3}\E\Big[e^{\mathrm{i}t\cdot\sum_{i=1}^nq_1(X_i)/(\sqrt{n}\xi_1)}\sum_{1\le i\neq j\le n}q_1(X_i)(q_{1}^2(X_{j})-\xi_{1}^2)\Big] \\
=&\frac{\mathrm{i}t(n-1)}{\sqrt{n}\xi_1^3}\varphi_n^{n-2}(t)\E\Big[e^{\mathrm{i}t\cdot(q_1(X_1)+q_1(X_2)/(\sqrt{n}\xi_1)}\cdot q_1(X_1)(q_{1}^2(X_{2})-\xi_{1}^2)\Big]\\
=&\frac{\mathrm{i}t(n-1)}{\sqrt{n}\xi_1^3}\varphi_n^{n-2}(t)\E\Big[q_1(X_1)(q_{1}^2(X_{2})-\xi_{1}^2)+\frac{\mathrm{i}t(q_1(X_1)+q_1(X_2))}{\sqrt{n}\xi_1}q_1(X_1)(q_{1}^2(X_{2})-\xi_{1}^2)\\
&\quad \quad \quad \quad \quad \quad \quad \quad \quad -\frac{t^2(q_1(X_1)+q_1(X_2))^2}{2n\xi_1^2}q_1(X_1)(q_{1}^2(X_{2})-\xi_{1}^2)+O\Big(n^{-3/2}t^3\Big)\Big]\\
=-&\frac{\mathrm{i}t^3}{\sqrt{n}\xi_1^3}e^{-t^2/2}\E\Big[q_1^3(X_1)\Big]+O\Big(n^{-1}\cdot \text{P}^{\ge 1}(t) e^{-t^2 / 12}\Big),\\
\E\Big[e^{\mathrm{i}tW_n^{(1)}}\cdot\mathrm{i}tL_{n,2}\Big]=& \frac{4\mathrm{i}t}{n^{3/2}\xi_1^3}\E\Big[e^{\mathrm{i}t\cdot\sum_{i=1}^nq_1(X_i)/(\sqrt{n}\xi_1)}\sum_{1\le i\neq j\le n}q_1(X_i)g_{\delta,1}(X_j)\Big] \\
=&\frac{4\mathrm{i}t(n-1)}{\sqrt{n}\xi_1^3}\varphi_n^{n-2}(t)\E\Big[e^{\mathrm{i}t\cdot(q_1(X_1)+q_1(X_2)/(\sqrt{n}\xi_1)}\cdot q_1(X_1)g_{\delta,1}(X_2)\Big]\\
=&\frac{4\mathrm{i}t(n-1)}{\sqrt{n}\xi_1^3}\varphi_n^{n-2}(t)\E\Big[q_1(X_1)g_{\delta,1}(X_2)+\frac{\mathrm{i}t(q_1(X_1)+q_1(X_2))}{\sqrt{n}\xi_1}q_1(X_1)g_{\delta,1}(X_2)\\
&\quad \quad \quad \quad \quad \quad \quad \quad \quad -\frac{t^2(q_1(X_1)+q_1(X_2))^2}{2n\xi_1^2}q_1(X_1)g_{\delta,1}(X_2)+O\Big(n^{-3/2}t^3\Big)\Big]\\
=&-\frac{4\mathrm{i}t^3}{\sqrt{n}\xi_1^3}e^{-t^2/2}\E\Big[q_1(X_1)q_1(X_2)q_2(X_1,X_2)\Big]+O\Big(n^{-1}\cdot \text{P}^{\ge 1}(t) e^{-t^2 / 12}\Big),\\
\E\Big[e^{\mathrm{i}tW_n^{(1)}}\cdot\mathrm{i}tL_{n,3}\Big]=& \frac{6\mathrm{i}t}{n^{3/2}\xi_1^3}\E[p_1(X_1)q_1(X_1)]\E\Big[e^{\mathrm{i}t\cdot\sum_{i=1}^nq_1(X_i)/(\sqrt{n}\xi_1)}\sum_{1\le i\neq j\le n}q_1(X_i)q_1(X_j)\Big] \\
=&\frac{6\mathrm{i}t(n-1)}{\sqrt{n}\xi_1^3}\E[p_1(X_1)q_1(X_1)]\varphi_n^{n-2}(t)\E\Big[e^{\mathrm{i}t\cdot(q_1(X_1)+q_1(X_2))/(\sqrt{n}\xi_1)}\cdot q_1(X_1)q_1(X_2)\Big]\\
=&\frac{6\mathrm{i}t(n-1)}{\sqrt{n}\xi_1^3}\E[p_1(X_1)q_1(X_1)]\varphi_n^{n-2}(t)\E\Big[q_1(X_1)q_1(X_2)+\frac{\mathrm{i}t(q_1(X_1)+q_1(X_2))}{\sqrt{n}\xi_1}q_1(X_1)q_1(X_2)\\
&\quad \quad \quad \quad \quad \quad \quad \quad \quad -\frac{t^2(q_1(X_1)+q_1(X_2))^2}{2n\xi_1^2}q_1(X_1)q_1(X_2)+O\Big(n^{-3/2}t^3\Big)\Big]\\
=&-\frac{6\mathrm{i}t^3}{\sqrt{n}\xi_1}e^{-t^2/2}\E\Big[p_1(X_1)q_1(X_1)\Big]+O\Big(n^{-1}\cdot \text{P}^{\ge 1}(t) e^{-t^2 / 12}\Big),
\end{align*}
and, by combining the above results together, we have
\begin{align*}
	&\E\Big[e^{\mathrm{i}tW_n^{(1)}}\Big\{1+\mathrm{i}t\Big(\alpha_n+W_n^{(2)}-\dfrac{L_{n,1}+L_{n,2}- L_{n,3}}{2}\Big)\Big\}\Big]\\
	= & e^{-t^2/2}\Big(1-\frac{\mathrm{i}t}{\sqrt{n}\xi_1^3}\cdot\Big(\frac{1}{2}\E\Big[q_1^3(X_1)\Big]+2\E\Big[q_1(X_1)q_1(X_2)q_2(X_1,X_2)\Big]-3\xi_1^2\E\Big[p_1(X_1)q_1(X_1)\Big]\Big)\\
&\quad \quad \quad +\frac{\mathrm{i}t^3}{\sqrt{n}\xi_1^3}\cdot\Big(\frac{1}{3}\E\Big[q_1^3(X_1)\Big]+\E\Big[q_1(X_1)q_1(X_2)q_2(X_1,X_2)\Big]{-3\xi_1^2\E\Big[p_1(X_1)q_1(X_1)\Big]}\Big)\Big)\\
&+O\Big(n^{-1}\cdot \text{P}^{\ge 1}(t) e^{-t^2 / 12}\Big).
\end{align*}

Following the same procedure, it can be shown that the quadratic term $\E[e^{\mathrm{i}tW_n^{(1)}}t^2(\alpha_n+W_n^{(2)}-L_n/2)^2]$ can be absorbed into $O\Big(n^{-1}\cdot \text{P}^{\ge 1}(t) e^{-t^2 / 12}\Big)$. Thus, by plugging the above results into~(\ref{eq: Ttilde leading term}), we have
\begin{align*}
\E\Big[e^{\mathrm{i}t\widetilde{T}_n}\Big]=Ch.f.(\mathcal{G}_n;t)+O\Big(n^{-1}\cdot \text{P}^{\ge 1}(t) e^{-t^2 / 12} + t^3 n^{-3/2}\log^3n\Big),
\end{align*}
where
\begin{align*}
\mathcal{G}_n(x)=\Phi(x)+\frac{\phi(x)}{\sqrt{n}\xi_1^3}&\Big((\frac{\E\Big[q_1^3(X_1)\Big]}{3}+\E\Big[q_1(X_1)q_1(X_2)q_2(X_1,X_2)\Big]{-3\xi_1^2\E\Big[p_1(X_1)q_1(X_1)\Big]})x^2\\
&+\frac{\E\Big[q_1^3(X_1)\Big]}{6}+\E\Big[q_1(X_1)q_1(X_2)q_2(X_1,X_2)\Big]\Big).
\end{align*}
Therefore, for $upper\asymp n^\epsilon$ with $\epsilon<1/12$, we have
\begin{align*}
	& \int_0^{upper}\left|\frac{\E\big[e^{\mathrm{i}t\widetilde{T}_n}\big]-Ch.f.(\mathcal{G}_n;t)}{t}\right| dt\\
	= & O\left( n^{-1}\int_0^{upper}\mid \text{P}^{\ge 0}(t) e^{-t^2 / 12}\mid dt + n^{-3/2}\log^3n \int_0^{upper} t^2 dt\right)\\
	= & O(n^{-1} +n^{3\epsilon-3/2}\log^3n)= O(n^{-1}).
\end{align*}
It remains to bound the quadratic and remainder terms in (\ref{eq: chf Gn}). For the quadratic term, by (\ref{Hoeffding.sigma_W^2}), we have 
\begin{align*}
&\E\Big[e^{\mathrm{i}t\widetilde{T}_n}t^2\var(\wtD_n\mid W)\Big]\\
=&(n\rho_n/s_n)^{-1}t^2\E\Big[e^{\mathrm{i}t\widetilde{T}_n}\Big(\E\Big[\sigma_W^2\Big]+\frac{1}{n}\sum_{ i=1}^n g_{\sigma_W^2}(X_i)+\wtO_{p,1}(n^{-1}\log{n})\Big)\Big]\\
=&(n\rho_n/s_n)^{-1}t^2\E[\sigma_W^2]\E\Big[e^{\mathrm{i}t\widetilde{T}_n}\Big]+(n\rho_n/s_n)^{-1}t^2\E\Big[e^{\mathrm{i}t\widetilde{T}_n}g_{\sigma_W^2}(X_1)\Big]+O\Big((n^2\rho_n/s_n)^{-1}t^2\log{n}\Big).
\end{align*}
Note that, by using the same procedure, we have  
\begin{align*}
\Big|\E\Big[e^{\mathrm{i}t\widetilde{T}_n}g_{\sigma_W^2}(X_1)\Big]\Big|=O\Big(n^{-1/2} \cdot \text{P}^{\ge 1}(t) e^{-t^2 / 12}\Big). 
\end{align*}
It follows that 
\begin{align*}
&\int_0^{upper}\mid\frac{\E\Big[e^{\mathrm{i}t\widetilde{T}_n}t^2\var(\wtD_n\mid W)\Big]}{t}\mid dt=O(n^{-1}\rho_n^{-1}s_n).
\end{align*}
We also have 
\begin{align*}
&\int_0^{upper}\mid\frac{\E\Big[e^{\mathrm{i}t\widetilde{T}_n}t^2\var(\delta_T)\Big]}{t}\mid dt\asymp n^{-1}\log n\int_0^{upper}\mid t\E\Big[e^{\mathrm{i}t\widetilde{T}_n}\Big]\mid dt=O(n^{-1}\log n).
\end{align*}
For the last remainder term in (\ref{eq: chf Gn}), we have for $\epsilon<{\min(\dfrac{1}{4},\dfrac{1}{4}+\dfrac{\log\rho_n-\log s_n}{4\log n})}$
\begin{align*}
\int_0^{upper}\mid\frac{t^4(\E[\var(\wtD_n\mid W)]\vee \var(\delta_T))^2}{t}\mid dt&\asymp n^{4\epsilon-2}(\rho_n^{-2}s_n^{2}\vee\log^2n)\prec n^{-1}\rho_n^{-1}s_n
\end{align*}
because of Condition (c).

By combining the above bounds together, we have  that 
\[\int_0^{upper}\left|\frac{Ch.f.(\mathcal{F}_{\widetilde{T}_n+\widetilde{\Delta}_n{+\delta_T}}; t)-Ch.f.(\mathcal{G}_n;t)}{t}\right| dt = O(n^{-1}\log n +   n^{-1}\rho_n^{-1}s_n),\]
which completes the proof of Lemma~\ref{lemma.integral} (d).
\end{proof}

\section{Proof of Theorem~\ref{thm2} and Corollary~\ref{thm3}}
\label{sec:proof-thm2-thm3}

With the empirical estimates of $p_1(X_i)$, $q_1(X_i)$ and $q_2(X_i,X_j)$ as follows:

\begin{align*}
\widehat{p}_1(X_i)=\sum_{\substack{1\le i_1 < i_2 \le n\\ i \neq i_1, i_2}}\frac{h_{\Delta}(A_{i,\,i_1,i_2})-\widehat{V}_n}{\binom{n-1}{2}\widehat{V}_n},\quad\widehat{q}_1(X_i)=\sum_{\substack{1\le i_1 < i_2 \le n\\ i \neq i_1, i_2}}\frac{\widehat{V}_n h_{\blacktriangle}(A_{i,\,i_1,i_2})-\widehat{U}_n h_{\Delta}(A_{i,\,i_1,i_2})}{\binom{n-1}{2}\widehat{V}^2_n},
\end{align*}
and
\begin{align*}
\widehat{q}_2(X_i,X_j)=\frac{1}{n-2}\sum_{\substack{1 \le i_1 \le n\\ i,j \neq i_1}}\frac{h_{\blacktriangle}(A_{i_1,\,i,\,j})}{\widehat{V}_n}-\frac{h_{\Delta}(A_{i_1,\,i,\,j})}{\widehat{V}^2_n}\widehat{U}_n-\widehat{q}_1(X_i)-\widehat{q}_1(X_j),
\end{align*}
we have
\begin{align*}
\widehat{\xi}_1^2=\frac{1}{n}\sum_{i=1}^n\widehat{q}_1^2(X_i),\, \widehat{\E}\{q_1^3(X_1)\}=\frac{1}{n}\sum_{i=1}^n\widehat{q}_1^3(X_i),\,\widehat{\E}\{q_1(X_1)p_1(X_1)\}=\frac{1}{n}\sum_{i=1}^n\widehat{q}_1(X_i)\widehat{p}_1(X_i),
\end{align*}
and 
\begin{align*}
\widehat{\E}\{q_1(X_1)q_1(X_2)q_2(X_1,X_2)\}=\frac{1}{\binom{n}{2}}\sum_{1 \le i < j \le n}\widehat{q}_1(X_i)\widehat{q}_1(X_j)\widehat{q}_2(X_i,X_j).
\end{align*}
For an observed network, we can determine the coefficients in the empirical Edgeworth expansion $$
\widehat{\mathcal{G}}_n(x)=\Phi(x)+\frac{\phi(x)}{\sqrt{n}}\left\{\widehat{a}(\frac{1}{3}x^2+\frac{1}{6})+\widehat{b}(x^2+1)-3\widehat{c}x^2\right\}
$$ as
\begin{align}\label{defineabchat}
\widehat{a}=\widehat{\E}\{q_1^3(X_1)\}/\widehat{\xi}_1^3,\,\widehat{b}=\widehat{\E}\{q_1(X_1)q_1(X_2)q_2(X_1,X_2)\}/\widehat{\xi}_1^3,\,\widehat{c}=\widehat{\E}\{q_1(X_1)p_1(X_1)\}/\widehat{\xi}_1.
\end{align}
Notice that calculating these statistics for network of size $n$ only requires a few $n\times n$ matrix multiplications, which is computationally efficient.
\begin{proof}[of Theorem~\ref{thm2}]
By Theorem~\ref{thm1}, we have $||\mathcal{F}_{\widehat{T}_n + \delta_T}(x)-\mathcal{G}_n(x)||_{\infty}=O(\mathcal{E}(n,s_n,\rho_n))$. We have also proved $\widehat{\xi}_1-\xi_1=\wtO_{p,1}(n^{-\frac{1}{2}}\,\log n)$ in Lemma~\ref{lemma}, which leads to
\[|\frac{1}{\sqrt{n}\xi^3_1}-\frac{1}{\sqrt{n}\widehat{\xi}^3_1}|=\wtO_{p,1}(n^{-1}\,\log n)\quad\text{and}\quad|\frac{1}{\sqrt{n}\xi_1}-\frac{1}{\sqrt{n}\widehat{\xi}_1}|=\wtO_{p,1}(n^{-1}\,\log n).\]
To proceed, we only need to demonstrate
\begin{align*}
\max&\Big\{\Big|\widehat{\E}[q_1^3(X_1)]-\E[q_1^3(X_1)]\Big|, \,\Big|\widehat{\E}[q_1(X_1)q_1(X_2)q_2(X_1,X_2)]-\E[q_1(X_1)q_1(X_2)q_2(X_1,X_2)]\Big|,\\
&\Big|\widehat{\E}[q_1(X_1)p_1(X_1)]-\E[q_1(X_1)p_1(X_1)]\Big|\,\Big\}=\wtO_{p,1}(n^{\frac{1}{2}}M(n, s_n, \rho_n)).
\end{align*}

We first bound $|\widehat{\E}[q_1^3(X_1)]-\E[q_1^3(X_1)]|$. By (\ref{hatA-Abound2}), (\ref{A-abound}), and (\ref{lemma4ahat-a}) in the proof of Lemma~\ref{lemma}, we have $\widehat{A}_i-A_i=\wtO_{p,2}(n^{-\frac{1}{2}}\,\rho_n^{-\frac{1}{2}}\,\log^{\frac{1}{2}}n)$, $A_i-\widehat{a}_i=\wtO_{p,2}(n^{-1}\,\rho_n^{-\frac{1}{2}}\,\log^{\frac{1}{2}}n + n^{-1/2} M(n, s_n, \rho_n))$ and $\widehat{a}_i-a_i=\wtO_{p,2}(n^{-\frac{1}{2}}log^{\frac{1}{2}}n)$ holds uniformly for $1\le i\le n$. Consequently, $\widehat{A}_i - a_i = \wtO_{p,2}(n^{-\frac{1}{2}}\rho_n^{-\frac{1}{2}}\log^{\frac{1}{2}}n)$ holds uniformly for $1\le i\le n$, which implies \[\dfrac{1}{n}\sum_{i=1}^n\widehat{A}_i^3-\dfrac{1}{n}\sum_{i=1}^n a_i^3 = \wtO_{p,1}(n^{-\frac{1}{2}}\rho_n^{-\frac{1}{2}}\log^{\frac{1}{2}}n).\] 
We also have $a_i - \widetilde{a}_i = \wtO_{p,2}(n^{-\frac{1}{2}} \log^{\frac{1}{2}}n)$ holds uniformly for $1\le i\le n$ because of (\ref{aitildea1a2}), which leads to \[\frac{1}{n}\sum_{i=1}^n a_i^3-\frac{1}{n}\sum_{i=1}^n\widetilde{a}_i^3 = \wtO_{p,1}(n^{-\frac{1}{2}} \log^{\frac{1}{2}}n).\] By Lemma~\ref{applemma2}, we have \[\dfrac{1}{n}\sum_{i=1}^n \widetilde{a}_i^3-\E[\widetilde{a}_i^3] = \wtO_{p,1}(n^{-\frac{1}{2}} \log^{\frac{1}{2}}n).\]
Therefore,
\begin{align*}
&|\widehat{\E}[q_1^3(X_1)]-\E[q_1^3(X_1)]|\\
=&|\frac{1}{n}\sum_{i=1}^n\widehat{A}_i^3-\E[\widetilde{a}_i^3]| \\
\le &|\frac{1}{n}\sum_{i=1}^n\widehat{A}_i^3-\frac{1}{n}\sum_{i=1}^n a_i^3|+|\frac{1}{n}\sum_{i=1}^n a_i^3-\frac{1}{n}\sum_{i=1}^n \widetilde{a}_i^3|+|\frac{1}{n}\sum_{i=1}^n \widetilde{a}_i^3-\E[\widetilde{a}_i^3]|\\
=&\wtO_{p,1}(n^{-\frac{1}{2}}\,\rho_n^{-\frac{1}{2}}\,\log^{\frac{1}{2}}n)=\wtO_{p,1}(n^{\frac{1}{2}}M(n, s_n, \rho_n)).
\end{align*}

Next, we bound $\widehat{\E}[q_1(X_1)q_1(X_2)q_2(X_1,X_2)]-\E[q_1(X_1)q_1(X_2)q_2(X_1,X_2)]$, which is
\begin{align*}
&\frac{1}{\binom{n}{2}}\sum_{1 \le i < j \le n}\widehat{q}_1(X_i)\widehat{q}_1(X_j)\widehat{q}_2(X_i,X_j)-\E[q_1(X_1)q_1(X_2)q_2(X_1,X_2)]\\
=&\frac{1}{\binom{n}{2}}\sum_{1 \le i < j \le n}(\widehat{q}_1(X_i)\widehat{q}_1(X_j)\widehat{q}_2(X_i,X_j)-q_1(X_i)q_1(X_j)q_2(X_i,X_j))\\
+&\frac{1}{\binom{n}{2}}\sum_{1 \le i < j \le n}(q_1(X_i)q_1(X_j)q_2(X_i,X_j)-\E[q_1(X_1)q_1(X_2)q_2(X_1,X_2)])\\
=&\frac{1}{\binom{n}{2}}\sum_{1 \le i < j \le n}(\widehat{q}_1(X_i)\widehat{q}_1(X_j)\widehat{q}_2(X_i,X_j)-q_1(X_i)q_1(X_j)q_2(X_i,X_j))+\wtO_{p,1}(n^{-1}\,\log n)\\
=&\frac{1}{\binom{n}{2}}\sum_{1 \le i < j \le n}\widehat{q}_1(X_i)\widehat{q}_1(X_j)[\widehat{q}_2(X_i,X_j)-q_2(X_i,X_j)]\\
+&\frac{1}{\binom{n}{2}}\sum_{1 \le i < j \le n}[\widehat{q}_1(X_i)\widehat{q}_1(X_j)q_2(X_i,X_j)-q_1(X_i)q_1(X_j)q_2(X_i,X_j)]+\wtO_{p,1}(n^{-1}\,\log n).
\end{align*}
Note that
\begin{align*}
&\widehat{q}_1(X_i)-q_1(X_i)=\widehat{A}_i - q_1(X_i)=(\widehat{A}_i-A_i)+(A_i-q_1(X_i))\\
=&(\widehat{A}_i-A_i)+\frac{1}{\binom{n-1}{2}}\sum_{\substack{1 \le i_1 < i_2 \le n\\ i_1, i_2 \neq i}}h_{\blacktriangle}(X_i,X_{i_1},X_{i_2})(\frac{1}{V_n}-\frac{1}{v_n})-h_{\Delta}(X_i,X_{i_1},X_{i_2})(\frac{U_n}{V_n^2}-\frac{u_n}{v_n^2})\\
=&\wtO_{p,2}(n^{-\frac{1}{2}}\,\rho_n^{-\frac{1}{2}}\,\log^{\frac{1}{2}}n)+\wtO_{p,2}(n^{-\frac{1}{2}}\,\log^{\frac{1}{2}}n)\\
=&\wtO_{p,2}(n^{-\frac{1}{2}}\,\rho_n^{-\frac{1}{2}}\,\log^{\frac{1}{2}}n).
\end{align*}
Since $q_2(X_i, X_j) \preceq 1$, we have $\widehat{q}_1(X_i)\widehat{q}_1(X_j)q_2(X_i,X_j)-q_1(X_i)q_1(X_j)q_2(X_i,X_j) = \wtO_{p,2}(n^{-\frac{1}{2}}\,\rho_n^{-\frac{1}{2}}\,\log^{\frac{1}{2}}n)$ holds uniformly for $1\le i\le n$, so $$\frac{1}{\binom{n}{2}}\sum_{1 \le i < j \le n}[\widehat{q}_1(X_i)\widehat{q}_1(X_j)q_2(X_i,X_j)-q_1(X_i)q_1(X_j)q_2(X_i,X_j)]=\wtO_{p,2}(n^{-\frac{1}{2}}\,\rho_n^{-\frac{1}{2}}\,\log^{\frac{1}{2}}n).$$
Now we need to bound $\dfrac{1}{\binom{n}{2}}\displaystyle\sum_{1 \le i < j \le n}\widehat{q}_1(X_i)\widehat{q}_1(X_j)[\widehat{q}_2(X_i,X_j)-q_2(X_i,X_j)].$ Let $$\widehat{A}_{ij}=\frac{1}{n-2}\sum\limits_{\substack{1 \le i_1 \le n\\i_1 \neq i,j}}\frac{h_{\blacktriangle}(A_{i_1,i,j})}{\widehat{V}_n}-\frac{h_{\Delta}(A_{i_1,i,j})\widehat{U}_n}{\widehat{V}^2_n},$$
$$ a_{ij}=\frac{1}{n-2}\sum\limits_{\substack{1 \le i_1 \le n\\i_1 \neq i,j}}\frac{h_{\blacktriangle}(X_{i_1},X_i,X_j)}{v_n}-\frac{h_{\Delta}(X_{i_1},X_i,X_j)u_n}{v^2_n},$$
and
$$
\widetilde{a}_{ij}=\frac{\E[h_{\blacktriangle}(X_{i_1},X_i,X_j)|X_i,X_j]}{v_n}-\frac{u_n}{v_n^2}\E[h_{\Delta}(X_{i_1},X_i,X_j)|X_i,X_j].
$$
We have 
\begin{align*}
&\widehat{q}_2(X_i,X_j)-q_2(X_i,X_j)=\widehat{A}_{ij} - \widetilde{a}_{ij}\\
=&(\widehat{A}_{ij}-\widehat{q}_1(X_i)-\widehat{q}_1(X_j))-(\widetilde{a}_{ij}-q_1(X_i)-q_1(X_j))\\
=&(\widehat{A}_{ij}-a_{ij})+(a_{ij}-\widetilde{a}_{ij})-(\widehat{q}_1(X_i)-q_1(X_i))-(\widehat{q}_1(X_j)-q_1(X_j)).
\end{align*}
Recall that $\widehat{q}_1(X_i)-q_1(X_i)= \wtO_{p,2}(n^{-\frac{1}{2}}\,\rho_n^{-\frac{1}{2}}\,\log^{\frac{1}{2}}n)$ holds uniformly for $1\le i\le n$. By Lemma~\ref{applemma2}, $a_{ij} - \widetilde{a}_{ij} = \widetilde{O}_{p,3}(n^{-\frac{1}{2}} \log^{\frac{1}{2}}n)$ holds uniformly for $1\le i<j\le n$. As a result, there exist constants $C_1,C_2,N_1>0$ such that $\pr(|(\widehat{q}_2(X_i,X_j)-q_2(X_i,X_j)) - (\widehat{A}_{ij}-a_{ij})| \le C_1 n^{-\frac{1}{2}}\,\rho_n^{-\frac{1}{2}}\,\log^{\frac{1}{2}}n, \, 1 \le i < j \le n) \ge 1 - C_2 n^{-1}$ when $n > N_1$. Thus we have
\begin{align*}
&\frac{1}{\binom{n}{2}}\sum_{1 \le i < j \le n}\widehat{q}_1(X_i)\widehat{q}_1(X_j)[\widehat{q}_2(X_i,X_j)-q_2(X_i,X_j)]\\
=&\frac{1}{\binom{n}{2}}\sum_{1 \le i < j \le n}\widehat{q}_1(X_i)\widehat{q}_1(X_j)[\widehat{A}_{ij}-a_{ij}]+\wtO_{p,1}(n^{-\frac{1}{2}}\,\rho_n^{-\frac{1}{2}}\,\log^{\frac{1}{2}}n)\\
=&\frac{1}{\binom{n}{2}}\sum_{1 \le i < j \le n}\widehat{A}_i\widehat{A}_j(\widehat{A}_{ij}-a_{ij})+\wtO_{p,1}(n^{-\frac{1}{2}}\,\rho_n^{-\frac{1}{2}}\,\log^{\frac{1}{2}}n)\\
=&\frac{1}{\binom{n}{2}}\sum_{1 \le i < j \le n} (\widehat{A}_i - a_i) \widehat{A}_j (\widehat{A}_{ij}-a_{ij}) + a_i (\widehat{A}_j - a_j)(\widehat{A}_{ij}-a_{ij}) + a_i a_j (\widehat{A}_{ij}-a_{ij})\\
&\quad+\wtO_{p,1}(n^{-\frac{1}{2}}\,\rho_n^{-\frac{1}{2}}\,\log^{\frac{1}{2}}n).
\end{align*}
Without loss of generality, we assume $\widehat{U}_n \asymp \widehat{V}_n \asymp \rho_n^3$ and $\widehat{A}_i \preceq 1$, $1 \le i \le n$, since these events occur with a probability of at least $1 - C_1 n^{-1}$. We define
$$ A_{ij}=\frac{1}{n-2}\sum\limits_{\substack{1 \le i_1 \le n\\i_1 \neq i,j}}\frac{h_{\blacktriangle}(X_{i_1},X_i,X_j)}{\widehat{V}_n}-\frac{h_{\Delta}(X_{i_1},X_i,X_j)\widehat{U}_n}{\widehat{V}^2_n}.
$$
Recall that $\widehat{U}_n - u_n = \wtO_{p,2}(n^{-\frac{1}{2}}\rho_n^3 \log^{\frac{1}{2}}n)$ and $\widehat{V}_n - v_n = \wtO_{p,2}(n^{-\frac{1}{2}}\rho_n^3 \log^{\frac{1}{2}}n)$. Without loss of generality, we assume $|A_{ij} - a_{ij}| \le C_2 n^{-1/2 }\log^{1/2}n$, $1 \le i < j \le n$ for some constant $C_2$ because this event occurs with at a probability of at least $1 - C_3 n^{-1}$ when $n$ is sufficiently large. 
By Lemma~\ref{applemma2}, $\widehat{A}_{ij} - A_{ij} = \widetilde{O}_{p,3}(\rho_n^{-1} + n^{-1/2} \rho_n^{-2} \log^{1/2} n)$ holds uniformly for $1\le i<j\le n$. Therefore, $\widehat{A}_{ij} - a_{ij}= \widetilde{O}_{p,3}(\rho_n^{-1} + n^{-1/2} \rho_n^{-2} \log^{1/2} n)$ holds uniformly for $1\le i<j\le n$. 
Given that $\widehat{A}_i - a_i = \wtO_{p,2}(n^{-1/2}\rho_n^{-1/2} \log^{1/2}n)$ holds uniformly for $1\le i\le n$, there exist constants $C_4,C_5,N_2>0$ such that $P ((\widehat{A}_i - a_i)(\widehat{A}_{ij} - a_{ij})\le C_4 n^{-1/2} \rho_n^{-3/2} \log^{1/2} n + n^{-1} \rho_n^{-5/2} \log n, 1 \le i < j\le n) \ge 1 - C_5n^{-1}$ when $n > N_2$. Thus we have
\begin{align*}
&\frac{1}{\binom{n}{2}}\sum_{1 \le i < j \le n}\widehat{q}_1(X_i)\widehat{q}_1(X_j)[\widehat{q}_2(X_i,X_j)-q_2(X_i,X_j)]\\
=&\frac{1}{\binom{n}{2}}\sum_{1 \le i < j \le n} a_i a_j (\widehat{A}_{ij}-a_{ij}) + \wtO_{p,1}(n^{-1/2} \rho_n^{-3/2} \log^{1/2} n+n^{-1} \rho_n^{-5/2} \log n)\\
=&\frac{1}{\binom{n}{2}}\sum_{1 \le i < j \le n} a_i a_j (\widehat{A}_{ij}-A_{ij}) + \wtO_{p,1}( n^{-1/2} \rho_n^{-3/2} \log^{1/2} n+n^{-1} \rho_n^{-5/2} \log n).
\end{align*}
Following the similar proof procedure of Lemma~\ref{apppro1}, we have
\begin{align*}
&\frac{1}{\binom{n}{2}}\sum_{1 \le i < j \le n} a_i a_j (\widehat{A}_{ij}-A_{ij})\\
=&\frac{1}{6 \binom{n}{3} \widehat{V}_n} \sum_{\substack{1 \le i, j, k \le n\\i \neq j, j \neq k, k \neq i}}a_i a_j(h_{\blacktriangle}(A_{i, j, k}) - h_{\blacktriangle}(X_i, X_j, X_k))\\
&-\frac{\widehat{U}_n}{6 \binom{n}{3} \widehat{V}^2_n} \sum_{\substack{1 \le i, j, k \le n\\i \neq j, j \neq k, k \neq i}}a_i a_j(h_{\Delta}(A_{i, j, k}) - h_{\Delta}(X_i, X_j, X_k))\\
=&\wtO_{p,2}(n^{-1} \rho_n^{-1/2} \log^{1/2}n).
\end{align*}
Consequently,
\begin{align*}
&\widehat{\E}[q_1(X_1)q_1(X_2)q_2(X_1,X_2)]-\E[q_1(X_1)q_1(X_2)q_2(X_1,X_2)]\\
=&\wtO_{p,1}(n^{-1/2} \rho_n^{-3/2} \log^{1/2} n + n^{-1} \rho_n^{-5/2} \log n)=\wtO_{p,1}(n^{\frac{1}{2}}M(n, s_n, \rho_n)).
\end{align*}

For the last term $\widehat{\E}[q_1(X_1)p_1(X_1)]-\E[q_1(X_1)p_1(X_1)]$, by Lemma~\ref{applemma2}, we have
\begin{align*}
&\widehat{\E}[q_1(X_1)p_1(X_1)]-\E[q_1(X_1)p_1(X_1)]\\
=&\frac{1}{n}\sum_{i=1}^n(\widehat{q}_1(X_i)\widehat{p}_1(X_i)-q_1(X_i)p_1(X_i))+(q_1(X_i)p_1(X_i)-\E[q_1(X_1)p_1(X_1)])\\
=&\frac{1}{n}\sum_{i=1}^n(\widehat{q}_1(X_i)\widehat{p}_1(X_i)-q_1(X_i)p_1(X_i))+\wtO_{p,1}(n^{-\frac{1}{2}}\,\log^{\frac{1}{2}}n).
\end{align*}
Because
$$
\widehat{q}_1(X_i)-q_1(X_i)=\wtO_{p,2}(n^{-\frac{1}{2}}\,\rho_n^{-\frac{1}{2}}\,\log^{\frac{1}{2}}n)\quad\text{and}\quad\widehat{p}_1(X_i)-p_1(X_i)=\wtO_{p,2}(n^{-\frac{1}{2}}\,\rho_n^{-\frac{1}{2}}\,\log^{\frac{1}{2}}n),
$$
we have $\widehat{q}_1(X_i)\widehat{p}_1(X_i)-q_1(X_i)p_1(X_i) = \wtO_{p,2}(n^{-\frac{1}{2}}\,\rho_n^{-\frac{1}{2}}\,\log^{\frac{1}{2}}n)$. Again, by Lemma~\ref{applemma2},
$$
|\widehat{\E}[q_1(X_1)p_1(X_1)]-\E[q_1(X_1)p_1(X_1)]|=\wtO_{p,1}(n^{-\frac{1}{2}}\,\rho_n^{-\frac{1}{2}}\,\log n)=\wtO_{p,1}(n^{\frac{1}{2}}M(n, s_n, \rho_n)).
$$
Here, we finish the proof of Theorem~\ref{thm2}. The proof of Corollary~\ref{thm3} follows the same approach as \citet[Theorem 4.3]{DBLP:journals/corr/abs-2004-06615}.
\end{proof}

\section{Technical Lemmas and their Proofs}

To establish the error rate of the proposed Edgeworth expansion approximation, we require the following technical lemmas. Some of these lemmas are existing results from the literature, which we restate here for ease of reference. For those not directly derived from existing literature, we provide the proofs immediately following each lemma.

\begin{lemma}
    \label{techbernsteinthreom1}
    (Theorem 1 in \cite{major2007multivariate}) Let $X_{1},\dots,X_{n}$ be a sequence of $i.i.d.$ random variables on a space (X, $\mathcal{X}$) with some distribution $\mu$. Suppose that $k$ is a fixed positive integer and $\varphi(x_{1},\dots,x_{k})$ is a function that is canonical with respect to the measure $\mu$
on the space $(X^k, \mathcal{X}^k)$, i.e.
$$
\int \varphi(x_{1},\dots,x_{j-1}, u, x_{j+1}, x_{k})\mu(du)=0,
$$
for all $1\le j \le k$ and $x_s \in X$, $s \in \{1,\dots, k\}\backslash \{j\}$.
Consider a U-statistic $I_{n, k}(\varphi)$ with order $k$:
$$
I_{n, k}(\varphi)=\frac{1}{k!}\sum_{\substack{ 
1 \le j_{s} \le n ,\,\, s=1,\dots,k \\
j_s\neq j_{s'} \,\, if \, s\neq s'
}}\varphi(X_{j_1},\dots,X_{j_k}).
$$
If $\varphi$ satisfies the following condition in~(\ref{lemma2cond}) with some $0 < \sigma^{2} \le 1$,
\begin{equation}
\label{lemma2cond}
\sup_{x_j \in X,\, 1 \le j \le k}|\varphi(x_{1},\dots,x_{k})| \le 1
,\quad
\int \varphi^2(x_{1},\dots,x_{k})\mu(dx_{1})\cdots\mu(dx_{k})\le \sigma^2,
\end{equation}
then there exist constants
$A = A(k) > 0$ and $B = B(k) > 0$ depending only on the order k of the U-statistic $I_{n, k}(\varphi)$ such
that
\begin{equation}\label{technewbern2}
\pr(k!\,n^{-k/2}|I_{n, k}(\varphi)|> u)\le A \cdot \exp\left\{-\frac{u^{2/k}}{2\sigma^{2/k}\{1+B[un^{-k/2}\sigma^{-(k+1)}]^{1/k}\}}\right\}
\end{equation}
for all $0 \le u \le n^{k/2}\sigma^{k+1}$.
\end{lemma}

\medskip

\begin{lemma}\label{applemma1}
Let $X_{1},\dots,X_{n}$ be a sequence of $i.i.d.$ random variables on a space (X, $\mathcal{X}$) with some distribution $\mu$. Suppose that  $k$ is a fixed positive integer and $\varphi(x_{1},\dots,x_{k})$ is a function that is canonical with respect to the measure $\mu$
on the space $(X^k, \mathcal{X}^k)$. If there exists a constant $C$ such that, for any $x_i \in X$, $|\varphi(x_{1},\dots,x_{k})|\le Cq_n$, then for any constant $h \ge 1$, we have $C_1$, $C_2$ and $N$ such that
$$
\pr(|\frac{1}{n^k}I_{n,k}(\varphi) \ge C_1 q_n\,n^{-\frac{k}{2}}\,\log^{\frac{k}{2}}n) \le C_2 \frac{1}{n^h}
$$
for $n>N$.
\end{lemma}
\begin{proof}[of Lemma~\ref{applemma1}]
Without loss of generality, we consider $n^{-2}log^2 n \prec q_n \preceq 1$. Otherwise, we can rescale $q_n$. We prove the result by applying Lemma~\ref{techbernsteinthreom1} for the function $\varphi \cdot q_n^{-3/4}$. To meet the condition in (\ref{lemma2cond}) and $0 \le u \le n^{k/2}\sigma^{k+1}$, we assign the values of $\sigma$ and u as follows. Note that there exists a constant $C$ such that $|\varphi(\cdot)| \le C q_{n}$. We set $\sigma = C\,q_n^{1/4} $, which satisfies the condition in (\ref{lemma2cond}). Then we set $u=C_1\,k!\,q_n^{1/4}\,\log^{k/2}n$ that meets the condition $0 \le u \le n^{k/2}\sigma^{k+1}$ for large $n$ and $n^{-2}log^2 n \prec q_n$. Here, the constant $C_1$ will be specified later. 
By plugging them into (\ref{technewbern2}), we have 
\begin{align*}
&\pr(n^{-k}|I_{n, k}(\varphi)|\ge C_1\,q_n\,n^{-k/2}\,\log^{k/2}n)\\
=&\pr(k!\,n^{-k/2}|I_{n, k}(\varphi)|\ge C_1\,k!\,q_n\,\log^{k/2}n)\\
=&\pr(k!\,n^{-k/2}|I_{n, k}(\varphi\cdot q_n^{-3/4})|\ge C_1\,k!\,q_n^{1/4}\,\log^{k/2}n)\\
=&\pr(k!\,n^{-k/2}|I_{n, k}(\varphi\cdot q_n^{-3/4})|\ge u)\\
\le& C_2\,exp\{-(\frac{C_1\,k!}{C})^{2/k}\,\log\,n\,(1+B)^{-1}\}\\
\le& \frac{C_2}{n^h}.
\end{align*}
The last inequality holds by choosing the constant $C_1$ sufficiently large, which completes the proof.
\end{proof}

\medskip

\begin{lemma}\label{applemma2}
Consider independent random variables $\{\eta_i\}_{i=1}^n$ with $|\eta_i| \le 1$ and $\E(\eta_i)=0$. Assuming there exist a constant $C_1 > 0$ such that $\E [\eta_i^2] \le C_1 p_n$, where $p_n \succ n^{-1}logn$.  And $|\Theta_i| \le C_2 q_n$ with a constant $C_2 > 0$. Then for any constant $h\ge 1$, we have $C_3$, $C_4$ and $N$ such that
$$
\pr(|\frac{1}{n}\sum_{i=1}^{n}\Theta_i \eta_i| \ge C_3 n^{-\frac{1}{2}}\,\log^{\frac{1}{2}}n\,q_n\,p^{\frac{1}{2}}_n ) \le C_4 \frac{1}{n^h}
$$
for $n > N$.
\end{lemma}
\begin{proof}[of Lemma~\ref{applemma2}] 
Note that $\Theta_i\eta_i$'s are independent zero-mean random variables with $|\Theta_i\eta_i| \le C_2 q_n $  and $\E [\Theta_i^2\eta_i^2]\le C_1C_2^2p_n q_n^2$ for all $i$. 
By applying the Bernstein inequality, we have
\begin{align*}
&\pr(\frac{1}{n}\sum\limits_{i=1}^{n}\Theta_i \eta_i \ge C_3\,n^{-\frac{1}{2}}\,\log^{\frac{1}{2}}n\,q_n\,p^{\frac{1}{2}}_n)\\
\le & \exp(-\frac{\frac{C_3^2}{2}n\,\log\,n\,p_n q_n^2}{\sum\limits_{i=1}^n \E[\Theta_i^2 \eta_i^2]+\frac{C_3C_2}{3}n^{\frac{1}{2}}\,\log^{\frac{1}{2}}n\,p^{\frac{1}{2}}_n q_n^2}) \text{\quad (by Bernstein Inequality)}\\
\le & \exp(-\frac{\frac{C_3^2}{2}n\,\log\,n\,p_n q_n^2}{C_1 C_2^2\,n\,p_nq_n^2+\frac{C_3C_2}{3}n^{\frac{1}{2}}\,\log^{\frac{1}{2}}n\,p^{\frac{1}{2}}_n q_n^2}) \quad (\text{by} \ \E[\Theta_i^2\eta_i^2]\le C_1C_2^2p_n q_n^2)\\
= & \exp(-\frac{\frac{C_3^2}{2}\,\log\,n}{C_1C_2^2+\frac{C_3C_2}{3}n^{-\frac{1}{2}}\,\log^{\frac{1}{2}}n\,p^{-\frac{1}{2}}_n })\\
\le & \frac{C_4}{n^h}.
\end{align*}
Since $p_n \succ n^{-1}\log n$, there exists constants $C_3$ and $C_4$ such that the inequality holds.
\end{proof}

\medskip
\begin{lemma} \label{hyperth}
(Theorem 1.3 in \cite{schudy2011bernstein}) 
A hypergraph is defined by a node set $\mathcal{V}(H):=\{1,\dots,N\}=[N]$ and a set $\mathcal{H}(H)$ of hyperedges. Denote the node set in a hyperedge $\mathfrak{h}$ by $\mathcal{V}(\mathfrak{h})\subset \mathcal{V}(H)$ and the number of nodes in a hyperedge by $|\mathcal{V}(\mathfrak{h})|$.  Consider a multilinear polynomial:
$$
f(Y):=\sum_{\mathfrak{h} \in \mathcal{H}(H): |\mathcal{V}(\mathfrak{h})| \le q}\mathfrak{W}_{\mathfrak{h}}\prod_{\mathfrak{b}\in \mathcal{V}(\mathfrak{h})}Y_{\mathfrak{b}},
$$
where $Y=(Y_1,\ldots , Y_N)$ are the random variables associated with the nodes and $\mathfrak{W}_{\mathfrak{h}}$ is an weight multiplier on each hyperedge $\mathfrak{h}$. For a natural number $\mathfrak{r} \ge 0$, we define
$$
\Xi_{\mathfrak{r}}:=\max_{S\subset[N]:|S|=\mathfrak{r}}
\sum_{\substack{\mathfrak{h}\in \mathcal{H}(H): \\ S\subseteq \mathcal{V}(\mathfrak{h})}}\,
|\mathfrak{W}_{\mathfrak{h}}|
\prod_{\mathfrak{v}\in 
\mathcal{V}(\mathfrak{h})\backslash
S}\E[|Y_{\mathfrak{v}}|].
$$
If $Y_1,\dots, Y_N$ are independent and they all satisfy
the following condition for a common parameter $L>0$:
\begin{align}
\label{hyperineqcondition}
\E\left[|Y_k-\E[Y_k]|^i\right]\le iL\cdot \E\left[|Y_k-\E[Y_k]|^{i-1}\right], \quad i\ge1, \quad 1 \le k \le N,
\end{align}
then we have 
$$
\pr(|f(Y)-\E[f(Y)]|\ge u)\le e^2\cdot max\left\{exp\left\{-\frac{u^2}{C^{\mathfrak{q}}\cdot \var(f(Y))}\right\},\max\limits_{1\le \mathfrak{r} \le \mathfrak{q}}exp\left\{-(\frac{u}{\Xi_{\mathfrak{r}}L^{\mathfrak{r}}C^{\mathfrak{q}}})^{1/\mathfrak{r}}\right\}\right\},
$$
where C is a universal constant.
\end{lemma}

\medskip
We provide a corollary of Lemma~\ref{hyperth} to facilitate its more efficient application. 
\begin{lemma} \label{hyperthquick}
Building on the definitions in Lemma~\ref{hyperth}, we further define $O_{a_n}$ as the upper-bound order of $a_n$. Let $O_{\Xi_{1:2}} = \max (O_{\Xi_1}, O_{\Xi_2})$ and $O_{\Xi_{\ge r_0}} = \max_{\mathfrak{r} \ge r_0} O_{\Xi_{\mathfrak{r}}}$. Then, if either of the following two conditions holds, 
\begin{align}
\label{hyperquickcon1}
(a) \quad O_{var^{1/2}(f(Y))} \succeq O_{\Xi_1}, \quad O_{var^{1/2}(f(Y))} \succ O_{\Xi_{\ge 2}} \log^{\mathfrak{q}-2} n;
\end{align}
\begin{align}
\label{hyperquickcon2}
(b) \quad O_{var^{1/2}(f(Y))} \succeq O_{\Xi_{1:2}}, \quad O_{var^{1/2}(f(Y))} \succ O_{\Xi_{\ge 3}} \log^{\mathfrak{q}-2} n,
\end{align}
we have 
\begin{align*}
\pr(|f(Y)-\E[f(Y)]|\ge C_1  O_{var^{1/2}(f(Y))} \log^2 n )\le e^2 n^{-h},
\end{align*}
for any constant $h \ge 1$ and sufficiently large constant $C_1$.
\end{lemma}
\begin{proof}[of Lemma~\ref{hyperthquick}]
We only prove this lemma under Condition (\ref{hyperquickcon2}), as the proof under Condition  (\ref{hyperquickcon1}) is similar.
We set $u = C_1  O_{var^{1/2}(f(Y))} \log^2 n$. 
When $\mathfrak{r} = 1$, for sufficiently large $C_1$, we have  \[-(\frac{C_1 O_{var^{1/2}(f(Y))} \log^2 n}{\Xi_{1} L^{1} C^\mathfrak{q} }) \le -h \log^2 n \le -h \log n,\] where the first inequality holds as $O_{var^{1/2}(f(Y))} \succeq O_{\Xi_{1}}$.
Similarly, when $\mathfrak{r} = 2$, for sufficiently large $C_1$ and $O_{var^{1/2}(f(Y))} \succeq O_{\Xi_{2}}$, we have $-(\frac{C_1 O_{var^{1/2}(f(Y))} \log^2 n}{\Xi_{2} L^{2} C^\mathfrak{q} })^{1/2} \le -h \log n$.
When $\mathfrak{r} \ge 3$, we have $-(\frac{C_1 O_{var^{1/2}(f(Y))} \log^2 n}{\Xi_{\mathfrak{r}} L^{\mathfrak{r}} C^\mathfrak{q} })^{1/\mathfrak{r}} \le -h \log n$, because $O_{var^{1/2}(f(Y))} \succ O_{\Xi_{\ge 3}} \log^{\mathfrak{q}-2} n$. Combining them together, by Lemma~\ref{hyperth}, it follows that
\begin{align*}
\pr(|f(Y)-\E[f(Y)]|\ge C_1  O_{var^{1/2}(f(Y))} \log^2 n )\le e^2 n^{-h},
\end{align*}
which completes the proof of Lemma~\ref{hyperthquick}.
\end{proof}

\medskip
\begin{lemma}\label{berryessentheroem}(Theorem 2.1 in \cite{chen2001non}) Let $\{X_i\}_{i=1}^n$ be independent but not necessarily identically distributed random variables with zero means and finite variance. Denote the cumulative distribution functions for $\sum_{i=1}^nX_i$ and standard normal distribution as $F$ and $\Phi$ respectively. If $Var(\sum_{i=1}^nX_i)=1$, we have 
    $$
    \Vert F-\Phi\Vert_{\infty}=\sup_{x\in \mathbb{R}}|F(x)-\Phi(x)| \le 4.1\left\{ \sum_{i=1}^n \E[X_i^2 I(|X_i|>1)] + \E[|X_i|^3 I(|X_i| \le 1)]\right\}.
    $$
\end{lemma}

\medskip 
\begin{lemma}[Esseen’s smoothing lemma in \cite{feller1991introduction}]\label{Esseen's smoothing lemma}
For (1) any distribution function $F$, (2) a general function $G$ whose derivative is universally bounded by $M>0$ and satisfies $G(-\infty)=0$ and $G(\infty)=1$, and (3) an arbitrary constant $\gamma>0$, there exist constants $C_1$ and $C_2$ such that
\begin{equation}
\label{eq:Esseen}
\Vert F-G\Vert_\infty\le C_1\int_{-\gamma}^\gamma\Big| \frac{Ch.f.(F; t)-Ch.f.(G; t)}{t}\Big|dt+\frac{C_2M}{\gamma},
\end{equation}
where $Ch.f.(G;t)$ is the characteristic function of $G$, \textit{i.e.}, $Ch.f.(G;t)=\int_{-\infty}^\infty e^{\mathrm{i}tx} dG(x).$
\end{lemma}

\medskip
\begin{lemma}\label{lemma.expop}
For $X=\wtO_{p,1}(a_n)$ with a positive sequence $\{a_n:n \ge 1\}$ that converges to zero, we have $e^{X}=1+\wtO_{p,1}(a_n).$
\end{lemma}

\begin{proof}[of Lemma~\ref{lemma.expop}]
The first-order Taylor expansion of $e^X$ is given by $e^X=1+X+R_X$, where the remainder term is denoted as $R_X=O(X^2)$. By definition, there exist $C>0$ and $\epsilon\in(0,1)$ such that when $|X|<\epsilon$, we have $|R_X|\le CX^2$. For this $\epsilon$, there exist constants $N,C_1,C_2$ such that when $n>N$, $\pr(|X|\ge C_1 a_n)\le C_2/n$ and $|a_n|<\epsilon$. Thus, $$\pr(|R_X|\ge CC_1^2|a_n|)\le \pr(|R_X|\ge CC_1^2a_n^2)\le \pr(X^2\ge C_1^2a_n^2)\le C_2/n,$$ which implies $e^X=1+\wtO_{p,1}(a_n)$.
\end{proof}

\medskip
\begin{lemma}[Lemma 8.2 in \cite{DBLP:journals/corr/abs-2004-06615}]\label{smoothing lemma}
Suppose the random variables $X$, $Y$ and $Z=X-Y$ satisfy (1) $F_Y$ is smooth, (2) there exist a universal $M>0$ and a positive sequence $\{c_n\}_{n=1}^\infty$ such that for any $u\in\mathbb{R}$ and positive sequence $\{a_n\}_{n=1}^\infty$, $F_Y(u+a_n)-F_Y(u)\le Ma_n+O(c_n)$, and (3) $Z=\wtO_{p,1}(d_n)$ for some positive sequence $\{d_n\}_{n=1}^\infty$, then we have
$$\Vert F_X-F_Y\Vert_\infty=O(c_n+d_n+n^{-1}).$$
\end{lemma}

\medskip
\begin{lemma}\label{lemma: phin n-k}
Let $X$ be a random variable, with $\E X^2=\xi_1^2>0$ and $\beta_4=\E[X^4]<\infty$. Define $\varphi_n(t)=\E\Big[e^{\mathrm{i}t\cdot X/(\sqrt{n}\xi_1)}\Big]$. In the interval $|t|<\dfrac{\xi_1^2}{\sqrt{\beta_4}}\sqrt{n}$, we have
\begin{align*}
\Big|\dfrac{d^m}{dt^m}\left(\varphi_n^{n}(t)-e^{-t^2/2}\Big(1-\frac{\mathrm{i}t^3}{6\sqrt{n}\xi_1^3}\E[X^3]\Big)\right)\Big|\le c(4)\frac{\beta_4}{\xi_1^4}n^{-1}e^{-t^2/12}(t^4+\vert t\vert^9),
\end{align*}
where $c(4)$ is a positive constant. 
\end{lemma}
\begin{proof}[ of Lemma~\ref{lemma: phin n-k}]
This is a special case of \citet[Section VI, Lemma 4]{petrov2012sums} by taking $m=0$ and $s=4$.
\end{proof}

\section{Results for $w_{n,t}$}
\label{sec: results for wnt}

The inference procedures for $w_{n,t}(t=1,2,3,4)$ are analogous to those for $w_n$. We only need to replace $h_{\blacktriangle}(A_{i_1,i_2,i_3})$ with $h_{t}(A_{i_1,i_2,i_3})$. The variance estimator becomes
\begin{align*}
n\widehat{S}^2_{n, t}=\frac{9}{n}\sum_{i=1}^n\left[
\sum_{\substack{1\le i_{1}< i_{2}\le n\\i_{1},i_{2}\neq i}}\left\{h_{t}(A_{i,i_{1},i_{2}})/\widehat{V}_n-\widehat{U}_{n, t} h_{\Delta}(A_{i,i_{1},i_{2}})/\widehat{V}^2_n \right\}/\binom{n-1}{2}
\right]^2,
\end{align*}
where \[\widehat{U}_{n,t}=\sum_{1 \le i_1 < i_2 < i_3 \le n}h_{t}(A_{i_1,i_2,i_3}) / \binom{n}{3}.
\]
And we write the empirical estimates of $p_{1,t}(X_i)$, $q_{1,t}(X_i)$ and $q_{2,t}(X_i,X_j)$ as follows: 
\begin{equation*}
\label{thm2defineqp}
\widehat{p}_{1,t}(X_i)=\sum_{\substack{1\le i_1 < i_2 \le n\\ i \neq i_1, i_2}}\frac{h_{\Delta}(A_{i,\,i_1,i_2})-\widehat{V}_n}{\binom{n-1}{2}\widehat{V}_n},\quad\widehat{q}_{1,t}(X_i)=\sum_{\substack{1\le i_1 < i_2 \le n\\ i \neq i_1, i_2}}\frac{\widehat{V}_n h_{t}(A_{i,\,i_1,i_2})-\widehat{U}_{n,t} h_{\Delta}(A_{i,\,i_1,i_2})}{\binom{n-1}{2}\widehat{V}^2_n},
\end{equation*}
and
\begin{equation*}
\label{thm2defineq2}
\widehat{q}_{2,t}(X_i,X_j)=\frac{1}{n-2}\sum_{\substack{1 \le i_1 \le n\\ i,j \neq i_1}}\frac{h_{t}(A_{i_1,\,i,\,j})}{\widehat{V}_n}-\frac{h_{\Delta}(A_{i_1,\,i,\,j})}{\widehat{V}^2_n}\widehat{U}_{n,t}-\widehat{q}_{1,t}(X_i)-\widehat{q}_{1,t}(X_j).
\end{equation*}
It follows that 
\begin{equation*}
\label{thm2definexi1ee}
\widehat{\xi}_{1,t}^2=\frac{1}{n}\sum_{i=1}^n\widehat{q}_{1,t}^2(X_i),\quad \widehat{\E}\{q_{1,t}^3(X_1)\}=\frac{1}{n}\sum_{i=1}^n\widehat{q}_{1,t}^3(X_i),\quad \widehat{\E}\{q_{1,t}(X_1)p_{1,t}(X_1)\}=\frac{1}{n}\sum_{i=1}^n\widehat{q}_{1,t}(X_i)\widehat{p}_{1,t}(X_i),
\end{equation*}
\begin{equation*}
\label{thm2definee}
\widehat{\E}\{q_{1,t}(X_1)q_{1,t}(X_2)q_{2,t}(X_1,X_2)\}=\frac{1}{\binom{n}{2}}\sum_{1 \le i < j \le n}\widehat{q}_{1,t}(X_i)\widehat{q}_{1,t}(X_j)\widehat{q}_{2,t}(X_i,X_j).
\quad 
\end{equation*}
The coefficients in the empirical Edgeworth expansion can be expressed as 
\begin{align*}
\label{defineabcthat}
\widehat{a}_t=\widehat{\E}\{q_{1,t}^3(X_1)\}/\widehat{\xi}_{1,t}^3, \widehat{b}_t=\widehat{\E}\{q_{1,t}(X_1)q_{1,t}(X_2)q_{2,t}(X_1,X_2)\}/\widehat{\xi}_{1,t}^3, \widehat{c}_t=\widehat{\E}\{q_{1,t}(X_1)p_{1,t}(X_1)\}/\widehat{\xi}_{1,t}.
\end{align*}
In order to obtain the error bound for $w_{n,t}$, results similar to Lemma~\ref{apppro1} and Lemma~\ref{apppro2} for $h_2(A_{i_1,i_2,i_3})$ and $h_4(A_{i_1,i_2,i_3})$ are required, as shown in Lemma~\ref{appproaddition}.
\begin{lemma}\label{appproaddition} We present the following decompositions.
\begin{enumerate}
\item By substituting the condition $\rho_n^3 s_n^2 \succeq n^{-2} \log^8 n$ in Lemma~\ref{lemma} with $\rho_n^3 s_n \succeq n^{-2} \log^8 n$ while preserving the other conditions, we derive the decomposition for $h_2$:
\begin{align*}
&\dfrac{1}{\binom{n}{3} }\sum\limits_{1 \le i_1 < i_2 < i_3 \le n}[h_2(A_{i_1, i_2, i_3})-h_2(X_{i_1},X_{i_2},X_{i_3})]\\
=&\frac{1}{\binom{n}{2}}\sum_{1 \le k_1 < k_2 \le n} \mathring{B}_{k_1 k_2}^{(1)} \eta_2(A_{k_1 k_2})+ \frac{1}{\binom{n}{2}}\sum_{1 \le k_1 < k_2 \le n} \mathring{C}_{k_1 k_2}^{(3)} \eta_3(A_{k_1 k_2}) +\wtO_{p,2}(n^{-3/2} \rho_n^{3/2} s_n^{1/2} \log^2 n),
\end{align*}
and for any fixed $1 \le i \le n$, 
\begin{align*}
&\frac{1}{\binom{n-1}{2}}\sum_{\substack{1 \le i_1 < i_2 \le n\\i_1,i_2\neq i}}\left(h_{2}(A_{i,i_1, i_2})-h_{2}(X_i,X_{i_1}, X_{i_2})\right)\\
=&\frac{1}{n-1}\sum_{\substack{1\le j \le n:\\j \neq i}}\widehat{\Theta}_{ij}^{(5)} \eta_3(A_{ij})+\frac{1}{n-1}\sum_{\substack{1\le j \le n:\\j \neq i}}\widehat{\Theta}_{ij}^{(6)}\eta_2(A_{ij})+\wtO_{p,2}(n^{-1} \rho_n^{3/2} s_n^{1/2} \log^2 n),
\end{align*}
where $\mathring{B}_{k_1 k_2}^{(1)}$, $\mathring{C}_{k_1 k_2}^{(3)}$, $\widehat{\Theta}_{k_1 k_2}^{(5)}$ and $\widehat{\Theta}_{k_1 k_2}^{(6)}$ are functions of $X_k$, $1 \le k \le n$. Additionally, 
$\mathring{B}_{k_1 k_2}^{(1)} \asymp \rho_n^2$, $\mathring{C}_{k_1 k_2}^{(3)} \asymp \rho_n^2 s_n$, $\widehat{\Theta}_{k_1 k_2}^{(5)} \asymp \rho_n^2 s_n$ and $\widehat{\Theta}_{k_1 k_2}^{(6)} \asymp \rho_n^2$.
\item By substituting the condition $\rho_n^3 s_n^2 \succeq n^{-2} \log^8 n$ in Lemma~\ref{lemma} with $\rho_n^3 s_n^3 \succeq n^{-2} \log^8 n$ while preserving the other conditions, we derive the decomposition for $h_4$:
\begin{align*}
&\dfrac{1}{\binom{n}{3} }\sum\limits_{1 \le i_1 < i_2 < i_3 \le n}[h_4(A_{i_1, i_2, i_3})-h_4(X_{i_1},X_{i_2},X_{i_3})]\\
=&\frac{1}{\binom{n}{2}}\sum_{1 \le k_1 < k_2 \le n} \mathring{B}_{k_1 k_2}^{(2)} \eta_2(A_{k_1 k_2})+\wtO_{p,2}(n^{-3/2} \rho_n^{3/2} s_n^{3/2} \log^2 n),
\end{align*}
and for any fixed $1 \le i \le n$, 
\begin{align*}
&\frac{1}{\binom{n-1}{2}}\sum_{\substack{1 \le i_1 < i_2 \le n\\i_1,i_2\neq i}}\left(h_{4}(A_{i,i_1, i_2})-h_{4}(X_i,X_{i_1}, X_{i_2})\right)\\
=&\frac{1}{n-1}\sum_{\substack{1\le j \le n:\\j \neq i}}\widehat{\Theta}_{ij}^{(7)}\eta_2(A_{ij})+\wtO_{p,2}(n^{-1} \rho_n^{3/2} s_n^{3/2} \log^2 n),
\end{align*}
where $\mathring{B}_{k_1 k_2}^{(2)}$ and $\widehat{\Theta}_{k_1 k_2}^{(7)}$ are functions of $X_k$. Additionally, 
$\mathring{B}_{k_1 k_2}^{(2)} \asymp \rho_n^2 s_n^2$ and $\widehat{\Theta}_{k_1 k_2}^{(7)} \asymp \rho_n^2 s_n^2$.
\end{enumerate}
\end{lemma}
The Edgeworth expansion approximation error bound for $w_{n,t} (1\le t\le 4)$ in Theorem~\ref{thm1} and Theorem~\ref{thm2} is given by $\mathcal{E}_t(n, s_n, \rho_n) = M_t(n, s_n, \rho_n) \log^{1/2} n + n^{-1}\rho_n^{-1/2} s_n^{-1/2}$,
where $M_1(n, s_n, \rho_n) = n^{-1}\rho_n^{-3/2} \log n $ and $M_t(n, s_n, \rho_n) = n^{-1}\rho_n^{-3/2} \log n (1 + s_n^{(t-1)/2}\log n)$ for $t \ge 2$. Notably, the exponent of $s_n$ is associated with the number of negative edges in the type-$t$ triangle. Condition~(c1) is revised from $\rho_n^3 s_n^2 \succeq n^{-2}\log^8 n$ to $\rho_n^3 s_n^t \succeq n^{-2}\log^8 n$. The confidence interval for $w_{n,t}$ is provided in Corollary~\ref{wnt CI}.
\begin{corollary}\label{wnt CI}
For any $\alpha \in (0,1)$, we define 
$$
\widehat{q}_{\widehat{T}_{n,t;\alpha}}:=z_{\alpha}-\frac{1}{\sqrt{n}}\{\widehat{a}_{t}(\frac{1}{3}z_{\alpha}^2+\frac{1}{6})+\widehat{b}_{t}(z_{\alpha}^2+1)-3\widehat{c}_{t}z_{\alpha}^2\} - \delta_T,
$$
where $z_{\alpha}:=\Phi^{-1}(\alpha)$. 
Assuming the modified conditions hold, the $1-\alpha$ two-sided symmetric CI for $w_{n,t}$,
\begin{equation*}
(\frac{\widehat{U}_{n,t}}{\widehat{V}_n}-\widehat{q}_{\widehat{T}_{n,t;1-\alpha/2}}\widehat{S}_{n,t} , \frac{\widehat{U}_{n,t}}{\widehat{V}_n}-\widehat{q}_{\widehat{T}_{n,t;\alpha/2}}\widehat{S}_{n,t}),
\end{equation*}
has a $1-\alpha+O(\mathcal{E}_t(n, s_n, \rho_n))$ coverage probability.
\end{corollary}

\section{Additional Details for Numerical Studies}
\subsection{Details for Node Resampling}\label{sec: node resampling}

We extend the node resampling procedure developed for the binary network in \citet{green2022bootstrapping} to derive confidence intervals for $w_{n,t}(t=1,2,3,4)$ in signed networks. Given an observed signed network $A_n$, we estimate $\rho_nF$ and $s_nG$ under the proposed graphon model for signed networks (see Definition~\ref{def: graphon}) using Algorithm~\ref{algo: estimate graphon}. With the empirical estimates $\hat{r}$ for $\rho_nF$ and $\hat{s}$ for $s_nG$, we generate $B$ signed networks $\{{A_n^*}^{(b)}\}_{b=1}^B$ from the graphon model. For a signed network ${A^*}$ generated from the graphon model associated with $(\hat{r}, \hat{s})$, we define the probability $\bar{s}_n(\hat{r},\hat{s})=\pr(A_{ij}^*=-1\mid|A_{ij}^*|=1; \hat{r}, \hat{s})$ and $\hat{s}_n$ as its empirical estimate of the proportion of negative edges among all edges in~$A^*$. Similarly, we define $w_{n,t}(\hat{r},\hat{s})=\pr(h_t(A_{1,2,3})=1\mid h_\Delta(A_{1,2,3})=1; \hat{r}, \hat{s})$ and  $\hat{w}_{n,t}(A^*)$ as its empirical estimate based on $A^*$. Inspired by \citet[Theorem 2]{green2022bootstrapping}, we construct the confidence interval for $w_{n,t}$ as $\hat{I}_{t}(A_n)=(\hat{q}_{\alpha/2}, \hat{q}_{1-\alpha/2})$, where $\hat{q}_{\alpha/2}$ and $\hat{q}_{1-\alpha/2}$ are the  $\alpha/2$ and upper $1-\alpha/2$ quantiles, respectively, of the empirical distribution of $\hat{w}_{n,t}(A_n)-(\hat{w}_{n,t}({A_n^*}^{(b)}) -w_{n,t}(\hat{r},\hat{s}))\hat{s}_n^{t-1}/\bar{s}_n^{t-1}$.

\begin{algo}\label{algo: estimate graphon}
Empirical graphon for a signed network
\begin{tabbing}
    \qquad Input observed signed network $A_n$\\
    \qquad Set $\hat{\theta}_r=|A_n|$, $\hat{\theta}_s=(A_n == -1)$\\
    \qquad Set $\hat{r}(x,y)=(\hat{\theta}_r)_{\lceil nx\rceil\lceil ny\rceil}$, $\hat{s}(x,y)=(\hat{\theta}_s)_{\lceil nx\rceil\lceil ny\rceil}$\\
    \qquad Output $\hat{r}$ and $\hat{s}$ 
\end{tabbing}
\end{algo}
\subsection{Full Results for Simulation Studies}\label{sec: full simulation results}
\begin{table}[h]
\centering
\begin{tabular}{|l|c|c|c|}
    \hline
    CDF distance & Our method & Normal & Resampling \\
    \hline
    Type-1    & \bf{0.0121}  & 0.0520  & 0.0722  \\
    \hline
    Type-2    & \bf{0.0028}  & 0.0171  & 0.0227  \\
    \hline
    Type-3    & \bf{0.0056}  & 0.0309  & 0.0362  \\
    \hline
    Type-4    & \bf{0.0015}  & 0.0043  & 0.0052  \\
    \hline
    Balanced  & \bf{0.0019}  & 0.0116  & \diagbox{}{} \\
    \hline
\end{tabular}
\caption{Kolmogorov--Smirnov (sup-norm) distances between the true cumulative distribution function and its approximations.}
\label{tab: CDF distance appendix}
\end{table}
\begin{figure}
\includegraphics[width=0.8\textwidth]{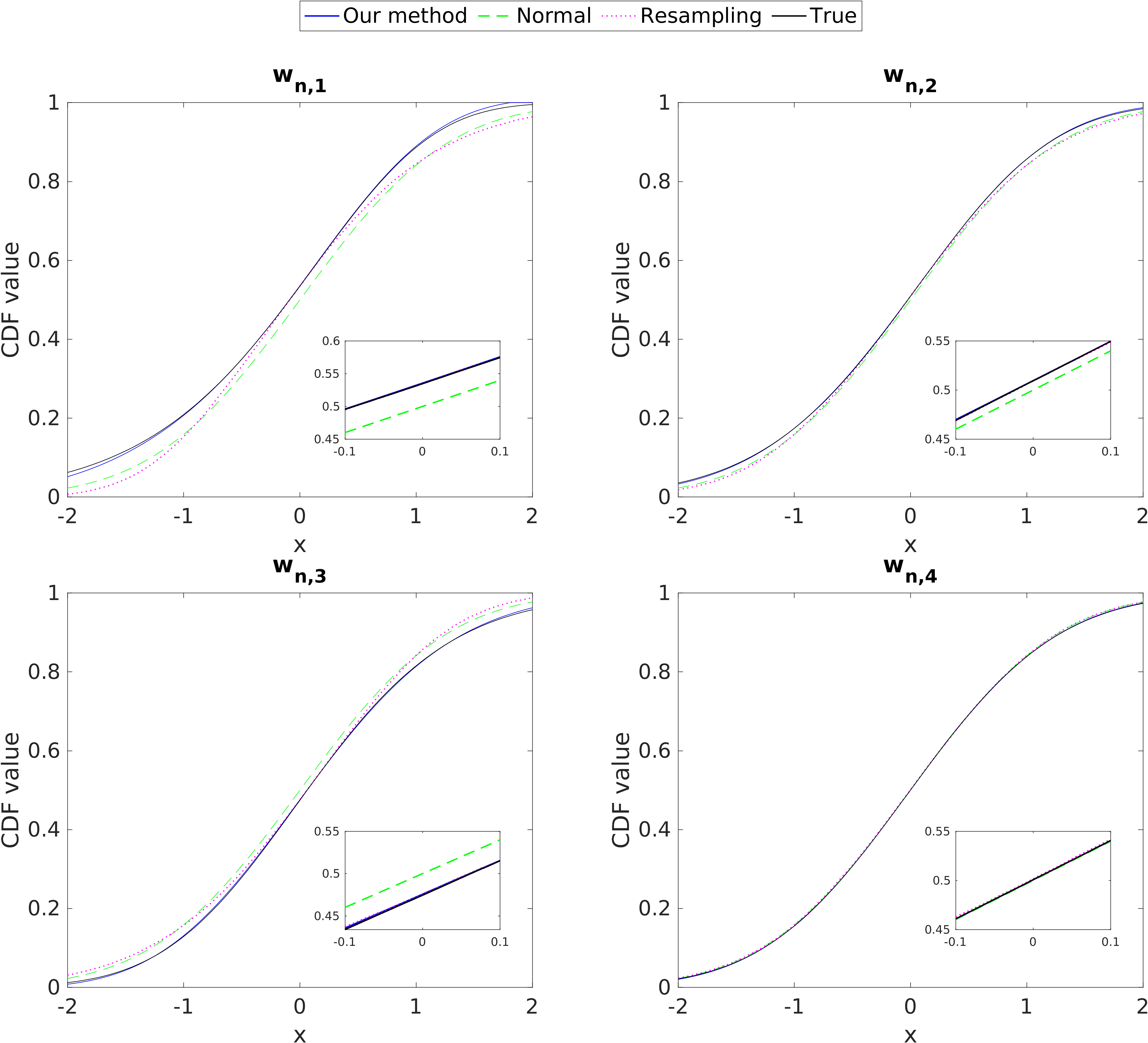}
\caption{The true CDF of the studentized ratio of network moments, along with its approximations (including our empirical Edgeworth expansion formula, standard normal approximation, and the bootstrap-based node-resampling method), is displayed on the top-left for type-1 triangles, on the top-right for type-2 triangles, on the bottom-left for type-3 triangles, and on the bottom-right for type-4 triangles.}
\label{fig.CDF_appendix}
\end{figure}

\begin{figure}[!htb]
\centering
\begin{subfigure}{0.39\textwidth}
    \includegraphics[height=4.9cm, keepaspectratio]{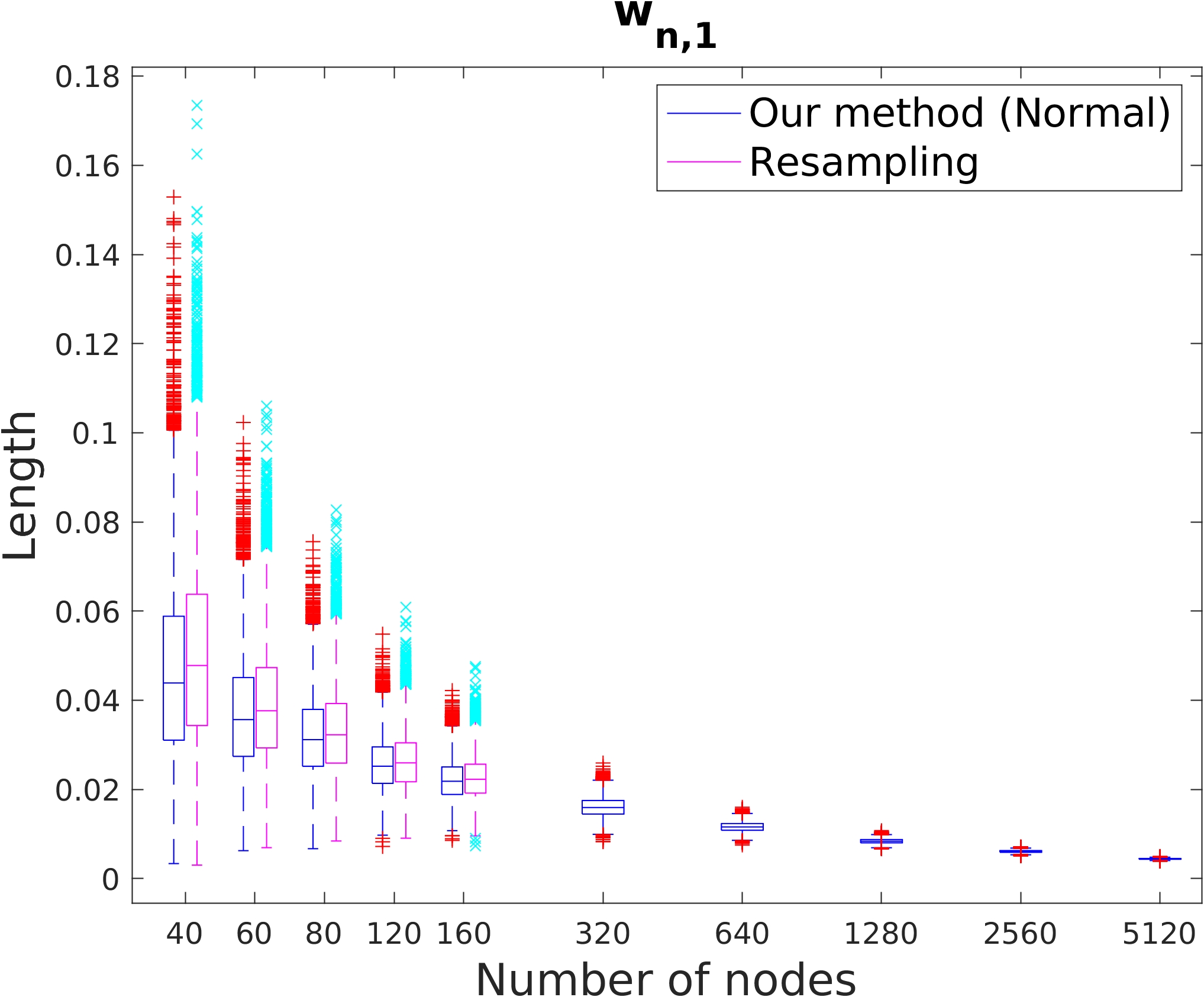}
\end{subfigure}
\hspace{0.02\textwidth}
\begin{subfigure}{0.39\textwidth}
    \includegraphics[height=4.9cm, keepaspectratio]{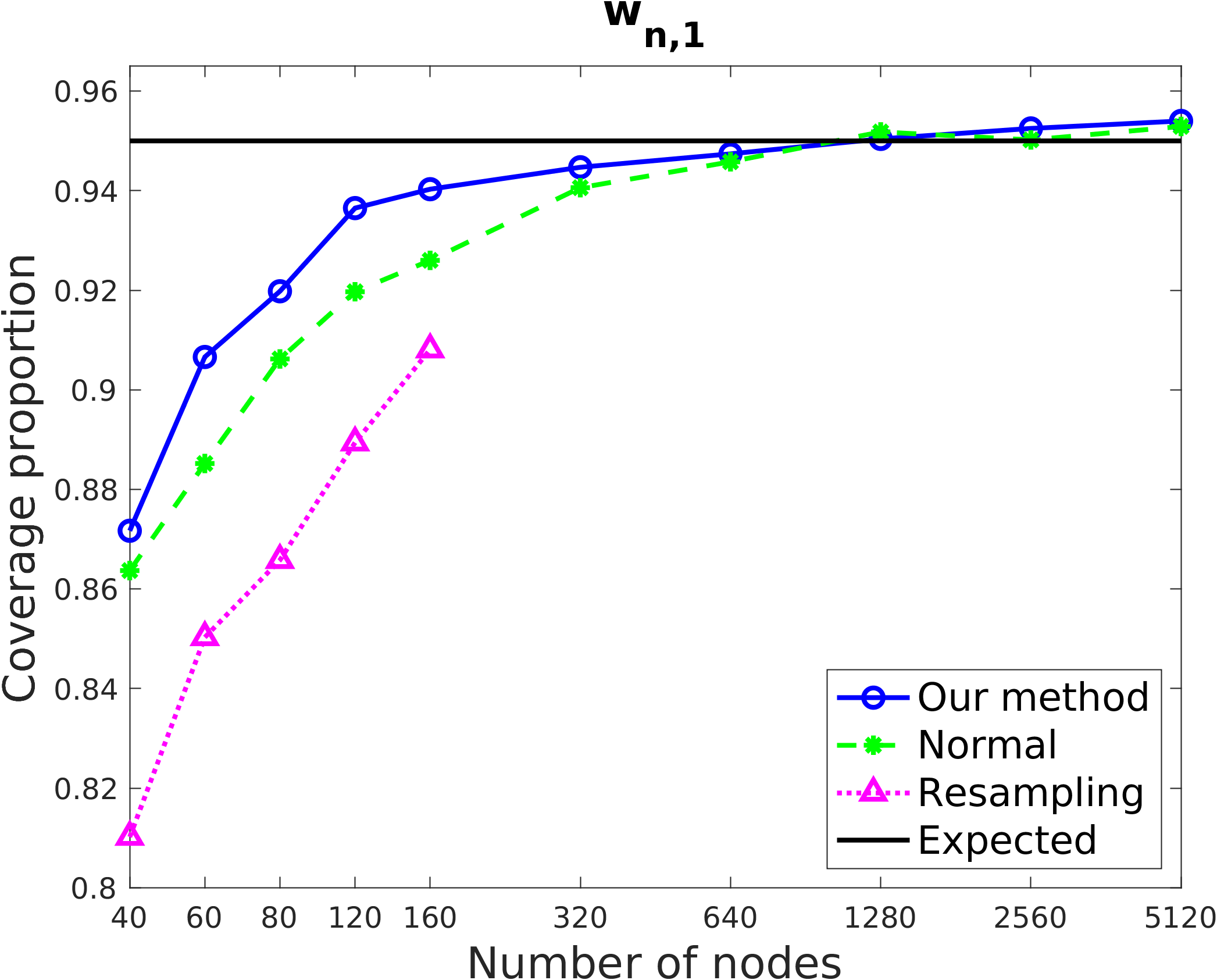}
\end{subfigure}
\vspace{0.01\textwidth}

\begin{subfigure}{0.39\textwidth}
    \includegraphics[height=4.9cm, keepaspectratio]{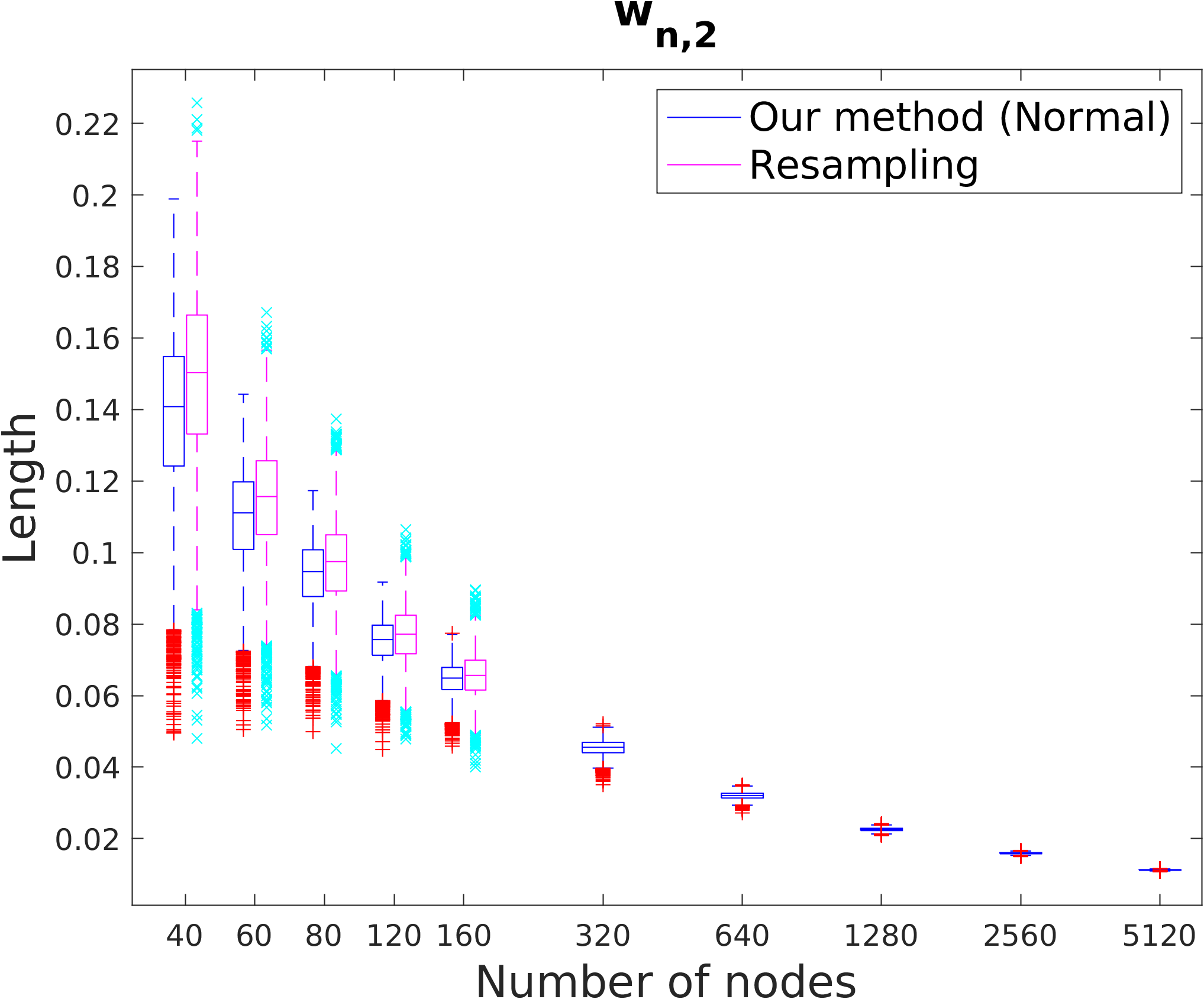}
\end{subfigure}
\hspace{0.02\textwidth}
\begin{subfigure}{0.39\textwidth}
    \includegraphics[height=4.9cm, keepaspectratio]{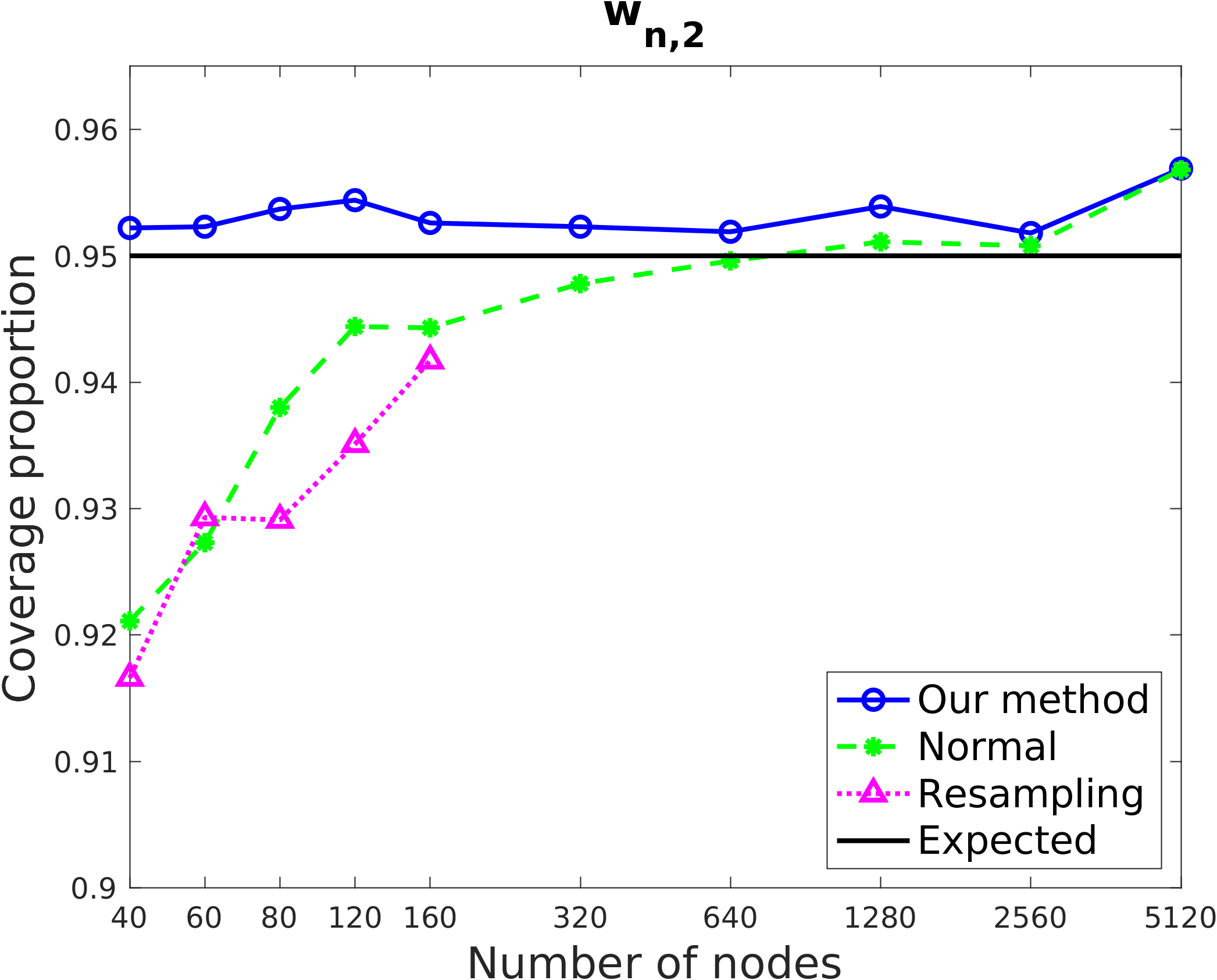}
\end{subfigure}
\vspace{0.01\textwidth}

\begin{subfigure}{0.39\textwidth}
    \includegraphics[height=4.9cm, keepaspectratio]{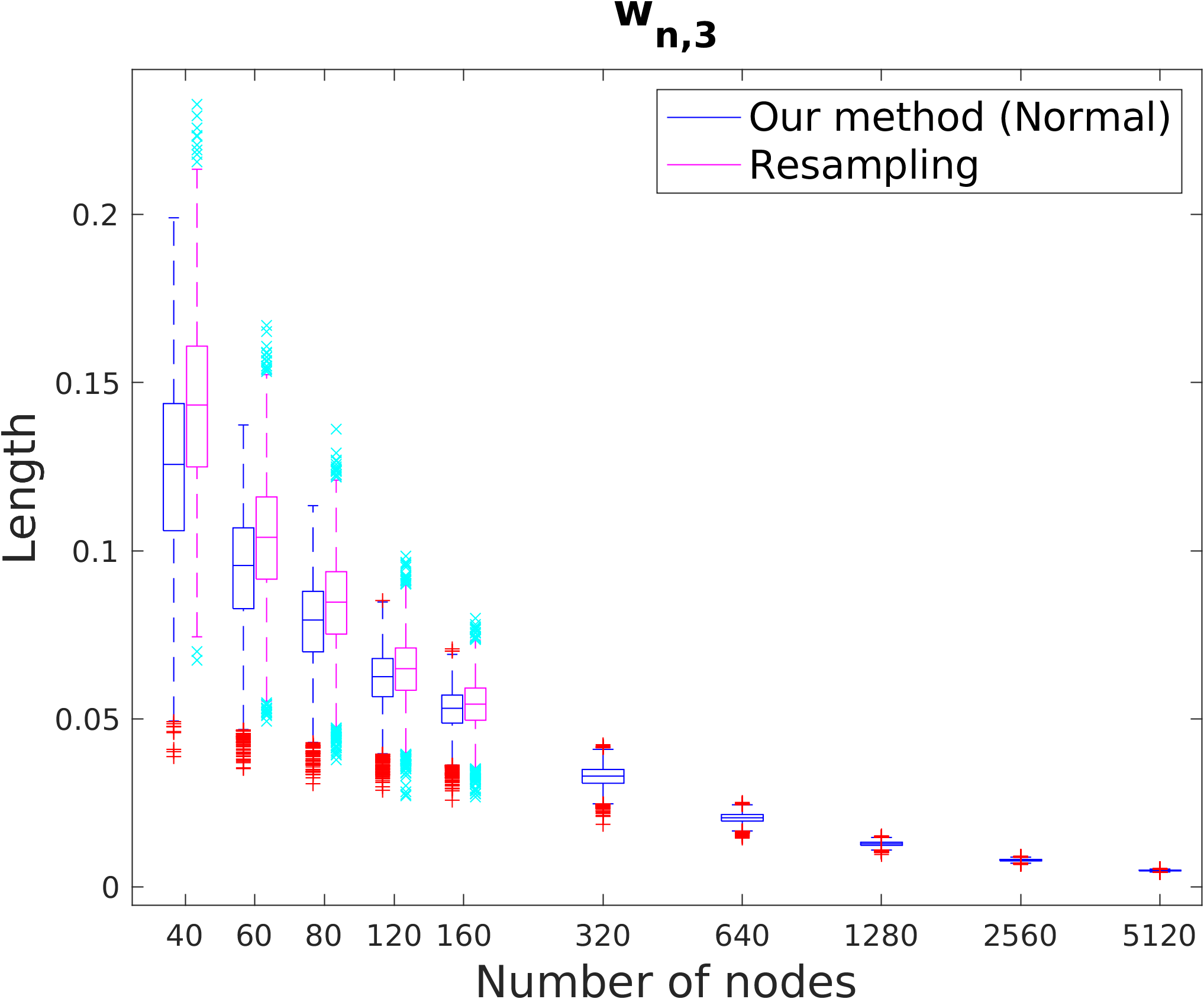}
\end{subfigure}
\hspace{0.02\textwidth}
\begin{subfigure}{0.39\textwidth}
    \includegraphics[height=4.9cm, keepaspectratio]{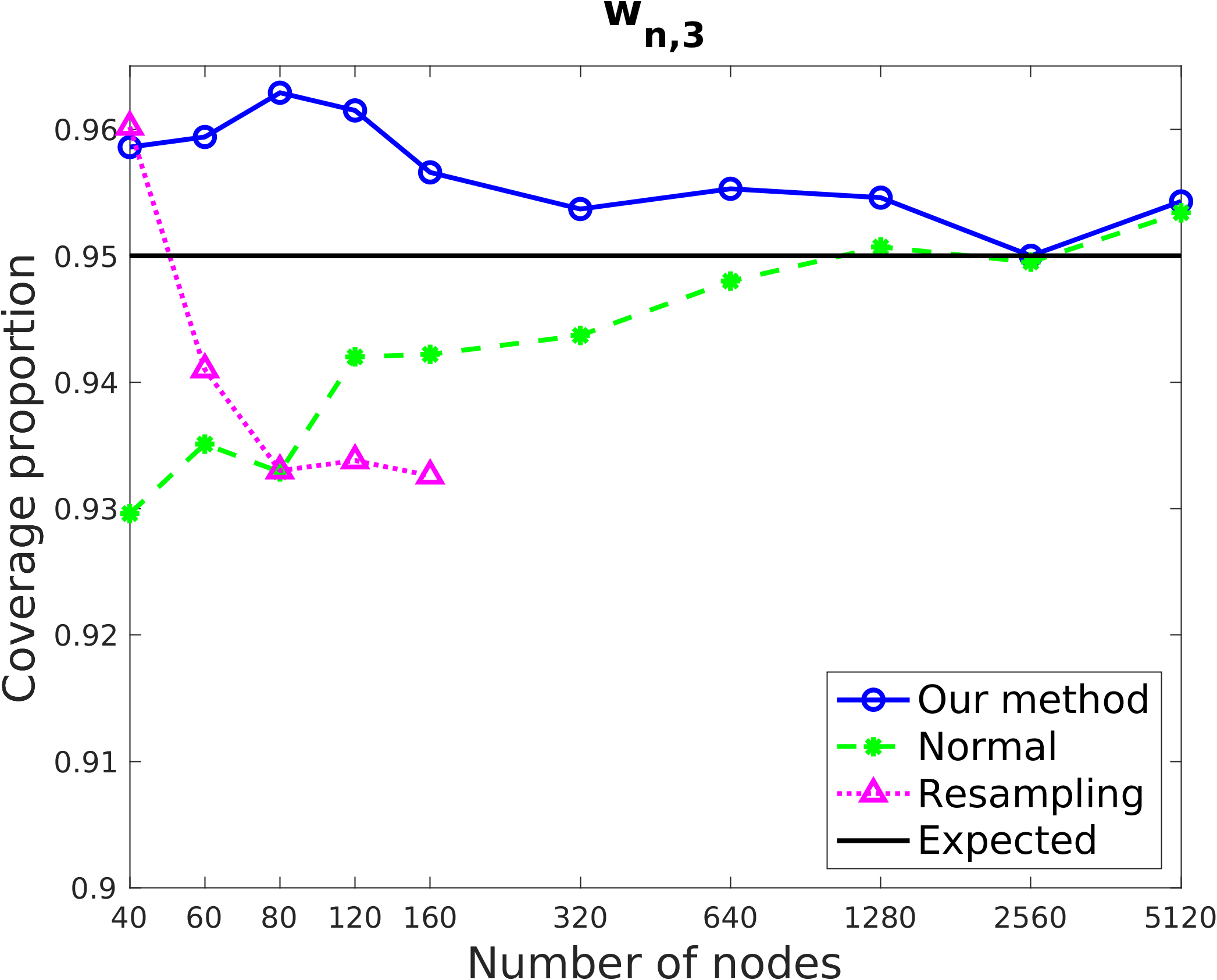}
\end{subfigure}
\vspace{0.01\textwidth}

\begin{subfigure}{0.39\textwidth}
    \includegraphics[height=4.9cm, keepaspectratio]{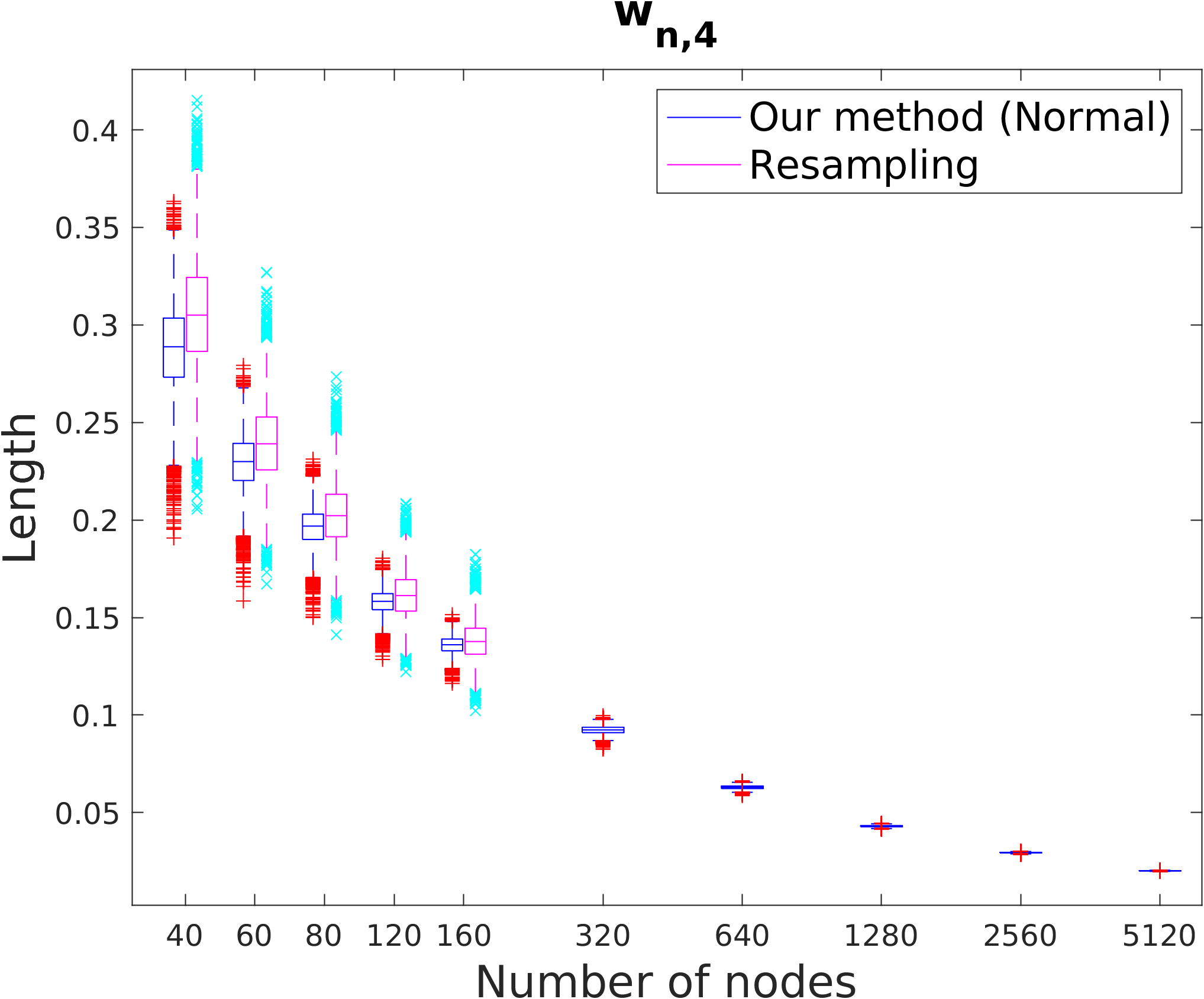}
\end{subfigure}
\hspace{0.02\textwidth}
\begin{subfigure}{0.39\textwidth}
    \includegraphics[height=4.9cm, keepaspectratio]{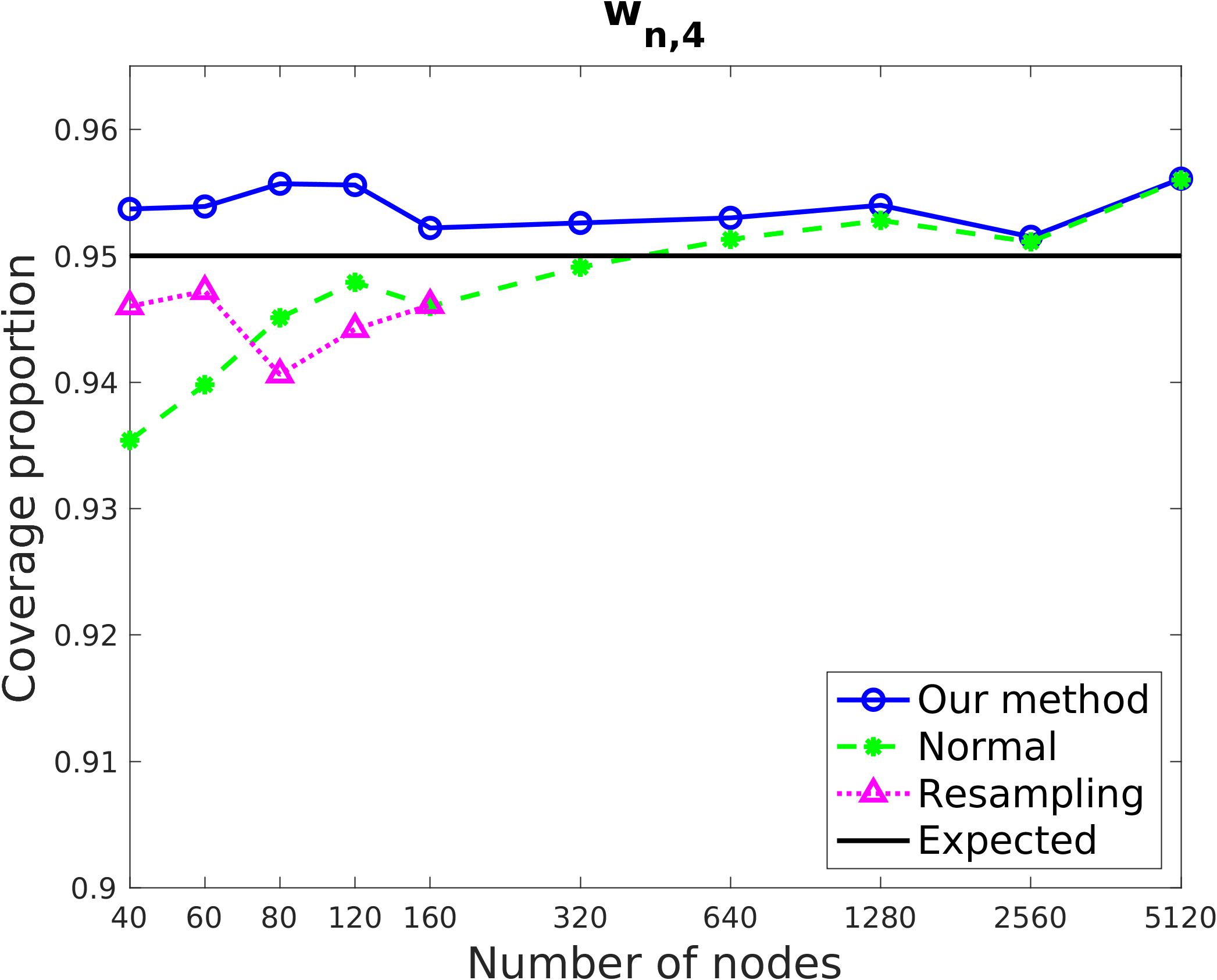}
\end{subfigure}
\vspace{0.02\textwidth}
\caption{The length and coverage proportion of 95\% confidence intervals in relation to varying network size are presented from left to right, respectively. From top to bottom, the details for the expected proportion of type-t triangles are provided.}
\label{fig.n_merged_appendix}
\end{figure}

\begin{figure}[!htb]
\centering
\begin{subfigure}{0.4\textwidth}
    \includegraphics[height=4.9cm, keepaspectratio]{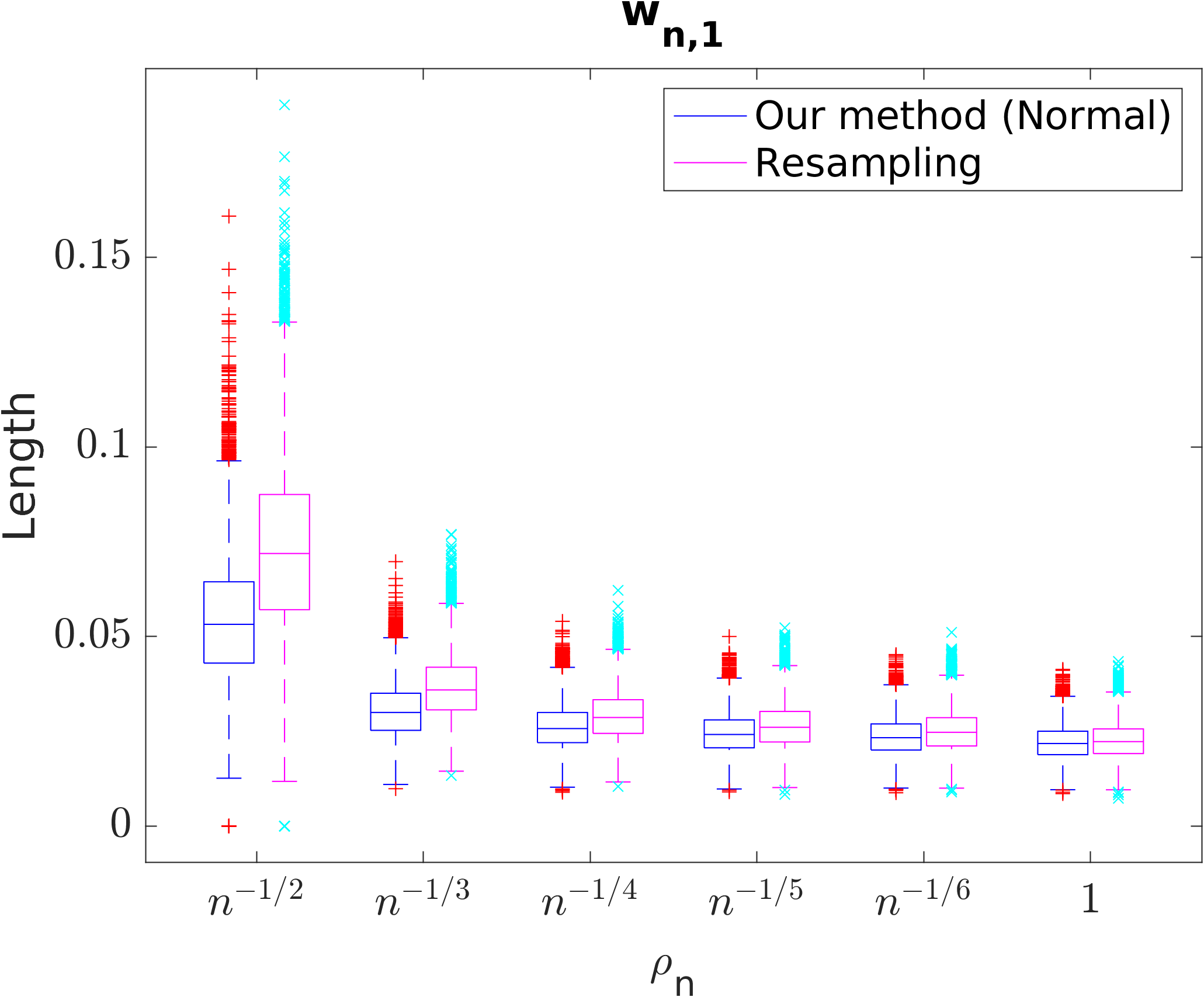}
\end{subfigure}
\hspace{0.02\textwidth}
\begin{subfigure}{0.4\textwidth}
    \includegraphics[height=4.9cm, keepaspectratio]{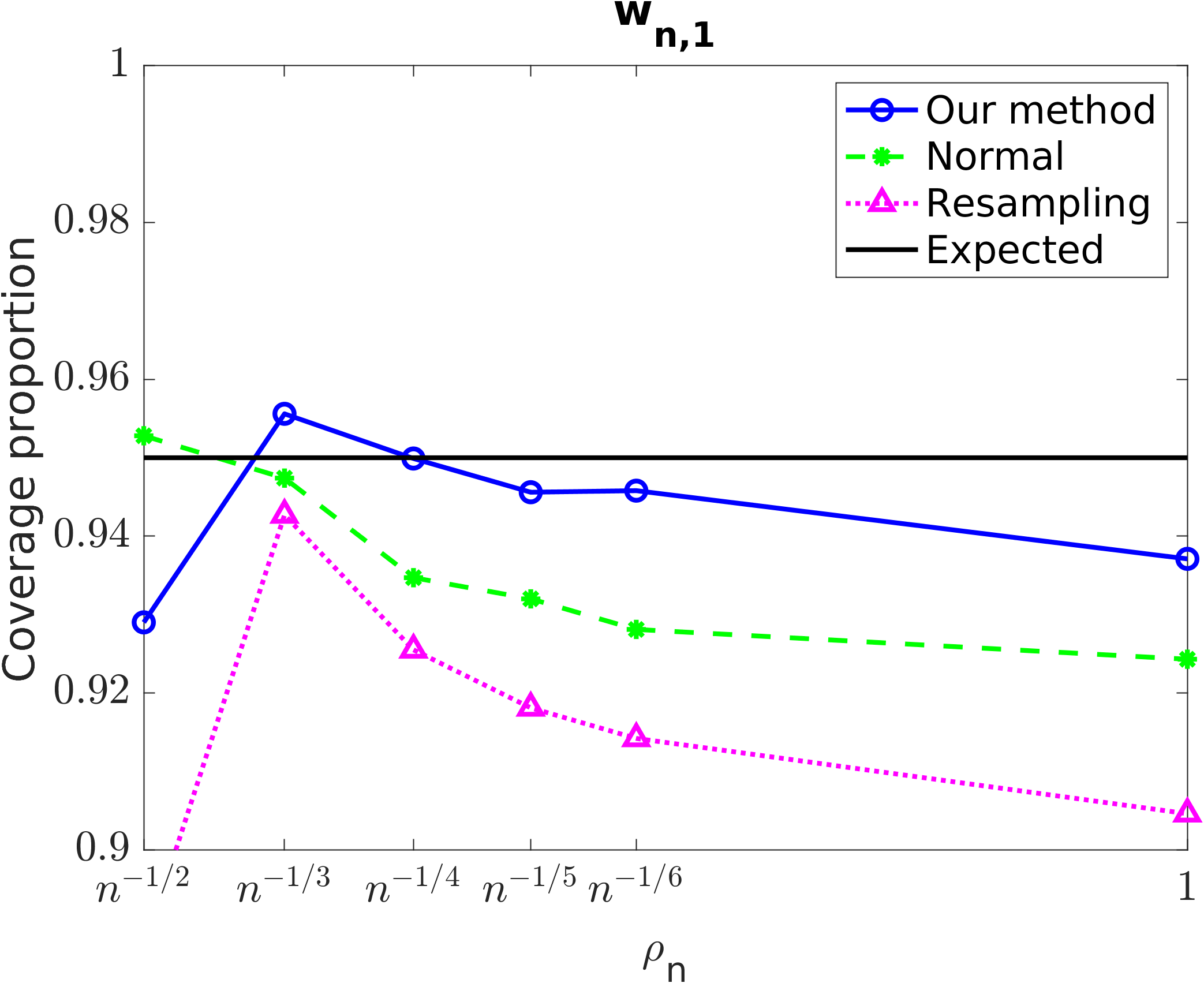}
\end{subfigure}
\vspace{0.01\textwidth}
\begin{subfigure}{0.4\textwidth}
    \includegraphics[height=4.9cm, keepaspectratio]{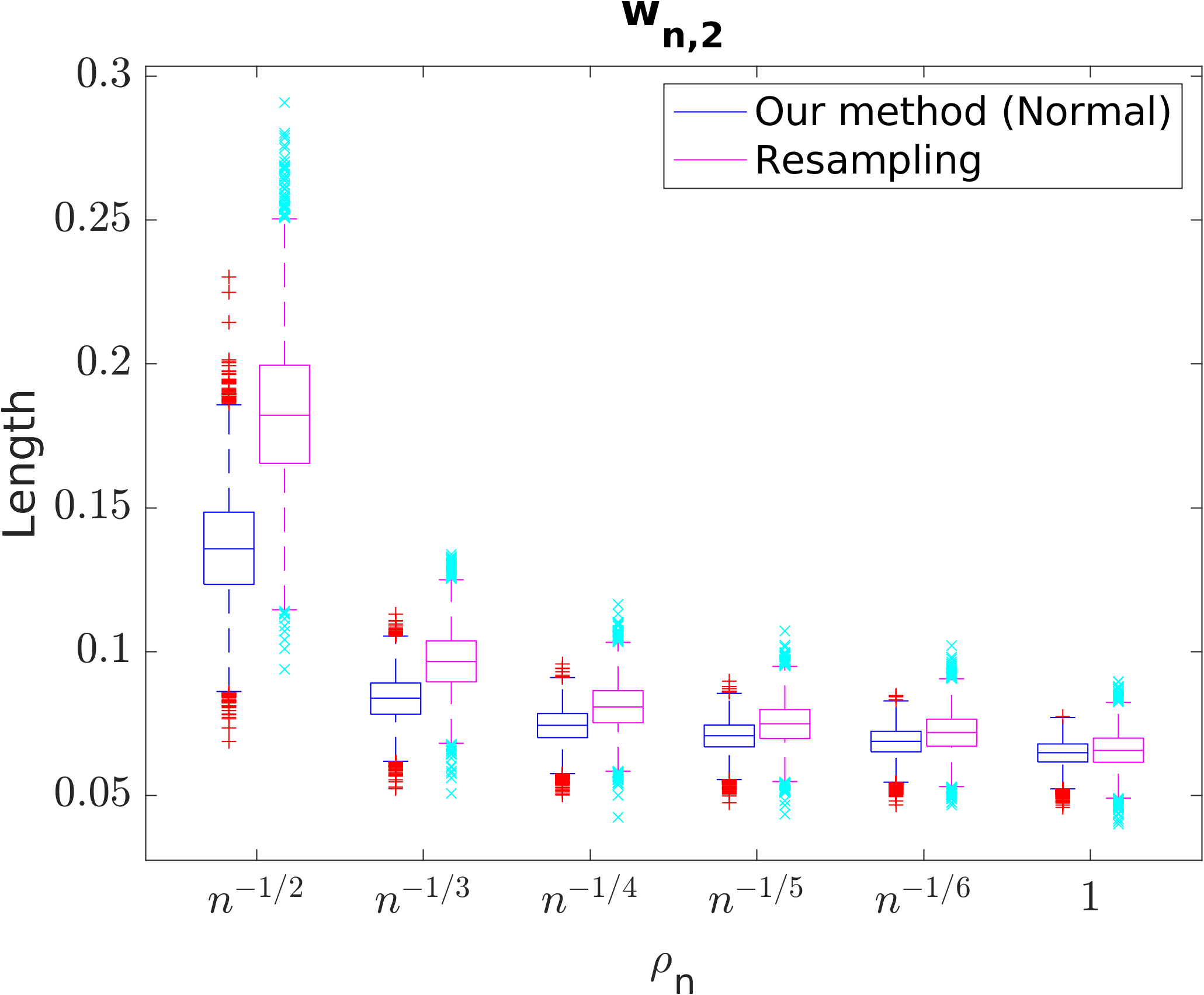}
\end{subfigure}
\hspace{0.02\textwidth}
\begin{subfigure}{0.4\textwidth}
    \includegraphics[height=4.9cm, keepaspectratio]{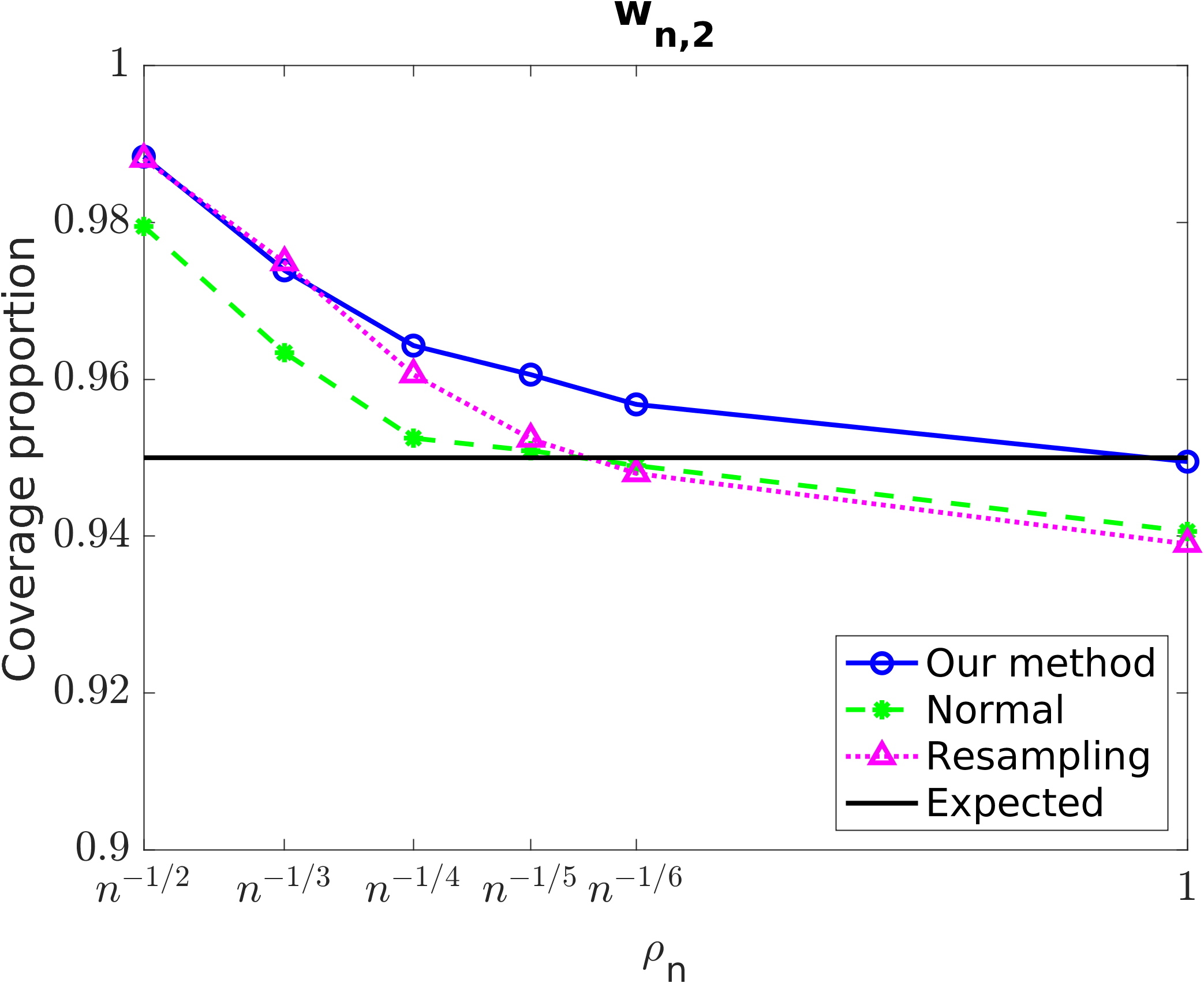}
\end{subfigure}
\begin{subfigure}{0.4\textwidth}
    \includegraphics[height=4.9cm, keepaspectratio]{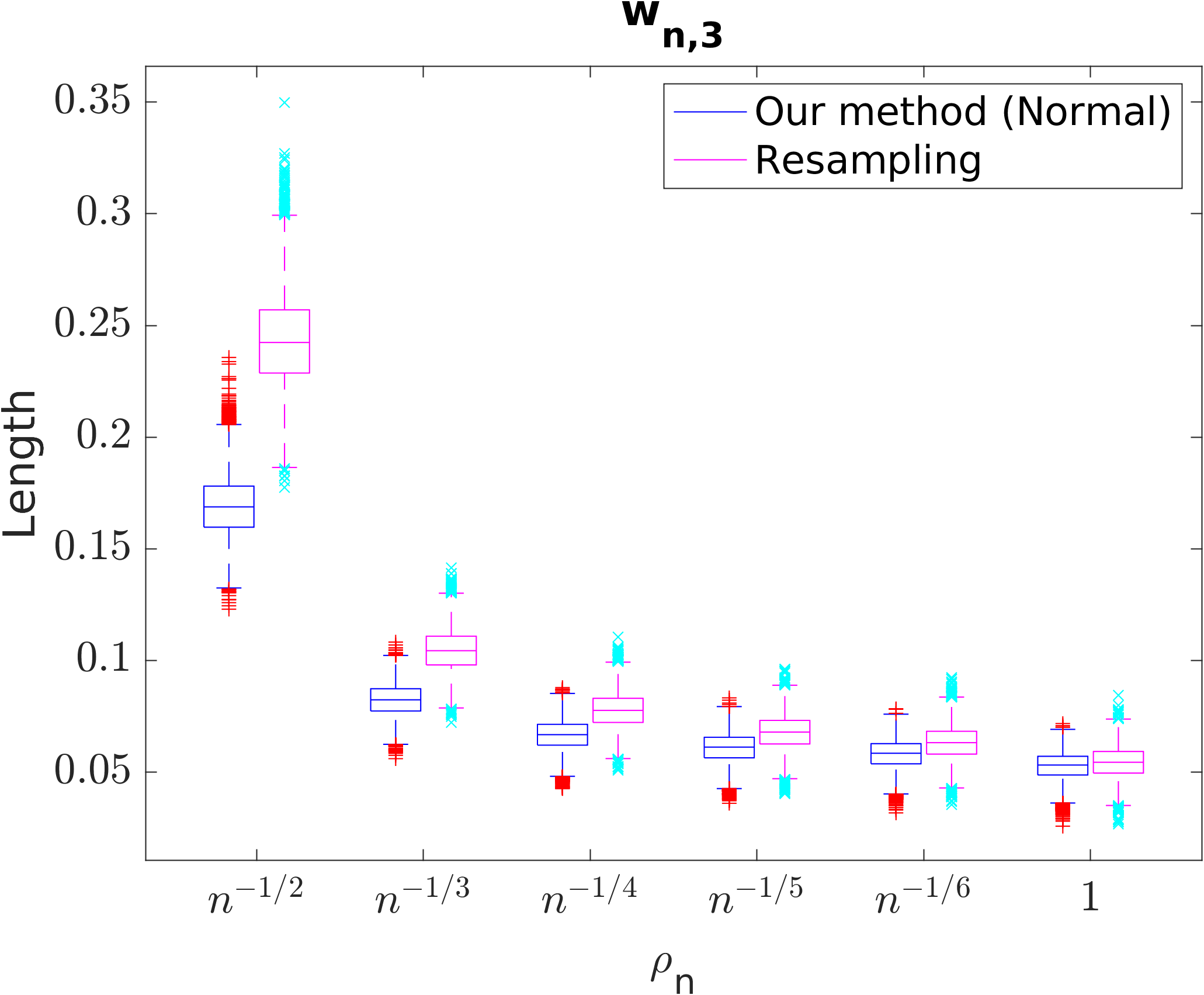}
\end{subfigure}
\vspace{0.01\textwidth}
\begin{subfigure}{0.4\textwidth}
    \includegraphics[height=4.9cm, keepaspectratio]{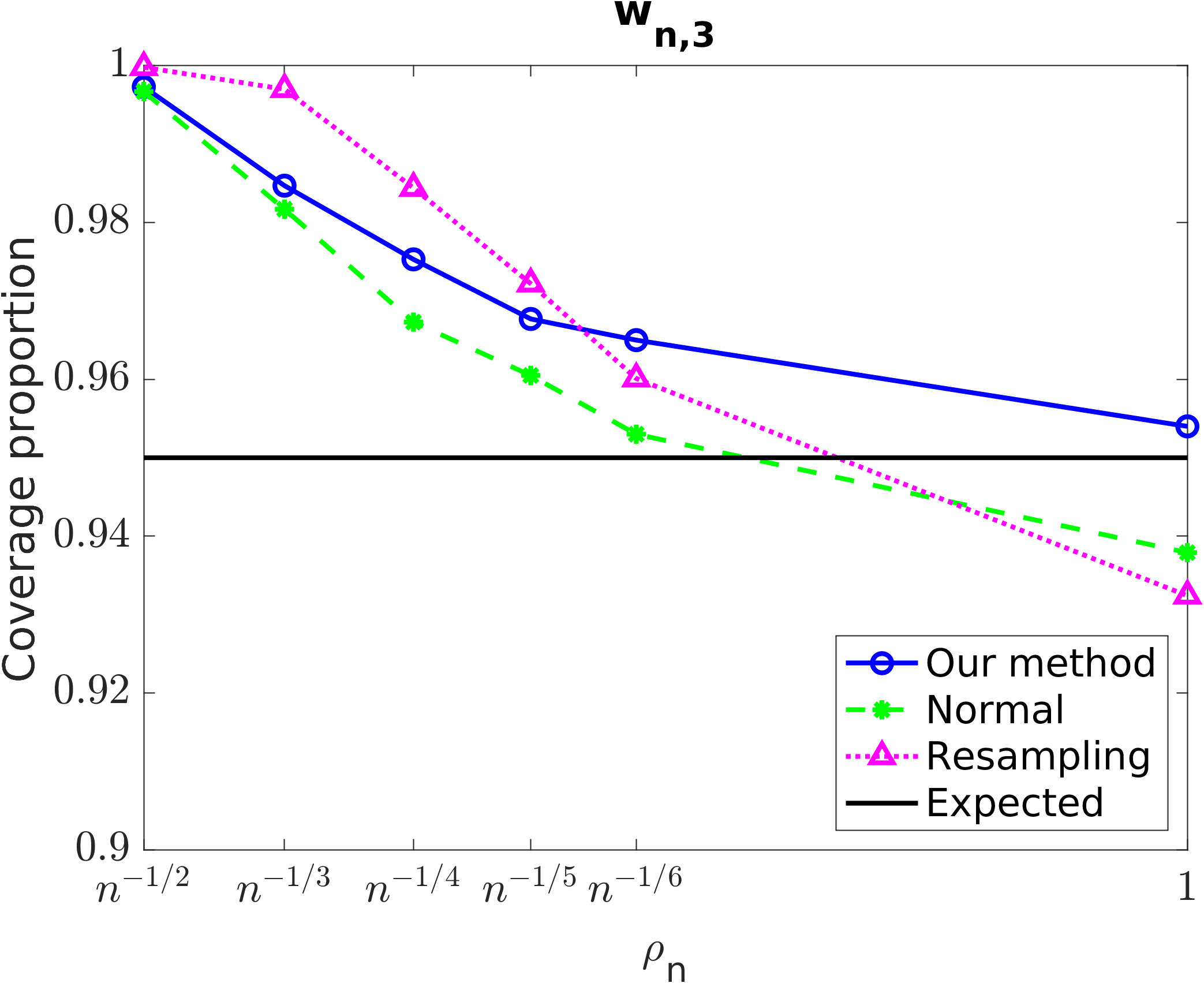}
\end{subfigure}
\begin{subfigure}{0.4\textwidth}
    \includegraphics[height=4.9cm, keepaspectratio]{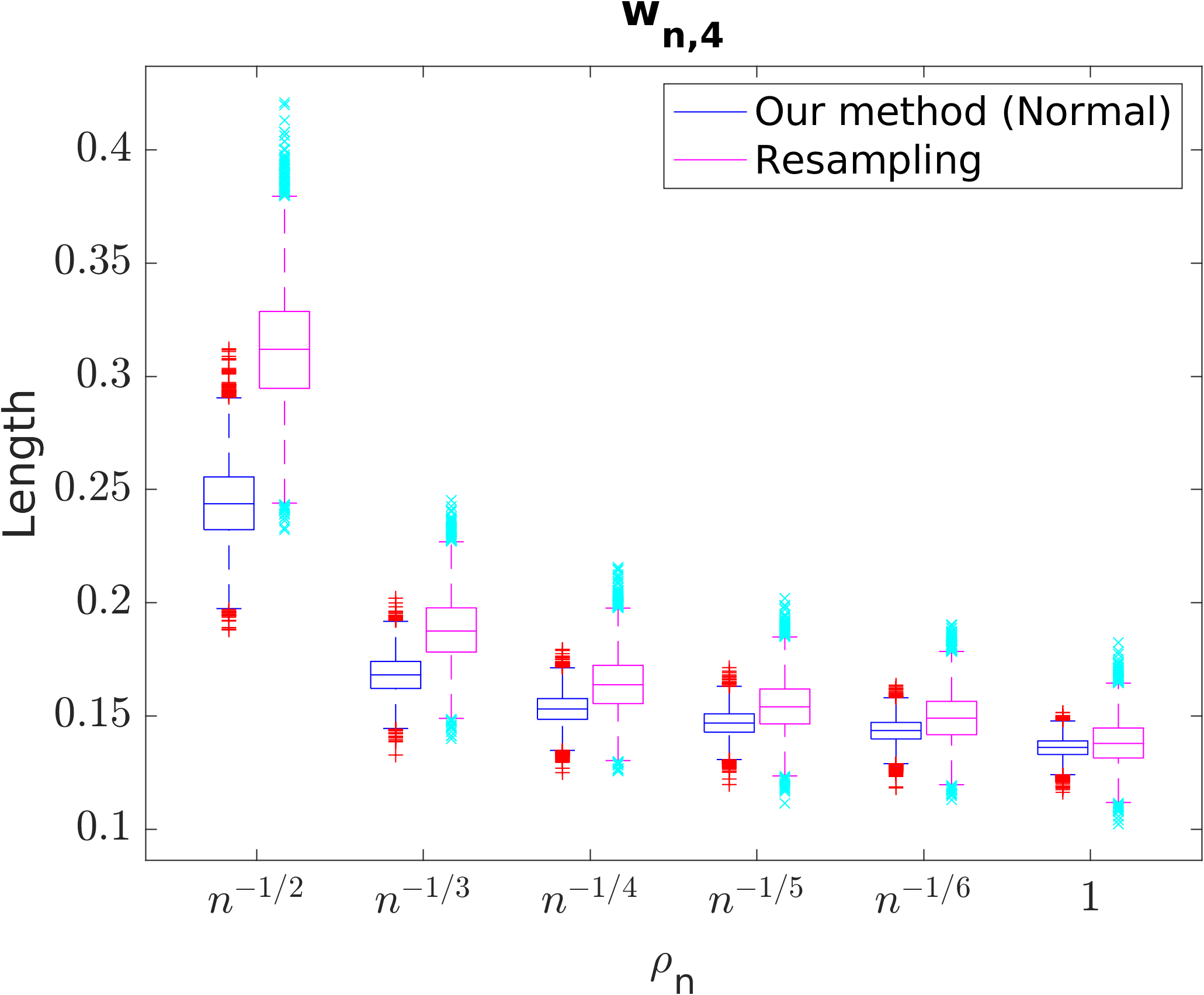}
\end{subfigure}
\hspace{0.02\textwidth}
\begin{subfigure}{0.4\textwidth}
    \includegraphics[height=4.9cm, keepaspectratio]{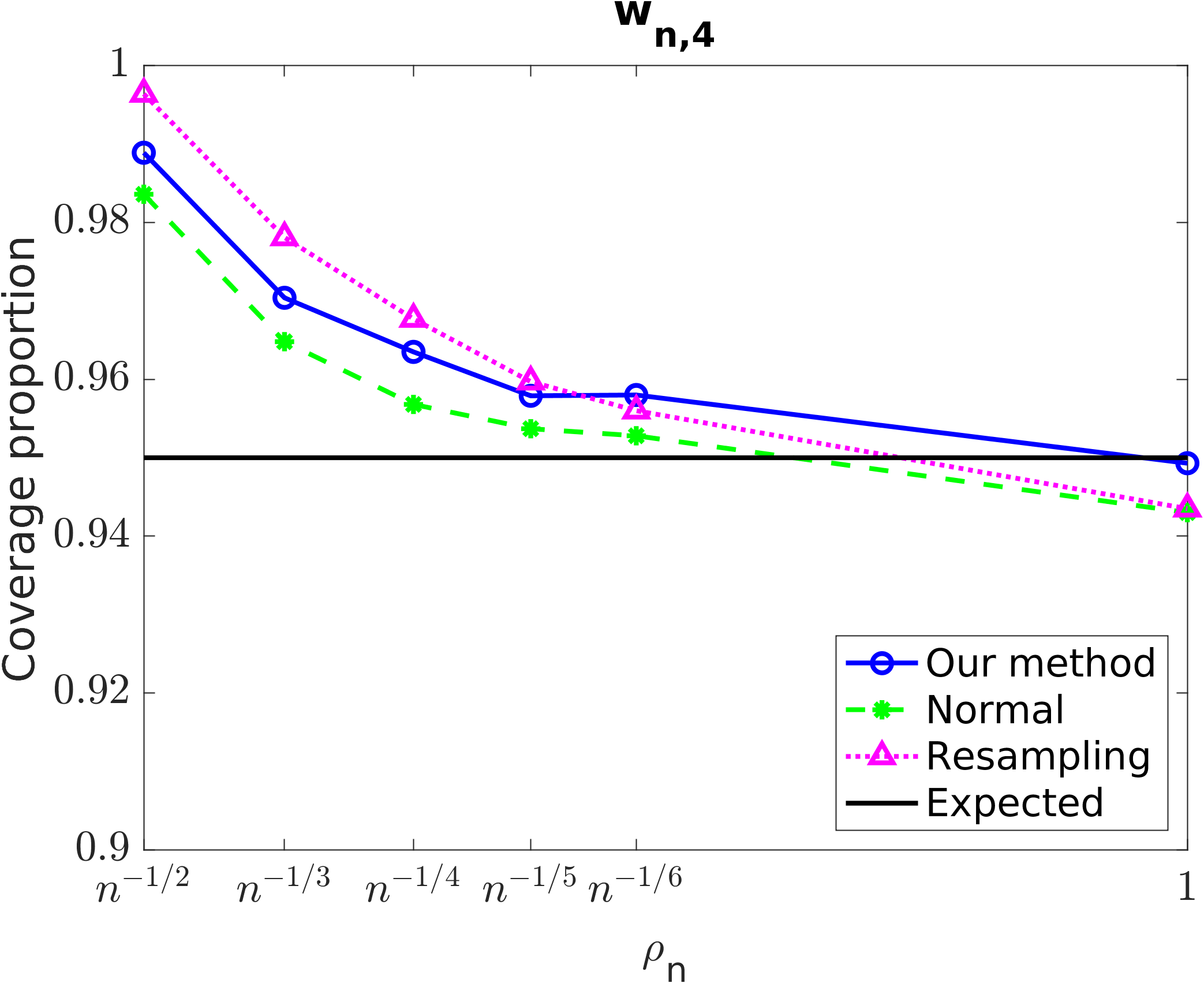}
\end{subfigure}
\caption{The length and coverage proportion of 95\% confidence intervals in relation to varying network sparsity are presented from left to right, respectively. From top to bottom, the details for the expected proportion of type-t triangles are provided.}
\label{fig.rho_n_appendix}
\end{figure}

\begin{figure}[!htb]
\centering
\begin{subfigure}{0.4\textwidth}
    \includegraphics[height=4.9cm, keepaspectratio]{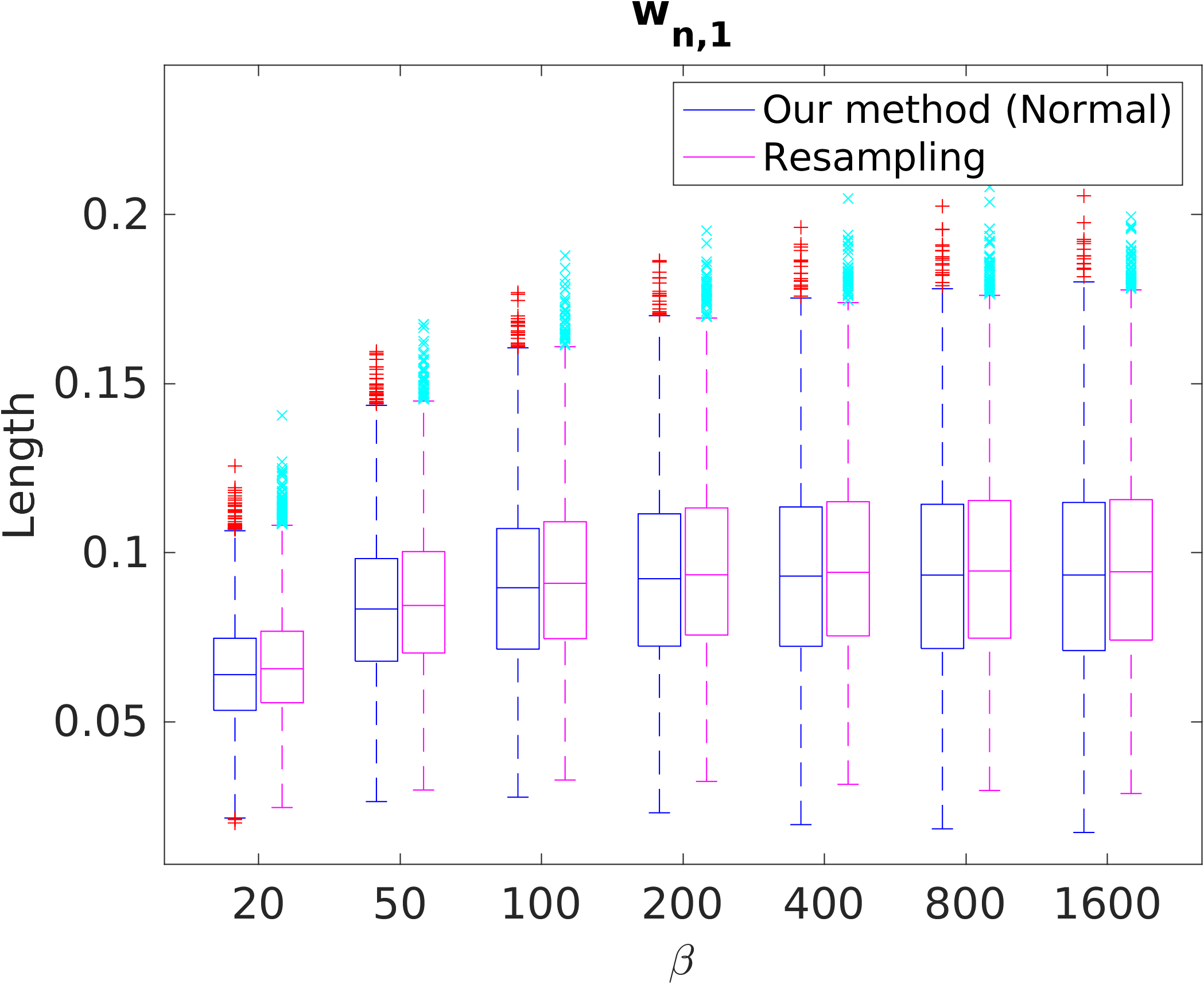}
\end{subfigure}
\hspace{0.02\textwidth}
\begin{subfigure}{0.4\textwidth}
    \includegraphics[height=4.9cm, keepaspectratio]{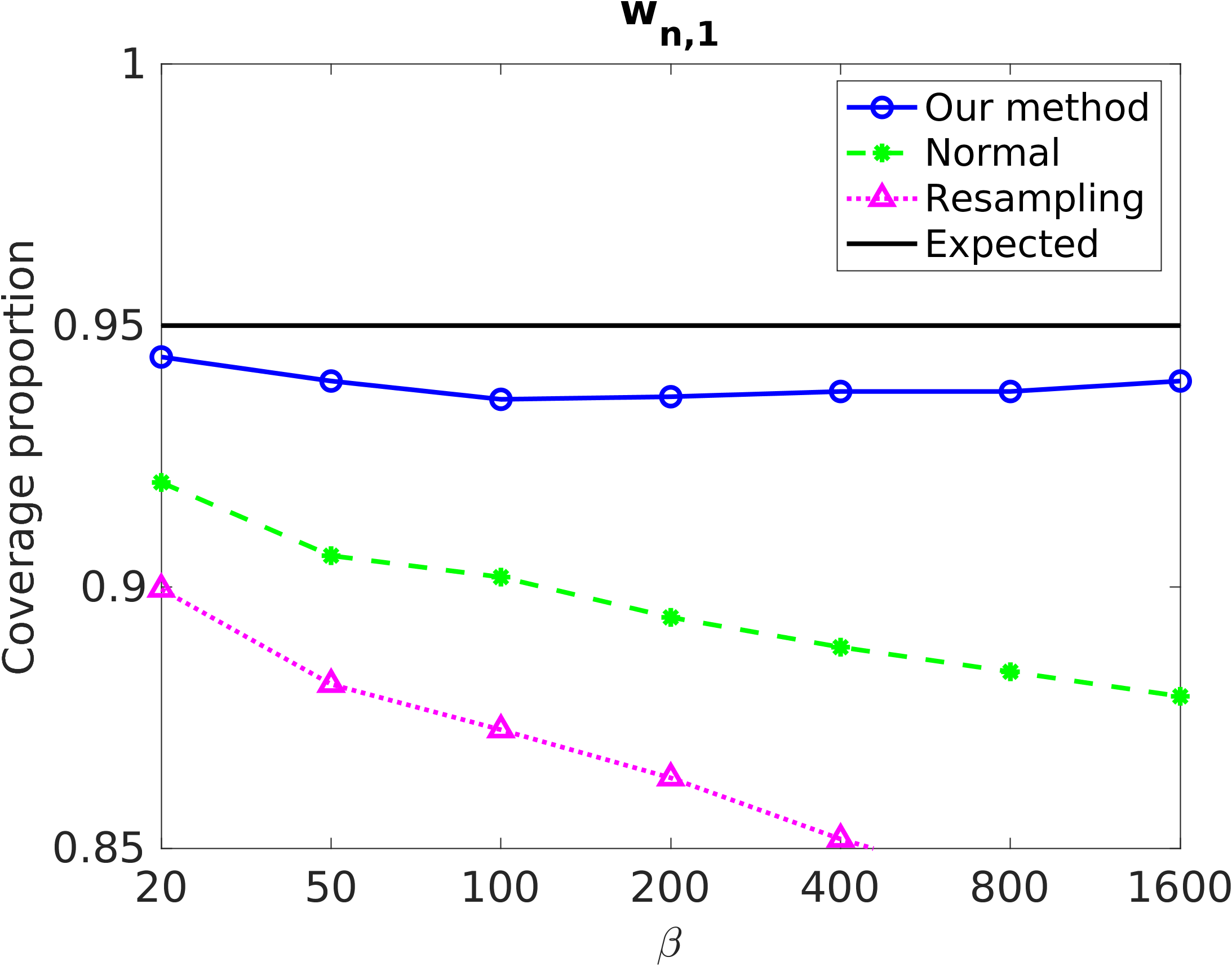}
\end{subfigure}
\vspace{0.01\textwidth}

\begin{subfigure}{0.4\textwidth}
    \includegraphics[height=4.9cm, keepaspectratio]{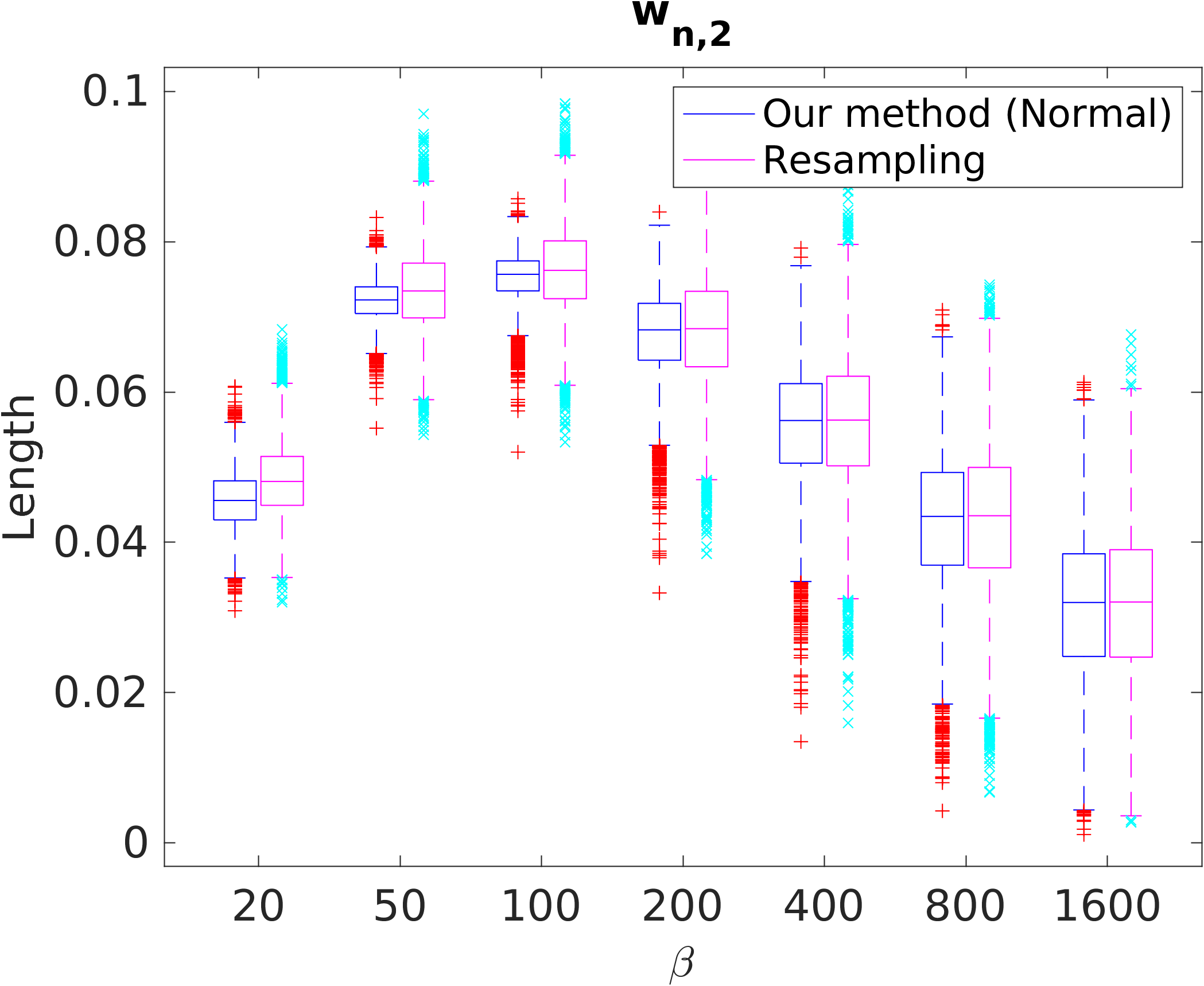}
\end{subfigure}
\hspace{0.02\textwidth}
\begin{subfigure}{0.4\textwidth}
    \includegraphics[height=4.9cm, keepaspectratio]{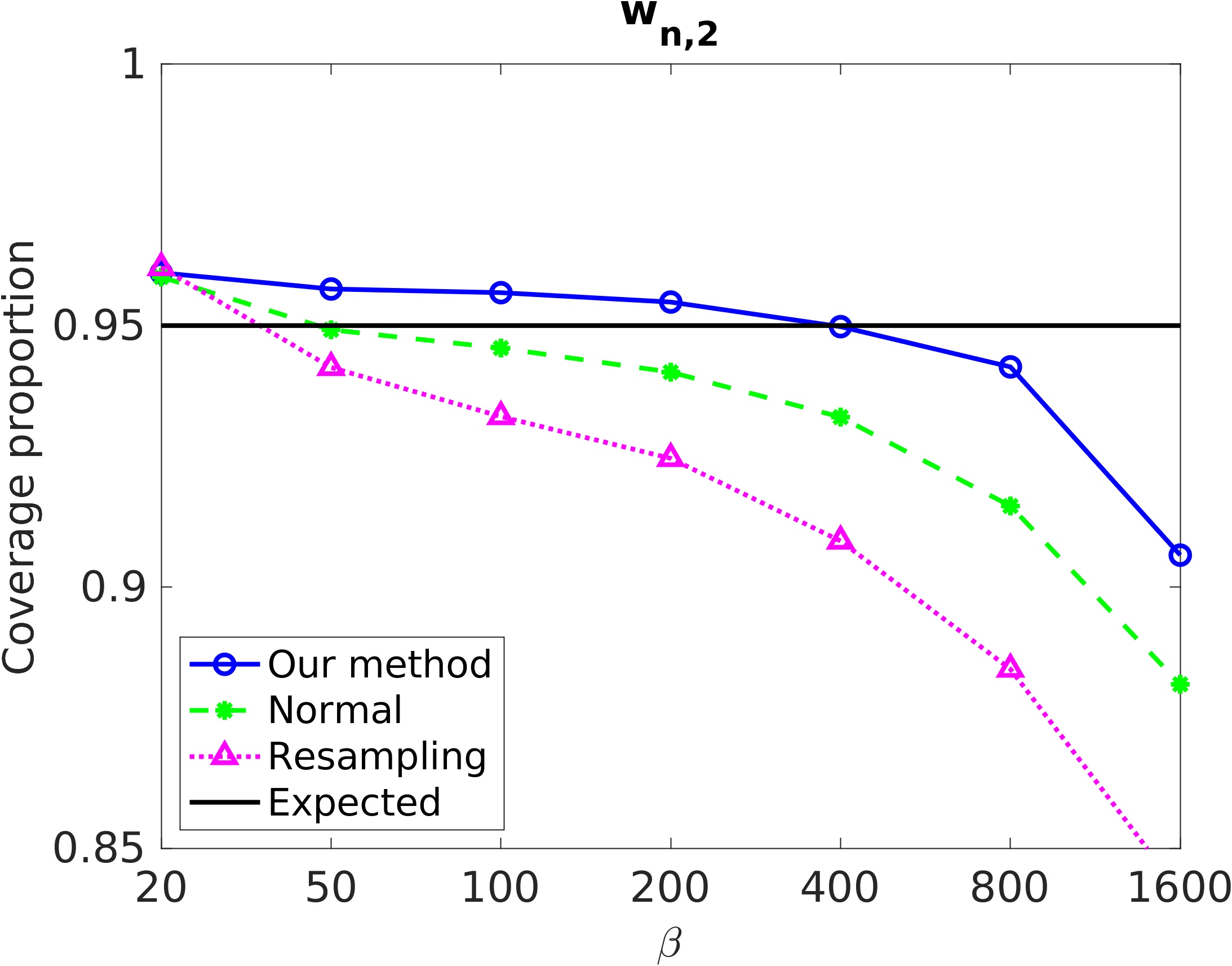}
\end{subfigure}
\vspace{0.01\textwidth}

\begin{subfigure}{0.4\textwidth}
    \includegraphics[height=4.9cm, keepaspectratio]{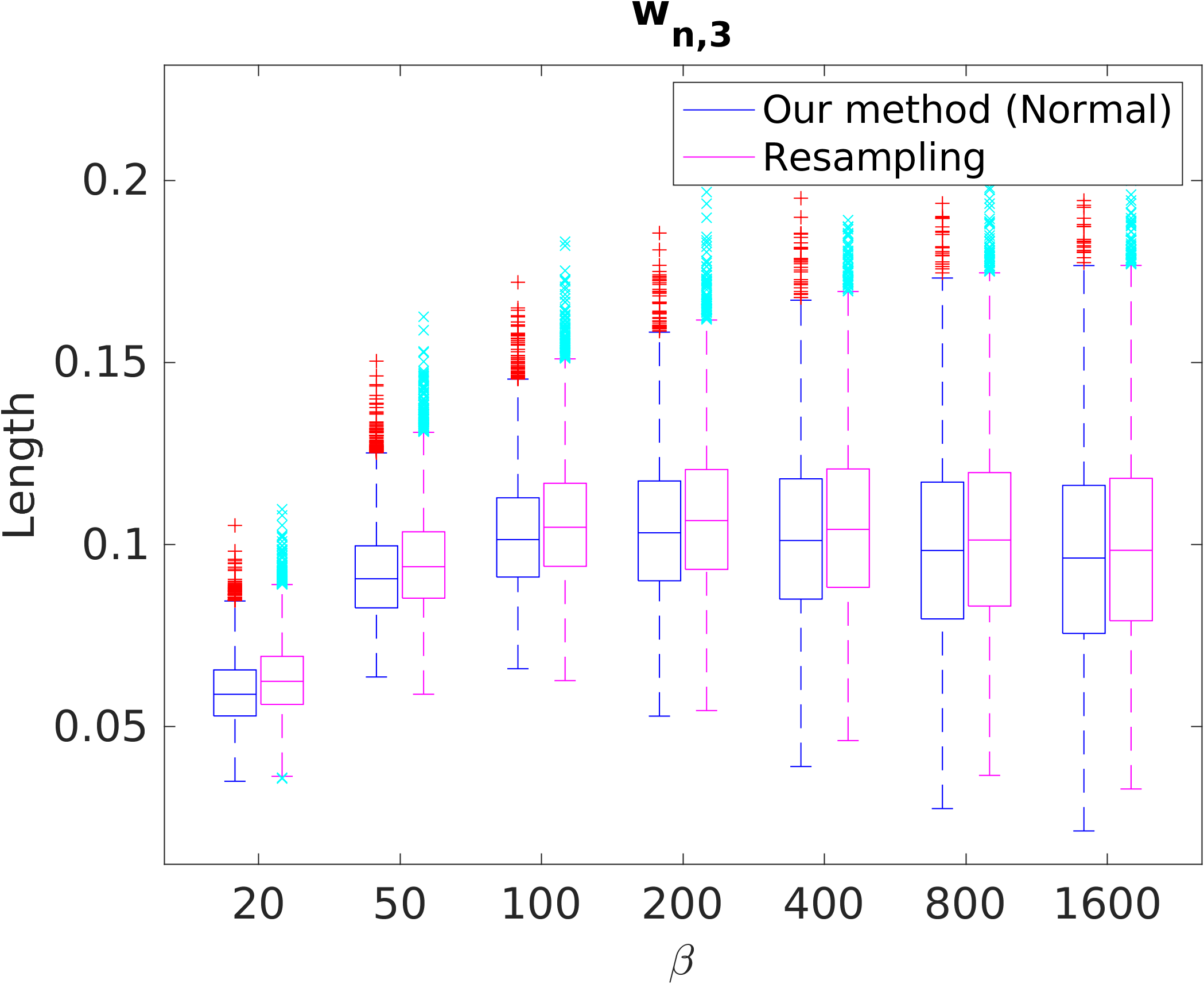}
\end{subfigure}
\hspace{0.02\textwidth}
\begin{subfigure}{0.4\textwidth}
    \includegraphics[height=4.9cm, keepaspectratio]{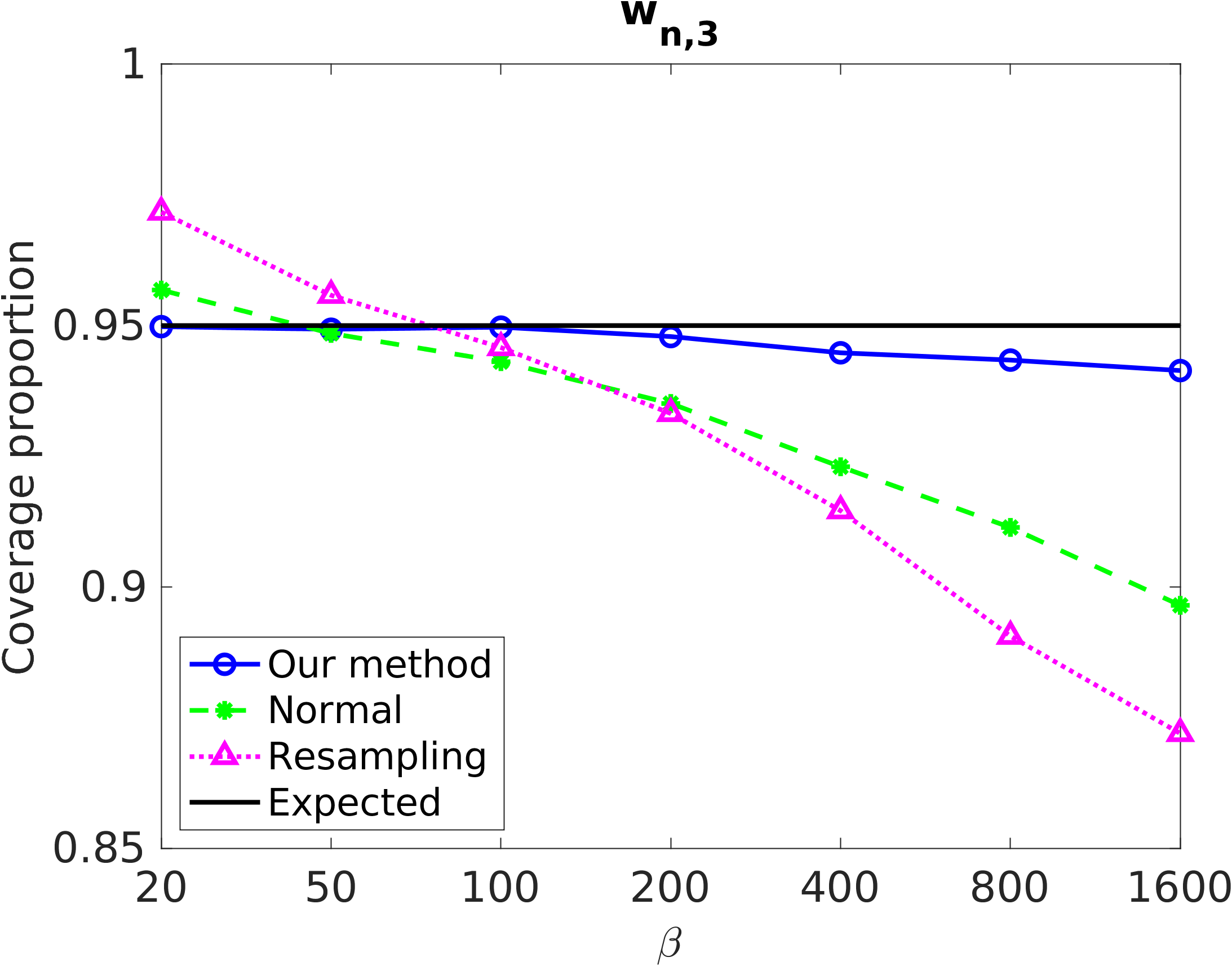}
\end{subfigure}
\vspace{0.01\textwidth}

\begin{subfigure}{0.4\textwidth}
    \includegraphics[height=4.9cm, keepaspectratio]{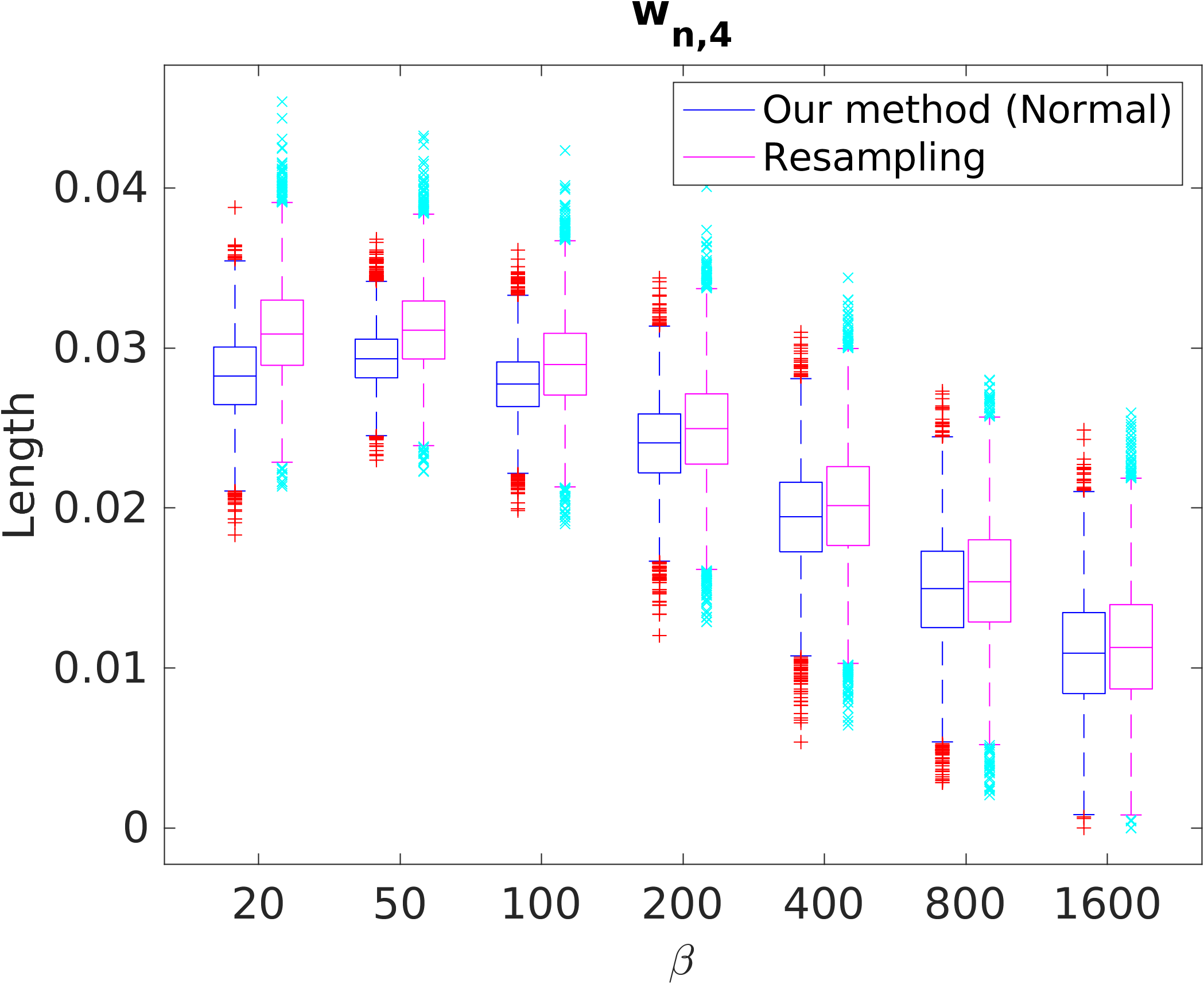}
\end{subfigure}
\hspace{0.02\textwidth}
\begin{subfigure}{0.4\textwidth}
    \includegraphics[height=4.9cm, keepaspectratio]{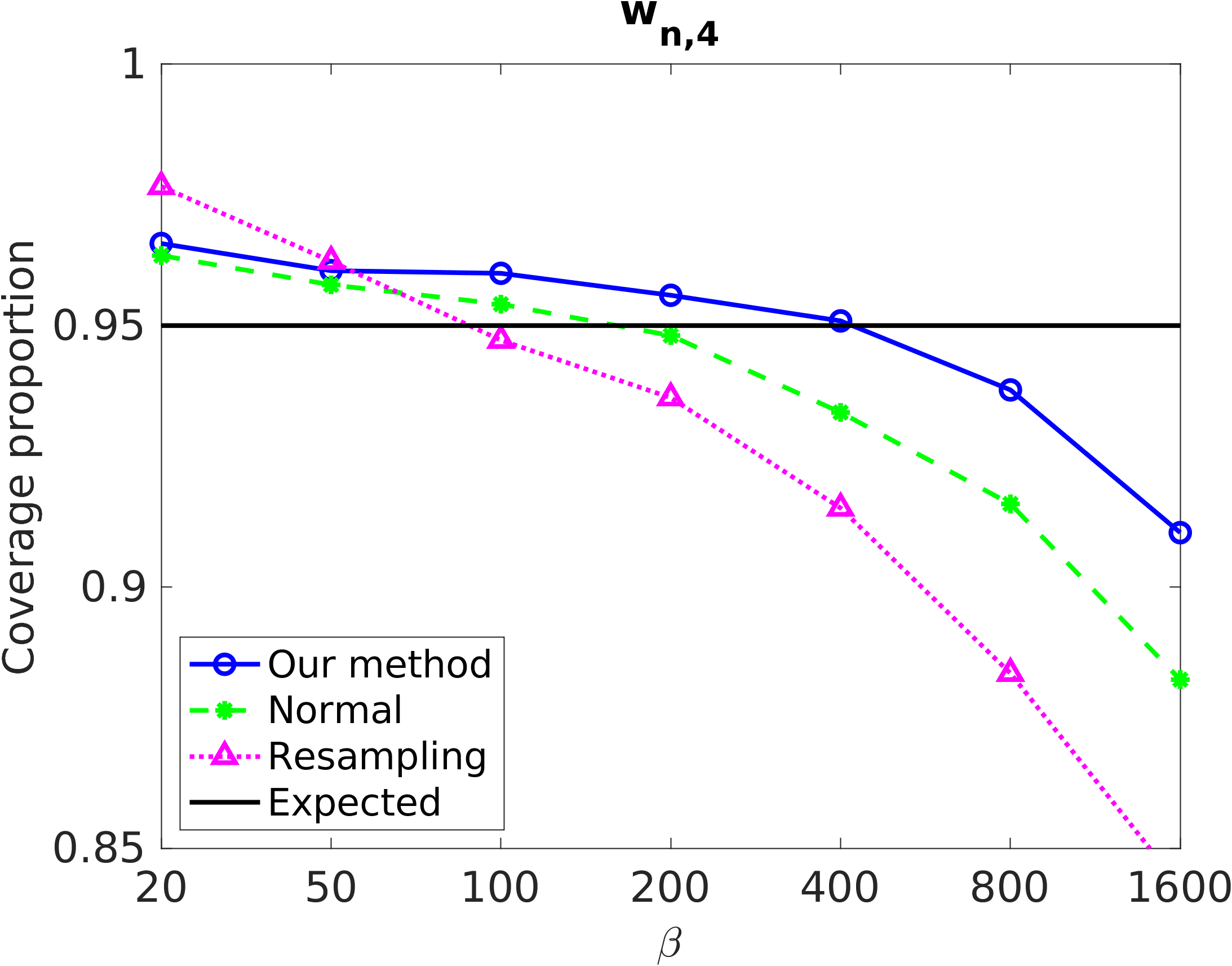}
\end{subfigure}
\caption{The length and coverage proportion of 95\% confidence intervals in relation to varying degree of balance are presented from left to right, respectively. From top to bottom, the details for the expected proportion of type-t triangles are provided.}
\label{fig.balance_appendix}
\end{figure}

\begin{figure}[!htb]
\centering

\begin{subfigure}{0.45\textwidth}
  \centering
  \includegraphics[width=\textwidth]{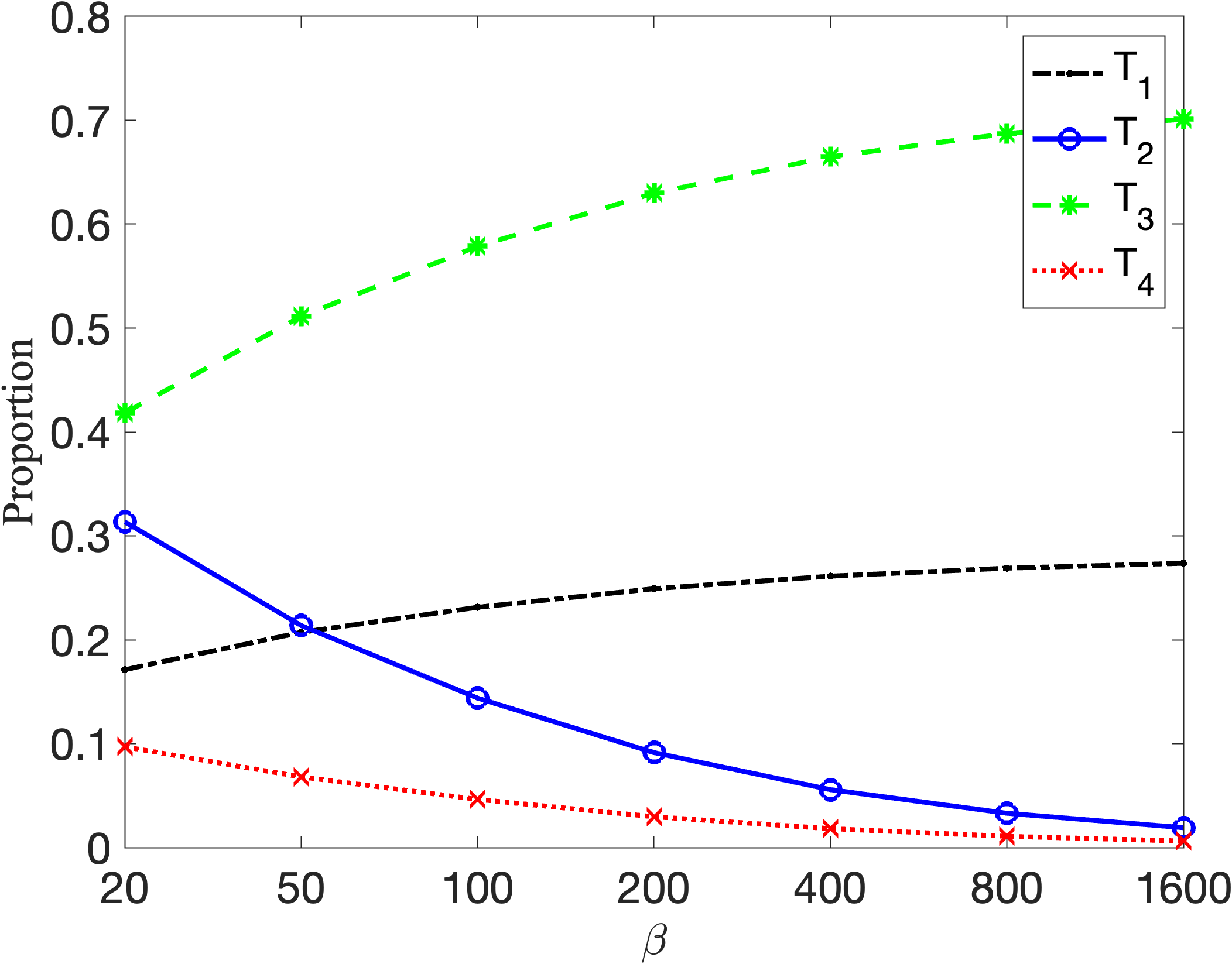}
\end{subfigure}
\hfill
\begin{subfigure}{0.51\textwidth}
  \centering
  \includegraphics[width=\textwidth]{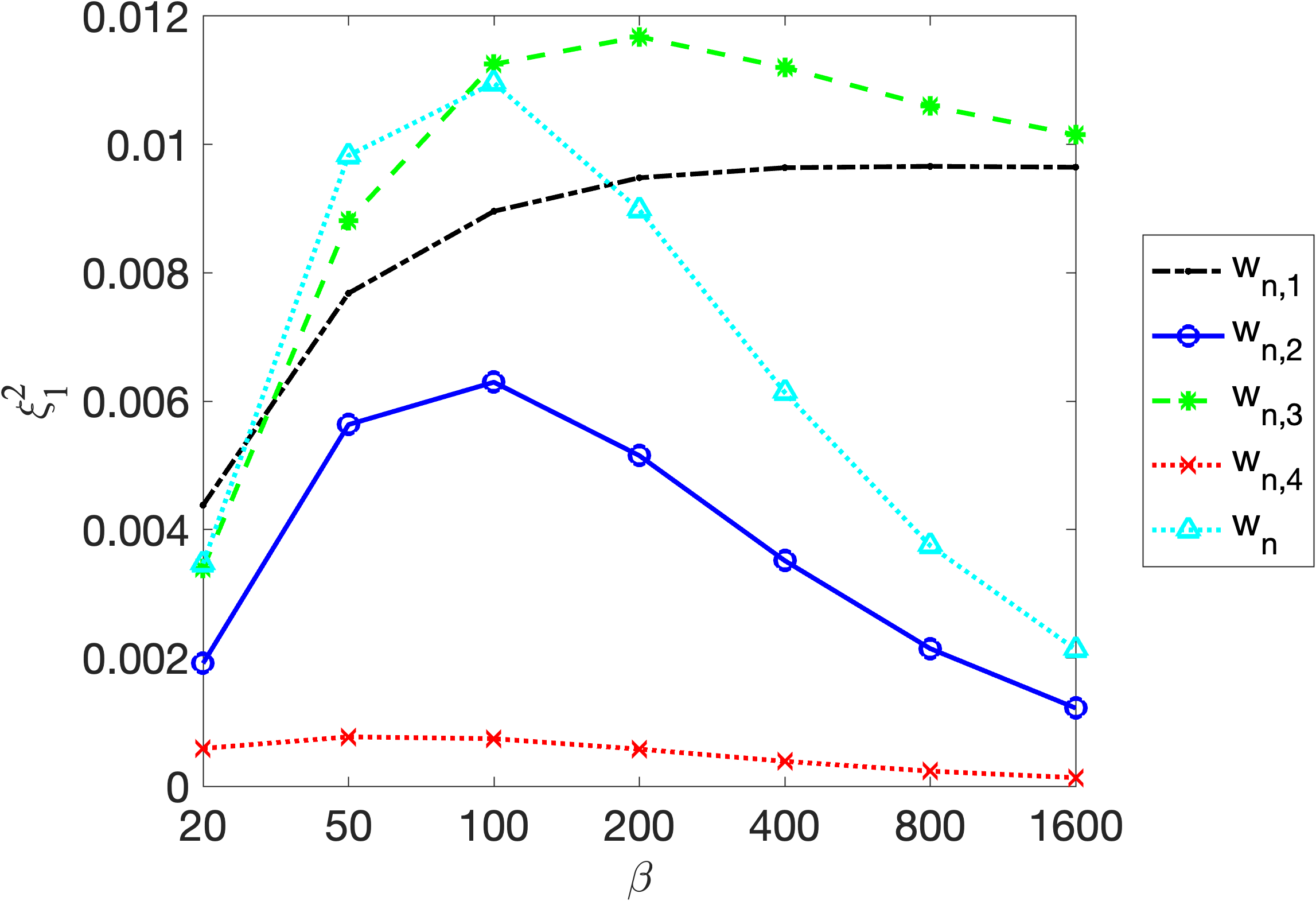}
\end{subfigure}

\caption{Triangle statistics versus balance degree $\beta$. The left panel shows the proportion of each triangle type, where $T_i$ denotes type-$i$ triangle. The right panel shows $\xi_1^2$ versus $\beta$ for all triangle types.}
\label{fig:balance_summary}
\end{figure}

In this subsection, we present simulation results for each type of triangles, i.e., $w_{n,t}$ for $1\le t \le 4$. Table~\ref{tab: CDF distance appendix} and Figure~\ref{fig.CDF_appendix} provide a comparative analysis of CDF approximations. Our method yields the closest estimates to the true CDFs. In contrast, both the standard normal approximation and the node resampling method show notable deviations from the true CDFs. 

Here we provide the coverage proportion and the length of confidence intervals for each type of triangles. We summarize the results as follows.

\textbf{Trend with respect to network size $n$.} As network size or density increases, Figures~\ref{fig.n_merged_appendix} demonstrate that our method outperforms the node resampling method in terms of shorter CI lengths while maintaining reasonable coverage proportions. Across all methods, CI lengths decrease and the coverage proportions of our method approach the nominal level as the network size increases.

\textbf{Trend with respect to network density $\rho_n$.} We include both a sparser regime $\rho_n\asymp n^{-1/2}$ and a denser regime $\rho_n\asymp 1$ in Figure~\ref{fig.rho_n_appendix}. As $\rho_n$ increases, more triangles are observed, estimation and inference becomes more precise, and the coverage proportion approaches the nominal level of 0.95. This behavior is expected and aligns with our theoretical error bound, which decreases as $\rho_n$ increases. Importantly, in denser regimes the coverage stabilizes near the nominal level.

\textbf{Trend with respect to balance degree $\beta$.} We investigate larger values of the balance parameter $\beta\in \{800,1600\}$ in Figure~\ref{fig.balance_appendix}.  In contrast to $\rho_n$, increasing $\beta$ leads to a different phenomenon. For very large $\beta$, the network becomes nearly perfectly balanced: unbalanced triangles occur with vanishing probability (see the proportion of each type of triangles in Figure~\ref{fig:balance_summary} left panel). In this extreme case, the asymptotic variance of the estimator degenerates (see Figure~\ref{fig:balance_summary} right panel), violating Condition (b), which requires the leading variance term to be bounded away from zero. As a result, the normal approximation underlying the confidence intervals breaks down, and the coverage probability deteriorates.

\begin{table}[ht]
    \centering
    \begin{minipage}{0.45\textwidth}
        \centering
        \begin{tabular}{|c|c|c|c|c|c|}\hline
            \diagbox{$c_{\delta}$}{$n$}& 160 & 320 & 640 & 1280 \\ \hline
            0    & 0.955 & 0.954 & 0.954 & 0.953 \\ \hline
            0.1  & 0.953 & 0.952 & 0.955 & 0.953 \\ \hline
            1    & 0.949 & 0.948 & 0.953 & 0.954 \\ \hline
            5    & 0.942 & 0.947 & 0.951 & 0.950 \\ \hline
        \end{tabular}
    \end{minipage}
    \hfill
    \begin{minipage}{0.45\textwidth}
        \centering
        \begin{tabular}{|c|c|c|c|c|c|}\hline
            \diagbox{$c_{\delta}$}{$n$} & 160 & 320 & 640 & 1280 \\ \hline
            0    & 0.951 & 0.953 & 0.952 & 0.953 \\ \hline
            0.1  & 0.952 & 0.950 & 0.952 & 0.954 \\ \hline
            1    & 0.946 & 0.947 & 0.951 & 0.954 \\ \hline
            5    & 0.940 & 0.946 & 0.949 & 0.950 \\ \hline
        \end{tabular}
    \end{minipage}
    \caption{Coverage proportions of balanced (left) and weakly balanced (right) statistics for $c_{\delta}\in\{0,0.1,1,5\}$ with network size $n\in\{160, 320, 640, 1280\}$.}
    \label{tab: c_delta_appendix}
\end{table}

\textbf{Finite-sample impact of $\delta_T\sim N(0, c_\delta n^{-1}\log n)$.} We introduce an artificial Gaussian perturbation $\delta_T\sim N(0, c_\delta n^{-1}\log n)$ to establish our approximation results, the variance of which vanishes as $n$ grows, so its effect is asymptotically negligible. To assess its finite-sample impact, we consider $c_\delta \in \{0, 0.1, 1, 5\}$ and network sizes $n \in \{160, 320, 640, 1280\}$. Table~\ref{tab: c_delta_appendix} reports coverage proportions and confidence interval lengths for both balanced and weakly balanced statistics. The results show only minor variation across different values of $c_\delta$, with the variation diminishing as $n$ increases. In particular, setting $c_\delta = 0$, that is introducing no artificial perturbation, already achieves reasonable coverage. This suggests that adding $\delta_T$ mainly serves a technical role in the theoretical analysis and does not materially affect practical performance.

\subsection{Details for Real-world Data Examples}
We first provide details on the construction of a signed network among U.S.\ senators. The original cosponsorship data represents a bipartite network characterizing 100 senators' sponsorship of 7,804 bills. Following ~\citet{neal2014backbone}, we fit a scobit model \citep{nagler1994scobit} to predict the sponsorship probabilities between senators and bills, as functions of their node degrees. Then we randomly sample multiple bipartite networks based on the fitted probabilities.
These networks and the original co-sponsorship network are, respectively, projected into one-mode networks, where edges represent the numbers of co-sponsored bills between senators. 
The signed network among senators is defined as follows: a positive edge is assigned between two senators if they co-sponsored bills significantly more frequent, and a negative edge is assigned if they co-sponsored bills significantly less frequent. Thus, a positive edge suggests political alliance, and a negative edge suggests political antagonism.

Next, we provide the derivation of the adjusted baseline, which accounts for the proportion of negative edges in the network. The balance-free model assumes that the edge signs are independently and identically distributed Bernoulli variables, with the probability equal to the observed fraction of negative edges. Specifically, the adjusted baseline for $w_{n, 1}$, $w_{n_2}$, $w_{n, 3}$, $w_{n, 4}$ and $w_n$ are given by $(1-s_n)^3$, $3s_n(1-s_n)^2$, $3s_n^2(1-s_n)$, $s_n^3$, and $(1-s_n)^3+3s_n^2(1-s_n)$, respectively, where $s_n$ represents the proportion of negative edges.

\begin{figure}[!t]
\centering
\begin{subfigure}{0.45\textwidth}
    \includegraphics[width=\textwidth]{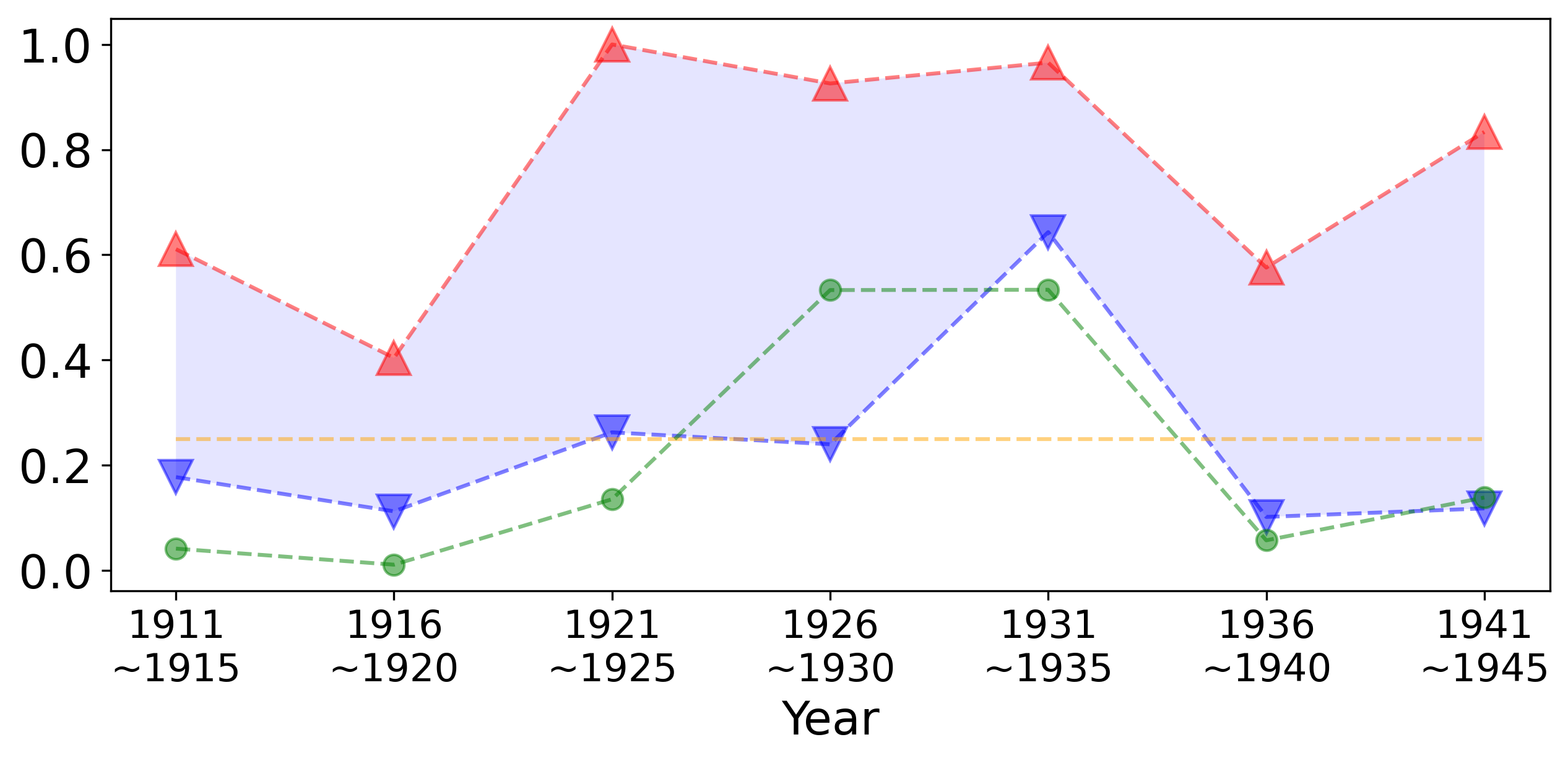}
    \caption{95\% confidence intervals for $w_{n, 1}$}
    \label{woc:w1}
\end{subfigure}
\begin{subfigure}{0.45\textwidth}
    \includegraphics[width=\textwidth]{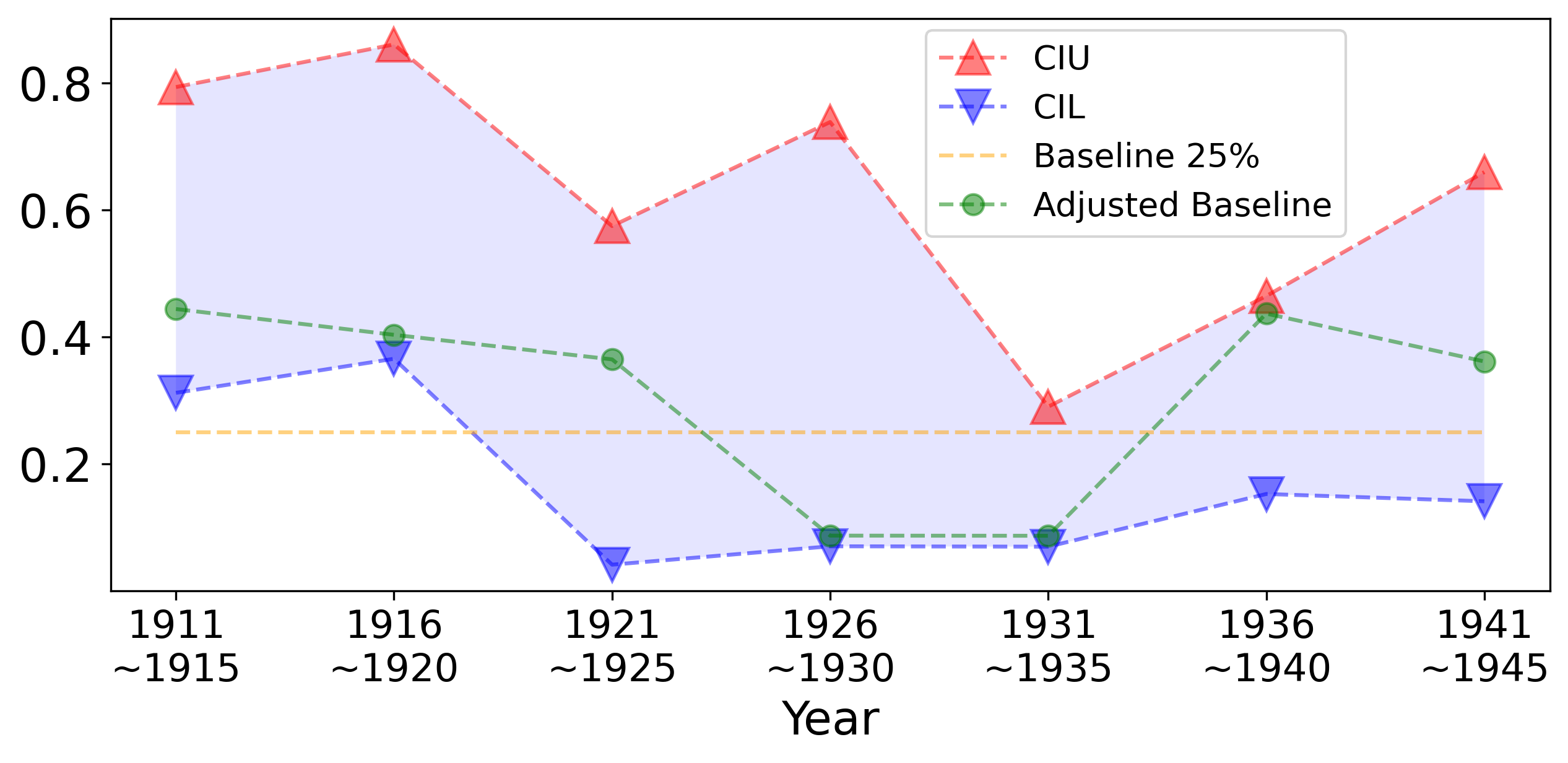}
    \caption{95\% confidence intervals for $w_{n, 3}$}
    \label{woc:w3}
\end{subfigure}
\caption{(a) The 95\% confidence intervals for the type-1 triangle, i.e.,  $w_{n,1}$; (b) The 95\% confidence intervals for the type-3 triangle, i.e., $w_{n,3}$. For (a)-(b), the red upper triangle represents the upper bound of confidence intervals; the blue lower triangle represents the lower bound of confidence intervals; and the yellow and green dashed lines represent baselines under two balance-free models. }
\label{wocremaining}
\end{figure}

Finally, we provide the full results for real-world data studies. Specifically, Figure~\ref{wocremaining} shows the results for type-1 triangles and type-3 triangles in the international relations network dataset. We find that the lower bound for $w_{n, 1}$ is almost always greater than the adjusted baseline. Table~\ref{appendix: ppi and social networks} presents the complete summary data and results of the protein-protein interaction network and social network.

\begin{table}[!ht]

    \centering
    \small
    \begin{tabular}{|l|c|c|c|}
    \hline
        Datasets  & Protein-Protein & U.S.Senate & Wikipedia Elections\\ \hline
        Nodes & 8777 & 100 & 7115 \\ \hline
        Number of edges  & 23313 & 1999 & 100693 \\ \hline
        Edge density  & $6.05\times 10^{-4}$ & $0.40$ & $3.98\times 10^{-3}$ \\ \hline
        Negative edge density  & 0.31 & 0.42 & 0.22 \\ \hline
        Number of Type-1 & 23468& 7195 & 400303 \\ \hline
        Proportion of Type-1 & 0.891& 0.490 & 0.659 \\ \hline
        Number of Type-2 & 500& 448 & 139769 \\ \hline
        Proportion of Type-2 & 0.019& 0.031 & 0.230 \\ \hline
        Number of Type-3 & 2337& 6019 & 58294  \\ \hline
        Proportion of Type-3 & 0.089& 0.410 & 0.096  \\ \hline
        Number of Type-4 &  26& 1017 & 8913 \\ \hline
        Proportion of Type-4 &  0.001& 0.069 & 0.015 \\ \hline
        CI for $w_{n}$ & (0.968,0.989) &(0.792, 0.943) & (0.728, 0.780)\\ \hline
        CI for $w_{n,1}$ & (0.841,0.931) & (0.320, 0.689)& (0.615, 0.698)\\ \hline
        CI for $w_{n,2}$ & (0.011,0.030) & (0.019, 0.065)&(0.208, 0.254)\\ \hline
        CI for $w_{n,3}$ & (0.057,0.132)& (0.245, 0.556)& (0.081, 0.114)\\ \hline
        CI for $w_{n,4}$ & (0.001,0.002) & (0.034, 0.175)& (0.011, 0.020)\\ \hline
        Baseline for $w_n$ & 0.529 & 0.502& 0.587\\ \hline
    \end{tabular}
\caption{Characteristics of real-world signed networks, including the number of nodes, the number of edges, the proportion of edges and negative signs, the number and proportion of type-$t$ triangles for $1 \le t \le 4$, 95\% confidence intervals for $w_{n,t}$, and the adjusted balance-free baseline for $w_n$.
}
\label{appendix: ppi and social networks}
\end{table}

\end{document}